\documentclass[a4paper,12pt,twoside]{report} 

\usepackage{setspace}

\usepackage[a4paper,top=30mm,bottom=30mm,inner=25mm,outer=30mm,bindingoffset=15mm]{geometry}

\usepackage{hyperref}  

\hypersetup{
    colorlinks=true,
    linkcolor=chaptercolor,
    citecolor=chaptercolor,
    urlcolor=chaptercolor
}

\usepackage{doi}
\usepackage{graphicx}
\usepackage{eso-pic}
\usepackage{listings}
\usepackage[table]{xcolor}
\usepackage{multirow}
\usepackage{array}
\usepackage{caption}
\usepackage{subcaption}
\usepackage{mathtools}
\usepackage{amsmath}
\usepackage{amssymb}

\usepackage{tcolorbox} 
\tcbuselibrary{skins,breakable,theorems} 

\usepackage{mathspec}

\setmathsfont(Digits,Latin)[
    Path = ./font/yrsa/,
    Extension = .ttf,
    UprightFont = Yrsa-Light,
    ItalicFont = Yrsa-LightItalic,
    BoldFont = Yrsa-SemiBold,
    BoldItalicFont = Yrsa-SemiBoldItalic,
]{Yrsa}

\usepackage{microtype}
\usepackage{booktabs}
\usepackage{rotating}
\usepackage{adjustbox}
\usepackage{quoting} 
\usepackage[absolute,overlay]{textpos} 
\usepackage{siunitx}
\usepackage{tikz}  
\usepackage{empheq}  

\newcommand\BackgroundPic{%
\put(0,0){%
\parbox[b][\paperheight]{\paperwidth}{%
\vfill
\centering
\includegraphics[width=\paperwidth,keepaspectratio]{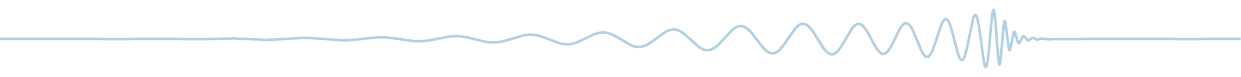}%
\vfill
}}}

\newtcolorbox{keyeqn}[1][]{%
  enhanced,
  colback=lightaccent,
  colframe=chaptercolor,
  boxrule=1pt,
  arc=3pt,
  left=8pt, right=8pt, top=8pt, bottom=8pt,
  before skip=12pt,
  after skip=12pt,
  #1
}

\newtcolorbox{keyeqntitled}[2][]{%
  enhanced,
  colback=lightaccent,
  colframe=chaptercolor,
  boxrule=1pt,
  arc=3pt,
  left=8pt, right=8pt, top=4pt, bottom=8pt,
  before skip=12pt,
  after skip=12pt,
  fonttitle=\bfseries\sffamily\color{chaptercolor},
  title=#2,
  attach boxed title to top left={yshift=-2mm, xshift=4mm},
  boxed title style={colback=white, colframe=white, boxrule=0pt},
  #1
}

\usepackage{printlen} 

\usepackage[numbers,sort&compress]{natbib}

\usepackage{arydshln}
\usepackage{makecell}
\usepackage{pdflscape} 

\usepackage{xspace}

\definecolor{uopgold}{HTML}{D2A61E}       
\definecolor{uopblue}{HTML}{00639B}       
\definecolor{uopslate}{HTML}{3F4444}      

\definecolor{chaptercolor}{named}{uopblue} 
\definecolor{accentgold}{named}{uopgold}   
\definecolor{darkslate}{named}{uopslate}   

\definecolor{warmgray}{HTML}{6B6B6B}
\definecolor{lightaccent}{HTML}{E8EEF2}
\definecolor{rulecolor}{HTML}{B8C5D0}
\definecolor{pomlightpurp}{HTML}{1E3A5F}
\definecolor{pomdarkpurp}{HTML}{142A45}
\definecolor{pomcyan}{rgb}{0, 0.63, 1}
\definecolor{pomtech}{rgb}{0.77, 0, 0.29}
\definecolor{lightgray}{gray}{0.9}
\definecolor{mediumgray}{gray}{0.5}

\usepackage[explicit]{titlesec}

\titleformat{\section}
  {\Large\color{chaptercolor}\bfseries\sffamily}
  {\textcolor{chaptercolor}{\thesection}}
  {0.8em}
  {#1}

\titleformat{\subsection}
  {\large\color{chaptercolor}\bfseries\sffamily}
  {\textcolor{chaptercolor}{\thesubsection}}
  {0.8em}
  {#1}

\titleformat{\subsubsection}
  {\normalsize\color{chaptercolor}\bfseries\sffamily}
  {\textcolor{chaptercolor}{\thesubsubsection}}
  {0.8em}
  {#1}

\usepackage{tabularx}
\usepackage{array}
\newcommand{\chapternumberfont}{\fontsize{50}{60}\selectfont\bfseries\color{chaptercolor}}
\newcommand{\chaptertitlefont}{\Huge\bfseries\color{chaptercolor}}

\titleformat{\chapter}[block]
  {\normalfont}
  {}
  {0pt}
  {%
    \noindent
    \begin{tabular}{ @{} b{0.70\textwidth} @{\hspace{0.5cm}} !{\tikz[overlay]\draw[accentgold, line width=1.5pt] (0,-10pt) -- (0, \paperheight);} @{\hspace{0.5cm}} l @{} }
       \raggedleft\chaptertitlefont #1 & \chapternumberfont\thechapter
    \end{tabular}%
  }

\titlespacing*{\chapter}{0pt}{-60pt}{30pt}

\titleformat{name=\chapter,numberless}[display]
  {\normalfont\centering}
  {}
  {0pt}
  {\LARGE\bfseries\color{chaptercolor} #1}

\titlespacing*{name=\chapter,numberless}{0pt}{-80pt}{10pt}

\newcommand{\chapterquote}[2]{%
  \vspace{1.5em}
  \hfill 
  \begin{minipage}{0.55\textwidth} 
    {\color{chaptercolor}\itshape #1}\\[-0.1em] 
    {\color{rulecolor}\rule{\textwidth}{0.4pt}}\\[0.4em] 
    \hspace*{\fill}{\small\color{chaptercolor}\textemdash\ \textsc{#2}} 
  \end{minipage}
  \par\vspace{2.5em} 
}

\newcommand{\gw}{gravitational wave\xspace}
\newcommand{\gws}{gravitational waves\xspace}
\newcommand{\gwadj}{gravitational-wave\xspace}

\newcommand{\Gws}{Gravitational waves\xspace}
\newcommand{\Gwadj}{Gravitational-wave\xspace}

\newcommand{\GR}{General Relativity\xspace}

\newcommand{\scl}{scattered light\xspace}
\newcommand{\Scl}{Scattered light\xspace}
\newcommand{\scladj}{scattered-light\xspace}
\newcommand{\Scladj}{Scattered-light\xspace}

\newcommand{\cbcs}{compact binary coalescences\xspace}
\newcommand{\Cbcs}{Compact binary coalescences\xspace}

\makeatletter
\renewcommand{\cleardoublepage}{%
  \clearpage%
  \if@twoside
    \ifodd\c@page
    \else
      \hbox{}%
      \thispagestyle{empty}
      \newpage
      \if@twocolumn\hbox{}\newpage\fi
    \fi
  \fi
}
\makeatother

\newcounter{ghostpage}
\newcommand{\frontmatterchapter}[1]{
    \cleardoublepage             
    \stepcounter{ghostpage}      
    \renewcommand{\thepage}{\roman{ghostpage}} 
    \phantomsection              
    \chapter*{#1}                
    \addcontentsline{toc}{chapter}{#1} 
    \markboth{#1}{#1}            
}

\let\oldchapter\chapter
\renewcommand{\chapter}{\cleardoublepage\oldchapter}

\usepackage{fancyhdr}
\usepackage{pdfpages}
\usepackage{etoolbox}
\newcommand{\customsymbol}[1]{%
  \raisebox{-0.047\height}{%
    \includegraphics[width=1.5cm]{images/header/#1.pdf}%
  }%
}

\newcommand{\arabicheader}{%
  \renewcommand{\headrule}{%
    \vspace{0pt}%
    {\color{accentgold}\hrulefill}%
    \raisebox{-10.0pt}{%
      \hspace{0.5em}%
      \customsymbol{\thepage}%
    }%
    {\color{accentgold}\hrulefill}%
  }
  \fancyhf{}
  \fancyhead[LE, RO]{{\color{chaptercolor}\bfseries\thepage}}
  \fancyhead[CO]{{\color{chaptercolor}\small\nouppercase{\leftmark}}}
  \fancyhead[CE]{{\color{chaptercolor}\small\nouppercase{\rightmark}}}
  \fancyfoot{}
}

\newcommand{\romanheader}{%
  \fancyhf{}
  \renewcommand{\footrulewidth}{0.0pt}
  \renewcommand{\headrulewidth}{0.4pt}
  \renewcommand{\headrule}{\hbox to\headwidth{\color{accentgold}\leaders\hrule height \headrulewidth\hfill}}

  \fancyhead[LO, RE]{{\color{chaptercolor}\bfseries\thepage}}
  \fancyhead[CO]{{\color{chaptercolor}\small\nouppercase{\leftmark}}}
  \fancyhead[CE]{{\color{chaptercolor}\small\nouppercase{\rightmark}}}
  \fancyfoot{}
}

\begin{document}

\pagenumbering{roman} 
\romanheader

\pagenumbering{gobble} 

\AddToShipoutPicture*{\BackgroundPic} 

\begin{titlepage}
    \begin{tikzpicture}[remember picture, overlay]
        \fill[chaptercolor] (current page.north west) rectangle ([xshift=1.5cm]current page.south west);

        \fill[accentgold] ([xshift=1.5cm, yshift=-8cm]current page.north west) rectangle ([xshift=1.65cm, yshift=-12cm]current page.north west);

        \node[anchor=north east, inner sep=0pt] at ([xshift=-2cm, yshift=-1.5cm]current page.north east) 
            {\includegraphics[width=4.5cm]{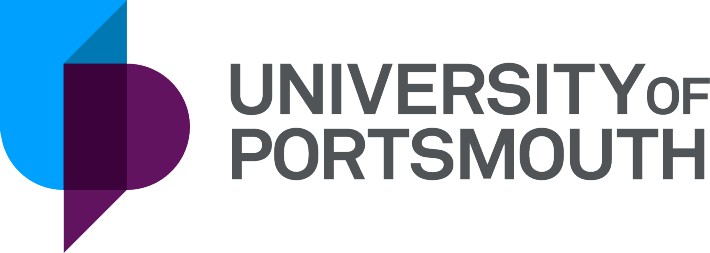}};
    \end{tikzpicture}

    \begin{center}
        \vspace*{1.5cm}

        {\fontsize{27}{27}\selectfont\bfseries\color{darkslate}%
        Improving the Detection of\\
        Gravitational-Wave Signals\\
        in Real Time\par}

        \vspace{0.8cm}

        {\color{accentgold}\rule{8cm}{2pt}}

        \vspace{0.8cm}

        {\LARGE\color{chaptercolor}\textsc{Arthur Tolley}}

        \vspace{1cm}

        \includegraphics[width=0.5\textwidth]{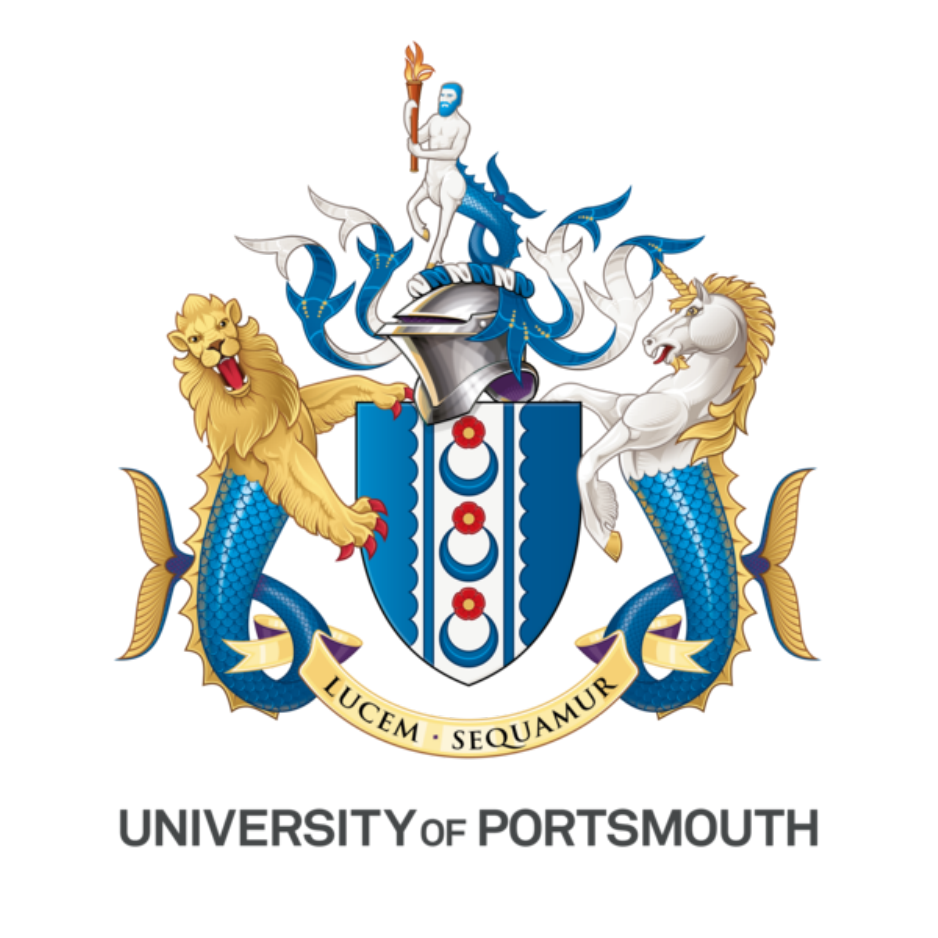}

        \vspace{0.8cm}

        {\large\color{warmgray}October 2024}

        \vspace{0.8cm}

        {\color{darkslate}
        This thesis is submitted in partial fulfilment of the requirements\\[0.3em]
        for the award of the degree of\\[0.5em]
        {\large\itshape Doctor of Philosophy}\\[0.5em]
        of the\\[0.3em]
        {\large\textsc{University of Portsmouth}}}

    \end{center}
\end{titlepage}

\cleardoublepage

\thispagestyle{empty}
\vspace*{\fill}

\begin{center}
    \begin{minipage}{0.9\textwidth} 
        \centering
        {\color{warmgray}\rule{3cm}{0.5pt}}\\[2em]

        {\fontsize{18}{26}\selectfont\color{darkslate}\itshape
        ``Yesterday is history, tomorrow is a mystery,\\[0.5em] 
        but today is a gift---that is why it is called the present.''}\\[2em]

        {\color{warmgray}\rule{3cm}{0.5pt}}\\[1.5em]

        {\color{warmgray}\textsc{Grand Master Oogway}}\\[3em]

        \makebox[\linewidth][c]{\includegraphics[width=0.75\linewidth]{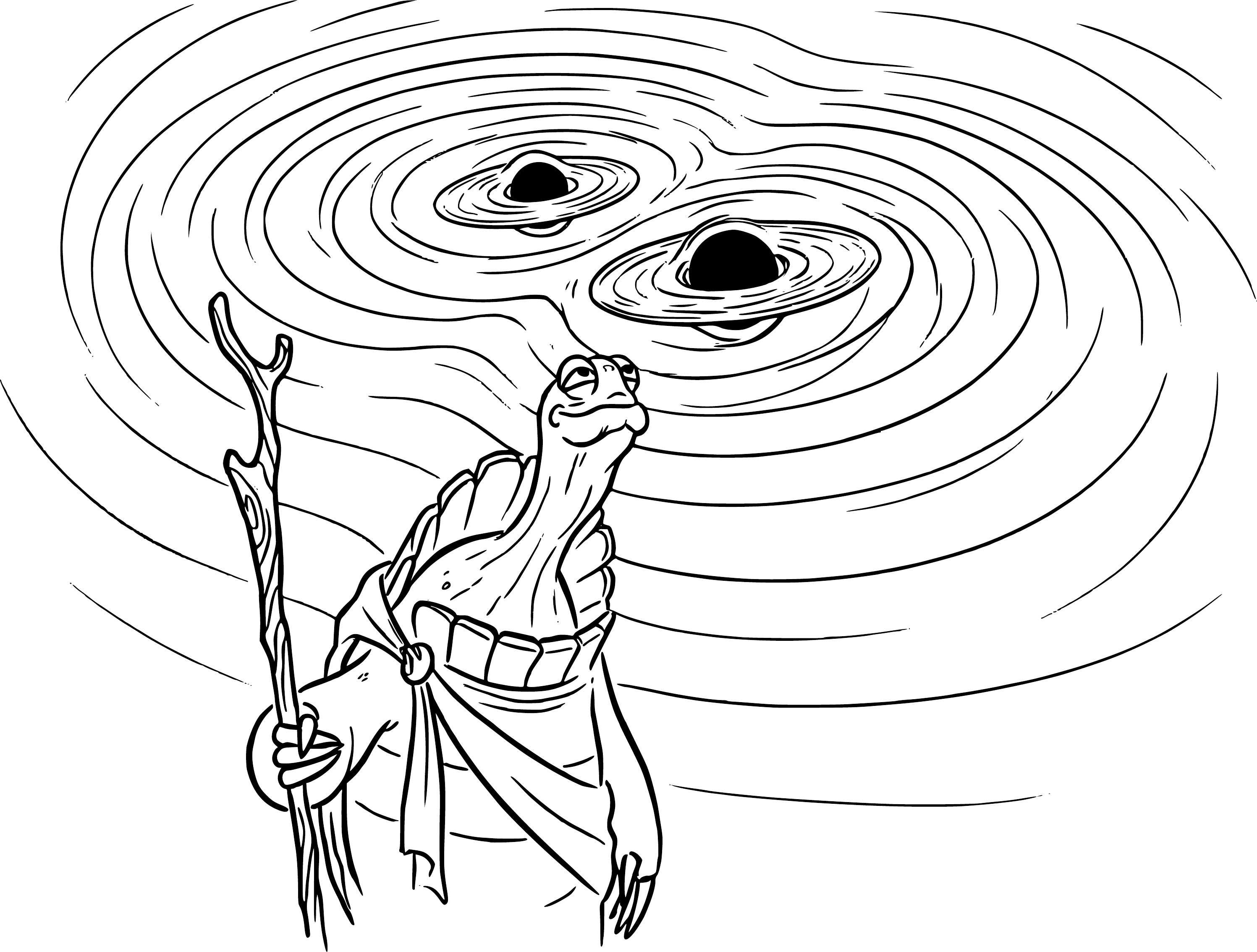}}

    \end{minipage}
\end{center}

\vspace*{\fill}

\pagestyle{fancy}
\romanheader
\frontmatterchapter{Declaration}
\begingroup
\setlength{\parindent}{0pt}
\setlength{\parskip}{1em}

\vfill 

\noindent{\bfseries\color{chaptercolor} Declaration of Authorship}

\noindent Whilst registered as a candidate for the above degree, I have not been registered for any other research award. The results and conclusions embodied in this thesis are the work of the named candidate and have not been submitted for any other academic award.

\noindent{\bfseries\color{chaptercolor} Examination \& Ethics}

\noindent The contents of this thesis were examined in a viva voce held on 3$^{rd}$ December 2024 by \textbf{Prof Graham Woan} and \textbf{Prof Kazuya Koyama}, to whom special thanks are owed. All research underlying this thesis has passed ethics review at the University of Portsmouth with certification code: ETHICS-10374.

\noindent{\bfseries\color{chaptercolor} Word Count \& Submission}

\noindent This thesis has an estimated word count of \textbf{\textcolor{teal!70}{42,069}}, this is calculated by the Overleaf word count tool. This thesis was originally submitted on the 30$^{th}$ September 2024 by \textbf{Arthur Tolley}.

\noindent{\bfseries\color{chaptercolor} Version 2:} Generated 25th January 2026. Formatting updates only; content unchanged.

\vspace{\baselineskip}
\begin{center}
    \begin{minipage}{0.35\textwidth} 
        \includegraphics[width=\textwidth]{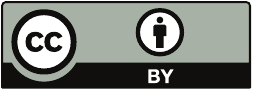}
    \end{minipage}%
    \hspace{0.02\textwidth} 
    \begin{minipage}{0.6\textwidth} 
        This work by Arthur Tolley is licensed under \\
        \textbf{Creative Commons Attribution 4.0 International.} \\
        (\href{https://creativecommons.org/licenses/by/4.0/}{https://creativecommons.org/licenses/by/4.0/}).
    \end{minipage}
\end{center}

\endgroup

\frontmatterchapter{Abstract}
\vfill
\begin{center}
\begin{minipage}{0.85\textwidth}
This thesis presents advancements in the detection of \gws from compact binary coalescences, utilising the most sensitive observatories constructed to date. The research focuses on enhancing \gwadj signal searches through the development of new tools and the application of existing methodologies to increase the sensitivity of live \gwadj searches. 

We introduced a novel noise artefact model, which enabled the identification and removal of glitches, thereby facilitating the recovery of previously missed \gwadj injections. This pioneering approach established a glitch search pipeline that adapted techniques typically used in \gwadj searches to address the unique characteristics of glitches. Additionally, we implemented an exponential noise model within the PyCBC Live search framework, significantly improving the detection ranking statistics for \gwadj signals and demonstrating the potential for substantial increases in detection sensitivity.

Furthermore, we analysed and proposed enhancements for the PyCBC Live Early Warning search to maximise the detection of \gwadj events in the early warning regime. Our findings highlighted deficiencies in the current ranking statistic and led to recommendations for optimising coincidence timing windows and refining phase-time-amplitude histograms. These adjustments aim to increase the detection of \gwadj signals, particularly binary neutron star events, in early warning scenarios.

The results underscore the importance of advancing search techniques in \gwadj astronomy, which can operate independently of detector improvements. By refining search methodologies, we enhance the capacity to detect a greater number of events, contributing significantly to our understanding of the Universe.

\end{minipage}
\end{center}
\vfill


\frontmatterchapter{Acknowledgements}
\begingroup
\setlength{\parindent}{0pt}
\setlength{\parskip}{1em}


Firstly, I owe special thanks to Graham and Kazuya for examining my thesis and conducting my viva voce.
%
Thank you to my academic supervisors: Andy Lundgren, you gave me the first opportunity to perform some `proper' research; Ian Harry, you supported me throughout the PhD as my main supervisor, ensuring I was suitably challenged and consequently over-prepared all the way to the end; Gareth and Xan, you have both helped me immeasurably with my projects and saved me from countless hours of debugging.

I required a great deal of personal support during the PhD, and for that the bulk of thanks goes to my amazing partner, Vickie. Without you by my side, I'm sure I wouldn't have made it to the end. There were many moments I was ready to quit and get a job that paid me a living wage. Thankfully, you were there to help me out and maintain my lavish lifestyle, I love you forever.

I am lucky enough to have many many friends who have made the PhD journey far more enjoyable than it could have been. Firstly, to my childhood friends in Ipswich: Billy, George, Ricky and Muntazim, I cherish the memories we've made together throughout Primary School, High School and Sixth Form.

To the friends I made during my undergraduate degree: Chris, Akash, Nick, Alex and Jake, we've had great adventures and living together made University feel like a breeze, I truly wish I could experience it all again. I want to say special thanks to Chris for being an amazing flatmate during the initial years of the PhD, I look back on those times fondly.

Thanks to the many PhD students I have shared a department with, you made this journey far less daunting than it could have been: Connor, Rafaela, Dan, Joe, Molly, Elena, Sergi, Emily, Rahma and many others too numerous to mention. I have to make a special note to Connor; we spent hours ranting about the PhD, you were the best support a man could've asked for and knowing we weren't alone in how we felt meant we both made it to the end, and it wasn't all for nothing. Love you buddy.

The root cause of my success is my family. These people have moulded me into the person I am, and I love you all more than anything. Jack and India, thank you for the childhood we shared. Mum and Lee, thank you for your constant support and safety net. Dad, I am thankful for who you are today and I know you're always trying your best, Jo, thank you for forcing him to try his best! Claire, thank you for always doing the best for me. I am who I am because of you all, and I hope I have made you proud.

\endgroup

\cleardoublepage
\renewcommand{\thepage}{\romannumeral\numexpr\value{page}-7\relax}
\addcontentsline{toc}{chapter}{Contents}
\tableofcontents



\cleardoublepage
\pagenumbering{arabic}
\arabicheader

\chapter*{Introduction}
\markboth{Introduction}{Introduction}
\label{chapter:introduction}
\addcontentsline{toc}{chapter}{Introduction}
In the 20th century, Albert Einstein introduced the theory of \GR~\cite{Einstein_1:1914, Einstein_2:1914, Einstein_3:1915, Einstein_4:1916, Einstein_5:1917, Einstein_6:1936}, which describes gravity as the curvature of spacetime caused by massive objects, rather than as a force. In this framework, objects move along geodesics in curved spacetime. A key prediction of the theory is the existence of \gws~\cite{Einstein_7:1937, Einstein_8:1938, Einstein_9:1939, Einstein_10:1948}, which are perturbations in spacetime generated by the acceleration and collisions of massive bodies, such as black holes or neutron stars. These waves propagate at the speed of light and induce small changes in the proper distance between freely falling objects. The detector strain caused by \gws is extremely small, typically on the order of $10^{-22}$ for the most energetic sources, making direct detection highly challenging. Despite this, the observation of \gws has become a crucial tool for studying astrophysical phenomena and testing the predictions of \GR.

The first direct detection of \gws occurred in 2015 with the observation of GW150914~\cite{GW150914:2016}, a \gwadj signal produced by the merger of two black holes. This event was detected by the Laser Interferometer Gravitational-Wave Observatory (LIGO) and the Virgo~\cite{aVirgo:2015} \gwadj observatory using ground-based interferometers. The observed signal matched the predictions of \GR~\cite{GW150914_TGR:2016} for the inspiral, merger, and ringdown phases of a binary black hole coalescence. This detection marked the beginning of \gwadj astronomy, providing the first direct evidence of black hole mergers and opening a new observational window to study the Universe.

Following the first detection of \gws, the global \gwadj detector network expanded with the inclusion of the KAGRA detector~\cite{KAGRA:2021}. The current International Gravitational-Wave Observatory Network (IGWN) collaboration has enabled more comprehensive and sensitive observations of \gwadj events. With a running total of 221 \gwadj events~\cite{gwtc1:2019, gwtc2:2021, gwtc21:2024, gwtc3:2023, 1OGC:2018, 2OGC:2020, 3OGC:2021, 4OGC:2021, Princeton_1:2019, Princeton_2:2019, Princeton_3a:2022, Princeton_3b:2023, gracedb_superevents:2024} from \cbcs, including two mergers of binary neutron star systems~\cite{GW170817:2017, GW190425:2020} and two neutron star–binary black hole mergers~\cite{nsbh:2021}, across four observing runs as of the drafting of this thesis.

The work performed in this thesis focuses on improvements to the current search methods for \gws from data obtained by ground-based \gwadj observatories. Chapter~\ref{chapter:1-gravitational-waves} provides a simple derivation of \gws, a brief discussion of how the current generation of detectors operates, and the expected \gwadj emissions from a compact binary coalescence. This discussion is intended to give the reader a basic understanding of the background required to appreciate the motivation for the research conducted in later chapters; for texts with more detailed discussions, please refer to~\cite{Moore_book:2012, Maggiore_book:2007, Schutz_book:2009, Creighton_book:2009}. Chapter~\ref{chapter:2-searches} reviews the detection methods used for \gws, focusing on the PyCBC~\cite{PyCBC:2016} search pipeline. Chapter~\ref{chapter:3-detchar} concentrates on Detector Characterisation and the techniques employed to characterise and calibrate ground-based interferometer noise, producing high-quality data for further study.

Chapter~\ref{chapter:4-archenemy} investigates the feasibility of modelling \gwadj detector data artefacts, specifically the \scladj artefact, and subtracting these artefacts to clean the data before the search for \gws. In Chapter~\ref{chapter:5-pycbc-live}, we describe the improvements to the PyCBC Live search's ranking statistic aimed at enhancing detection sensitivity. Chapter~\ref{chapter:6-earlywarning} details the PyCBC Live Early Warning search, examining its deficiencies and suggesting improvements to ensure the identification of all potentially electromagnetically bright events. In Chapter~\ref{chapter:7-snr-optimiser}, we discuss the investigation and enhancements made to the PyCBC Live rapid signal-to-noise ratio optimiser. Finally, in Chapter~\ref{chapter:8-industry}, we briefly discuss the industrial placements undertaken as a requirement of this doctoral programme.

\chapter[Gravitational Waves]{Gravitational Waves}
\label{chapter:1-gravitational-waves}
\chapterquote{This little manoeuvre's gonna cost us 51 years.}{Joseph Cooper}
In this thesis, we focus on improving the capabilities of \gwadj search pipelines to detect \gwadj signals from compact binary coalescences (CBCs). As an introduction to the research chapters later in this thesis, we will briefly discuss gravitational radiation, its detection using laser interferometry and, the \gwadj signals we expect to see from CBCs.
In this chapter, we construct a simple description of \gws as a consequence of \GR. In Section~\ref{1:sec:gravitational-radiation} we derive \gws as solutions of the Einstein field equations in a linearised gravity regime under the weak-field approximation. In Section~\ref{1:sec:gravitational-wave-detection} we will discuss the ground-based \gwadj observatories and how \gws will create physical effects that we can measure using these observatories. Finally, in Section~\ref{1:sec:modelling_CBC}, we model the gravitational waveform from the primary source of \gws---CBCs---we see with our detectors. For greater detail on any of these sections, readers are referred to the texts~\cite{Moore_book:2012, Schutz_book:2009, Maggiore_book:2007, Creighton_book:2009}.

\section{\label{1:sec:gravitational-radiation}Gravitational radiation}

Gravitational radiation refers to the propagation of perturbations in spacetime, predicted by \GR, that manifest as \gws. These waves are generated by the accelerated motion of massive objects and propagate at the speed of light. In this section, we will derive \gws as solutions to the Einstein field equations within the weak-field approximation, where the gravitational field is considered a small perturbation to flat spacetime. This approach, known as linearised gravity, allows for a simplified treatment of \gws and provides the foundation for understanding their key properties. We will also discuss the general characteristics of gravitational radiation, such as their polarisation and energy carried by \gws.

Einstein's field equations (commonly referred to as Einstein's equations) are a set of $10$\footnote{Due to the symmetry of $G_{\mu\nu}$ and $T_{\mu\nu}$.} non-linear coupled equations which describe how the gravitational field is generated by matter and can be written eloquently using \GR notation as
\begin{keyeqntitled}{Einstein's Field Equations}
\begin{eqnarray}
    G_{\mu\nu} = 8\pi T_{\mu\nu},
    \label{1:eqn:EFE}
\end{eqnarray}
\end{keyeqntitled}
where we use natural units, $G \!=\! c \!=\! 1$.

Einstein's equations relate the curvature of spacetime using the Einstein tensor, $G_{\mu\nu}$, to the stress-energy tensor, $T_{\mu\nu}$, which describes the density and flux of matter, energy, and momentum at each point in spacetime.

\subsection{\label{1:sec:perturbations}Perturbations in spacetime}

\Gws are a natural consequence of \GR and we show this by making a number of simplifications to Einstein's equations to reveal valid plane wave solutions. We begin in a flat-Minkowski spacetime, represented by the spacetime metric, $\eta_{\mu\nu}$, and make a \textit{weak} linear perturbation, $h_{\mu\nu}$ due to a gravitational field. In this scenario we can write the metric tensor, $g_{\mu\nu}$, as
\begin{equation}
    g_{\mu\nu} = \eta_{\mu\nu} + h_{\mu\nu},
    \label{1:eq:metric_perturbation}
\end{equation}
and to ensure our perturbation is small we enforce the condition $|h_{\mu\nu}| \ll 1$. In this regime, we are able to raise and lower indices using the Minkowski metric
\begin{equation}
    h^{\mu\nu} = \eta^{\mu\alpha}\eta^{\nu\beta}h_{\alpha\beta}.
    \label{1:eq:raise_lowering_indices}
\end{equation}
Terms of $h_{\mu\nu}$ greater than linear-order are negligible and are ignored, this is called taking the \textit{weak-field limit}. This simplification derives \gws in a regime where Einstein's equations are linearised, which significantly reduces the complexity of the calculations while still capturing the essential physical behaviour of \gws.

We can write the Einstein tensor, $G_{\mu\nu}$, in terms of the Riemann tensor, $R_{\mu\nu}$, the Ricci scalar, $R$, and the metric tensor
\begin{align}
    G_{\mu\nu} &= R_{\mu\nu} - \frac{1}{2}g_{\mu\nu}R,
    \label{1:eq:g_munu_riemann}
\end{align}
and we can introduce a quantity known as the \textit{trace reverse tensor}
\begin{equation}
    \Bar{h}_{\mu\nu} = h_{\mu\nu} - \frac{1}{2} \eta_{\mu\nu}h,
\end{equation}
where $h = \eta^{\mu\nu}h_{\mu\nu}$ is the trace of the metric perturbation.

We can then express the linearised Einstein tensor in terms of the linearised Riemann tensor and our trace reverse metric tensor~\cite{Moore_book:2012}
\begin{equation}
    R_{\mu\nu} - \frac{1}{2}g_{\mu\nu}R = \frac{1}{2}\left(\partial^{\sigma}\partial_{\mu}\Bar{h}_{\sigma\nu} - \partial^{\sigma}\partial_{\sigma}\Bar{h}_{\mu\nu} + \partial_{\nu}\partial^{\alpha}\Bar{h}_{\mu\alpha} - \eta_{\mu\nu}\partial^{\alpha}\partial^{\beta}\Bar{h}_{\alpha\beta}\right).
\end{equation}
We can then express our fully linearised Einstein field equations as
\begin{equation}
    - \Box\Bar{h}_{\mu\nu} + \partial^{\sigma}\partial_{\mu}\Bar{h}_{\sigma\nu}  + \partial_{\nu}\partial^{\alpha}\Bar{h}_{\mu\alpha} - \eta_{\mu\nu}\partial^{\alpha}\partial^{\beta}\Bar{h}_{\alpha\beta} = 16\pi T_{\mu\nu},
    \label{1:eq:linearised_efe}
\end{equation}
where $\Box$ is the flat space d'Alembertian operator,
\begin{equation}
    \Box = \partial^{\sigma}\partial_{\sigma} = -\frac{\partial^{2}}{dt^{2}} + \frac{\partial^{2}}{dx^{2}} + \frac{\partial^{2}}{dy^{2}} + \frac{\partial^{2}}{dz^{2}}.
\end{equation}

\subsection{\label{1:sec:gauge-transformations}Gauge transformations}

Inspecting Equation~\ref{1:eq:linearised_efe} we can identify the presence of the d'Alembertian operator $\Box$ along with terms involving divergences of $\bar{h}_{\mu\nu}$. Performing a gauge transformation to the harmonic gauge will eliminate the divergences,
\begin{equation}
    \partial^{\nu} \Bar{h}_{\mu\nu} = 0,
    \label{1:eq:divergence_gone}
\end{equation}
further simplifying our linearised Einstein equations.

In the weak-field limit we can make this gauge transformation by applying small coordinate translations to the perturbed spacetime that ensure the metric perturbations remain small,
\begin{align}
    x &\rightarrow x^{\prime}, \\
    x^{\prime\mu} &= x^{\mu} + \xi^{\mu}(x),
    \label{1:eq:gauge_transform}
\end{align}
where the metric in the new coordinate system is
\begin{equation}
    g^{\prime}_{\mu\nu} = \eta_{\mu\nu} + \bar{h}^{\prime}_{\mu\nu},
    \label{1:eq:gauge_metric}
\end{equation}
and remains valid as long as the weak-field approximation holds in the new coordinates $|h^{\prime}_{\mu\nu}| \ll 1$.

We have chosen a coordinate system where Equation~\ref{1:eq:divergence_gone} is satisfied, and in this gauge, the Einstein field equations simplify to
\begin{equation}
    \Box \Bar{h}_{\mu\nu} = -16 \pi T^{\mu\nu} .
    \label{1:eq:lorentz_gauge_efe}
\end{equation}

\subsection{\label{1:sec:gw_in_vacuum}\Gws in vacuum}

In the weak-field approximation, where we are far from any sources of mass or energy ($T_{\mu\nu} \!=\! 0$), Equation~\ref{1:eq:lorentz_gauge_efe} simplifies to
\begin{equation}
    \Box \Bar{h}_{\mu\nu} = 0,
\end{equation}
which is the wave equation and has plane wave solutions for $\Bar{h}_{\mu\nu}$,
\begin{equation}
    \Bar{h}_{\mu\nu} = A_{\mu\nu} {\rm e}^{i(k_\alpha x^\alpha)},
\end{equation}
indicating the wave nature of gravitational perturbations. $A_{\mu\nu}$ is the amplitude tensor and $k_\alpha = (-\omega, \textbf{k})$, the wave vector that satisfies
\begin{equation}
    k^{\alpha} k_{\alpha} = 0 = -\omega^{2} + |\vec{k}|^{2}.
    \label{1:eq:wave_vector_trace}
\end{equation}
Equation~\ref{1:eq:wave_vector_trace} tells us that $\omega^{2} \!=\! |k|^{2}$ meaning the speed of gravity $v \!=\! \omega / |k_{i}| \!=\! 1$: \gws travel at the speed of light.

\subsection{\label{1:sec:TT_gauge}The transverse-traceless gauge}

We now have solutions to the Einstein field equations in the form of plane waves. $\bar{h}_{\mu\nu}$ is a tensor which contains $10$ independent components (as a symmetric tensor in $4$-dimensional spacetime) with redundancy between them. We can express another set of gauge transformations to reduce this to a tensor with only $4$ components and $2$ degrees of freedom.

First, we apply the gauge condition stated in Equation~\ref{1:eq:divergence_gone} to $\Bar{h}_{\mu\nu}$ to obtain,
\begin{align}
    0 &= k^{\mu} A_{\mu\nu},
\end{align}
this implies that \gws are \textit{transverse}. This removes the components of $\bar{h}_{\mu\nu}$ that are parallel to the direction of propagation of the \gw; hence, $\bar{h}_{0i} \!=\! 0$\footnote{Roman letter indices represent spatial dimensions ($x$, $y$, $z$).} and $A_{t\mu} \!=\! 0$. For instance, a \gw travelling in the $z$-direction will have purely spatial components in the $x$ and $y$ directions.

Second, we require $\bar{h}^{\mu}_{\mu} = 0$, the trace of the \gw to vanish. This removes the possibility of a volume change in spacetime due to \gwadj transmission. We say that our \gws are now \textit{traceless}. This gauge is called the \textit{transverse-traceless} (TT) gauge.

\subsection{\label{1:sec:gravitational-propagation}The propagation of \gws}

In the TT gauge, we can express the metric perturbation $h^{TT}_{\mu\nu}$ for a \gw propagating in the $z$-direction
\begin{keyeqn}
\begin{equation}
   h^{TT}_{\mu \nu} =
   \begin{pmatrix}
      0 & 0 & 0 & 0 \\
      0 & h_+ & h_\times & 0 \\
      0 & h_\times & -h_+ & 0 \\
      0 & 0 & 0 & 0
   \end{pmatrix}.
   \label{1:eqn:h_TT}
\end{equation}
\end{keyeqn}
We define the plus and cross polarisations as~\cite{Moore_book:2012}
\begin{equation}
    h_+ = A_+ \cos(\omega t - \omega z + \phi_{0}),
    \label{1:eq:h_+_cos}
\end{equation}
\begin{equation}
    h_{\times} = A_{\times} \cos(\omega t - \omega z + \phi_{0}),
\end{equation}
where $A_+$ and $A_\times$ are the amplitudes of the two polarisations, and $\phi_{0}$ is a constant phase offset. We have described the metric perturbation with only two independent components: the plus ($+$) and cross ($\times$) polarisations, representing the two degrees of freedom in gravitational radiation.

We consider the effect of a passing \gw on two particles in the $x$-$y$ plane placed at ($x_{1}, y_{1}, 0$) and ($x_{2}, y_{2}, 0$). The spacetime interval between the two particles can be written
\begin{align}
    \Delta s^{2} &= \left(\eta_{\mu\nu} + h^{TT}_{\mu\nu}\right) \Delta x^{\mu} \Delta x^{\nu}, \\
    &= -dt^{2} + \left(1 + h_{+}\right)(x_{1} - x_{2})^{2} \nonumber \\
    &\quad + \left(1 - h_{+}\right)(y_{1} - y_{2})^{2} \nonumber \\
    &\quad + 2h_{\times} (x_{1} - x_{2})(y_{1} - y_{2}).
\end{align}
and in the simple case where ($y_{2} - y_{1} = 0$) we can calculate the proper distance between the two particles at an arbitrary time, $t$, as
\begin{equation}
    \Delta s \approx \left( 1 + \frac{1}{2} h_{+}\right)(x_{1} - x_{2}).
    \label{1:eq:proper_dist_two_particles}
\end{equation}
We can see that the proper distance between the two particles is stretched by the factor $(1 + \frac{1}{2}h_{+})$. It can be shown that the simultaneous motion in the $y$-direction results in a squeezing of the proper distances to $(1 - \frac{1}{2}h_{+})$. This change in proper distance is called \gwadj ``strain'' and is what we want to measure to directly observe the effects of \gws.

\begin{figure}
   \includegraphics[width=\textwidth]{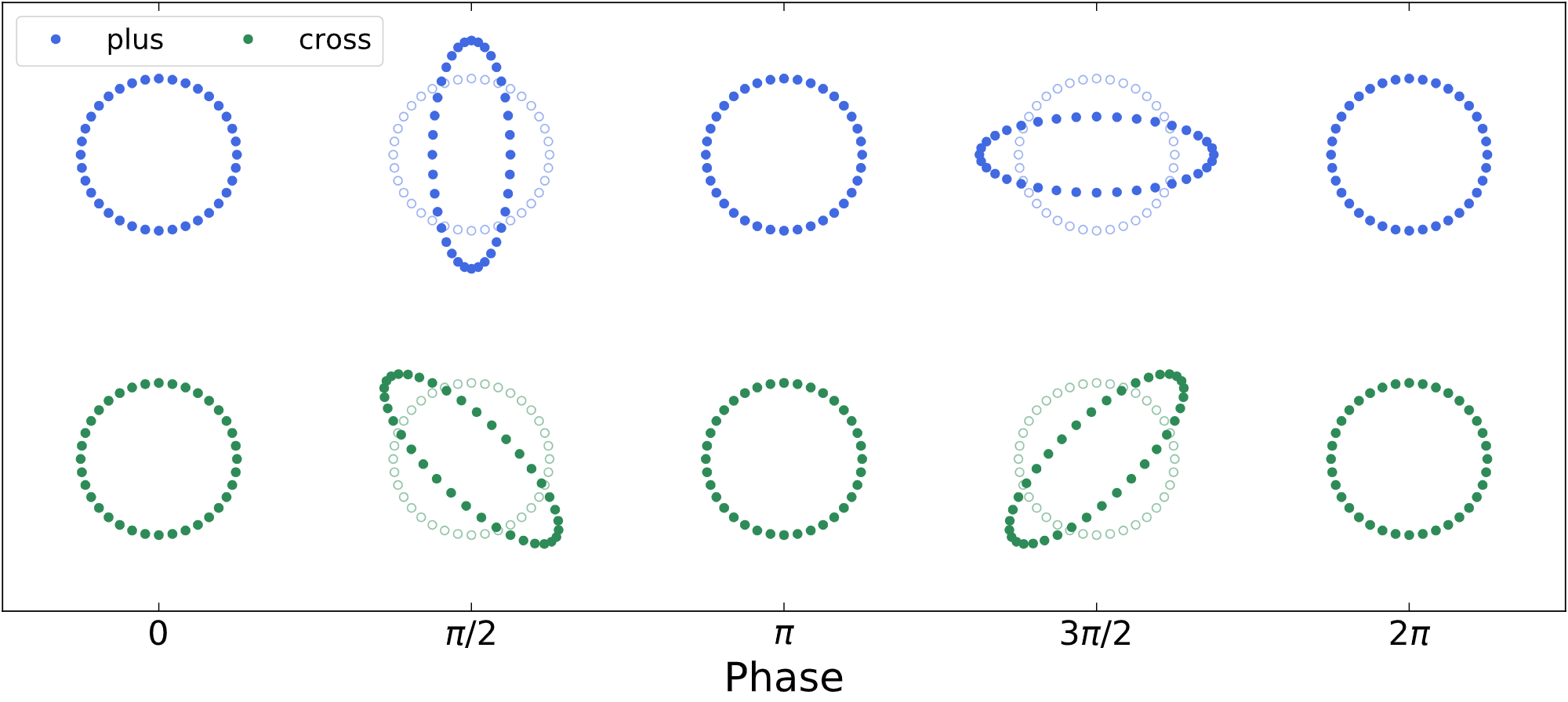}
   \caption{The effect of the two polarisations on a ring of test particles~\cite{gw_polarisation_plots}.}
   \label{1:fig:ring_of_particles}
\end{figure}

Figure~\ref{1:fig:ring_of_particles} provides a visual representation of how the proper distances in a ring of particles change under the influence of a \gw with pure $+$ or $\times$ polarisation.
It can be seen that the effect of the $\times$ polarisation is the same as the $+$ polarisation but with a $45^{\circ}$ rotation.

\subsection{\label{1:sec:gw-emission}\Gwadj emission in linearised gravity}

We now return to Equation~\ref{1:eq:lorentz_gauge_efe} and reconsider the stress-energy tensor. The solution to the Einstein equations can be obtained using the Green's function~\cite{Dhurkunde:2024},
\begin{equation}
    \Bar{h}_{\mu\nu}(t, \vec{x}) = 4 \int d^3 x^{\prime} \frac{T_{\mu\nu}\left(t - |\vec{x} - \vec{x}^{\prime}|, \vec{x}^{\prime}\right)}{|\vec{x} - \vec{x}^{\prime}|},
    \label{1:eq:greens_functions}
\end{equation}
where the observer is at $\vec{x}$ and the source is at $\vec{x}^{\prime}$. The solution can be approximated using the quadrupole formula under two assumptions: the observer is far away $\left(|\vec{x} - \vec{x}^{\prime}| \approx |\vec{x}| = r\right)$ and the source velocity is small compared to the speed of light. 

We expand the stress-energy tensor using a Taylor expansion around the retarded time ($t - r$), and Equation~\ref{1:eq:greens_functions} simplifies to
\begin{equation}
    \Bar{h}_{\mu\nu}(t, \vec{x}) = \frac{4}{r} \int d^3 x^{\prime} T_{\mu\nu}\left(t - r, \vec{x}^{\prime}\right),
    \label{1:eq:quadrupole_simplification}
\end{equation}
to leading order. We can see that the \gws are dependent on the integral of the stress-energy tensor over a volume containing the source, evaluated at the retarded time.

Using the laws of conservation of total energy and momentum, $\partial^{\mu}T_{\mu\nu}=0$, we can see that the integral of the $T_{t\nu}$ and $T_{\nu t}$ components will be constant in time. We choose to ignore these components and focus only on observable signals in the spatial dimensions moving forward.

The right-hand side of Equation~\ref{1:eq:quadrupole_simplification} can be expressed as
\begin{equation}
    \frac{4}{r} \int d^3 x^{\prime} T_{ij} = \frac{2}{r}\frac{\partial^{2}}{\partial t^{2}} \int T_{00} x_{i} x_{j} d^{3} x^{\prime},
\end{equation}
where $T_{00} \approx \rho$, the Newtonian mass density. This expression is referred to as the \textit{mass quadrupole moment tensor} of the mass distribution
\begin{equation}
    M_{ij} := \int T_{00} x_{i} x_{j} d^{3} x^{\prime},
    \label{1:eq:quadrupole_moment_tensor}
\end{equation}
and when substituted into Equation~\ref{1:eq:quadrupole_simplification}, we obtain the \textit{quadrupole formula}
\begin{keyeqntitled}{Quadrupole Formula}
\begin{equation}
    h_{ij} = \frac{2}{r} \ddot{M}_{ij}(t-r),
    \label{1:eq:quadrupole_formula}
\end{equation}
\end{keyeqntitled}
where each dot over a tensor represents on derivative with respect to time.

The key properties of this equation are the dependence of gravitational radiation on the second time derivative of the quadrupole moment tensor, indicating that \gws are produced by rapid changes in the quadrupole moment tensor, such as the accelerations of masses and their directional changes during orbital cycles. Additionally, the \gwadj strain falls off inversely with distance $r$ from the source.

To extract the plus and cross polarisations of the metric perturbation we must now transform into the transverse-traceless gauge, this is done using the projection tensor~\cite{McIsaac_Thesis:2023}
\begin{equation}
    P^{i}_{j}(\hat{n}) = \delta^{i}_{j} - n^{i}n_{j}
\end{equation}
where
\begin{equation}
    P^{i}_{j}P^{j}_{k} = P^{i}_{k}
\end{equation}
and $\hat{n}$ is a normal vector such that $|\hat{n}| = 1$. Applying the projection vector to a three-dimensional vector will project it down into the two-dimensional plane orthogonal to $\hat{n}$. To get the components of $h_{ij}$ which are transverse to $\hat{n}$ we need to apply the projection tensor once for each index
\begin{equation}
    h^{T}_{kl} = P^{i}_{k}P^{j}_{l}h_{ij},
\end{equation}
and then to remove the trace of the tensor we use the Minkowski metric
\begin{equation}
    h^{T} = \eta^{ij}h^{T}_{ij} = P^{ij}h_{ij}.
\end{equation}
From this we can construct the tensor $\frac{1}{2}P_{kl}h^{T}$ which will have a trace of $h^{T}$ and will be transverse to $\hat{n}$ allowing us to subtract it from the transverse metric perturbation to get the metric perturbation in it's transverse-traceless form
\begin{equation}
    h^{TT}_{kl} = \Lambda^{,ij}_{kl} h_{ij}
\end{equation}
where
\begin{equation}
    \Lambda^{,ij}_{kl} = \left(P^{i}_{k}P^{j}_{l} - \frac{1}{2}P_{kl}P^{ij}\right).
\end{equation}
Therefore, we can apply $\Lambda^{,ij}_{kl}$ to Equation~\ref{1:eq:quadrupole_formula} to get the transverse-traceless perturbation. For a gravitational wave travelling in the z-direction with $n^{i} = (0,0,1)$ (referred to as the "radiation" frame) we get plus and cross polarisations as functions of $\ddot{M}_{ij}$
\begin{equation}
    h_{+}^{TT} = \frac{G}{r}\left(\ddot{M}_{xx} - \ddot{M}_{yy}\right)
\end{equation}
\begin{equation}
    h_{\times}^{TT} = \frac{2G}{r}\left(\ddot{M}_{xy}\right).
\end{equation}

The total power radiated by a source is the \textit{luminosity flux}, and it must equal the energy carried by the \gws, due to the conservation of energy. We can obtain the luminosity flux by integrating the energy flux over a sphere with infinite radius from the source. The total energy loss rate is expressed as
\begin{equation}
    \frac{dE}{dt} = -\mathcal{L},
    \label{1:eq:de_dt}
\end{equation}
where $\mathcal{L}$ is the \gwadj luminosity flux. The luminosity flux at a large distance from the source is given by~\cite{Creighton_book:2009}
\begin{keyeqntitled}{Gravitational Wave Luminosity}
\begin{equation}
    L = \frac{1}{5} \left\langle \dddot{M}^{jk} \dddot{M}_{jk} \right\rangle,
    \label{1:eq:luminosity_flux}
\end{equation}
\end{keyeqntitled}
where the angular brackets represent averaging over all directions, and $\dddot{M}_{jk}$ refers to the third time derivative of the quadrupole moment. This shows that any source with a non-vanishing second derivative of its mass quadrupole moment will emit \gws.

Thus, through the quadrupole formula and \gwadj luminosity, we see that rapidly accelerating masses with a non-zero quadrupole moment are ideal sources of gravitational radiation.

\section{\label{1:sec:gravitational-wave-detection}\Gwadj detection}

Having described the effect of \gws on a ring of particles, we now describe how these principles are applied to detect \gws on Earth. The first direct detection of \gws was achieved on the 14th September 2015~\cite{GW150914:2016}, made possible by ground-based \gwadj observatories. The current network of \gwadj detectors includes the Advanced LIGO (aLIGO) detectors located in Hanford and Livingston~\cite{aLIGO:2015}; Virgo in Italy~\cite{aVirgo:2015}; KAGRA in Japan~\cite{KAGRA:2021}; and GEO600 in Germany~\cite{GEO600:2002}. While this section will primarily focus on the design and operation of the aLIGO interferometers, the underlying detection principles are similar across all observatories. The configuration and performance of aLIGO are described comprehensively in~\cite{aLIGO:2015}.

\subsection{\label{1:sec:laser_interferometry}Laser interferometry}

\begin{figure}
    \centering
    \includegraphics[width=0.9\linewidth]{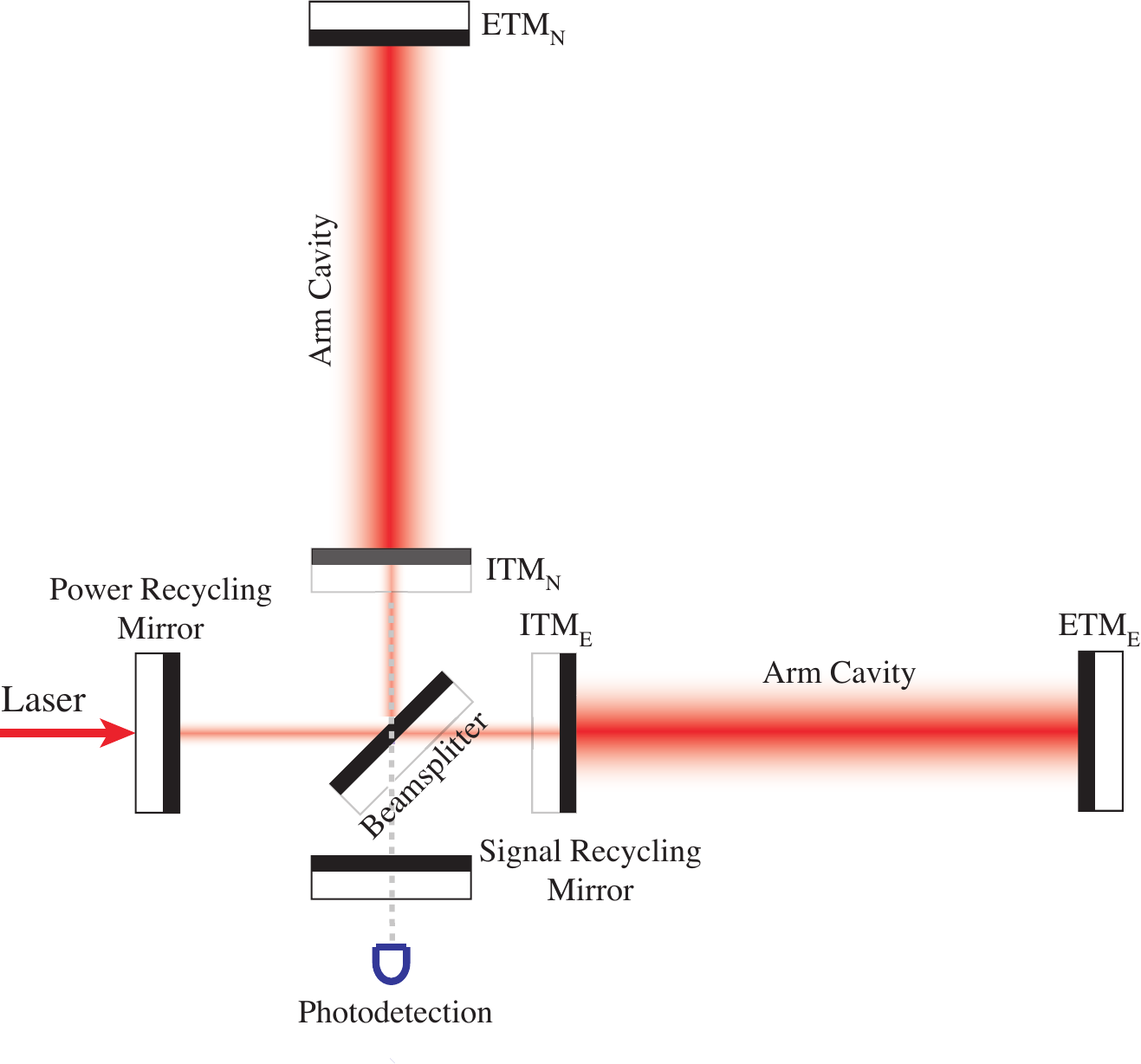}
    \caption{An example Michelson laser interferometer detailing the main components of the advanced LIGO \gwadj observatory~\cite{aLIGO:2015}. Half the laser is reflected to the north arm cavity, physically defined by the input test mass (ITM$_{\text{N}}$) and the end test mass (ETM$_{\text{N}}$), and the other half is transmitted through to the east arm. Laser power builds up in the arm cavities through reflection between test masses before returning to the beam splitter, where there will be a total destructive interference of light when no \gwadj signal is present and therefore no signal will be measured at the photo-detector output. The power recycling mirror reflects laser light back into the detector arms to build up more power. The signal recycling mirror reflects light back into the arms with frequencies like those expected from \gwadj signals which resonates inside the cavity and increase the signal sensitive frequency power. Taken from~\cite{IFO_diagram:2008}.}
    \label{1:fig:ifo}
\end{figure}

The physical effect of a passing \gw on a ring of particles (Figure~\ref{1:fig:ring_of_particles}) is described as a stretching and squeezing, changing the particles' proper distances depending on the polarisation of the \gw. The \gwadj observatories are Michelson interferometers, and we can visualise the L-shape of them as experiencing similar effects to the ring of particles. A basic diagram of the LIGO interferometer design can be seen in Figure~\ref{1:fig:ifo}.

Half the light from the laser incident on the beam splitter is transmitted to the end test mass in the $y$-direction (ETM$_{\text{N}}$) and half to the end test mass in the $x$-direction(ETM$_{\text{E}}$). The end test masses are mirrors which reflect the light back toward the beam splitter. When no \gw is present, the length of the $x$ and $y$ arm cavities are equal and the light returning to the beam splitter interferes destructively, meaning no light is detected by the photo-detector output of the detector.

Using Equation~\ref{1:eq:proper_dist_two_particles}, where we are treating the detector arms like two particles placed at ($L, 0, 0$) for the end test mass of the $x$ arm and ($0, L, 0$) for the $y$-arm, we can calculate the change in arm length (proper distance) with respect to the beam splitter at the origin under the influence of a \gw propagating in the $z$-direction
\begin{equation}
    ds_{x} \approx \left(1 + \frac{1}{2}h_+\right)L,
\end{equation}
where $L$ is the length of the detector arm ($4 \, \text{km}$ for advanced LIGO) and in the $y$ arm
\begin{equation}
    ds_{y} \approx \left(1 - \frac{1}{2}h_+\right)L.
\end{equation}
The strain experienced by the $x$ and $y$ arms is directly related to both the magnitude of the $h_{+}$ \gwadj polarisation and the length of the detector arm. The arm in the $x$-direction will experience stretching, while the arm in the $y$-direction will experience squeezing.

The coupling of motion in both arms leads us to require the use of the differential change in arm length to calculate the \gwadj strain,
\begin{keyeqn}
\begin{equation}
    h(t) = \frac{\delta L_1 - \delta L_2}{L},
    \label{1:eq:frac_length_diff_h_t}
\end{equation}
\end{keyeqn}
where $\delta L_{1}$ and $\delta L_{2}$ are the changes in the lengths of the arms.

We generalise $h(t)$ for any \gw of any polarisation. The strain induced in a single arm pointing in the $\vec{\mathbf{n}}$ direction, where $\vec{\mathbf{n}}$ is a unit normal vector, is
\begin{equation}
    \frac{\delta L}{L} = \frac{1}{2} h_{ij}n^{i}n^{j}
\end{equation}
and the difference in lengths for two interferometer arms is
\begin{equation}
    h(t) = \frac{1}{2} h_{ij}n_{1}^{i}n_{1}^{j} - \frac{1}{2} h_{ij}n_{2}^{i}n_{2}^{j},
\end{equation}
where crucially $h_{ij}$ is measured in the detector's reference frame. If our detector is located in the $x-y$ plane, the \gwadj signal is entirely polarised in the $h_{+}$ direction.

The \gw signal observed will depend on the orientation between source and detector. To transform between radiation frame and detector frame, we perform a set of rotations by angles ($\theta, \phi, \psi$) to adjust for the orientation and polarisation of the wave. Figure~\ref{1:eq:detector_frame_rotations} shows the required angles to describe effects on the \gwadj signal due to these transformations.
\begin{figure}
    \centering
    \includegraphics[width=0.75\linewidth]{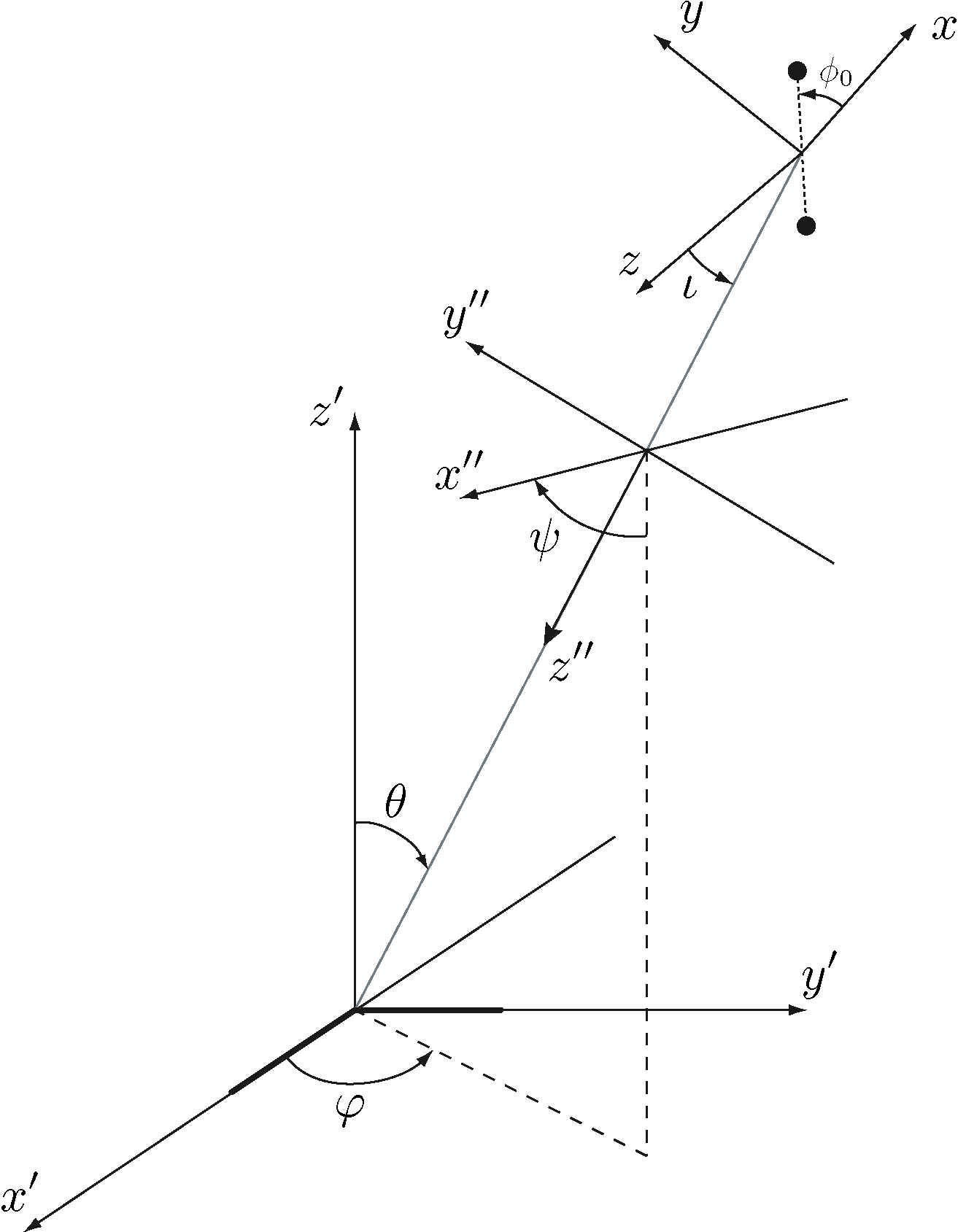}
    \caption{The angles of rotation to describe a \gw from the source from (unprimed coordinates) to the detector frame (single primed coordinates) through the radiation frame (double primed coordinates). Taken from~\cite{Brown_Thesis:2004}.}
    \label{1:fig:source_to_detector_frame}
\end{figure}
Suppose we have a \gwadj signal in the radiation frame, to transform into the detector frame we must first define a rotation tensor
\begin{equation}
    R^{i}_{j} = 
    \begin{pmatrix}
        \cos\phi & -\sin\phi & 0 \\
        \sin\phi &  \cos\phi & 0 \\
        0        &  0        & 1
    \end{pmatrix}
    \begin{pmatrix}
        1 & 0 & 0 \\
        0 & \cos\theta & -\sin\theta \\
        0 & \sin\theta & \cos\theta \\
    \end{pmatrix}
    \begin{pmatrix}
        \cos\psi & -\sin\psi & 0 \\
        \sin\psi &  \cos\psi & 0 \\
        0        &  0        & 1
    \end{pmatrix},
    \label{1:eq:detector_frame_rotations}
\end{equation}
such that when transforming the spatial components of $h_{ij}$ using this tensor we can evaluate the detector strain as a function of the $+$ and $\times$ polarisations in the radiation frame and the three angles relating the radiation frame to the detector frame. A detector with arms aligned in the $x$ and $y$ directions will give
\begin{keyeqntitled}{Detector Response}
\begin{equation}
    h(t) = F_{+}(\Theta, \Phi, \Psi)h_{+}(t) + F_{\times}(\Theta, \Phi, \Psi)h_{\times}(t) ,
    \label{1:eq:h_t_linear_combination}
\end{equation}
\end{keyeqntitled}
where,
\begin{equation}
    F_{+}((\Theta, \Phi, \Psi)) = \frac{1}{2}(1 + \cos^{2}(\Theta))\cos2\Phi\cos2\Psi - \cos\Theta\sin2\Phi\sin2\Psi,
\end{equation}
\begin{equation}
    F_{\times}((\Theta, \Phi, \Psi)) = -\frac{1}{2}(1 + \cos^{2}(\Theta))\cos2\Phi\cos2\Psi - \cos\Theta\sin2\Phi\sin2\Psi
\end{equation}
which are the antenna responses of the detector to the $+$ and $\times$ polarisations respectively, which describe the sensitivity of the detector to \gws at different sky positions.

\subsection{\label{1:sec:increase-det-sens}Increasing detector sensitivity}

With detector arm lengths on the order of ${\sim}10^{3} \, \text{m}$ and a laser wavelength of approximately $1 \, \mu\text{m}$, we are only able to measure a fractional change in the arm length on the order of $10^{-9} \, \text{m}$. However, our detectors are capable of measuring induced strain from passing \gws as small as $10^{-22}$. In this section, we discuss the detector improvements that enable these detections.

One technique used to effectively increase the arm length of the detectors is to allow the light to travel down the arms multiple times. The Fabry-Pérot cavities~\cite{aLIGO:2015} in the arms of the aLIGO detectors include an additional mirror on the input test mass (ITM), facing the end mirrors, which is highly reflective, allowing only a small fraction of light to transmit through. As a result, the light reflects multiple times between these two mirrors, increasing the optical path length to approximately $1000 \, \text{km}$. This also has the effect of building up laser power in the arms, sharpening the interference fringes to improve the ability to distinguish between noise and signal at the photo-detector~\cite{Meers:1988}.

Two additional components that increase detector sensitivity are: \textit{power recycling}, where a mirror at the laser input port redirects reflected light back into the interferometer, boosting the circulating power within the arms and enhancing the signal-to-noise ratio, enabling the detection of weaker \gwadj signals. The second improvement is the \textit{signal recycling mirror}, placed at the output port, which resonates with specific \gwadj frequencies, effectively increasing the detector's sensitivity to signals of interest. By tuning this mirror, the interferometer's sensitivity can be optimized for certain frequency bands. These improvements together allow the detector to reach sensitivities capable of observing strain on the order of $10^{-22}$ for short-duration signals up to a few seconds long.

\section{\label{1:sec:modelling_CBC}Modelling compact binary coalescences}

The primary sources of \gws seen by Earth-based \gwadj detectors are those from \cbcs. \Cbcs (CBC) are the inspiral and merger of two compact objects (black holes or neutron stars). The binary system will emit \gws and slowly dissipate the orbital energy of the system. The system is locked in the cycle of a decreasing orbit leading to greater acceleration, increasing the rate at which energy is being dissipated and the orbital radius decreases. The orbit will decay until an eventual merger between the two objects. We have observed three distinct CBC systems: binary black holes~\cite{GW150914:2016} (BBH), binary neutron stars~\cite{GW170817:2017} (BNS) and neutron star — black hole binaries~\cite{nsbh:2021}. 

In this section, we describe the \gws emitted by a CBC and the physical effects these would have on \gwadj detectors. First, we need to describe the intrinsic parameters of the \gwadj signal being created and the extrinsic, observational parameters which affect the signal based on how it is being observed. Next, we derive a gravitational waveform with a very simple binary system of two point-like masses using the Newtonian formalism.

\subsection{\label{1:sec:CBC-parameters}Waveform parameters}

\subsubsection{Intrinsic parameters}
The intrinsic parameters are those which have physical effects on the system and alter how the \gwadj signal is created,
\begin{itemize}
   \item Primary and secondary component masses, $m_{1}$ and $m_{2}$
   \item Primary and secondary three-dimensional spin vectors, $\vec{s_{1}}$ and $\vec{s_{2}}$
\end{itemize}
Here, the subscript $_1$ refers to the more massive object in the binary system, and $_2$ to the less massive one.

\subsubsection{Extrinsic parameters}
The extrinsic parameters affect how the \gwadj signal is observed,
\begin{itemize}
   \item Right ascension of the source location, $\alpha$ 
   \item Declination of the source location, $\delta$
   \item Luminosity distance to the source, $r$, 
   \item Inclination angle between the line of sight and the orbital angular momentum of the binary, $\iota$
   \item Polarisation angle of the \gw, $\psi$
   \item Time of coalescence, $t_{c}$
   \item Coalescence phase, $\phi_{c}$ 
\end{itemize}

There are additional parameters that describe other physical effects such as: tidal deformations, eccentricity, the neutron star equation of state, or potential deviations from \GR. In this work, we will ignore these additional effects and focus on the $15$ parameters listed above.

\subsection{\label{1:sec:keplerian_derivation}Simple CBC inspiral}

We can derive $h_{+}$ and $h_{\times}$ for a binary system modelled as a decaying quasi-circular orbit of two point masses around a shared centre of mass. These two point masses are in the `inspiral' regime, where they are radiating \gws, travelling with speeds much less than $c$ and, are not close to merging. We choose a frame of reference where the plane of the binary is in the $x$-$y$ direction and the binary is situated at a distance $r$ from the observer.

Assuming a circular orbit, we obtain the orbital velocity in Kepler's laws of planetary motion by equating the Newtonian gravitational force and the orbital centripetal force
\begin{align}
    \frac{m_{1} m_{2}}{a^{2}} = \frac{\mu v^{2}}{a}, \\
    v = \sqrt{\frac{m_{1} + m_{2}}{a}},
\end{align}
where $m_{1}$ and $m_{2}$ are the masses of the objects, $\mu$ is the reduced mass of the system ($\mu = \frac{m_1m_2}{(m_1+m_2)}$) and $a$ is the distance between the two objects.
Using $\omega = v/a$, we can obtain the orbital frequency
\begin{equation}
    \omega_{orbit} = \sqrt{\frac{m_{1} + m_{2}}{a^{3}}},
    \label{1:eq:omega_orbit}
\end{equation}
and relate it to the rate of change in the orbital phase
\begin{equation}
    \omega_{orbit} = \frac{d \Phi_{orbit}}{dt}.
\end{equation}
Using the quadrupole formula (Equation~\ref{1:eq:quadrupole_formula}), we can derive the metric perturbations caused by this system
\begin{equation}
    h_{\mu\nu} = \frac{4}{r} \mu a^{2} \omega^{2}_{orbit}
    \begin{pmatrix}
      0 & 0 & 0 & 0 \\
      0 & -\cos\left(2\omega_{orbit}t\right) & -\sin\left(2\omega_{orbit}t\right) & 0 \\
      0 & -\sin\left(2\omega_{orbit}t\right) & \cos\left(2\omega_{orbit}t\right) & 0 \\
      0 & 0 & 0 & 0
   \end{pmatrix}.
\end{equation}
The line of sight between the binary system and the observer is defined by the inclination angle, $\iota$, and when projecting the \gwadj emission in the line of sight we get the waveform polarisations,
\begin{equation}
    h_{+}(t) = \frac{4}{r} M\eta v^{2} \left(\frac{1 + \cos^{2}\iota}{2}\right)\cos\left(2\omega_{orbit}t+2\Phi_{c}\right),
\end{equation}
\begin{equation}
    h_{\times}(t) = \frac{4}{r} M\eta v^{2} \left(\frac{\cos\iota}{2}\right)\sin\left(2\omega_{orbit}t+2\Phi_{c}\right),
\end{equation}
where $M = m_1 + m_2$ is the total mass of the system, $\eta = \frac{m_{1}m_{2}}{(m_{1} + m_{2})^{2}}$ is the symmetric mass ratio and $\Phi_{c}$ is the coalescence phase of the system. A key observation from these equations is that the frequency of the \gws is twice that of the binary orbital frequency.

We have an embedded assumption that the separation distance of the two objects will remain constant. We know that the radiation of \gws will decrease the orbit of the binary system, and so by solving equations~\ref{1:eq:de_dt} and~\ref{1:eq:luminosity_flux}, we can find the power radiated by the binary system. The energy loss will increase the binary orbital frequency, from this we can calculate the change in phase and amplitude as a function of time.

In the two point source example we relate the orbital velocity, $\omega_{orbit}$ to the particle separation, $a$, using Equation~\ref{1:eq:omega_orbit} and then when introducing the \textit{chirp mass}, $\mathcal{M}$, of the binary system
\begin{equation}
    \mathcal{M} = \frac{(m_{1}m_{2})^{\frac{3}{5}}}{(m_{1} + m_{2})^\frac{1}{5}},
\end{equation}
and solving equations~\ref{1:eq:de_dt} and~\ref{1:eq:luminosity_flux} we are able to determine the power radiated by this system as
\begin{equation}
    \frac{dE}{dt} = -\frac{32}{5}(\mathcal{M}\omega_{orbit})^\frac{10}{3}.
    \label{1:eq:cbc_eg_de_dt}
\end{equation}
The total energy of the system can be expressed as
\begin{align}
    E &= -\frac{m_{1}m_{2}}{2a}, \\
      &=-\frac{1}{2}(\mathcal{M}^{5} \omega^{2}_{orbit})^\frac{1}{8},
      \label{1:eq:total_energy_cbc}
\end{align}
and when substituted into Equation~\ref{1:eq:cbc_eg_de_dt} and solving we obtain the orbital frequency as a function of time
\begin{align}
    \omega_{orbit}(t) &= \left(-\frac{256}{5}\mathcal{M}^{\frac{5}{3}}(t - t_{0})\right)^{-\frac{3}{8}}, \\
                      &= \frac{1}{8}\left(\frac{\tau}{5}\right)^{-\frac{3}{8}}\mathcal{M}^{-\frac{5}{8}},
    \label{1:eq:omega_orbit_f_of_tau}
\end{align}
where $t_{0}$ is the fiducial starting time for integration which is used to map the time to define a new time variable $\tau = t_{c} - t$.

We can integrate Equation~\ref{1:eq:omega_orbit_f_of_tau} over time to find the orbital phase of the \gwadj signal,
\begin{align}
    \Phi(\tau) &= \int^{t}_{t+{0}} \omega_{orbit} dt, \\
    &= -2\left(\frac{5G\mathcal{M}}{c^{3}}\right)^{-\frac{5}{8}} \tau^{\frac{5}{8}} + 2\Phi_{c},
\end{align}
where $\Phi_{c}$ is the coalescence phase.

Finally, combining the previous equations, we have derived
\begin{keyeqn}
\begin{equation}
    h_{+}(\tau) = \frac{1}{r}\left(\frac{G\mathcal{M}}{c^{2}}\right)^{\frac{5}{4}}\left(\frac{5}{c\tau}\right)^{\frac{1}{4}}\left(\frac{1+\cos^{2}\iota}{2}\right)\cos\Phi(\tau),
\end{equation}
\begin{equation}
    h_{\times}(\tau) = \frac{1}{r}\left(\frac{G\mathcal{M}}{c^{2}}\right)^{\frac{5}{4}}\left(\frac{5}{c\tau}\right)^{\frac{1}{4}}\cos\iota\sin\Phi(\tau),
\end{equation}
\end{keyeqn}
the \gwadj polarisations for a metric perturbation originating from our simple two-point particle binary system. From these equations, it's easy to show that both the amplitude and frequency of the system will increase rapidly and ``chirp'' as it approaches coalescence. Also clear to see is that these \gwadj signals are dependent only on the chirp mass of the system.

\subsection{\label{1:sec:fourier_transform_chirp}Fourier transform of the chirp signal}

In \gwadj analysis, the Fourier transform of a chirp signal is crucial for understanding its frequency domain characteristics. We consider a gravitational waveform given by
\begin{equation}
    h(t) = A(t) \cos \Phi(t),
\end{equation}
where \( A(t) \) represents the amplitude, and \( \Phi(t) \) is the phase of the signal. To simplify our analysis, we use the \textit{stationary phase approximation}, applicable when the amplitude \( A(t) \) varies slowly compared to the phase \( \Phi(t) \). Under this approximation, the Fourier transform of \( h(t) \) is given by
\begin{equation}
    \Tilde{h}(f) \simeq \frac{1}{2} A\left(t(f)\right) \left(\frac{dt(f)}{df}\right)^{\frac{1}{2}} {\rm e}^{i \left(2\pi f t(f) - \Phi(f) - \frac{\pi}{4}\right)},
\end{equation}
where \( t(f) \) and \( \Phi(f) \) are the time and phase at frequency \( f \). For a chirp signal, these quantities can be expressed as
\begin{equation}
    t(f) = t_c - \frac{5}{8\pi}\left(8\pi f\right)^{-\frac{8}{3}} \mathcal{M}^{-\frac{5}{3}},
\end{equation}
\begin{equation}
    \Phi(f) = \Phi_c - 2\left(8\pi \mathcal{M} f\right)^{-\frac{5}{3}},
\end{equation}
where \( t_c \) is the coalescence time, \( \mathcal{M} \) is the chirp mass, and \( \Phi_c \) is the phase at coalescence.

The Fourier transform of the chirp signal, considering the Newtonian approximation, can be expressed in the form of
\begin{keyeqntitled}{Frequency Domain Waveform}
\begin{equation}
    \Tilde{h}(f) = \sqrt{\frac{5}{24}} \frac{G^2 \mathcal{M}^2}{c^5 d_L} \left(\frac{\pi G \mathcal{M} f}{c^3}\right)^{-\frac{7}{6}} \left(\frac{1 + \cos^2 \iota}{2}\right) {\rm e}^{i \Phi(f; t_c)},
\end{equation}
\end{keyeqntitled}
where the phase term \( \Phi(f; t_c) \) is defined by
\begin{equation}
    \Phi(f; t_c) = \frac{\pi}{4} - \frac{3}{128} \left(\frac{\pi G \mathcal{M} f}{c^3}\right)^{-\frac{5}{3}} - 2\pi f t_c + \Phi_c.
\end{equation}
Here, \( d_L \) denotes the luminosity distance to the source, and \( \iota \) is the inclination angle of the orbital plane with respect to the line of sight.

\subsection{\label{1:sec:post_newtonian_treatment}Post-Newtonian approximation}

Our Newtonian model of the \gwadj signal from a CBC source has been derived with simplifications. However, the real signals detected by \gwadj detectors are far more complex, requiring corrections to account for the effects of \GR. The post-Newtonian (PN) expansion technique improves the accuracy of gravitational waveforms by introducing relativistic corrections to the waveform, accounting for deviations from Newtonian gravity due to \GR~\cite{2PN_1:1996, 2PN_2:1996, 2PN_3:1995}.

The PN expansion can be performed by expanding in powers of \( v/c \), where \( v \) is the velocity of the orbiting bodies, to obtain the corrections to the waveform which can then be summed to get waveforms up to the required PN order. For a CBC, the waveform could be written as
\begin{equation}
    h(t) = h_{\text{Newtonian}}(t) + h_{\text{1PN}}(t) + h_{\text{1.5PN}}(t) + h_{\text{2PN}}(t) + \cdots,
\end{equation}
where \( h_{\text{Newtonian}}(t) \) is the leading-order Newtonian term, and the following terms represent higher and higher PN order corrections~\cite{PN_models:2009}.

\subsection{\label{1:sec:full_waveform}Full waveform}

The post-Newtonian (PN) formalism is accurate during the inspiral phase of the waveform when the two objects are still distant from each other, and formalism does not cover the entire evolution of the system. The system continues to emit \gws throughout the merger and post-merger phases, including the ringdown phase, where the newly formed object settles into its final shape. The \gwadj emission peaks during the merger, making it crucial for detecting \gws. The ringdown phase is particularly valuable for testing General Relativity (GR)~\cite{GW150914_TGR:2016, GW170817_TGR:2019, O3_TGR:2021}.

\begin{figure}
    \centering
    \includegraphics[width=0.75\linewidth]{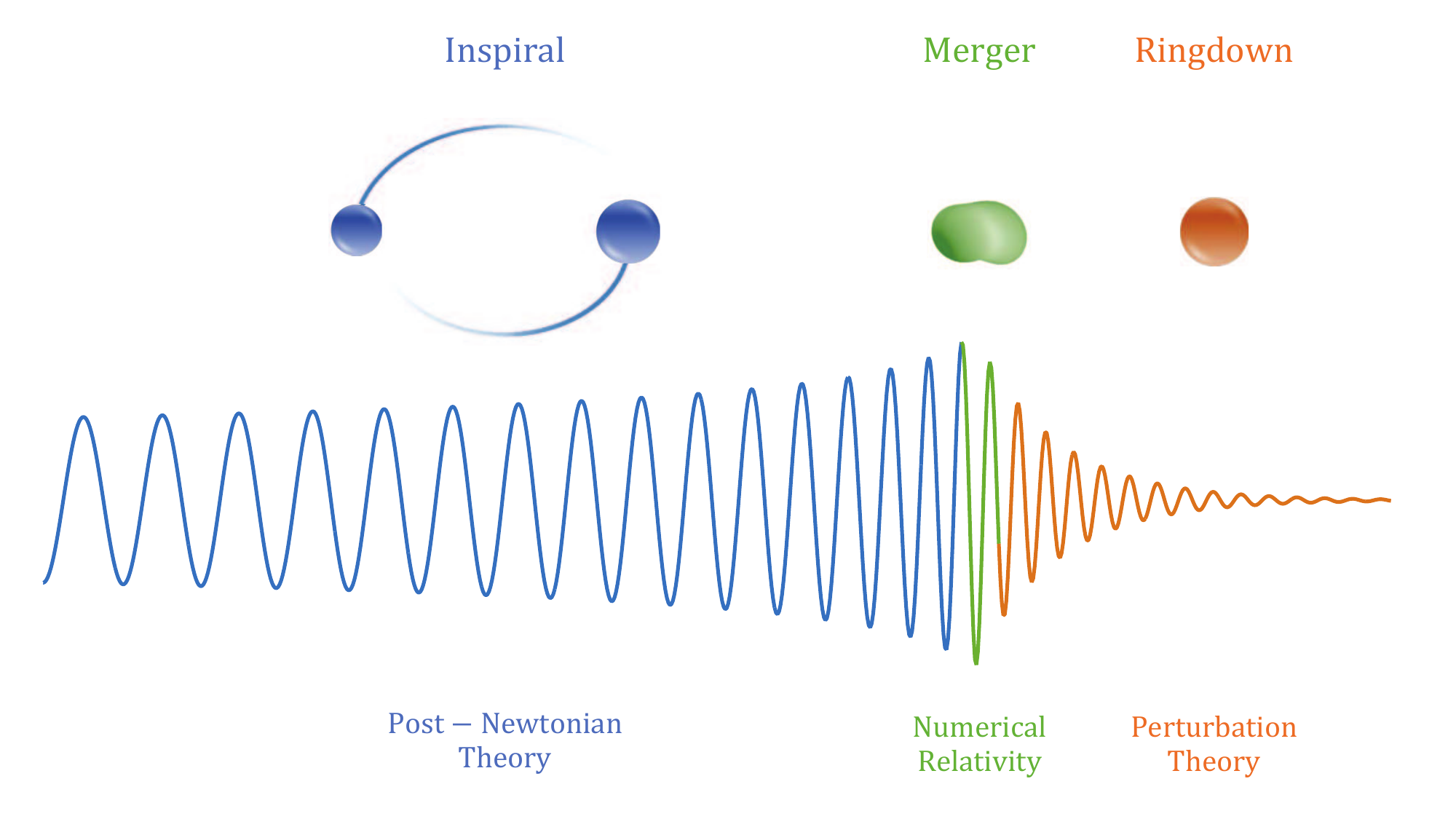}
    \caption{The evolution of a compact binary merger is illustrated through three distinct phases. The inspiral phase involves the two compact objects orbiting each other, with the orbit decaying as they emit \gws. This phase is modelled using post-Newtonian theory. The merger phase occurs when the orbital radius decreases sufficiently for the objects to merge into a new compact object, modelled using numerical relativity. The ringdown phase represents the newly formed object's vibration as it settles into its final shape, modelled using perturbation theory. This image is adapted from~\cite{IMR_plot:2016}.}
    \label{1:fig:IMR}
\end{figure}

Generating full waveforms requires solving the Einstein field equations using numerical relativity (NR), a process that can take weeks. To efficiently model the entire evolution of the waveform---comprising the inspiral, merger, and ringdown (IMR) phases (see Figure~\ref{1:fig:IMR})---waveform families use a combination of NR waveforms and analytical or semi-analytical approximations. This approach balances accuracy with computational efficiency. The following subsections briefly describe these waveform models.

\subsubsection{Effective-One-Body (EOB)}

The Effective-One-Body (EOB) model~\cite{EOB_1:1998} simplifies the binary system by transforming it into an equivalent single test particle moving in an external gravitational field, described by a Hamiltonian~\cite{EOB_1:1998, EOB_2:2000, EOB_3:2000, EOB_4:2001}. This Hamiltonian includes both leading-order Newtonian terms and higher-order post-Newtonian corrections~\cite{EOB_5:2008}.

EOB models tuned to numerical relativity (NR) data are known as ``EOBNR'' models~\cite{EOB_6:2007}. Some of these models, such as \texttt{SEOBNRv5}~\cite{SEOBNRv5:2023tna}, also incorporate the effects of anti-aligned spins. Further enhancements allow for precession effects and higher-order modes of \gwadj emission, as seen in the state-of-the-art model \texttt{SEOBNRv5\_PHM}~\cite{SEOBNRv5_PHM-Buades:2023ehm}, which represents the most advanced EOB waveform model as of the drafting of this thesis.

\subsubsection{Phenomenological models}

Phenomenological models, often abbreviated as `Phenom', use a phenomenological ansatz calibrated with NR simulations to produce full waveforms. These models are fully analytical and can generate both Fourier domain and time domain waveforms~\cite{IMR_1:2007, IMR_2:2020}. The ability to create waveforms directly in the Fourier domain is highly desirable for analyses that require generating millions of waveforms, such as offline searches or parameter estimation.

Phenom models that cover the inspiral, merger, and ringdown phases are prefixed with \texttt{IMRPhenom}~\cite{IMRPhenomD:2009}. The state-of-the-art model of this family is \texttt{IMRPhenomXPHM}~\cite{IMRPhenomXPHM:2020}, which can accurately model anti-aligned spins, precessing binaries (\texttt{P}), higher-order modes (\texttt{HM}), and extreme mass ratios (\texttt{X}).

\subsubsection{Surrogate models}

Surrogate models create waveforms by interpolating from NR simulations~\cite{Surr_1:2019, Surr_2:2022}. These models, such as \texttt{NRSur7dq4}~\cite{NRSur7dq4:2019}, provide greater accuracy than other model families but are limited by the parameter regions covered by the NR simulations used for interpolation. The current leading surrogate model, \texttt{NRSur7dq4}, is valid up to a mass ratio of four.

\chapter[Searching for Gravitational-Waves from Compact Binary Coalescences]{Searching for Gravitational Waves from Compact Binary Coalescences}
\label{chapter:2-searches}
\chapterquote{Even in death I serve the Omnissiah}{Magos Dominus Reditus}


\Gwadj search pipelines, such as PyCBC, play a critical role in analysing data from detectors to identify \gwadj signals, both in real time (live) and in post-processing (offline). This chapter develops the theoretical foundation for these search pipelines and explores the techniques employed in \gwadj detection. Emphasis is placed on the PyCBC pipelines---PyCBC Offline and PyCBC Live---both of which were extensively utilised and further developed as part of this thesis.

\section{\label{2:sec:gw-data}\Gwadj data}

\Gwadj observatories produce a dimensionless strain time-series, $s(t)$, which is composed of detector noise, $n(t)$, and, when present, an astrophysical \gwadj signal, $h(t)$
\begin{equation}
    s(t) =
    \begin{cases}
        n(t), & \text{if no signal is present}, \\
        n(t) + h(t), & \text{if a signal is present}.
    \end{cases}
\end{equation}
The primary objective of \gwadj search pipelines is to extract $h(t)$ from $s(t)$, detecting the astrophysical signal from within the background noise.

\section{\label{2:sec:search-methods}Searching through \gwadj data}

Detecting \gws from \cbcs requires sophisticated search methods capable of sifting through vast amounts of data collected by \gwadj detectors. This section details the search techniques used to identify and characterise \gwadj signals. There are many pipelines that search for \gws~\cite{pipelines, PyCBC:2017, GstLAL:2020, SPIIR:2020, MBTA:2021, cWB:2020, oLIB:2015, MLy:2020qax} and in this section we will focus on the PyCBC search~\cite{PyCBC:2016, PyCBC:2017, PyCBC_package:2021}

We begin with the single-detector search techniques to identify potential \gwadj events, and then detail the tests and methods used to combine single detector detections to identify coincidentally found \gwadj signals.

\subsection{\label{2:sec:matched-filter}Matched filtering}

Detecting \gws relies on being able to distinguish between data that contains only noise, \( s(t) \!=\! n(t) \), and data which contains a \gwadj signal, \( s(t) \!=\! n(t) + h(t) \). We must construct an optimal detection statistic, which expresses the value of the probability that the data contains a known signal. The Neyman-Pearson likelihood ratio forms the optimal detection statistic~\cite{Biswas:2012} and is found by computing the ratio of the probability that the data contains the signal (hypothesis $\mathcal{H}_{1}$) to the probability that the data is pure noise (the null hypothesis $\mathcal{H}_{0}$)
\begin{equation}
    \Lambda = \frac{P(s|\mathcal{H}_{1})}{P(s|\mathcal{H}_{0})},
\end{equation}
where \( \Lambda \) is the likelihood ratio that serves as the detection statistic, \( P(s|\mathcal{H}_{1}) \) is the probability that the data contains the signal, and \( P(s|\mathcal{H}_{0}) \) is the probability that the data is pure noise. It is natural to use probability densities due to the detection process involving continuous data and not discrete events
\begin{equation}
    \Lambda = \frac{p(s|\mathcal{H}_{1})}{p(s|\mathcal{H}_{0})},
    \label{2:eq:likelihood_ratio}
\end{equation}
where \( p(s|\mathcal{H}_{1}) \) is the probability density that the data contains the signal, and \( p(s|\mathcal{H}_{0}) \) is the probability density that the data is pure noise.

We first define the noise-weighted inner product between two time series $s(t)$ (data) and $h(t)$ (signal) as~\cite{PyCBC:2016}
\begin{keyeqntitled}{Noise-Weighted Inner Product}
\begin{equation}
  (s | h) = 4 \Re \int^{\infty}_{0} \frac{\tilde{s}(f) \tilde{h}^*(f)}{S_n(f)} df,
  \label{2:eqn:inner_product}
\end{equation}
\end{keyeqntitled}
where a tilde denotes a Fourier transformed version of the variable and where $S_n(f)$, represents the one-sided power spectral density (PSD) of the data, defined as
\begin{equation}
  \langle \tilde{s}(f) \tilde{s}(f^\prime) \rangle = \frac{1}{2} S_n(f) \delta(f - f^\prime) \;,
  \label{2:eqn:psd}
\end{equation}
where angle brackets denote an average over noise realisations and $\delta$ is the Dirac delta function. It is important to note, $p(a|b)$ denotes conditional probability and $(a|b)$ denotes the noise-weighted inner product.

If the detector noise is Gaussian, we can compute the probability densities and derive the matched filter:
\begin{align}
    p(s|\mathcal{H}_{0}) &\propto {\rm e}^{-(s|s)/2}, \\ 
    p(s|\mathcal{H}_{1}) &\propto {\rm e}^{-(s-h|s-h)/2},
\end{align}
and the likelihood ratio (Equation~\ref{2:eq:likelihood_ratio})
\begin{equation}
    \Lambda(\mathcal{H}_{1}|s) = \frac{{\rm e}^{-(s-h|s-h)/2}}{{\rm e}^{-(s|s)/2}} = {\rm e}^{(s|h)}{\rm e}^{-(h|h)/2}.
\end{equation}
The exponential scaling of the likelihood values can introduce numerical instability when computing the likelihood ratio. Therefore, we take the logarithm, resulting in the log-likelihood ratio
\begin{keyeqn}
\begin{equation}
    \log \Lambda(\mathcal{H}_{1}|s) = (s|h) - \frac{(h|h)}{2}.
    \label{2:eq:log_likelihood_ratio}
\end{equation}
\end{keyeqn}
From Equation~\ref{2:eq:log_likelihood_ratio} we can see that the likelihood ratio depends only on the data through the inner product of $s$ and $h$ and is the optimal detection statistic known as the \textit{matched filter}, effectively a noise-weighted correlation between the known signal and the data.

Since we are interested in evaluating the presence of a signal in the data, it is useful to define the \textit{signal-to-noise ratio} (SNR), \( \rho \), which quantifies how strong the signal is relative to the background noise. The SNR of a known signal in data can be derived as~\cite{FINDCHIRP:2012}
\begin{keyeqntitled}{Signal-to-Noise Ratio}
\begin{equation}
    \rho = \frac{(s|h)}{\sqrt{(h|h)}}.
    \label{2:eq:snr}
\end{equation}
\end{keyeqntitled}
This expression indicates how well the signal correlates with the data relative to the noise level in the detector. When the detector contains a real signal, a high SNR corresponds to a stronger, more detectable signal, whereas a low SNR indicates a weak signal buried in the noise. We will use the SNR value as the detection statistic moving forward.

\subsection{\label{2:sec:snr_timeseries}Signal-to-noise ratio over time}

We have assumed that we know all the parameters of the signal we are searching for; however, this is not likely for real \gwadj signals. We will begin to build up a more robust search, in which we know very few of the initial \gwadj signal parameters. First, we consider the case of a signal that has a known waveform but an unknown amplitude and arrival time. We can describe the true signal as
\begin{equation}
    h(t) = A g(t - t_{0}),
\end{equation}
where A is the unknown amplitude of the signal, $t_{0}$ is the unknown arrival time and $g(t)$ is the known waveform. We can take the Fourier transform of this signal
\begin{equation}
    \tilde{h}(f) = A \tilde{g}(f) {\rm e}^{-2\pi i f t_{0}},
\end{equation}
and obtain the matched filter using Equation~\ref{2:eqn:inner_product} to be
\begin{equation}
    (s|h) = 4 A \Re \int^{\infty}_{0} \frac{\tilde{s}(f) \tilde{h}^*(f)}{S_n(f)} {\rm e}^{2\pi i f t_{0}} df.
\end{equation}
From this we can define
\begin{align}
    (s|h) &= A x(t_{0}) \quad \text{where}, \\
    x(t) &= 4 \Re \int^{\infty}_{0} \frac{\tilde{s}(f) \tilde{g}^*(f)}{S_n(f)} {\rm e}^{2\pi i f t} df.
\end{align}
$x(t)$ is a time series representing the matched filter at any arrival time $t$. From this we can define an SNR time series $\rho(t)$ containing information about the SNR value at each point in time. To find the maximum likelihood detection statistic we simply find the largest value of $\rho(t)$ which will correspond to the amplitude and will reveal the previously unknown arrival time $t_{0}$.

\subsection{\label{2:sec:phase-maximisation}Phase maximisation}

The phase of the \gwadj signal is another unknown parameter. In Section~\ref{1:sec:fourier_transform_chirp} we give a signal of the form
\begin{equation}
    h(t) = A(t) \cos\left(\Phi(t)\right),
\end{equation}
and can include an additional phase offset, $\phi_{0}$, to account for the random orientation and sky position of the binary
\begin{equation}
    h(t) = A(t) \cos\left(\Phi(t) + \phi_{0}\right).
    \label{2:eq:phase_signal_model}
\end{equation}
We can maximise over this phase offset using the matched filter by rewriting our signal as
\begin{equation}
    h(t) = h_{0}(t) \cos(\phi_{0}) + h_{\pi/2}(t)\sin(\phi_{0}),
\end{equation}
where $h_{0}(t)$ and $h_{\pi/2}(t)$ are realisations of Equation~\ref{2:eq:phase_signal_model} where the phase offset has been set equal to $0$ and $\frac{\pi}{2}$ respectively~\cite{IHOPE:2012zx}.

We can then calculate $\rho^{2}$ using these two new signals to maximise over the phase
\begin{keyeqn}
\begin{equation}
    \underset{\Phi}{\text{max}}(\rho^{2}(t)) = \frac{(s|h_{0})^{2} + (s|h_{\pi/2})^2}{(h_{0}|h_{0})},
    \label{2:eq:phase_max}
\end{equation}
\end{keyeqn}
having made the assumption that $\tilde{h}_{\pi/2}(f) \!=\! i\tilde{h}_{0}(f)$ which is true for frequency domain waveforms with the stationary phase approximation~\cite{Droz:1999}. Conventionally we take the square root of Equation~\ref{2:eq:phase_max} to obtain the SNR, where phase has been maximised over
\begin{equation}
    \rho = \sqrt{\underset{\Phi}{\text{max}}(\rho^{2}(t))}.
\end{equation}

\subsection{\label{2:sec:template-bank}Template bank}

We have demonstrated that the matched filter can be used to analytically and efficiently maximise over the amplitude, time of arrival and phase of a known signal. We acknowledge that for a real search for \gws, we will not know the $15$ parameters of the signal. The search is performed by creating many realisations of the \gwadj signal and searching over the data with each of them. However, we need to discuss how the realisation parameter values are chosen to ensure a sufficiently covered parameter space.

When a signal is found by a template\footnote{A realisation of the \gwadj signal waveform.} with parameters not equal to the true values, we will expect to see a fractional loss in the expected SNR. The closer the template parameters are to the signal parameters, the closer to the maximum SNR we will see.

We can define a signal with parameters $\lambda$
\begin{equation}
    h(t) = \rho u(t;\lambda),
\end{equation}
where $\rho$ is the expected SNR value for the true template. If we have a template with parameters $u(t;\lambda + \Delta \lambda)$, the expected SNR when matched filtering the signal with this template will be
\begin{equation}
    \rho^{\prime} = (h|u(\lambda + \Delta \lambda)) = \rho(u(\lambda)|u(\lambda + \Delta \lambda)),
\end{equation}
and we can see therefore that the expected fractional loss in the expected SNR is
\begin{equation}
    \frac{\rho - \rho^{\prime}}{\rho} = 1 - (u(\lambda)|u(\lambda + \Delta \lambda)) = 1 - \mathcal{A},
\end{equation}
where we can define $\mathcal{A}$ as the ambiguity function
\begin{equation}
    \mathcal{A}(\lambda;\lambda + \Delta \lambda) := (u(\lambda)|u(\lambda + \Delta \lambda)),
\end{equation}
which tells us how well our nearby template matches to the true signal\footnote{Templates are assumed to be normalised such that $(u(\lambda)|u(\lambda)) = 1$.}. If the ambiguity value is large, then we have a small fractional loss, and the template is a good match to the true signal; if the value is small, then the template is a poor description of the signal.

$\lambda$ contains both the \textit{intrinsic} and \textit{extrinsic} parameters of the signal, we can define the \textit{overlap} between two templates $h_{1}$ and $h_{2}$ considering only the intrinsic parameters when we maximise over the extrinsic parameters,
\begin{equation}
    \mathcal{O}(h_{1}, h_{2}) := (h_{1} | h_{2}) = \frac{(h_{1} | h_{2})}{\sqrt{(h_{1} | h_{1})(h_{2} | h_{2})}}.
\end{equation}
The overlap for a signal $h_{1}$ represents the fraction of the SNR recovered by matched filtering with the template $h_{2}$. The match is then the overlap maximised over time of arrival and phase~\cite{Harry_Lundgren:2012}
\begin{equation}
    \mathcal{M}(h_{1}, h_{2}) = \underset{\phi_0, t_{c}}{max}(\hat{h}_{1}|\hat{h}_{2}(\phi_{c}, t_{c})).
\end{equation}
We then require that any signal can be recovered with a maximum $3\%$~\cite{Owen_Sathya:1999} SNR loss, corresponding to a mismatch, $1 - \mathcal{M}$, of $0.03$, at least one template must have a maximised overlap of at least $0.97$ with the true signal waveform. We can then construct a bank of templates with this requirement. Template banks can be constructed using two primary approaches: geometrically~\cite{geom_bank_1:1991, geom_bank_2:1992, geom_bank_3:1995, geom_bank_4:1995, Owen_Sathya:1999} or stochastically~\cite{Harry_sbank:2009, Stochastic_tb:2008}.

In a geometric approach, a lattice is defined within the template parameter space, with each lattice point corresponding to a unique template. In the stochastic approach, randomly generated templates are placed in the bank and their match to existing templates is calculated. A new template is retained if its maximum match with any existing template is less than $0.97$; it is discarded if the match is greater than or equal to $0.97$.

\subsection{\label{2:sec:signal-consistency}Signal consistency tests}

The matched filter is described as the optimal detection statistic in stationary Gaussian noise for searching for known signals. While we have dealt with the case of an unknown signal, we now consider the case where the detector noise is not Gaussian. Within the detector data, we have many instances of short duration bursts of excess power that are non-Gaussian, commonly called `glitches'~\cite{LIGO_data_quality:2015}.

Glitches produce large SNRs in the matched filter even when they do not share the same morphological characteristics. To combat this, we use signal consistency tests, which are able to discriminate between glitches and signal based on the distribution of the power present in the detector data.

We know the expected time and frequency evolution of a \gwadj signal using our waveform models. The time-frequency waveform consistency test described in~\cite{Allen_Chi:2005} (the Allen $\chi^{2}$ test) divides the signal template into $p$ sub-templates such that each sub-template contributes equally to the total SNR,
\begin{equation}
    4 \int^{f_{1}}_{0}\frac{\tilde{s}(f) \tilde{h}_{1}(p)^*(f)}{S_n(f)}df = 4 \int^{f_{2}}_{f_{1}}\frac{\tilde{s}(f) \tilde{h}_{2}(p)}{S_n(f)}df = ... =  4 \int^{\infty}_{f_{p-1}}\frac{\tilde{s}(f) \tilde{h}_{p}(f)}{S_n(f)}df ,
\end{equation}
where $\tilde{s}(f)$ is the data and $\tilde{h_{p}}(f)$ the sub-templates from discrete non-overlapping frequencies.

To calculate the divergence from the expected time-frequency distribution we can calculate the chi-squared value evaluated at some time, $t_{0}$, using
\begin{equation}
    \chi^{2}(t_{0}) = \sum^{p}_{i=1} \left(\frac{\rho}{\sqrt{p}} - \rho_{bin, i}\right),
\end{equation}
where $p$ is the number of sub-templates (bins), $\rho$ is the SNR for the full template matched-filtered with the data and $\rho_{bin}$ is the SNR found when matched filtering the sub-template and the data. The $\chi^{2}$ value will be small for data containing the true signal but will be large when the SNR found in some bins is different from the expected SNR, notably in the presence of a glitch with power distribution that does not match the signal.

If the data noise is Gaussian the $\chi^{2}$ value will be $\chi^{2}$ distributed with $2p - 2$ degrees of freedom, therefore, we use the reduced-$\chi^{2}$ value which will evaluate close to $1$ for the true signal template
\begin{keyeqntitled}{Reduced Chi-Squared Statistic}
\begin{equation}
    \chi_{r}^{2} = \frac{\chi^{2}}{2p-2}.
    \label{2:eq:reduced_chisq}
\end{equation}
\end{keyeqntitled}
The sine-Gaussian $\chi^{2}$ test is another test used by \gwadj searches to identify excess power in frequencies \textit{above} the expected frequency range of the signal. To do this, multiple sine-Gaussian functions are defined across the frequency range and matched filtered with the data to find deviations from the expected SNR at each frequency~\cite{PyCBC_sg:2018}. If a \gwadj signal merges at a specific frequency, then we do not expect a large SNR for sine-Gaussian functions at frequencies greater than the merger frequency.

These two $\chi^{2}$ tests are used to re-weigh the SNR of the matched filter, first the Allen $\chi^{2}$ test is applied~\cite{McIsaac_Chi:2022}
\begin{equation}
    \rho =
    \begin{cases}
        \hat{\rho}, & \text{if \,} \chi^{2}_{r} \le 1, \\
        \rho\left[(1 + (\chi^{2}_{r})^{3}) /2\right]^{-\frac{1}{6}}, & \text{if \,} \chi^{2}_{r} > 1,
    \end{cases}
\end{equation}
then the sine-Gaussian $\chi^{2}$ test is applied
\begin{equation}
    \hat{\rho}_{sg} =
    \begin{cases}
        \hat{\rho}, & \text{if \,} \chi^{2}_{r} \le 4, \\
        \hat{\rho}\left(\chi^{2}_{r, sg})/4\right)^{-\frac{1}{2}}, & \text{if \,} \chi^{2}_{r} > 4.
    \end{cases}
\end{equation}
We note that the coefficients used in the re-weighing are determined empirically and can be tuned for different populations (though this has not been done).

We define a detection threshold such that a template found at a time with SNR above the threshold will be considered for consideration as a \gwadj signal, we call these detections `triggers'. The re-weighted SNR is used to rank detector triggers from single detector matched filters.

\subsection{\label{2:sec:auto-gating}Auto-gating}

When matched filtering the template bank and the data, glitches will produce triggers which are suppressed by the signal-consistency tests. Very loud glitches which resemble delta-functions will produce very large SNR triggers and will cause a \textit{ringing} effect where the matched filter will continue to produce high SNR triggers for a short time.

\begin{figure}
    \centering
    \includegraphics[width=0.9\linewidth]{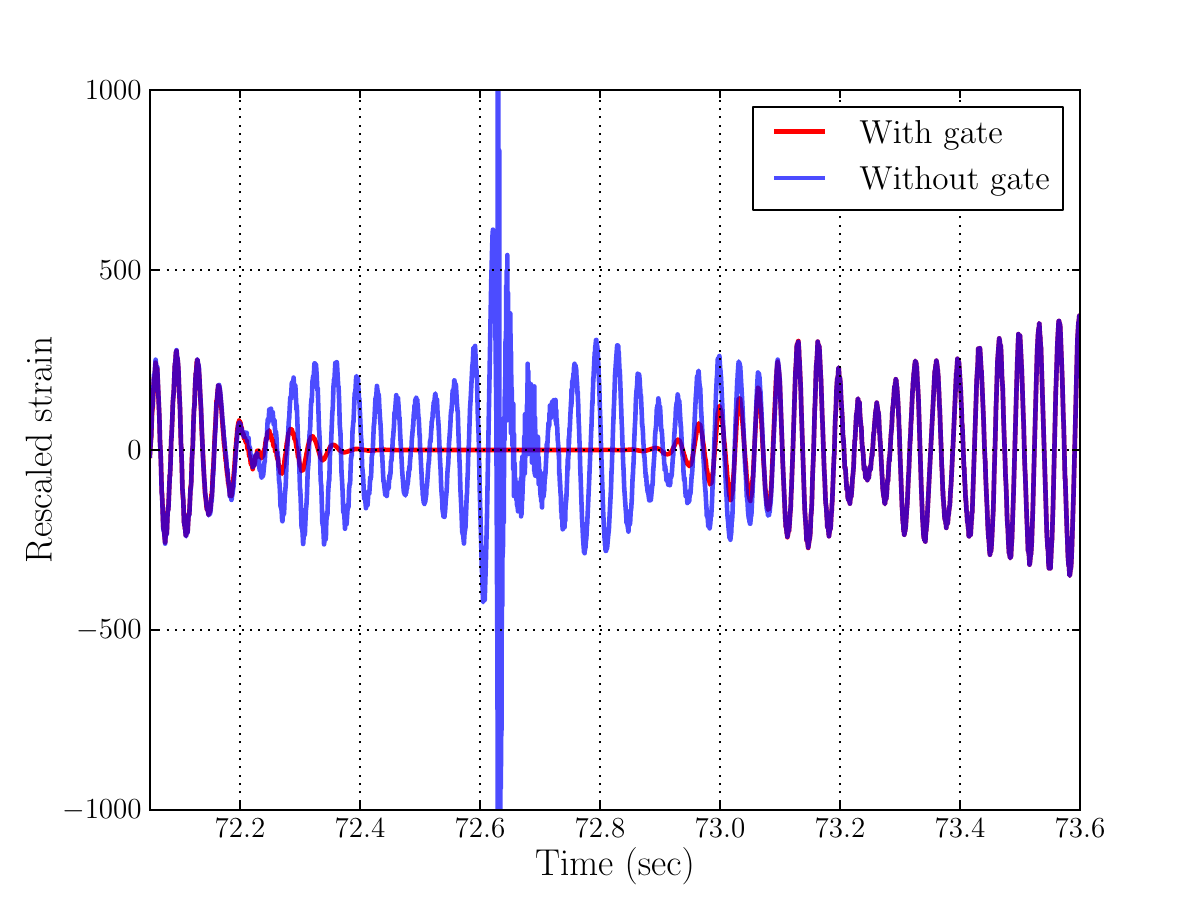}
    \caption{Gating a very loud noise transient. The detector strain has been rescaled by a factor of $10^{21}$ and the glitch has a peak magnitude over $5$,$000$. The blue line shows the data before applying the Tukey window and the red shows the data after applying the Tukey window. Note the smooth decrease in data amplitude at the edges of the windowing function. Image taken from~\cite{PyCBC:2016}.}
    \label{2:fig:autogating}
\end{figure}

These glitches are identified by looking for excess power in the whitened\footnote{Whitened data has a flat power spectral density with no frequency-dependent noise.} data and are handled by applying a windowing function around the data. This process is known as \textit{gating} and the window is applied such that a smooth transition between data and zeroing is made to avoid discontinuities in the data which can introduce more data artefacts. Figure~\ref{2:fig:autogating} shows an example of a glitch which has been gated.

Gating occurs automatically in \gwadj searches in a process called `auto-gating'. Auto-gating in PyCBC Live zeroes only $0.25$ seconds of data with a taper of $0.25$ seconds on either side. Auto-gating will completely remove the delta-function glitch and will prevent the ringing effect, removing the triggers caused by the glitch.

\subsection{\label{2:sec:coincidence-test}Coincidence tests}

Non-Gaussian transients in our detector data lead to greatly increased possibility for a \gwadj detector to report detections of \gwadj signals caused by non-astrophysical sources. If, after our signal-consistency tests, a glitch has a large SNR we are unable to make a distinction between it and a real event.

In terms of the optimal detection statistic, we must include further components which can reject glitches while ensuring we continue to detect all possible real events. The most powerful method for confirming the detection in one detector is the coincidental detection of the same signal in another detector. We call this a \textit{coincidence test}; the requirement that for a signal to be considered real, it must have been observed in multiple detectors.

\begin{figure}
    \centering
    \includegraphics[width=0.8\linewidth]{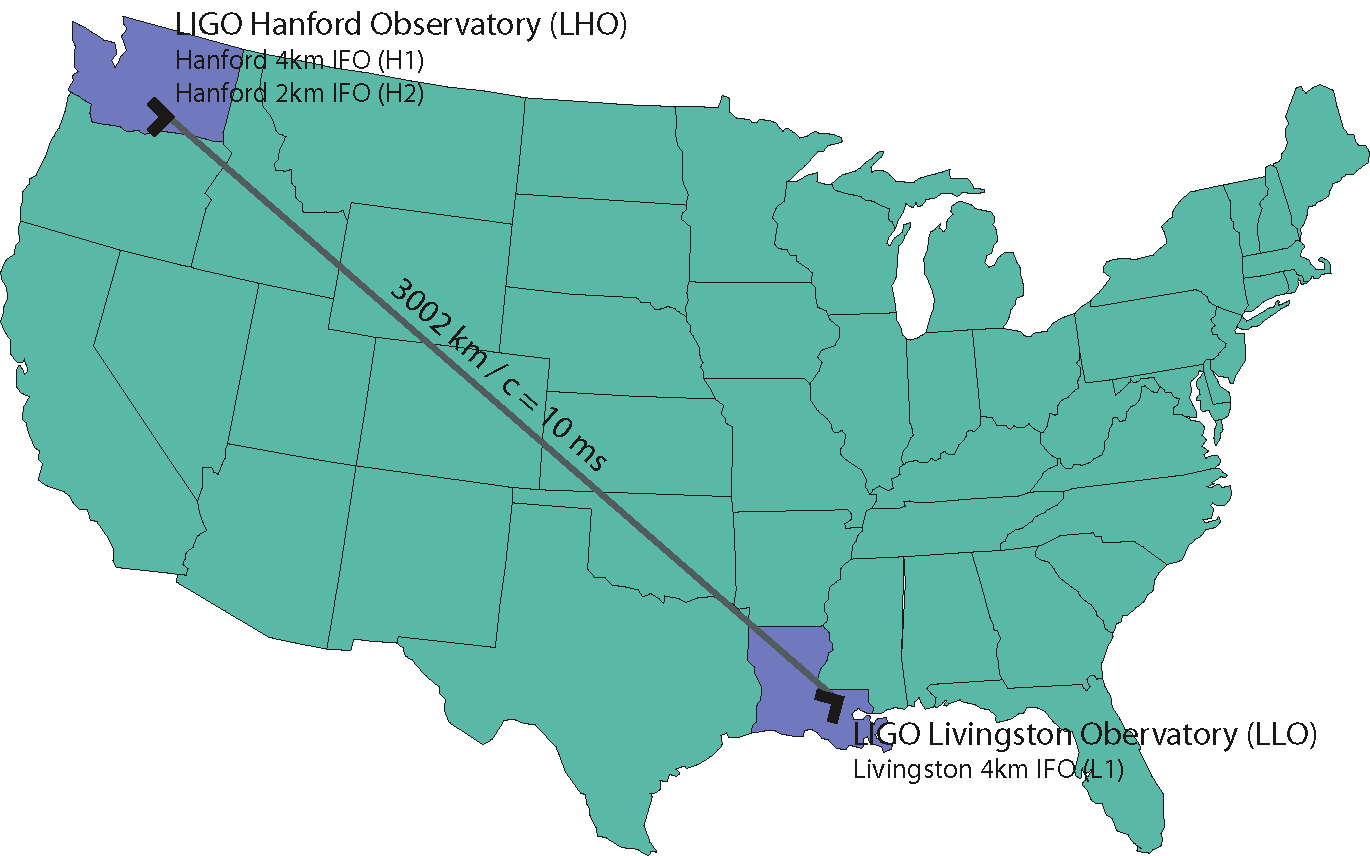}
    \caption{The two LIGO detectors locations and orientations, separated by approximately $3$,$000 \, \text{km}$. Taken from~\cite{Brown_Thesis:2004}.}
    \label{2:fig:observatories}
\end{figure}

In a two detector example, for example LIGO-Hanford and LIGO-Livingston, the signal seen by both detectors will not be exactly the same. The detectors are located approximately $3$,$000 \, \text{km}$ from one another, (Figure~\ref{2:fig:observatories}) and we know \gws propagate at the speed of light (Section~\ref{1:sec:keplerian_derivation}).
Therefore, we can expect a maximum delay between single-detector detections of $10 \, \text{ms}$. For the coincidence test, we must allow a window between single detector triggers of this light travel time. An additional coincidence requirement in PyCBC is the intrinsic parameters of the two single detector triggers being the same, i.e. they have seen the same gravitational waveform.

\subsection{\label{2:sec:ranking-statistic}Ranking statistic}

Finally, we can combine all these components to create a ranking statistic to use as the detection statistic for identifying \gwadj signals from \gwadj detector data. The likelihood ratio is calculated using the signal and noise event rate densities, defined as~\cite{PyCBC_global:2020}
\begin{equation}
    r_{s}(\kappa) = \frac{\text{Number of signal events}}{\text{Volume} \times \text{Time}},
\end{equation}
\begin{equation}
    r_{n}(\kappa) = \frac{\text{Number of noise events}}{\text{Volume} \times \text{Time}}.
\end{equation}
We compare the number of signal and noise events over the total Volume-Time (VT) observed, the number of signal events should increase with an increase in both volume and time, but the number of noise events will not change with an increase in volume, only time. We use the volume in the noise event rate density calculation to allow us to simply compare the two rates as the VT cancels in likelihood ratio
\begin{equation}
    \Lambda(\vec{\kappa}) = \frac{r_{s}(\vec{\kappa})}{r_{n}(\vec{\kappa})},
\end{equation}
where $\vec{\kappa}$ is a set of parameters
\begin{equation}
    \vec{\kappa} = \left\{ \left[\rho_{d}, \chi^{2}_{d}, \chi^{2}_{d, sg}, \sigma_{d}\right], \vec{\theta}, \left[\mathfrak{A}_{d_{1}d_{2}}, \delta t_{d_{1}d_{2}}, \delta\phi_{d_{1}d_{2}}\right] \right\}.
\end{equation}
$\vec{\kappa}$ contains three categories of parameter: first, the single detector trigger parameters for each detector $d$, SNR, Allen-$\chi^{2}$, sine-Gaussian $\chi^{2}$ and template sensitivity ($\sigma_{d}$); next, the intrinsic template parameters $\vec{\theta}$ and; the coincident trigger parameters, amplitude ratio ($\mathfrak{A}_{ab}$), time of arrival difference ($\delta t_{ab}$) and phase difference ($\delta \phi_{ab}$) between two detectors $d_{1}$ and $d_{2}$.

Again, it is natural to use the log-likelihood ratio due to the order of magnitude difference between the event rate densities
\begin{equation}
    R(\vec{\kappa}) = \log \Lambda(\vec{\kappa}) = \log r_{s}(\vec{\kappa}) - \log r_{n}(\vec{\kappa}).
\end{equation}
We call the log-likelihood detection statistic the ranking statistic, which is constructed from a noise model and signal model. The PyCBC noise and signal models have changed over the course of the \gwadj search history. We will describe the models used in~\cite{PyCBC:2016}, the first iteration of the \texttt{PyCBC} search for \cbcs.

\subsubsection{\label{2:sec:pycbc-2016}PyCBC}

The \texttt{PyCBC} search pipeline ranked single detector triggers by their SNR and $\chi^{2}$ values to create a re-weighted SNR, $\hat{\rho}_{d}$, for each detector. The \texttt{PyCBC} search pipeline ranking statistic noise event rate model provides information on how likely an event is to be caused by noise. The noise event rate model used by the \texttt{PyCBC} search pipeline was a simple quadrature sum of single detector re-weighted SNRs,
\begin{keyeqn}
\begin{equation}
    \hat{\rho}^{2} = \sqrt{\hat{\rho}^{2}_{d_{1}} + \hat{\rho}^{2}_{d_{2}}},
    \label{2:eq:PyCBC_noise_model}
\end{equation}
\end{keyeqn}
where a large $\hat{\rho}^{2}$ indicates a more significant event, less likely to be caused by noise.

The \texttt{PyCBC} search pipeline had no signal event rate model, and so the ranking statistic was simply the noise event rate model presented above. As detailed in~\cite{PyCBC:2016}, this ranking statistic was shown to `downrank' all triggers below a re-weighted SNR of $6$ in both a Gaussian noise simulation and the $6^{th}$ science run of LIGO~\cite{rw_snr_eq:2012}. We note that in this case there are $0$ real signals in the data and therefore this noise model has been applied to only noise triggers.

\subsection{\label{2:sec:background-estimation}Candidate event significance}

We can define some threshold at which a coincident trigger with ranking statistic value above this threshold can be preliminary considered to be a \gwadj event, we refer to these coincident triggers as \textit{candidates}. We define the \textit{significance} of the candidate as the \textit{false-alarm rate}. False alarms are coincident triggers that have been cause entirely by noise, with no astrophysical origin. The rate of false alarms depends on the search pipeline's response to detector noise and must be measured empirically.

\begin{figure}
    \centering
    \includegraphics[width=0.9\linewidth]{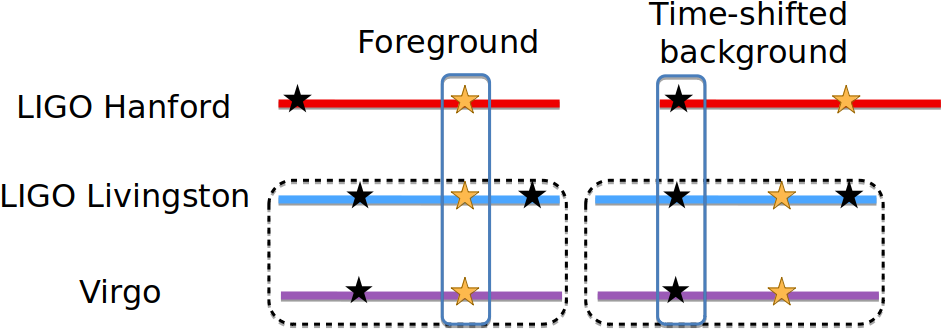}
    \caption{The time-sliding process carried out by the PyCBC search to generate a background distribution of candidate events. The foreground, real \gwadj event, is shown as a golden star and is the coincidence between three detectors with a single detector trigger in each which falls within the allowed light travel time of the detector network. The dashed black box indicates a time slide being made where two detectors have their data shifted by an interval greater than the allowed light travel time window to produce events which can only be caused by non-astrophysical noise sources. An example of a background event is shown as the three black stars in the blue box on the right image. Taken from~\cite{PyCBC_global:2020}.}
    \label{2:fig:timeslides}
\end{figure}

The PyCBC search pipeline measures the false-alarm rate using `time slides'. As discussed in Section~\ref{2:sec:coincidence-test}, a coincident trigger can only be formed if the triggers are within the light travel time window. Therefore, if a coincidence is made between two single detector triggers outside this window it \textbf{must} be caused by detector noise and not an astrophysical signal. To measure the background rate of coincident triggers, we can take the triggers from the first detector and count all the coincidences made with the triggers from the second detector after shifting the time in the second detector by greater than the light travel time window. This ensures that any coincidences made cannot possibly have been due to a real signal because the coincidences will have occurred outside the light travel time window. An example of a background coincidence by these time slides for the three detector case can be seen in Figure~\ref{2:fig:timeslides}.

The time sliding procedure is repeated many times to produce a large sample of false-alarm coincidences, which are used to compute the false-alarm rate as a function of the ranking statistic.

To measure the significance of each candidate event, we assign a p-value. In our pipeline, the p-value of a candidate event is the probability that there are one or more false-alarm events that have a ranking statistic value greater than or equal to the ranking statistic of the detector value, $p_{b} = P(R_{FA} \ge R_{CE})$. The p-values of candidate events are calculated under the null hypothesis that \textit{all} triggers are seen due to noise. To confidently claim the candidate event as real, we must demonstrate that the null hypothesis given the candidate event ranking statistic value is highly improbable (the p-value is small).

We can do this by measuring the number of noise background events, $n_{b}$, that have a ranking statistic value higher than the candidate event. If we do this for all ranking statistic values we can build up a mapping of ranking statistic to false-alarm rate. The function $n_{b}(R)$ gives the number of background events with a ranking statistic value higher than $R$, the ranking statistic value. The probability that one or more background events are found above $R$ given the observing time $T$ and the background time $T_{b}$ is~\cite{PyCBC:2016}
\begin{equation}
    p(\ge 1 \, \text{above} R|T, T_{b})_{0} = 1 - \exp \left[\frac{-T(1 + n_{b}(R))}{T_{b}}\right].
\end{equation}
The background time will equal $T_{b} = T^{2}/\delta$ where $\delta$ is the time-slide interval. We can produce a very large amount of background data from a relatively small period of observing data, fifteen days of coincident data with a time-slide interval of $0.1$ seconds allows us to measure false-alarm rates of $1$ in $200$,$000$ years.

Finally, we can express the mapping of false-alarm rate to ranking statistic as
\begin{equation}
    \text{FAR}(R^{*}) = \int r_{n}(\vec{\kappa}) \Theta(R(\vec{\kappa}) - R^{*}) d^{n}\vec{\kappa},
    \label{2:eq:far_mapping}
\end{equation}
where $\Theta$ is the Heaviside step function
\begin{equation}
    \Theta(x) =
    \begin{cases}
        0 & \text{if } x < 0 \\
        1 & \text{if } x \geq 0,
    \end{cases}
\end{equation}
and the false-alarm rate at the ranking statistic value $R^{*}$ is being calculated by integrating over all possible background events and summing up the events that have a ranking statistic greater than or equal to $R^{*}$, the ranking statistic threshold. The optimal ranking statistic will maximise the expected number of coincident events due to signals recovered above $R^{*}$.

\section{\label{2:sec:gw-pipelines}\Gwadj search pipelines}

We can develop the idea of a \gwadj search pipeline by describing the structure of the \texttt{PyCBC}~\cite{PyCBC:2016} search pipeline and how \gwadj events are identified from the initial \gwadj data using all the techniques described previously in this chapter.

\subsection{\label{2:sec:searching-for-gw-with-pycbc}\texttt{PyCBC}}
\begin{figure}
    \centering
    \includegraphics[width=1.0\linewidth]{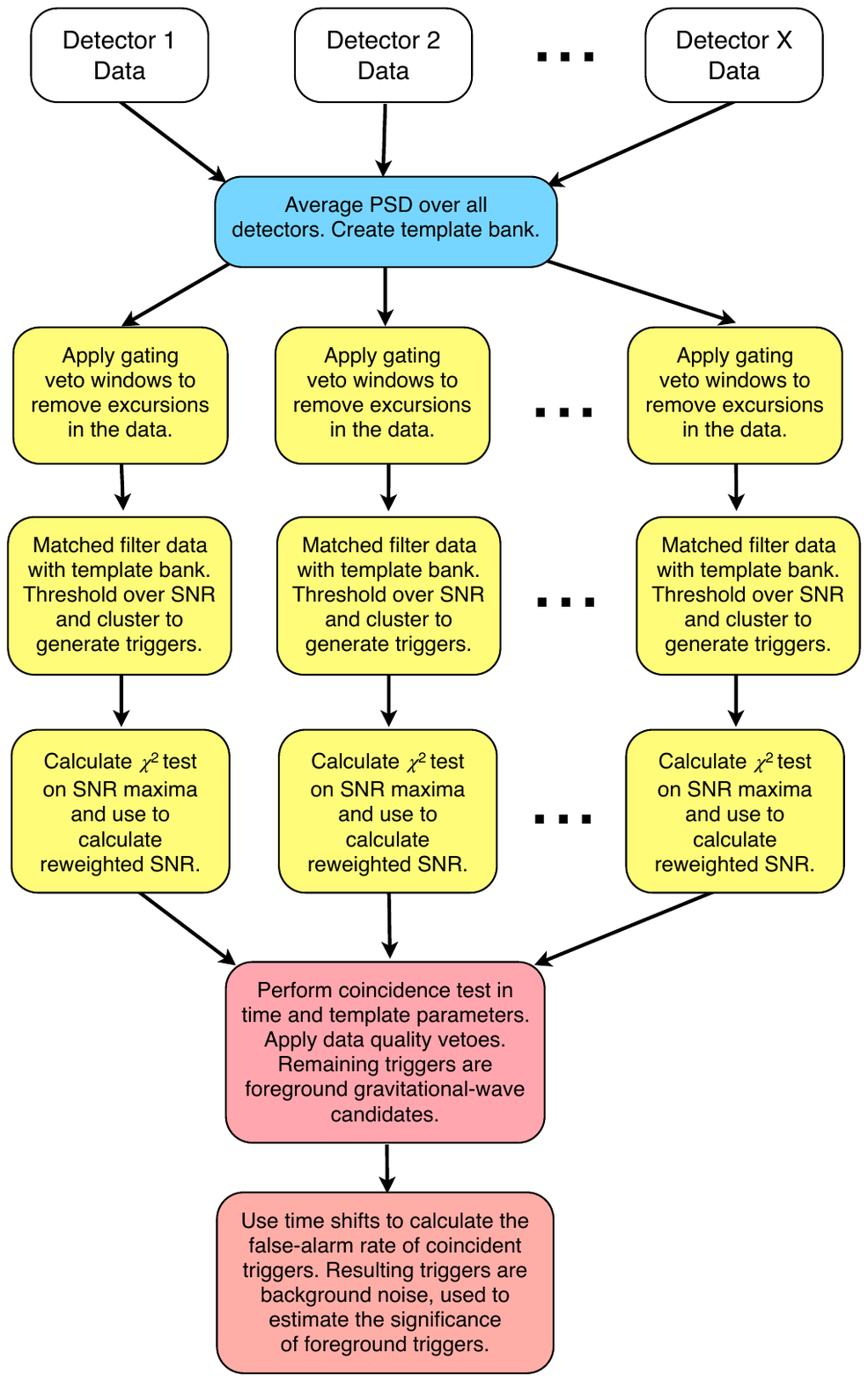}
    \caption{The structure of the \texttt{PyCBC} search pipeline for \gws. Taken from~\cite{PyCBC:2016}.}
    \label{2:fig:pycbc-flowchart}
\end{figure}

The flow chart describing the structure of the \texttt{PyCBC} pipeline has been taken from the \texttt{PyCBC} paper and can be seen in Figure~\ref{2:fig:pycbc-flowchart}. We follow this flowchart in our description of the search pipeline.

\subsubsection{Template bank generation}

A key consideration for \gwadj search pipelines is computational cost. A greater computational cost requires more time to analyse \gwadj data, alongside greater monetary cost and carbon emission cost. The dominant cost for modelled searches is matched filtering. Therefore, the number of templates in the template bank is proportional to the computational cost of the search. The number of templates is tuned to balance a minimal loss in matched filter SNR with the computational cost.

The template bank in the PyCBC search is shared between all detectors to allow a coincidence requirement of triggers sharing the same template between detectors. The template bank is generated given a power spectral density (PSD) and therefore we must create a PSD that is averaged over time for all detectors. The PSD is measured every $2048 \, \text{seconds}$ independently in each detector using Welch's method and taking the median average, this gives $N$ power spectra $S_{n}$ for each detector. The harmonic mean PSD for a single detector across all analysis time is then found
\begin{keyeqn}
\begin{equation}
    S_{n}^{\text{harmonic}}(f_{k}) = N / \sum^{N}_{i = 1} \frac{1}{S^{i}_{n}(f_{k})},
\end{equation}
\end{keyeqn}
and using the same method we average across all detector harmonic mean PSDs to get an estimated PSD across all time and detectors to use for template bank generation. This PSD estimate only need to be re-generated when the detector's noise PSD has changed significantly, which will typically only happen when the detectors are physically changed or upgraded.

The \texttt{PyCBC} template bank is placed with a minimum match of $0.97$ between any signal and the template bank. The loss in sensitivity due to this minimum match limit is equal to the minimum-match cubed and is ${\simeq}10\%$. The template bank covers a four dimensional parameter space of two component masses and aligned spins. The PyCBC template bank was generated using a combined geometric-based aligned-spin algorithm~\cite{Harry_Lundgren:2012, Harry_precession:2013} and a stochastic algorithm~\cite{Ajith:2012, Privitera:2013}, described in~\cite{pycbc_template_bank:2016}.

\begin{figure}
    \centering
    \includegraphics[width=0.75\linewidth]{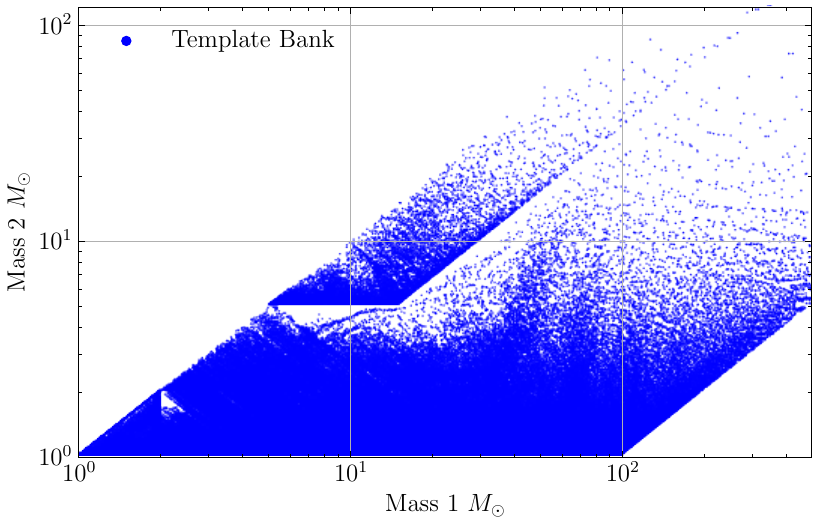}
    \caption{An example of a template bank used by the \texttt{PyCBC} search pipeline that has been generated from a single block of data and has been created in the four dimensional component mass and aligned spin parameter space. It consists of $428$,$725$ templates.}
    \label{2:fig:pycbc-o3-template-bank}
\end{figure}

A third observing run template bank can be seen plotted in the $m_{1}\text{--}m_{2}$ parameter space in Figure~\ref{2:fig:pycbc-o3-template-bank} which contains $428$,$725$ unique templates.

\subsubsection{Matched filtering and clustering}

The matched filter between template bank and data is performed on each block of data to create a list of single detector triggers. Triggers are only stored if their matched filter SNR is greater than $5.5$. Suppose we have a trigger with an SNR of $10.0$, according to our template bank minimum match we should have at least one additional trigger (and indeed have many) above the SNR threshold for a nearby template. To prevent the recording of multiple triggers from different templates for the same candidate event, we use a \textit{time clustering} algorithm which selects and keeps only the highest SNR trigger in a predefined time window. Another clustering algorithm, is employed to trigger across the template bank to prevent the triggering of a loud glitch in one region of the template bank from subduing a real signal trigger in a completely different region.

\subsubsection{Signal consistency test}

\begin{figure}
    \centering
    \includegraphics[width=0.75\linewidth]{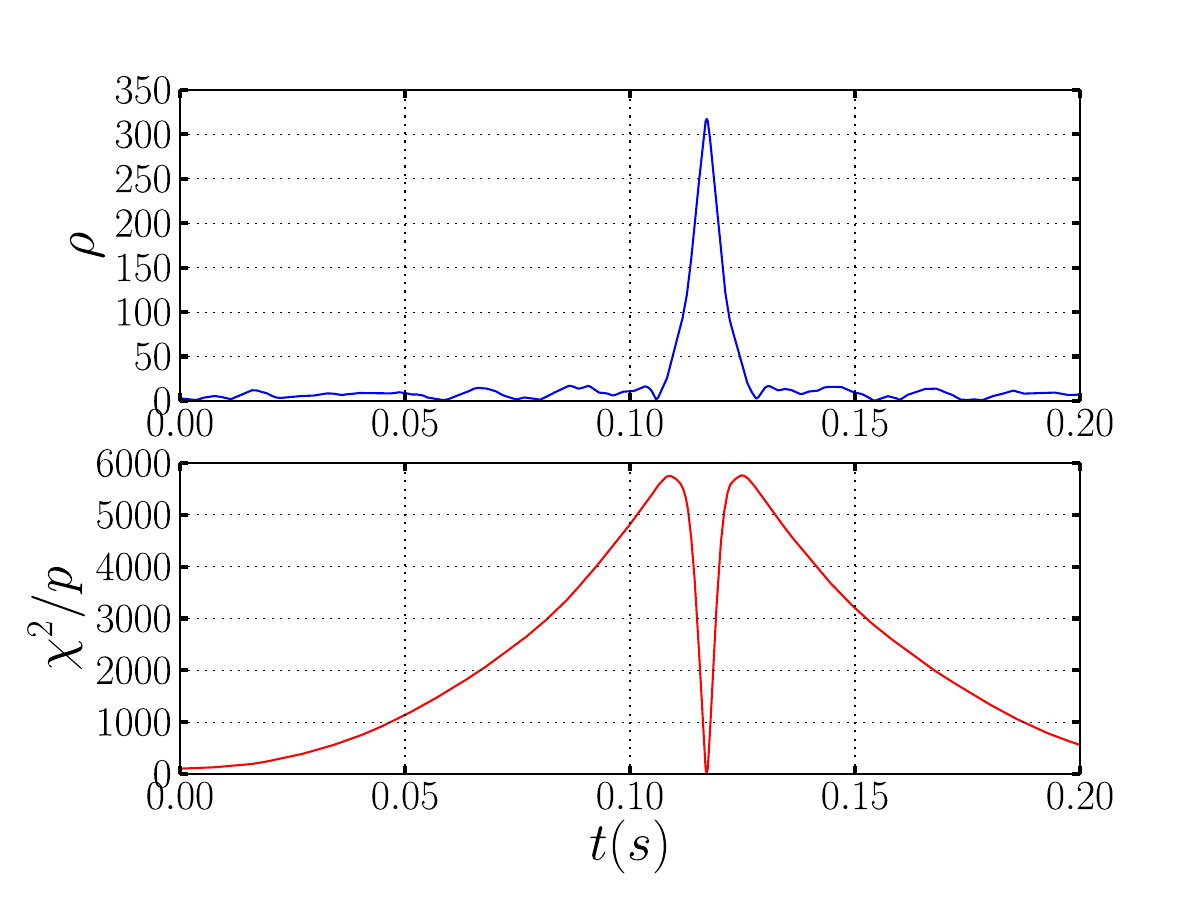}
    \caption{An SNR time series and $\chi^{2}$ time series for a simulated compact binary coalescence injected into S5~\cite{S5:2012} data with an SNR of $300$. Taken from~\cite{IHOPE:2012zx}.}
    \label{2:fig:snr-timeseries}
\end{figure}

The triggers then have the $\chi^{2}$ test~\cite{Allen_Chi:2005} applied to ensure consistency in the signal power distribution. The $\chi^{2}$ test values are calculated for each SNR trigger found by the template bank (Equation~\ref{2:eq:reduced_chisq}) and used to re-weight SNR values to obtain new SNR values for each trigger. An example of the SNR and $\chi^{2}$ time series from a loud signal can be seen in Figure~\ref{2:fig:snr-timeseries}.

\subsubsection{Coincidence tests}

Single detector triggers are combined and evaluated based on their trigger times and template parameters, triggers that have been found within the maximum light travel time between the two detectors ($10 \, \text{ms}$ for LIGO-Hanford and LIGO-Livingston) with an exactly matching template can be considered to originate from the same \gwadj signal~\cite{Robinson:2008}. The result of this process is a list of coincident triggers found with SNR above the threshold in two or more detectors, with exact template parameters across detectors.

\subsubsection{Generate background, rank triggers, establish significance and estimate search sensitivity}

The ranking statistic and false-alarm rates are calculated for each of these triggers as described in the previous section, using a quadrature sum of the reweighted SNR in each detector. The sensitivity of the search pipeline can be evaluated by injecting many simulated \gwadj signals into the data and recovering them with the search pipeline. The most realistic injections are hardware injections~\cite{Brown:2003, Biwer:2016} in which an actuator exerts a force on the end test mass to simulate the response of a real \gwadj signal. These injections are rarely performed because the data is now contaminated and cannot be used to search for real \gwadj signals. The injections used for large scale injection campaigns are software injections, in which the \gwadj signal is added to the data. These injections are added to add detectors with the correct time, phase, and amplitude differences and are placed so as to probe the full population of potential \gwadj events. The injection campaigns are a very powerful tool that can reveal search pipeline sensitivity, and we have used them extensively throughout this thesis to test the sensitivity increase of changes made to search pipelines.

\chapter[Detector Characterisation]{Detector Characterisation}
\label{chapter:3-detchar}
\chapterquote{Working on the weekend like usual.}{Drake}

In this thesis, we have made improvements to the search for \gws by improving the quality of the detector data and understanding the noise background of the detectors.

This chapter is laid out as follows: we repeat the state of the current \gwadj detector network in Section~\ref{3:sec:gw_detectors}, in Section~\ref{3:sec:detchar_calib} we define the research area of detector characterisation, the systematic sources of noise faced in the detectors are described in Section~\ref{3:sec:detector-analysis}, some common detector data noise transients in Section~\ref{3:sec:noise-transients} and finally, the detector characterisation tools most commonly used and referred to throughout this thesis in Section~\ref{3:sec:detchar-tools}.

\section{\label{3:sec:gw_detectors}\Gwadj detectors}

\Gwadj astronomy in the present day is performed by ground-based detectors at different locations globally: the LIGO-Livingston and LIGO-Hanford detectors are located in the United States of America, the Virgo interferometer in Italy, and the KAGRA telescope in Japan. We will primarily reference the LIGO detectors in the United States of America when discussing the challenges faced in detector characterisation.

As described in Chapter~\ref{chapter:1-gravitational-waves}, the LIGO detectors work on the principle of laser interferometry to detect \gwadj signals. The sensitivity~\cite{aLIGO_design_curve:2018} of \gwadj detectors is limited by fundamental sources of noise~\cite{PSD_var:2020} such as seismic noise~\cite{Glanzer:2023}, thermal noise~\cite{thermal_noise:2018} and quantum noise~\cite{quantum_noise:2003}. Additionally, we have non-Gaussian noise~\cite{Noise_Guide:2020}, which manifests as transient noise bursts in the data, which can contribute to false positives in \gwadj searches and the obscuring of real \gwadj signals~\cite{GW170817:2017, GW150914_noise:2016}.

\subsection{\label{3:sec:detector-analysis}Detector analysis}


Systematic sources of noise are the dominating limitation in the sensitivity of ground-based detectors. The sources of noise with the greatest contributions are
\begin{itemize}
    \item Seismic noise~\cite{Glanzer:2023}: the vibrations of the ground causes a coupled vibration motion in the mirrors of the interferometer. Higher frequency seismic noise ([1, 10]\,\text{Hz}) is often caused by local ground movements typically anthropogenic in nature, or greater magnitude earthquakes~\cite{Nuttall:2018}. Lower frequency seismic noise (<1\,\text{Hz}) is caused by larger-scale geophysical processes such as ocean waves, tidal effects, or smaller local earthquakes~\cite{aLIGO:2015}. Seismic noise can be mitigated with advanced suspension systems to isolate mirrors and end-station components from the ground to prevent coupling to seismic motion~\cite{seismic_isolation:2015}.
    \item Thermal noise~\cite{thermal_noise:2018}: the thermal energy of interferometer components will cause them to vibrate. This noise can be reduced with new mirror materials with resonant frequencies outside the sensitive regions or with cryogenically cooled mirrors such as those used by KAGRA~\cite{KAGRA:2021}.
    \item Shot noise~\cite{quantum_noise:2003}: the distribution in the number of photons observed by the interferometer photodetector in any time interval is Poisson distributed, and the error in the photon count places limits on the sensitivity. The error is proportional to the square-root of the count ($\sigma \propto \sqrt{\lambda}$) therefore, the simplest way to reduce this error is to increase the laser power and increase the photon count in the time interval.
    \item Radiation pressure~\cite{quantum_noise:2003}: the pressure exerted by the photons hitting the end mirrors will decrease search sensitivity. To reduce this noise, the laser power can be reduced; however, this will increase the shot noise.
\end{itemize}

\begin{figure}
    \centering
    \includegraphics[width=0.75\linewidth]{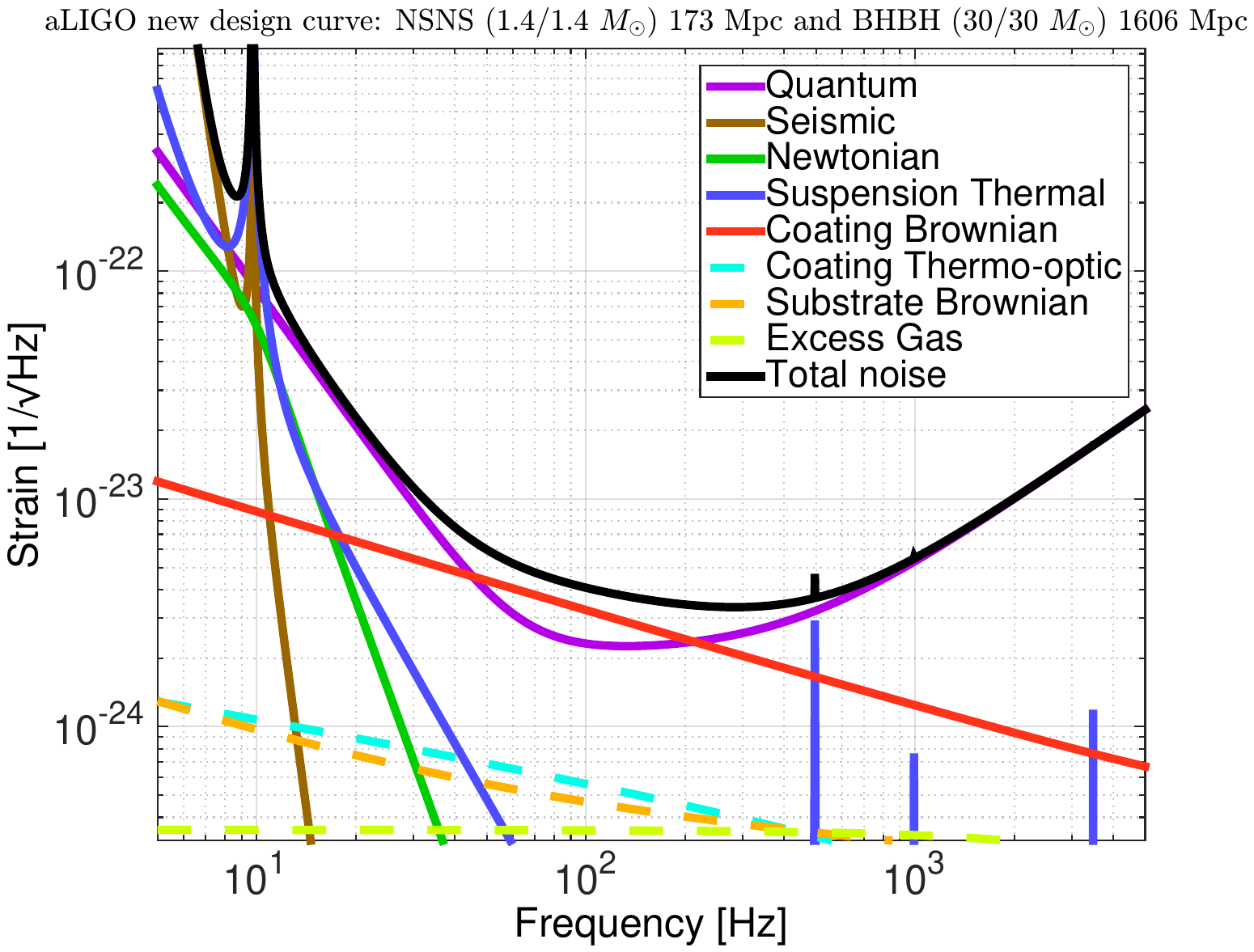}
    \caption{The Advanced LIGO~\cite{aLIGO:2015} strain sensitivity as a function of frequency (black solid line), accompanied by the systematic noise sources which limit the sensitivity of the detector. Taken from~\cite{aLIGO_design_curve:2018}.}
    \label{3:fig:aLIGO_noise}
\end{figure}

It can be seen that the laser power must be calibrated to balance the impact of increasing the photon count, which will decrease shot noise but simultaneously increase radiation pressure noise. A figure displaying these systematic noise sources and their limitations in the LIGO detector case can be seen in Figure~\ref{3:fig:aLIGO_noise}.
%

\begin{figure}
    \centering
    \includegraphics[width=1\linewidth]{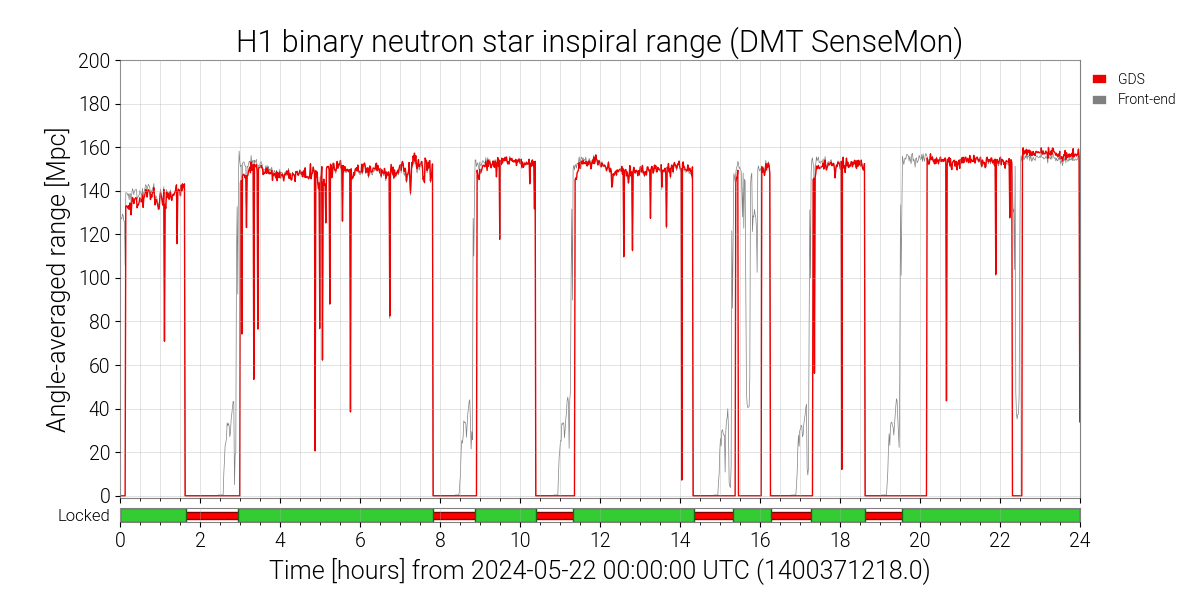}
    \caption{The LIGO-Hanford binary neutron star inspiral range for the 22nd May 2024, created using GWSumm~\cite{gwsumm:2024} and taken from the LIGO Summary Pages-\href{https://summary.ligo.org/}{https://summary.ligo.org/}, please see \href{https://gwosc.org/detector_status/}{https://gwosc.org/detector\_status/} for the public summary pages.}
    \label{3:fig:bns_range}
\end{figure}

To monitor detector sensitivity, the binary neutron star inspiral range is calculated. This is the range at which a binary neutron star signal, with both components having a mass of $1.4 \, \text{M}_{\odot}$, will be detected given the characteristic noise (PSD) of the data. The PSD is calculated by taking an average of the detector noise in the previous minute~\cite{range_calculation:2003, ota:2023}. Figure~\ref{3:fig:bns_range} shows an example of the binary neutron star range for a day of LIGO-Hanford during the fourth observing run.

\subsection{\label{3:sec:noise-transients}Noise transients}


Noise transients, commonly referred to as glitches, are short duration bursts of noise found in \gwadj data. The systematic sources of noise described in Section~\ref{3:sec:detector-analysis} limit the sensitivity of the detector to a certain frequency range, glitches appear within this sensitivity frequency range. The specific noise transients investigated by detector characterisation are non-Gaussian noise artefacts that have the ability to obscure~\cite{GW170817:2017} or mimic \gwadj signals~\cite{GWMimicking:2010}, producing false-alarms in our \gwadj search pipelines. Understanding these glitches is crucial for \gwadj detection; they have the potential to reduce the sensitivity and hinder the reliability of the detectors.
%
%


There are at least $27$~\cite{gravityspy:2023, gravityspy:2024} different classes of glitch which all manifest with different durations (typically in the millisecond to second range), amplitudes, and glitch morphology. Glitches populate the whole sensitive frequency range of the detectors, with some glitches being broadband, affecting up to the whole frequency bandwidth, to others being narrowband and affecting only specific frequency ranges. Glitches are commonly characterised and studied in 2-dimensional time-frequency representations of the one-dimensional strain time series that the detector outputs. The typical time-frequency representation used is the OmegaScan~\cite{qscan:2004} (referred to as `Omega scan' or `Q-scan') which is described later in this chapter.


Glitches originate from either environmental sources, instrumental sources, or a coupling of the two. As mentioned previously, seismic noise limits the sensitivity of the detectors below ${\sim}10 \, \text{Hz}$, but the coupling of seismic motion into interferometer components can cause one of the more common glitches---\scl. Other environmental noise sources are heavy winds, lightning, and human activity (traffic, construction, trains). A historical glitch identified at LIGO-Livingston was caused by an air conditioning compressor cycling, these glitches were seen by both a magnetometer and the \gwadj strain channel. Other instrumental glitches might be caused by mirror suspensions, electronics, control systems or laser fluctuations.

\begin{figure}
    \centering
    \includegraphics[width=1.0\linewidth]{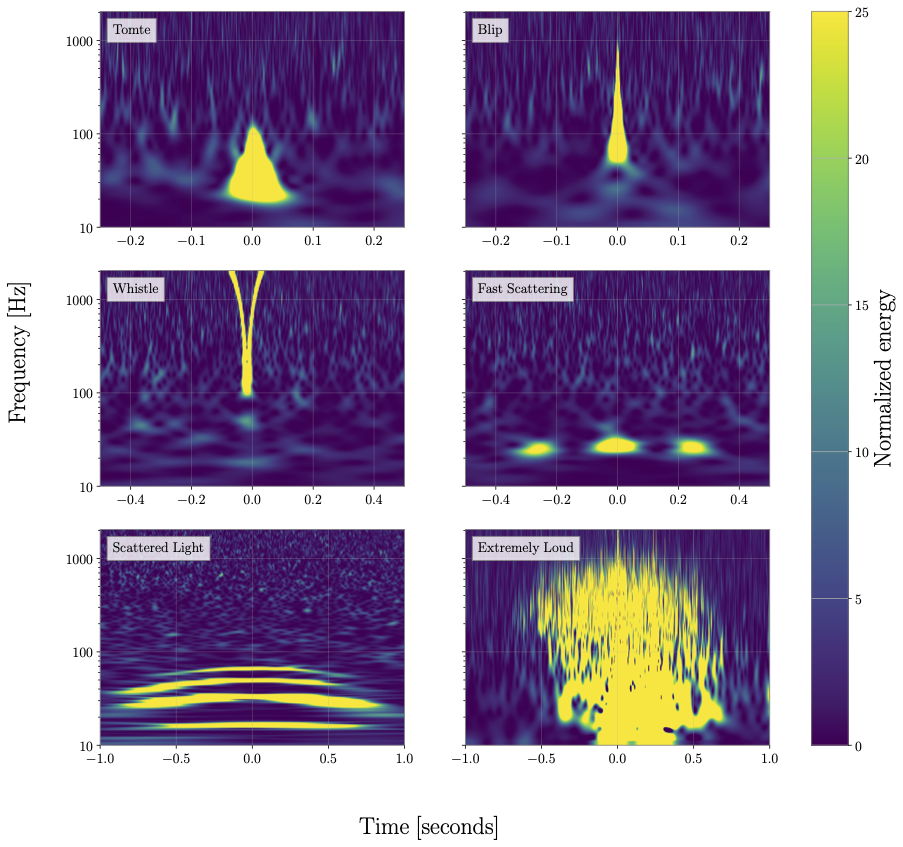}
    \caption{Six classes of glitch commonly found in LIGO detector data. Taken from~\cite{GlitchPlot:2024, gravityspy:2023}.}
    \label{3:fig:glitches_subset}
\end{figure}

The most common glitches are: blips~\cite{blips:2019}, \scl~\cite{ArchEnemy:2023} and whistles~\cite{glitschen:2021}, Omega scans of these glitches can be seen in Figure~\ref{3:fig:glitches_subset}. The sources of these glitches have been studied. For example, whistles are caused by the beating of radio frequencies in the detector; however, some glitches come from unknown sources, such as blips. \Scl has been found from many sources in the detector and great effort has been made to reduce the presence of \scl in the data~\cite{reducing_scattering:2020} but as of the fourth observing run some \scl still remains.
%


A number of algorithms and tools have been developed to detect and characterise noise transients in \gwadj data~\cite{ArchEnemy:2023, reducing_scattering:2020, Glanzer:2023, gravityspy:2017, gravityspy:2021, gravityspy:2023, glitschen:2021,  BayesWave:2015, gwadaptive:2022, O3_subtraction:2022, Powell:2016, glitschen:2021}. The primary aim of these tools is the identification of glitches and analysis of properties of the glitches (amplitude, frequency, duration, waveform shape) to determine the source of the noise. Once the noise source has been identified, improvements to the detector can be made to eliminate it. Other tools focus on modelling noise transients for the purpose of subtracting them from affected \gwadj data~\cite{ArchEnemy:2023, BayesWave:2015, glitschen:2021, antiglitch:2023}.
%
%


Another tool used in the identification and classification of glitches is the thousands of auxiliary channels alongside the main strain channel~\cite{iDQ:2020}, which monitor the many subsystems of the detector and can be used to find noise correlations between the different parts of the instrument~\cite{DQ_vetoes:2017}. The auxiliary channels which are not sensitive to \gws are very useful in identifying any correlations between noise in the gravitational strain channel and noise also seen in another channel which cannot observe \gws, an example of this is the magnetometer recording the same glitch as seen in the \gwadj strain channel when the air conditioning compressor was cycling at LIGO-Livingston~\cite{Nuttall:2018}.

\section{\label{3:sec:detchar_calib}Detector characterisation and calibration}

Detector characterisation is a field of study which assess and improves the performance of \gwadj detectors to ensure accurate and reliable detection of \gws. \Gwadj detector calibration determines how the response of the detector translates to astrophysical signals, ensuring accuracy in extracted astrophysical information. Detector characterisation is responsible for identifying and mitigating the many sources of noise that can obscure or mimic real signals. The optimisation of detector sensitivity is also studied, whereby different designs and components can be fine-tuned to increase sensitivity. Another responsibility of detector characterisation is the monitoring of detector performance in real-time to assess sensitivity changes.

Overall, effective detector characterisation ensures that \gwadj observatories can operate at maximum potential and provide accurate and reliable observations of \gws. The \gwadj detector organisations have dedicated working groups for detector characterisation and calibration~\cite{O2O3_DetChar:2021, VirgoDetChar:2023}. In this thesis, chapter~\ref{chapter:4-archenemy} is a direct contribution to the field of detector characterisation with a contribution to the LIGO detector characterisation working group.

The detector characterisation group has the important task of verifying data quality when a \gwadj signal is detected by live detection pipelines, giving the confirmation that the \gwadj signal isn't a false alarm potentially caused by noise transients or poor data quality immediately surrounding the \gwadj signal.

\subsection{\label{3:sec:detchar-tools}Detector characterisation tools}

\paragraph{Omega scans}

are two-dimensional time-frequency representations of \gwadj data used very commonly to display \gwadj data in a human interpretable fashion, which contains more information than a simple spectrogram and can reveal time-frequency relationships that are virtually impossible for humans to see in the time series. Example Omega scans can be seen in Figure~\ref{3:fig:glitches_subset} for a number of glitches.

Detector characterisation uses Omega scans to highlight glitches which can have sharp and localised features in the time-frequency plane. Omega scans calculate the amplitude of strain power at each pixel using the \textit{Q-transform}~\cite{qscan:2004}, where for each point in time and frequency a number of wavelets are created with a varying \textit{quality factor} ($Q$). $Q$ is calculated using
\begin{equation}
    Q = \frac{f_{c}}{\sigma_{f}},
\end{equation}
where $f_{c}$ is the central frequency and $\sigma_{f}$ is the bandwidth of a wavelet. The bandwidth has a uncertainty relation with the duration, $\sigma_{t}$ of the wavelet
\begin{equation}
    \sigma_{t} \sigma_{f} \ge \frac{1}{4\pi},
\end{equation}
and so a wider frequency bandwidth will lead to a shorter duration for the wavelet, balanced by the Q-factor. The Q-transform of all these wavelets with the \gwadj strain is taken
\begin{equation}
    H(\tau, f, Q) = \int^{\infty}_{-\infty} h(t) w(t - \tau, f, Q) dt,
\end{equation}
and for each pixel the Omega scan algorithm will calculate the power of every wavelet, $|H(\tau, f, Q)|^{2}$, and select the one with the greatest amount of power. As an example, a short-duration glitch (like a blip) would be better tiled with multiple low-frequency bandwidth, high-Q tiles which will resolve the glitch in greater definition than a low-Q tile which captures the whole glitch in one tile.

\paragraph{GravitySpy}

is a citizen science machine learning tool for classifying glitches found in \gwadj data~\cite{gravityspy:2017}. GravitySpy has been trained by volunteers at the project website:~\href{https://www.zooniverse.org/projects/zooniverse/gravity-spy}{https://www.zooniverse.org/projects/zooniverse/gravity-spy}, hosted on the Zooniverse~\cite{zooniverse}.

GravitySpy has uploaded more than $1.4$ million of Omega scans of \gwadj data which have been found to contain bursts of power which could be caused by glitches. The Omega scans are published on the GravitySpy project and volunteers are given these images and the option for which glitch they think is contained within the image. Once these images have been classified, the machine learning algorithm is trained, and it can go further to classify the images without the need of volunteers~\cite{gravityspy:2021}.

GravitySpy has been run on data from the second and third observing runs, where it was able to classify 24 different glitch categories. The downsides of GravitySpy are the constantly changing noise background of the \gwadj detectors and the new glitch types emerging which need to be trained on with volunteering again~\cite{gravityspy:2023}.

\paragraph{Data Quality Vetoes} are a flag applied to periods of data which contain potential data quality issues~\cite{DQ_vetoes:2017}. These periods of data are identified using the ${\sim}200$,$000$ auxiliary channels~\cite{DQ_vetoes:2017}, an example is a significantly elevated transient noise rate in the strain channel five days prior to GW150914~\cite{GW150914:2016} which was traced back to the $45 \, \text{MHz}$ electro-optic modulator driver system used to generate optical cavity control feedback signals~\cite{aLIGO:2015}. The noise caused by this channel was given a category 1 veto data quality flag and removed 2.62\% of the total coincident time from the analysis period. While there is a significant reduction in the number of high SNR triggers that would've been found during these times, these flags have fallen out of favour in the fourth observing run due to removing large amounts of data.

\chapter[ArchEnemy]{ArchEnemy}
\label{chapter:4-archenemy}
\chapterquote{One less enemy within.}{Swain}
This chapter reproduces the published work carrying the name \textit{"ArchEnemy: Removing scattered-light glitches from gravitational wave data"}, in CQG~\cite{ArchEnemy:2023} with changes for formatting consistency and no changes to content. I am the first author of this publication and was responsible for the development of the scattered-light glitch model, $\chi^{2}$ test and glitch search pipeline and I then tested the changes using a PyCBC Offline search injection study to evaluate the glitch search and subtraction tool ArchEnemy. 
\section{\label{4:sec:ArchEnemy-intro}Introduction}

The Laser Interferometer Gravitational-Wave Observatory (LIGO)~\cite{aLIGO:2015} and Virgo~\cite{aVirgo:2015} collaborations made the first observation of \gws in September 2015~\cite{GW150914:2016}. The detection established the field of \gwadj astronomy and a global network of \gwadj detectors, now joined by KAGRA~\cite{KAGRA:2021}, has allowed for the detection of approximately 100 \gwadj events~\cite{gwtc1:2019, gwtc2:2021, gwtc21:2024, gwtc3:2023, 1OGC:2018, 2OGC:2020, 3OGC:2021}.

The detection of \gws is made possible by both the sensitivity of the detectors and the search pipelines~\cite{PyCBC:2016, GstLAL:2020, SPIIR:2020, MBTA:2021, cWB:2020} which analyse raw strain data from the output of the detectors and identify observed \gwadj signals. One of the problems that these search pipelines must deal with is the fact the data contains both non-stationary noise and short duration `glitches'~\cite{Noise_Guide:2020, O2O3_DetChar:2021, VirgoDetChar:2023} where noise power increases rapidly. Glitches are caused by instrument behaviour or interactions between the instrument and the environment~\cite{GW150914_noise:2016, Glanzer:2023} and glitches reduce the sensitivity of the detectors~\cite{O3_sensitivity:2020}, can potentially obscure candidate \gwadj events~\cite{gwtc2:2021} and can even mimic \gwadj events~\cite{GWMimicking:2010, PyCBC_singles:2022}.

\begin{figure}
  \includegraphics[width=\textwidth]{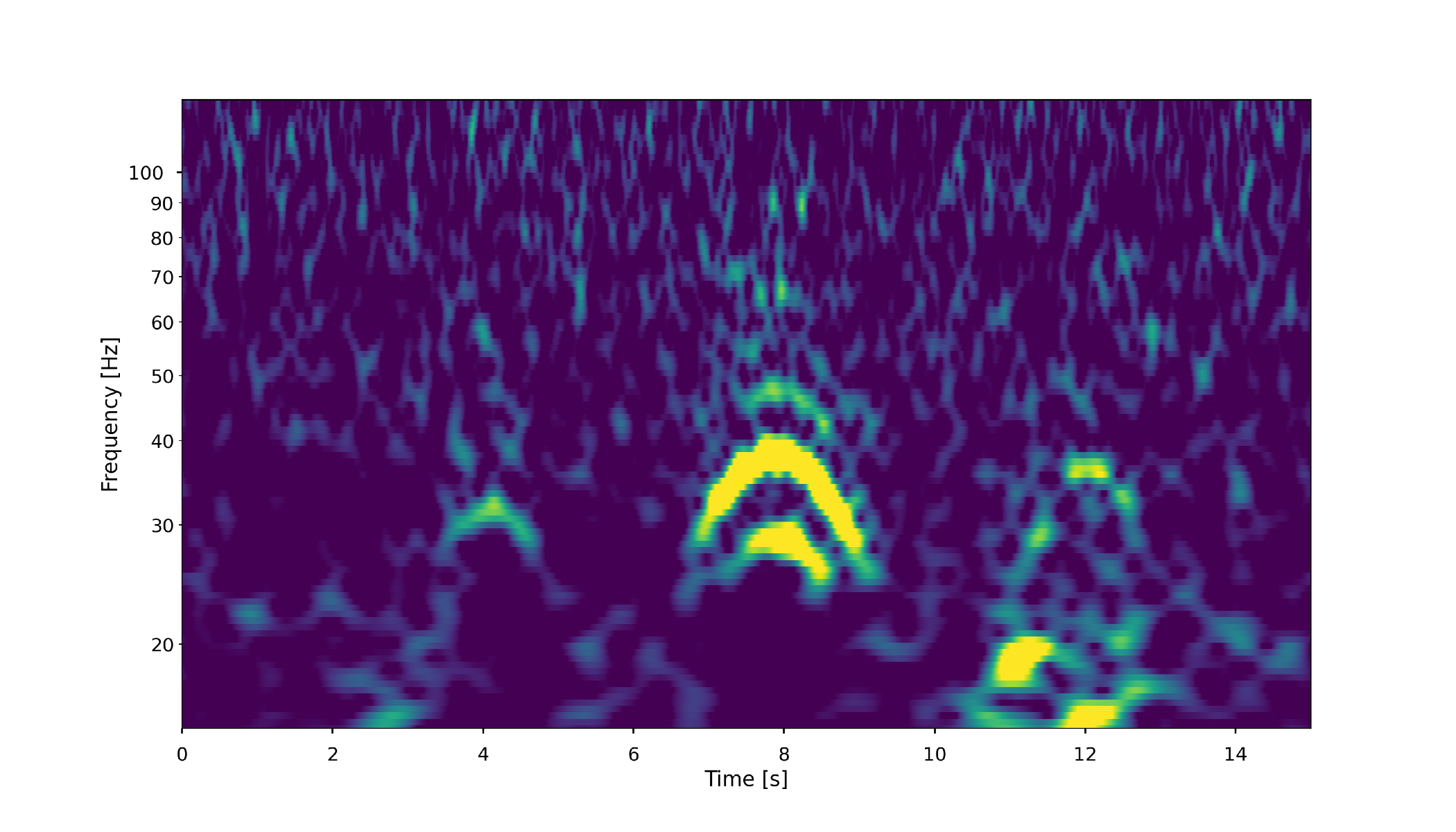}
  \caption{An Omega scan \cite{gwdetchar_tools:2021} of \gwadj data containing an example of a \scladj glitch. \Scladj glitches are characterised by a symmetric arch-like pattern. Multiple \scladj glitches can be seen at the 4, 8 \& 12 second marks as well as multiple harmonic glitches at 8 seconds.}
  \label{4:fig:scattered_light}
\end{figure}

Different classes of glitches have been characterised using tools such as Gravity Spy~\cite{gravityspy:2017, gravityspy:2023}. Of the 325,101 glitches classified by Gravity Spy in the third observing run of Advanced LIGO~\cite{gravityspy:2021} with a confidence of $90\%$ or higher, 120,733 ($32.1\%$) were classified as ``Scattered Light''. \Scladj glitches occur in the 10-120Hz frequency band~\cite{reducing_scattering:2020} which coincides with the frequency band where we observe the inspiral and merger signatures of compact binary coalescences (CBCs). \Scladj glitches are characterised by an arch-like pattern in a time-frequency spectrogram of the detector output, as seen in Figure~\ref{4:fig:scattered_light}.
\Scladj glitches occur when laser light in the interferometer is scattered from the main optical path by components within the detector. The motion of these components is coupled to seismic motion, inducing a phase shift on the light being scattered as the surface moves back and forth. This \scl then recombines with the main laser, producing \scladj glitches in the data. The surfaces from which \scladj glitches originate have been objects on optical benches such as lenses, mirrors and photo-detectors~\cite{TAccadia:2010}.

\begin{figure}
  \includegraphics[width=\textwidth]{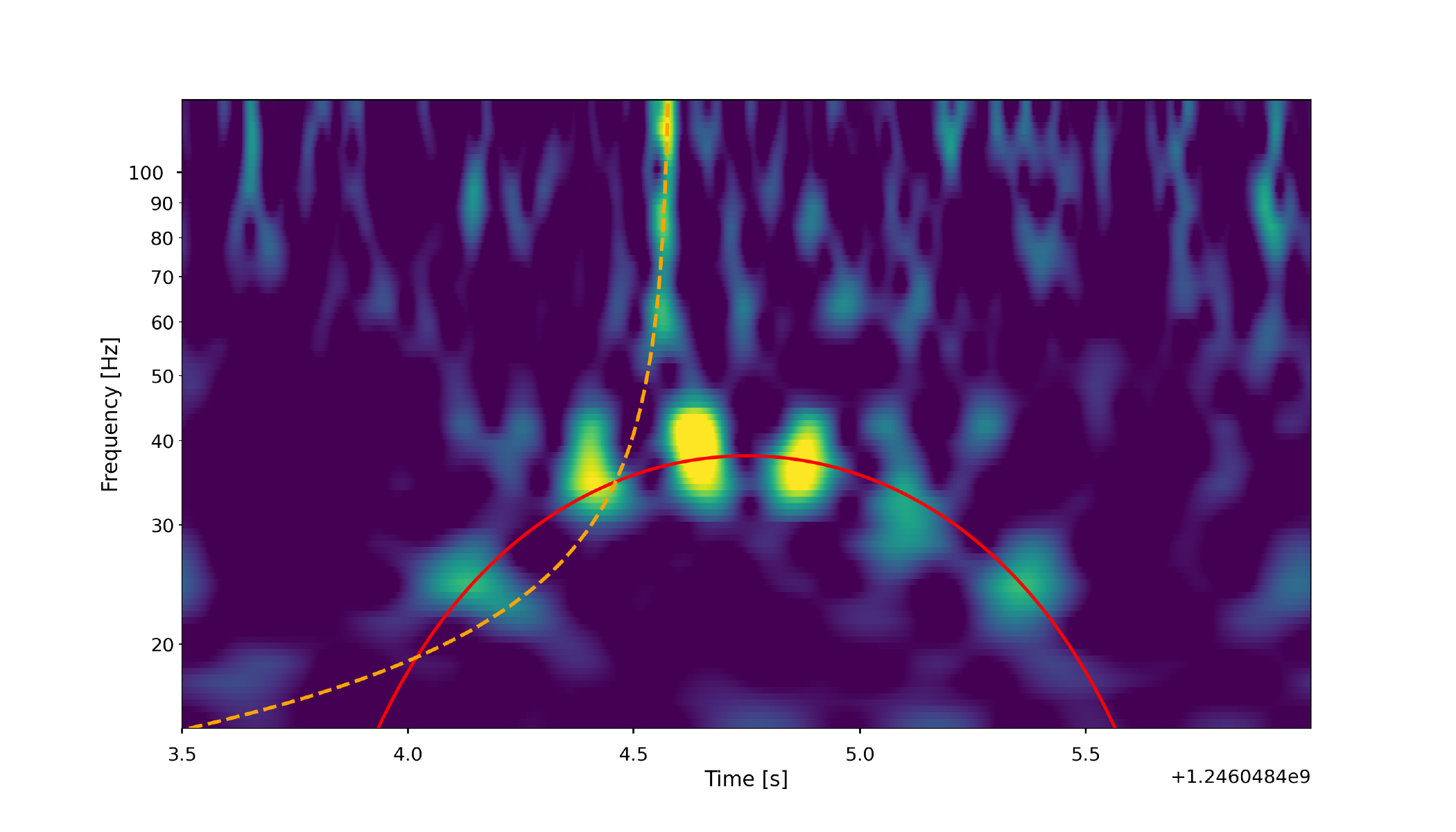}
  \caption{GW190701\_203306, a \gwadj event coincident with a \scladj glitch in the data from the LIGO Livingston observatory. The orange dashed track shows the inferred time-frequency evolution of a \gwadj event produced by a compact binary coalescence, the red solid line is an overlaid track of the approximate location of the coincident fast scattering glitches.}
  \label{4:fig:obscured_detection}
\end{figure}

\begin{figure}
  \includegraphics[width=\textwidth]{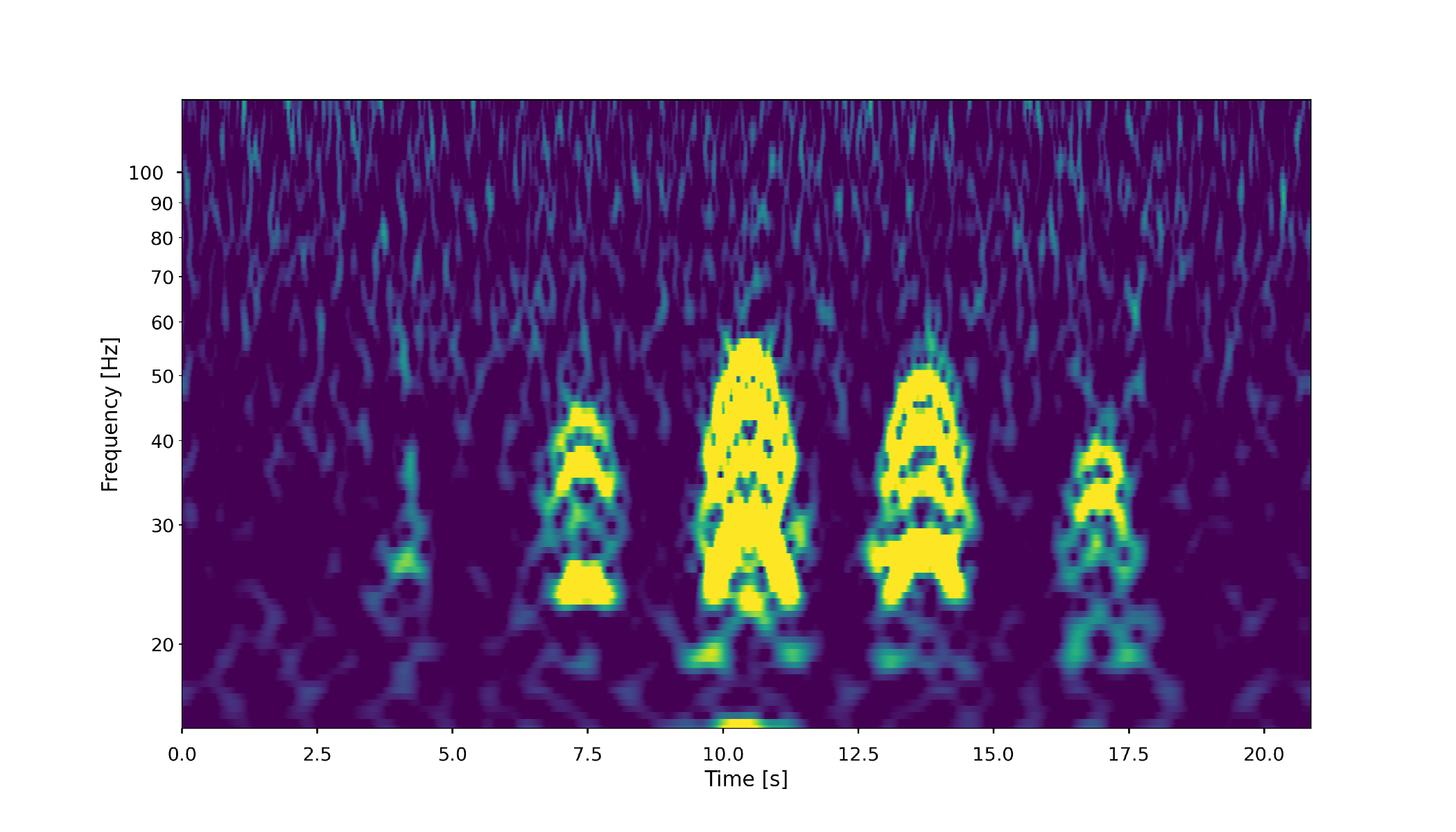}
  \caption{An Omega scan \cite{gwdetchar_tools:2021} of \gwadj data containing multiple examples of \scladj glitches. Here we see multiple \scladj glitches repeating periodically in a $20$ second period of time. Harmonic \scladj glitches are also seen in multiple stacks, the harmonic glitches share \emph{time period} values and the \emph{glitch frequency} values are $n-$multiples of the lowest frequency harmonic within the stack.}
  \label{4:fig:consec_scattered_light}
\end{figure}

\Scladj glitches have been a significant problem when observing CBCs. As an example, GW190701\_203306 was coincident with a \scladj glitch, as shown in figure \ref{4:fig:obscured_detection}~\cite{gwtc2:2021}, requiring subtraction from the data before the event could be properly characterised~\cite{O3_subtraction:2022}. A further 7 candidate events were found to be in coincidence with \scladj glitches in the third observing run~\cite{gwtc3:2023}. For this reason, it is important to reduce the effect of \scladj glitches in the detectors and \gwadj search pipelines. \Scladj glitches occur as single or multiple glitches and can appear rapidly in time and simultaneously in frequency (see figure \ref{4:fig:consec_scattered_light}), which we refer to as harmonic glitches.

The most obvious way to remove the impact of glitches is removing the mechanisms which produce the glitches in the observatories. This has been investigated in previous works~\cite{reducing_scattering:2020, TAccadia:2010, Nuttall:2018, gwadaptive:2022, HilbertHuang:2017, tvf-EMD:2020, Scattering_Monitoring:2022, Was_Subtract:2021}, which focus on identifying the surfaces in which light is being scattered from and then mitigating the scattering by reducing the reflectivity of the surface, seismically isolating it or relocating it. 

An alternative method for reducing \scladj glitches, known as ``RC tracking'', was implemented in the Advanced LIGO observatories in January 2020~\cite{reducing_scattering:2020}. The Advanced LIGO detectors employ a quadruple pendulum suspension for the test masses, where two chains suspend four masses in this suspension system, one for the test mass optic and the other for the reaction mass. The reaction mass is used to impose a force upon the test mass, and a significant source of \scladj glitches was the large relative motion between the test mass chain and the reaction chain. To mitigate this effect, the relative motion between the end test mass and the reaction mass needed to be reduced. This was achieved by ensuring the two chains are moving together by applying force to the top stage of the quadruple suspension system in Advanced LIGO. The implementation of RC tracking represented a decrease from $0.01 s^{-1}$ to $0.0001 s^{-1}$ and $0.0072 s^{-1}$ to $0.0012 s^{-1}$ in the number of \scladj glitches detected by Gravity Spy for LIGO-Hanford and LIGO-Livingston, respectively.

While methods for preventing \scladj glitches have been developed and have shown success, they have not been able to remove the problem of \scladj glitches from the data. Additionally, as the detectors continue to be upgraded and increase in sensitivity, new sources of \scladj glitches will continue to appear. Identifying these new sources and mitigating their effects can take many months, during which time the detectors are taking in data which might be affected by the presence of \scladj glitches. Therefore, it is not realistic to believe that analyses will be able to regularly run on data that does not contain \scladj glitches and this motivates us to develop a technique for mitigating the impact of these glitches when trying to identify CBCs in \gwadj data.

In this work, we present a new method for identifying and removing \scladj glitches from \gwadj data in advance of running searches to identify CBCs.
We first introduce a method for identifying when \scladj glitches are present in detector data, through the creation of a new modelled search for \scladj glitches, similar to how we search for \gws using matched filtering. We can model \scladj glitches, generate a suitable set of glitch waveforms and perform a matched filter search on detector data. We then subtract identified glitches from the data to increase detector sensitivity. The detector data isn't Gaussian and stationary, so the matched filter does have the potential to identify non-\scladj glitches, and potentially even \gwadj signals, as \scladj glitches. To prevent this we also demonstrate a new \scl $\chi^{2}$ test, which can distinguish between \scladj glitches, and other glitches---and \gwadj signals---in the data.

We begin by reviewing previous research and describing the formulation of the waveform model used for characterizing \scladj glitches in section \ref{4:sec:sc_li}. In section \ref{4:sec:search_techniques} we introduce the various techniques used in the search to identify \scladj glitches in \gwadj data and the results of the \scladj glitch search. In section \ref{4:sec:results} we describe the results of a ``glitch-subtracted'' \gwadj search and any increases in sensitivity. We conclude in section \ref{4:sec:conclusion} and discuss the implementation of this method in future observing runs.

\section{\label{4:sec:sc_li}\Scl}

To identify \scladj glitches in \gwadj data requires an accurate model of \scladj glitches. This section details the derivation of the model we will use for generating our \scladj glitch filter waveforms, along with its parameterisation.

\subsection{Modelling \scladj glitches — a review}

Our model for \scladj glitches draws heavily from~\cite{TAccadia:2010}, and we briefly review the main details of the model presented there. In~\cite{TAccadia:2010}, the authors construct a model to accurately predict the increase in noise due to \scl during periods of increased micro-seismic activity. The model in~\cite{TAccadia:2010} is constructed from parameters related to physically measurable properties of the detector such as the mirror transmission factor, $T$, the finesse of the Fabry–Pérot cavity, $F$, or the wavelength of the light, $\lambda$.

They define the amplitude of the additional beam produced by light scattering off of a surface as
\begin{equation}
    A_{sc} = A_{0} T \sqrt{\frac{2 F}{\pi}} \sqrt{f_{sc}} e^{i \phi_{sc}}\,,
    \label{4:eqn:accadia_amplitude}
\end{equation}

where $A_{0}$ is the amplitude of the light resonating in a Fabry–Pérot cavity, $f_{sc}$ is the fraction of the optical power scattered back into the main beam and $\phi_{sc}$ is the phase angle modulated by the displacement of the scattering optics, defined as
\begin{equation}
    \phi_{sc}(t) = \frac{4 \pi}{\lambda} ( x_{0} + \delta x_{opt}(t) ),
    \label{4:eqn:accadia_phase_noise}
\end{equation}
where $\delta x_{opt}$ is the displacement of the scattering surface and $x_0$ is the static optical path.

The total amplitude of the beam inside the arm is given by $A_{tot} \!=\! A_{0} + A_{sc}$, with a phase angle equal to the phase noise introduced by the \scl $\delta \Phi \!=\! \frac{A_{sc}}{A_{0}} \cdot \sin \phi_{sc}$. The noise introduced by the \scl, $h_{sc}$, can be approximated through the relationship $\sin(\phi+ \delta\phi) \approx \sin\phi + \cos\phi \cdot \delta\phi$ for small $\delta\phi$, and can be expressed as
\begin{equation}
    h_{sc}(t) = G \cdot \sin \left(\frac{4 \pi}{\lambda} (x_{0} + \delta x_{sc}(t) ) \right),
    \label{4:eqn:accadia_strain}
\end{equation}
where $G$ is the \textit{coupling factor}, defined as $K \cdot \sqrt{f_{sc}}$ where
\begin{equation}
K = \frac{\lambda}{4 \pi} \frac{T}{\sqrt{2 F \pi}}.
\end{equation}

The displacement of the scatterer when presenting with oscillatory motion is then given as
\begin{equation}
    \delta x_{sc} (t) \simeq A_{m} \sin(2 \pi f_{m} t),
    \label{4:eqn:accadia_oscillatory}
\end{equation}
where $f_{m}$ is the frequency-modulated signal with modulation index $m = A_{m} \frac{4 \pi}{\lambda}$ and where $A_{m}$ is the amplitude of the $n$th harmonic. Finally, Equation~\ref{4:eqn:accadia_strain_linearised} can be simplified when considering only small bench motion, according to
\begin{equation}
    h_{sc}(t) = G \cdot \cos\phi_{0} \cdot \frac{4 \pi}{\lambda} \cdot \delta x_{sc}(t) + h_{\text{offset}}.
    \label{4:eqn:accadia_strain_linearised}
\end{equation}

\subsection{Model}

The model introduced in~\cite{TAccadia:2010} for the \gwadj strain noise introduced by \scl uses a lot of knowledge about the detector state. The model used in this work will be more phenomenological in the parameterisation, allowing us to rely only on the characteristics of the glitches in the strain data and not knowledge of the detector configuration, especially in cases where this detector information might not be known. Each \scladj glitch, as viewed in a spectrogram (see figure \ref{4:fig:scattered_light}), has two easily identifiable features: the maximum frequency reached, \emph{glitch frequency} ($f_{gl}$); and the period of time the glitch exists in detector data, \emph{time period} ($T$).  In addition, we can fully describe an artifact by defining an \emph{amplitude} ($A$), \emph{phase} ($\psi$) and \emph{centre time} of the glitch ($t_0$).

To formulate a model of \scladj glitches in terms of these parameters, we simplify equation \ref{4:eqn:accadia_strain}, treating the strain noise caused by \scl as the sinusoidal function
\begin{equation}
  h_{sc}(t) = A \sin\left(\frac{4\pi}{\lambda}x(t) + \psi\right),
  \label{4:eqn:h_sc_initial}
\end{equation}
where $A$ is some constant, $x(t)$ is the displacement of the scattering surface and $\psi$ is an offset.

\Scladj glitches are caused by the physical increase in the distance travelled by the light as a consequence of being reflected off of a surface. We assume a simple harmonic oscillator for the surface displacement with frequency $f_{rep}$
\begin{equation}
    x(t) = B \sin\left(2\pi f_{rep}t\right),
    \label{4:eqn:harmonic_displacement}
\end{equation}
which is directly related to the \emph{time period}, $T$,
\begin{equation}
  f_{rep} = \frac{1}{2 T}.
  \label{4:eqn:f_rep}
\end{equation}
The \emph{time period} of a \scladj glitch only corresponds to half of a sinusoidal wave hence the multiplier of 2 on the denominator of equation \ref{4:eqn:f_rep}. We calculate the instantaneous frequency of the strain noise~\cite{TAccadia:2010}
\begin{equation}
    f(t) = \left|\frac{1}{2\pi} \frac{d}{dt}\left(\frac{4\pi}{\lambda}x(t) + \psi\right)\right|,
    \label{4:eqn:instant_freq}
\end{equation}
which we can use to predict the frequency of each scattered light glitch. We define the maximum frequency of each glitch (at the peak of the arch) as the glitch frequency, $f_{gl}$, which can be found by taking the maximum of Equation~\ref{4:eqn:instant_freq}
\begin{equation}
    f_{gl} = \text{max}\left[f(t)\right],
\end{equation}
giving an equation for the glitch frequency of a scattered light glitch
\begin{equation}
    f_{gl} = B \frac{4\pi}{\lambda} f_{rep}.
    \label{4:eqn:glitch_frequency}
\end{equation}
Rearranging Equation~\ref{4:eqn:glitch_frequency} yields the expression $\frac{4\pi}{\lambda} = \frac{f_{gl}}{f_{rep} B}$ and when combined with Equations~\ref{4:eqn:h_sc_initial} and~\ref{4:eqn:harmonic_displacement} gives the final form of the strain noise induced by scattered light
\begin{keyeqntitled}{Scattered Light Glitch Model}
\begin{equation}
  h_{sc}(t) = A \sin\left(\frac{f_{gl}}{f_{rep}} \sin(2 \pi f_{rep} t) + \psi\right),
  \label{4:eqn:model}
\end{equation}
\end{keyeqntitled}
where $A$ and $\psi$ are amplitude and phase parameters to be maximised over.

In~\cite{Was_Subtract:2021} they provide another term to describe \scladj glitches which uses the instantaneous frequency as a function of time, allowing the identification of the correct amplitude at all frequencies. This term is due to radiation pressure coupling and is thought to be dominant at low frequencies. The relationship between our amplitude and the new amplitude depends on the power in the arm cavities and the signal recycling mirror reflectivity, which we do not consider in our model and so disregard this extra term.

\subsection{Harmonics}

As described in~\cite{TAccadia:2010}, harmonic glitches appear at the same time with different \emph{glitch frequencies}. A harmonic glitch is a glitch with a \emph{glitch frequency} that is a positive integer multiple of the glitch frequency of the glitch in the stack with the lowest glitch frequency value. This lowest frequency glitch has the potential to appear below $15$Hz and will be masked by other sources of noise and therefore cannot be seen. An example of harmonics can be seen in figure \ref{4:fig:consec_scattered_light}.

\section{\label{4:sec:search_techniques}Identifying \scladj glitches in \gwadj strain data}

Equipped with our model for \scladj glitches from the previous section, we now discuss how we identify and parameterize \scladj glitches in a stretch of \gwadj data before we apply this to searches for CBCs in the next section.

\subsection{\label{4:subsec:MF}Matched filtering}

Given our model of \scladj glitches, we can consider a realisation with specific values of the parameters discussed above. To identify glitches with these values of the parameterisation we use matched filtering, described in Section~\ref{2:sec:matched-filter}. \Scladj glitches will take a range of values of the parameters describing them and we must be able to identify glitches anywhere in the parameter space. Following~\cite{FINDCHIRP:2012} we can analytically maximize over the \emph{amplitude}, \emph{phase} and the \emph{centre time} glitch parameters in equation \ref{4:eqn:model} exactly as described in Section~\ref{2:sec:matched-filter}.

\subsection{Template bank}

Our \scladj glitch model is parameterised by 5 variables. As shown above, by maximizing over \emph{phase}, \emph{amplitude} and \emph{time}, we can analytically maximize the SNR over 3 of these variables. However, the remaining two describe the intrinsic evolution of the glitch, and we must explicitly search over these parameters. To do this we create and use a ``template bank'' of \scladj glitch waveforms, created such that it would be able to identify glitches with any value of our 2 remaining parameters, \emph{glitch frequency} and \emph{time period}.

To generate this template bank, a stochastic template placement algorithm~\cite{Harry_sbank:2009} is used. This algorithm randomly generates a new template, places the template in the existing template bank and calculates the ``distance'' (how similar two templates appear) between the new template and existing templates. If the new template is too close to an existing template it is discarded, otherwise, it is accepted into the bank. The density of the bank is dependent on the allowed distance between two templates. The dominant cost of the search for \scladj glitches is matched filtering and is approximately linear in the number of templates, a larger template bank will find all the glitches with more accurate parameter values, but the computational cost of the search will be increased. The template bank generation alone does not constitute a significant computational cost in the search.

The distance function we have chosen to evaluate templates by is the match between two glitch templates. To calculate the match, we first normalize both templates such that the matched filter between a template and itself would be equal to 1. We can then compute the inner product between the two normalised templates to find their match
\begin{keyeqntitled}{Template Match}
\begin{equation}
    M = \max_{t, \psi} (\hat{a} | \hat{b}).
  \label{4:eqn:match}
\end{equation}
\end{keyeqntitled}
The value for the match, $M$,  is bounded between 0 and 1, where a value of 0 indicates orthogonal waveforms and a value of 1 indicates identical normalised waveforms. The maximum match allowed between any two templates in our bank is 0.97, which implies that a fully converged stochastic bank would have at least one waveform in it with $M \geq 0.97$ for any point in the space of parameters.

\begin{figure}
  \centering
  \includegraphics[width=1.0\textwidth]{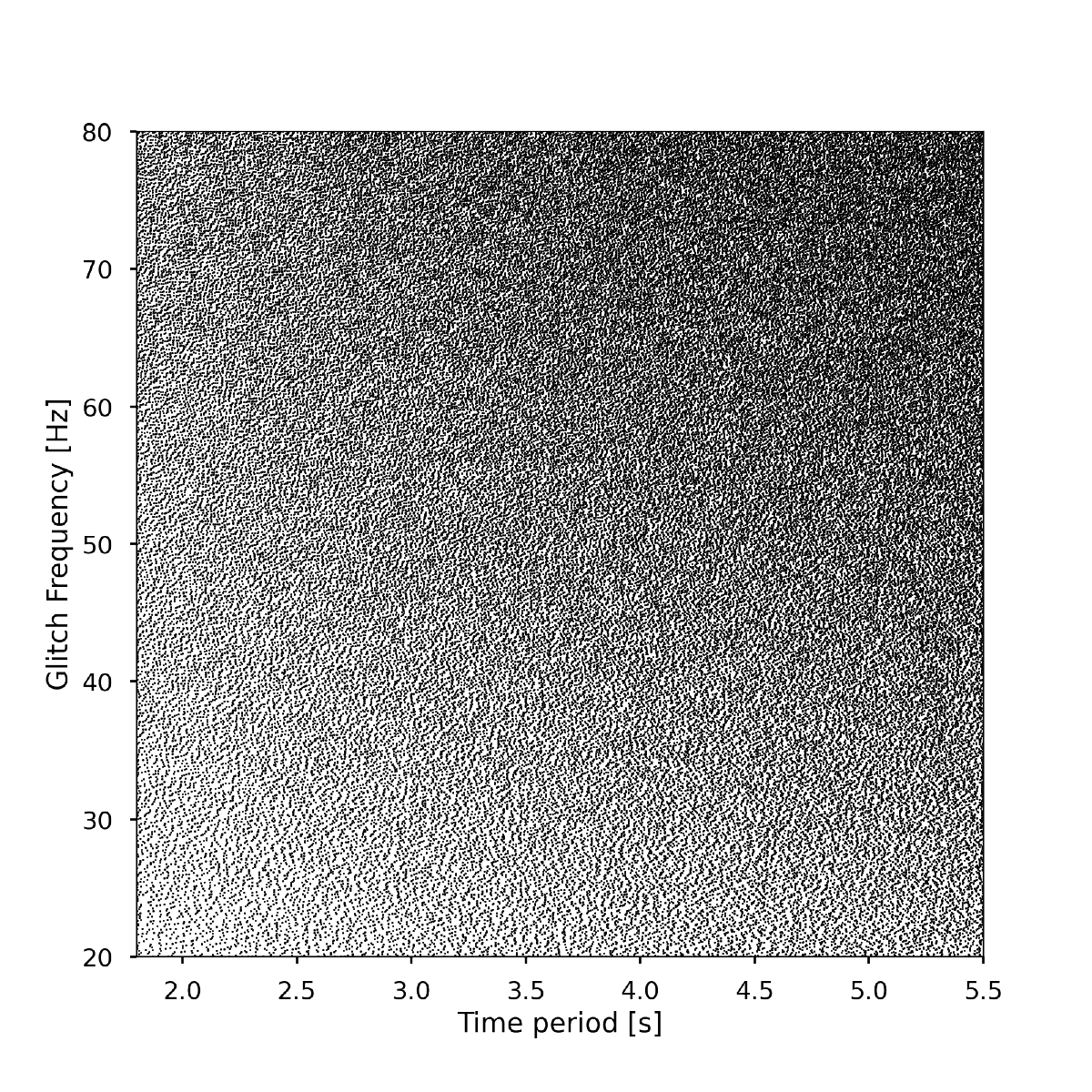}
  \caption{The template bank of \scladj glitch templates (black points) used in the search for \scladj glitches. The \emph{glitch frequency} parameter values range from $20 \, \text{Hz}\text{--} 80 \, \text{Hz}$ and the \emph{time period} parameter values range from $1.8 \, \text{s}\text{--} 5.5 \, \text{s}$. This bank was created with a maximum match allowed between two templates of $0.97$ and contains 117,481 templates.}
  \label{4:fig:sq_bank}
\end{figure}

For our \scladj glitch search, we generated a template bank with \emph{time periods} $\in$ $1.8 \, \text{s}\text{--}5.5 \, \text{s}$ and \emph{glitch frequencies} $\in$ $20 \, \text{Hz}\text{--} 80 \, \text{Hz}$. We chose to use the Advanced LIGO zero-detuned high-power sensitivity curve~\cite{aLIGO_design_curve:2018}\footnote{The zero-detuned high-power sensitivity curve is a broader noise curve than the O3 Advanced LIGO data that we identify \scladj glitches in. However, this broader noise curve will result in \textit{more} template waveforms than we need, and will therefore overcover the parameter space, rather than potentially miss \scladj glitches.}. This allowed us to generate a template bank that contains 117,481 templates. We visualize the distribution of the templates as a function of \emph{time period} and \emph{glitch frequency} in Figure~\ref{4:fig:sq_bank}, observing a greater density of templates at higher frequencies and longer durations, indicating that templates in this region look less similar and therefore, more templates are needed to fully describe the parameter space.

\subsection{Identifying potential \scladj glitches in the data}

We test our method by searching through \gwadj data from 2019-11-18 16:13:05--2019-11-25 22:11:09 for the LIGO-Hanford and LIGO-Livingston detectors. This time corresponds to the 25th period of data used in the LVK analysis of O3 data for CBCs~\cite{gwtc3:2023} and is prior to the implementation of RC tracking~\cite{reducing_scattering:2020} in these detectors. We only analyse data that is flagged as ``suitable for analysis" on the Gravitational Wave Open Science Center~\cite{GWOSC:2021}\footnote{We note that in O3 \emph{only} data suitable for analysis is released, so we simply analyse all the publicly available data.}. This corresponds to 5.96 days of analysable data for LIGO-Hanford and 5.93 days for LIGO-Livingston.

Equipped with our template bank, we now identify potential \scladj glitches in the data. We matched filter all the data against all the templates, producing a SNR time series for every template in the bank. These SNR time series will contain peaks which, when above a certain limit, indicate a potential \scladj glitch at a particular time. We retain any maxima within the time series that have a SNR greater than 8. However, as we do this independently for every template, we will identify multiple peaks for any given glitch, and we will also find peaks that correspond to other types of glitch, or even \gwadj signals. We discuss how we reduce this to a list of identified \scladj glitches in the following subsections.

\subsection{Scattered-light signal consistency test}

To prevent the search for \scladj glitches from misclassifying other classes of glitches, or \gwadj signals, we employ a $\chi^2$ consistency test. The $\chi^2$ discriminator introduced in \cite{Allen_Chi:2005} divides \gwadj templates into a number of independent frequency bins. These bins are constructed so as to contain an equal amount of the total SNR of the original matched filter between template and data. The $\chi^{2}$ value is obtained by calculating the SNR of each bin, subtracting this from the expected SNR value for each bin and squaring the output. These values are summed for all bins and this value forms the $\chi^{2}$ statistic,
\begin{keyeqn}
\begin{equation}
  \chi_{r}^{2} = \frac{\chi^{2}}{\textrm{DOF}} = \frac{n}{2n - 2} \sum_{i=1}^n \left(\frac{\rho}{\sqrt{n}} - \rho_{bin,i}\right)^2.
  \label{4:eqn:chi_squared}
\end{equation}
\end{keyeqn}
Here $n$ is the number of bins, $\rho$ is the SNR of the original matched filter between template and data, and $\rho_{bin}$ is the value of the SNR found when matched filtering one of the divided templates and the data. This test is constructed so as to produce large values when the data contains a glitch, or astrophysical signal, that is not well described by the template, but to follow a $\chi^2$ distribution if a glitch that matches well to the template is present, or if the data is Gaussian and stationary.

\begin{figure}
  \centering
  \begin{minipage}[t]{1.0\linewidth}
  \includegraphics[width=0.49\textwidth]{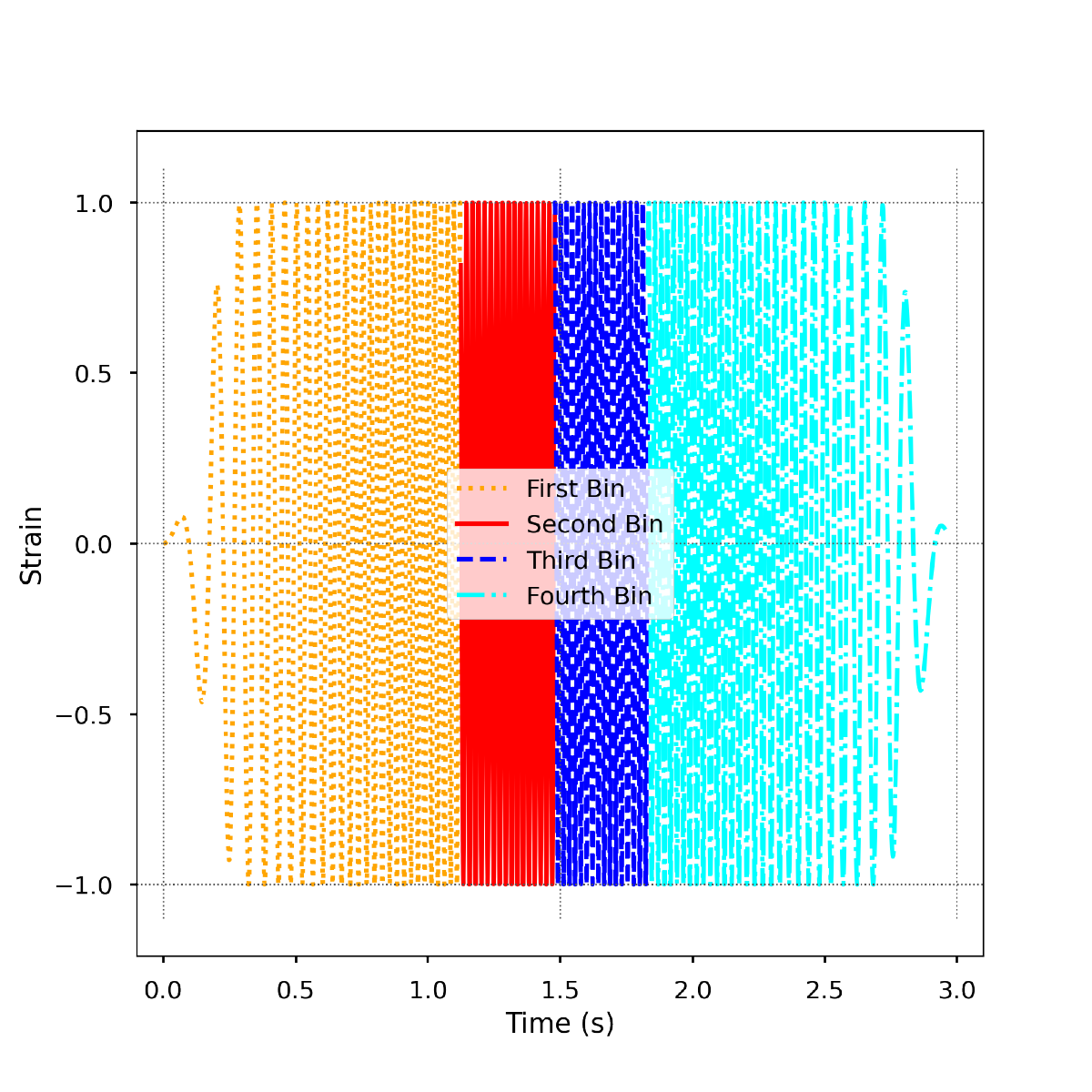}
  \hspace{0.01\linewidth}
  \includegraphics[width=0.49\linewidth]{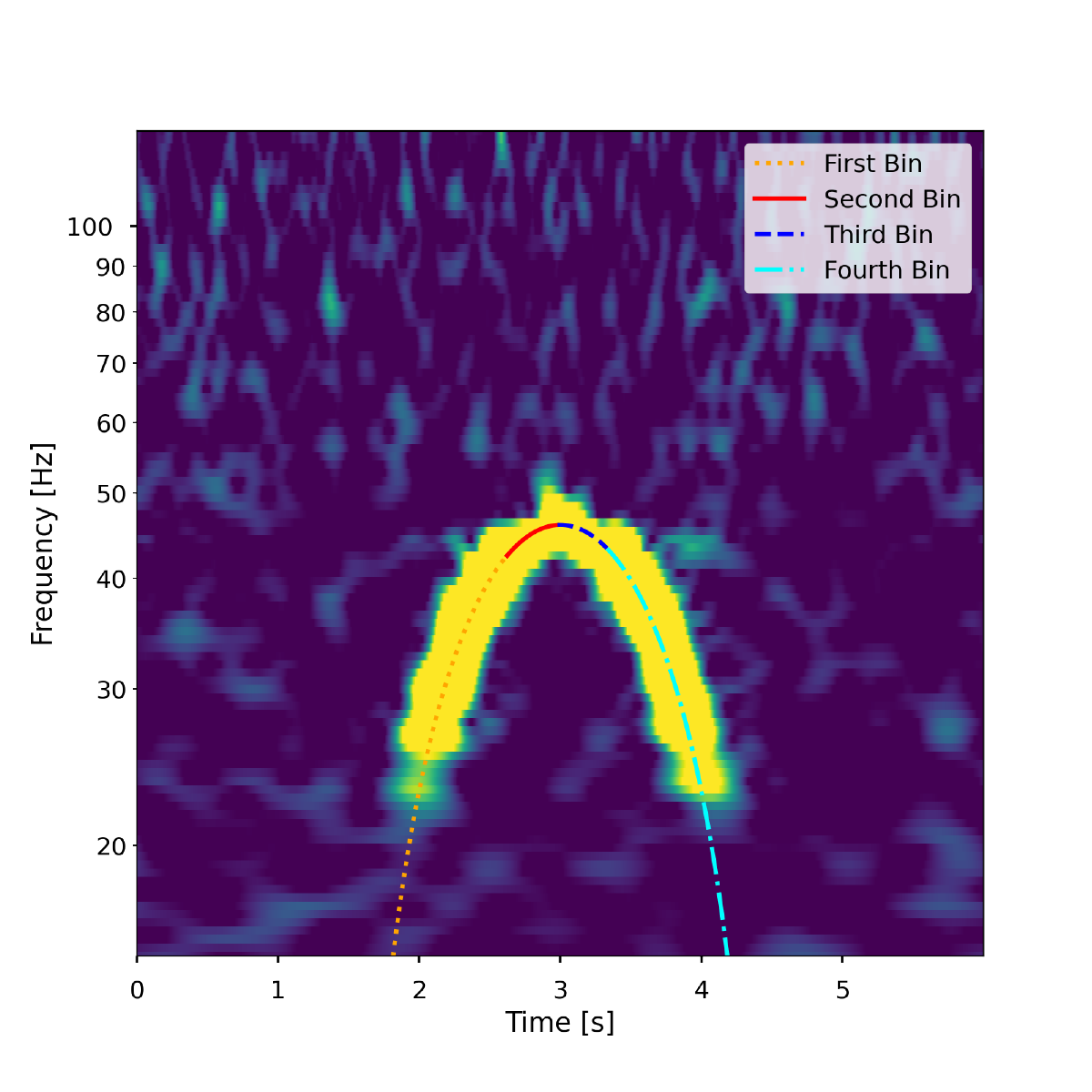}
  \end{minipage}
  \caption{A \scladj glitch template (left) where the colours and line-styles are indicative of the four equal SNR time bins to be used in calculating the $\chi_{r}^{2}$ value and re-weighting the SNR. The same \scladj glitch template bins overlaid on an injection of the \scladj glitch (right). The inner two bins are considerably shorter than the outer two bins, which informs us that the centre---higher frequency---region of the \scladj glitch contributes a larger amount to the SNR per unit time than the lower frequency regions.}
  \label{4:fig:split_temp_subplot}
\end{figure}

The $\chi^2$ test that we employ is similar to that of \cite{Allen_Chi:2005}, however CBC waveforms increase in frequency over time whereas \scladj glitch templates are symmetric about their centre. We therefore choose to construct our $\chi^2$ test with four non-overlapping bins \emph{in the time domain}, each of which contributes equally to the SNR, an example of the split template can be seen in figure \ref{4:fig:split_temp_subplot}. The matched filter between the bins and data is computed, and the $\chi_{r}^{2}$ value is calculated using equation \ref{4:eqn:chi_squared} (where $n \!=\! 4$). Any glitch, or astrophysical signal, which does not exhibit symmetric morphology should not fit with this deconstruction of the template, and should result in elevated $\chi_{r}^{2}$ values.

\begin{figure}
  \makebox[\textwidth][c]{\includegraphics[width=\textwidth]{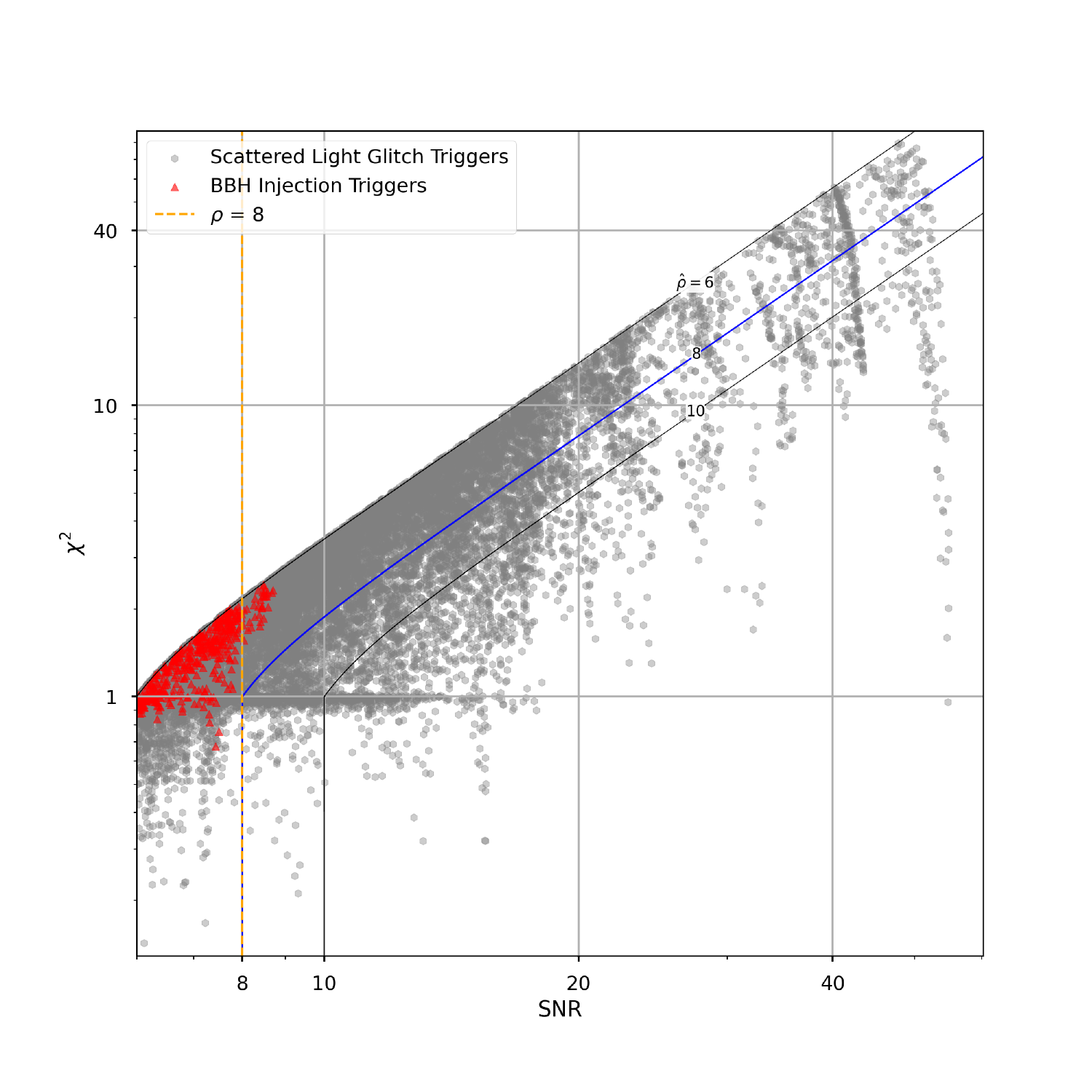}}
  \caption{The signal-to-noise ratio (SNR) and $\chi_{r}^{2}$ values for the triggers identified by the matched filtering and clustering of the \scladj glitch template bank with data containing only \scladj glitches (grey hexagons) and data containing a binary black hole \gwadj injection (red triangles). The black contour lines represent the re-weighted SNR values the trigger will take when equation \ref{4:eqn:reweighting} is applied. The orange dashed vertical line indicates the SNR value limit of 8, above which we decide to perform the $\chi^{2}$ test and calculate the re-weighted SNR. The blue solid contour line indicates a re-weighted SNR value of 8, which is the limit at which we consider the trigger to be real. Different re-weighting parameter values will produce different contour line shapes. It can be seen that no triggers for the data containing the \gwadj injection lie beneath the contour line, and therefore no \scladj glitches are found on the \gwadj signal.}
  \label{4:fig:chi_snr}
\end{figure}

After computing the $\chi_{r}^{2}$ value for potential \scladj glitches, we follow~\cite{rw_snr_eq:2012} to compute a ``re-weighted SNR'', which is an empirically tuned statistic depending on the SNR and the $\chi_{r}^{2}$ value.  The re-weighting function we use matches that presented in~\cite{rw_snr_eq:2012},
\begin{keyeqn}
\begin{equation}
\rho_{rw} =  \left\{  \begin{array}{l@{\quad}cr}
\rho & \mathrm{if} & \chi_{r}^{2} \leq 1, \\
\rho [(1 + (\chi_{r}^{2})^3)/2]^{-\frac{1}{6}} &  \mathrm{if} & \chi_{r}^{2} \ge 1,
\end{array}\right.
\label{4:eqn:reweighting}
\end{equation}
\end{keyeqn}
where $\rho_{rw}$ represents the re-weighted SNR of the \scladj glitch template calculated using the SNR, $\rho$, and the $\chi_{r}^{2}$ value of that template.
We do note that this re-weighting function has been tuned for CBC waveforms, and we did not repeat that tuning here with \scladj glitches. One could retune this statistic, specifically targeting \scladj glitches, using (for example) the automatic tuning procedure described in~\citep{McIsaac_Chi:2022}. However, we demonstrate the suitability of the $\chi^{2}$ test for our purposes in Figure~\ref{4:fig:chi_snr} where we show the $\chi_{r}^{2}$ vs SNR distribution of the triggers found by our \scladj glitch search when performed on data containing only \scladj glitches and data containing a binary black hole \gwadj injection.

We note that the $\chi^{2}$ test increases the number of matched filters required by the search and therefore the computational cost of the search. Each template would require the matched filtering of an additional $4$ ``binned'' templates to calculate the $\chi_{r}^{2}$ value to re-weight the SNR time series of that template, increasing computational cost by a maximum factor of $5$. However, we reduce this increase by only computing the $\chi^{2}$ where it is needed, specifically for any template where the SNR time series has values above the threshold of 8.

\subsection{Identifying all \scladj glitches in the data}

\begin{figure}
    \centering
    \begin{minipage}[t]{1.0\linewidth}
        \centering
        \includegraphics[width=0.48\linewidth]{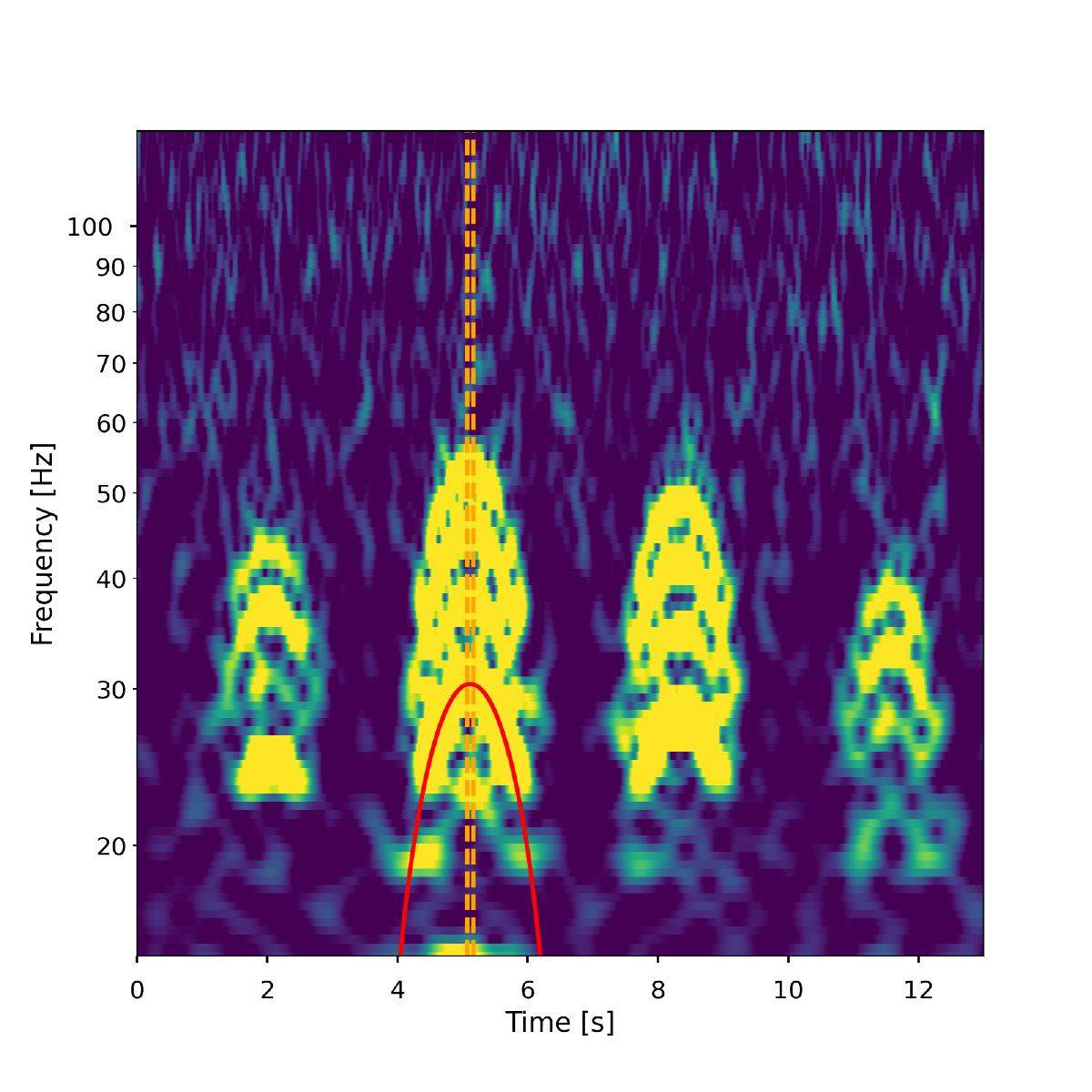}%
        \hfill%
        \includegraphics[width=0.48\linewidth]{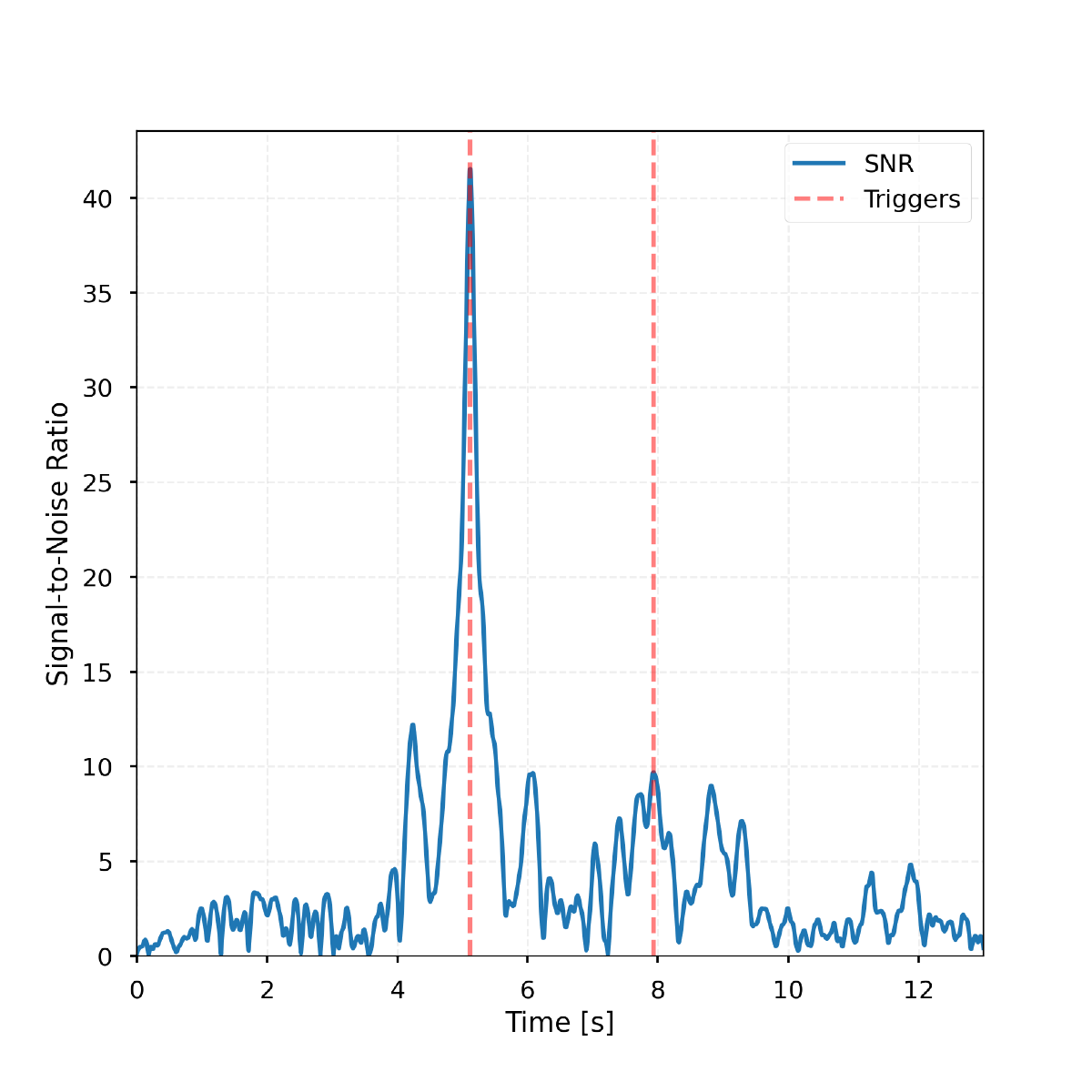}
    \end{minipage}

    \vspace{0.2cm} 

    \begin{minipage}[t]{1.0\linewidth}
        \centering
        \includegraphics[width=0.48\linewidth]{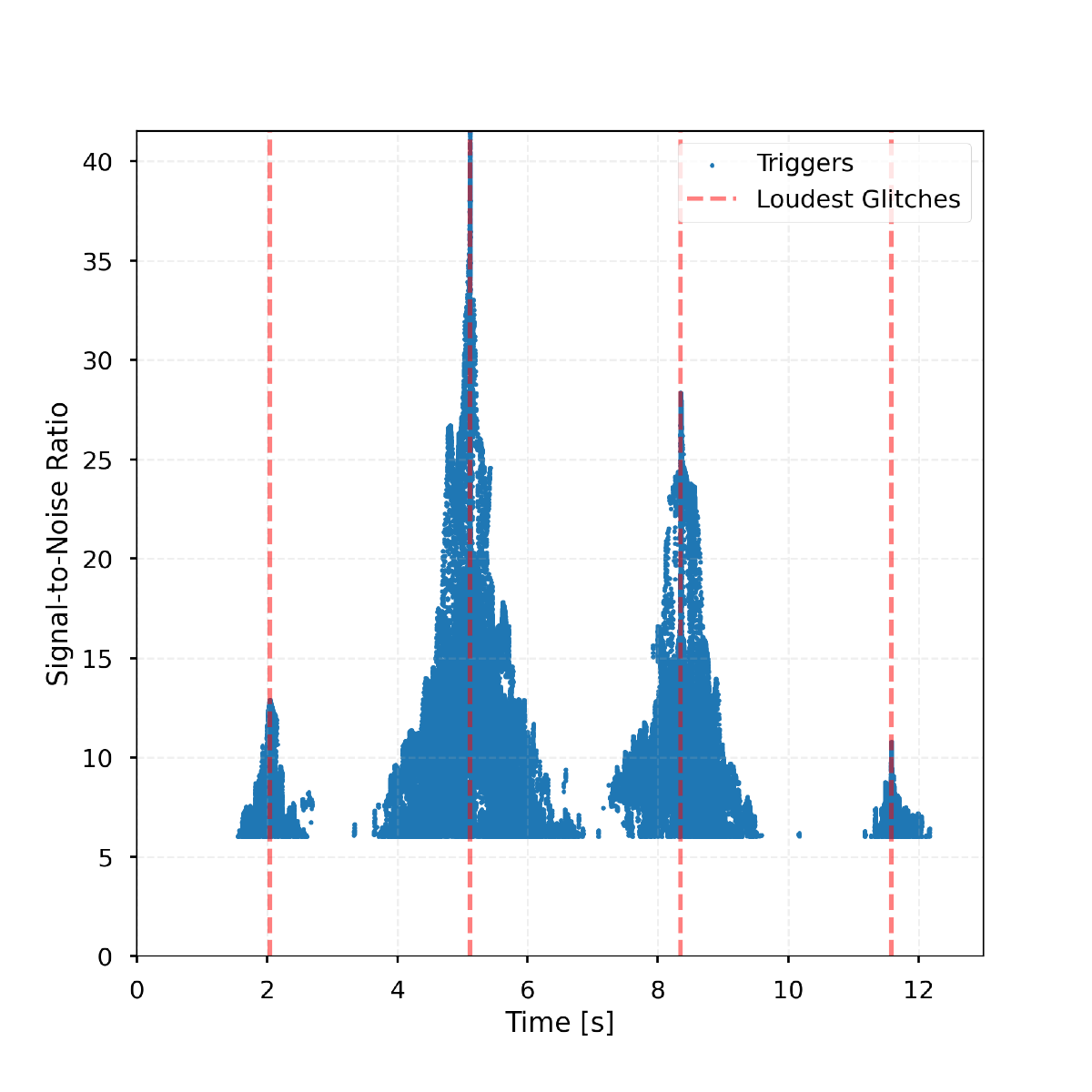}%
        \hfill%
        \includegraphics[width=0.48\linewidth]{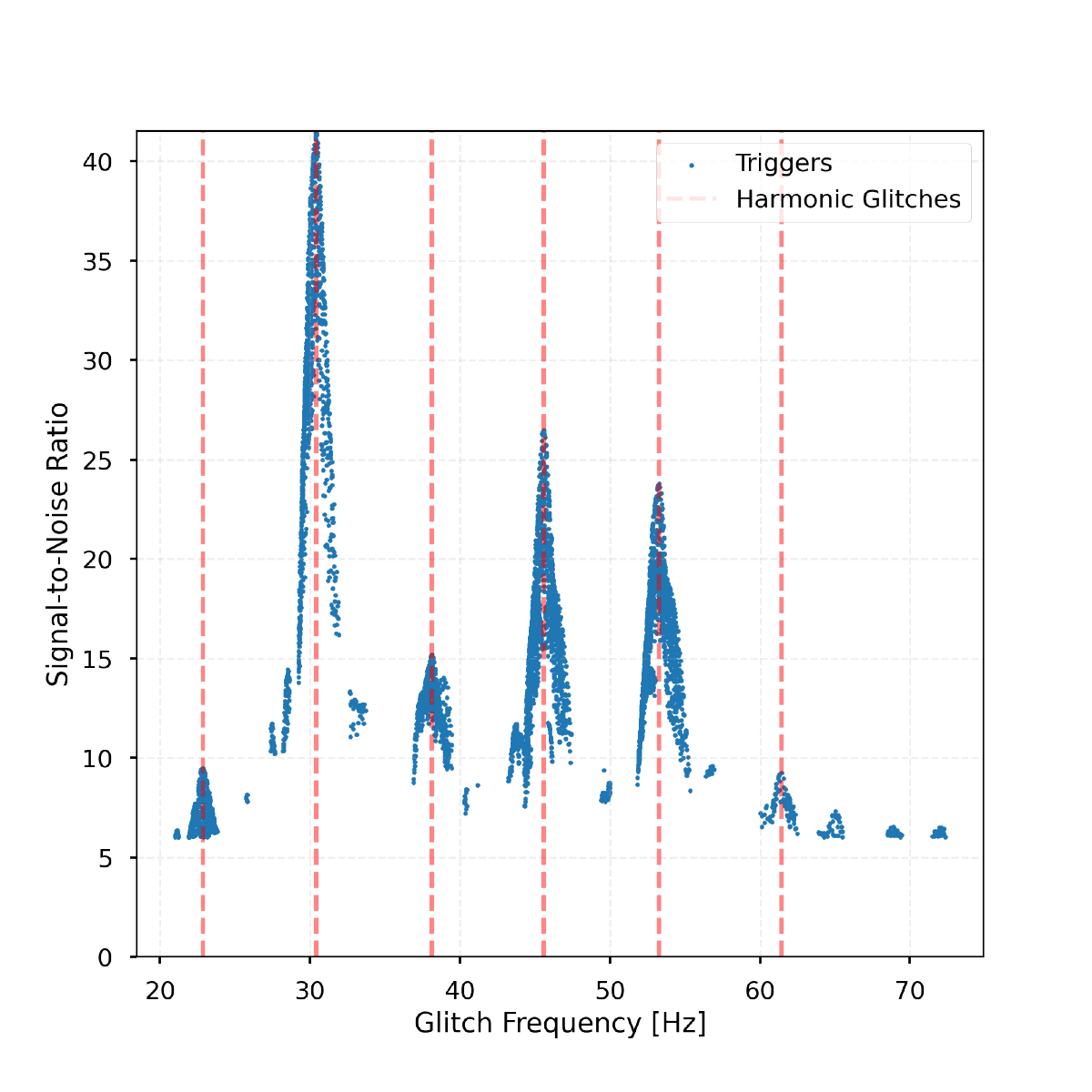}
    \end{minipage}

    \caption{The process for identifying all the \scladj glitches in a period of \gwadj data. The red overlay in the \gwadj data used in this example (top left) indicates the highest re-weighted signal-to-noise trigger found, the dashed vertical lines represent the time slice window around this trigger. The re-weighted SNR time series resulting from the matched filter of this trigger's template and the data (top right) is clustered in time to identify the triggers found above a signal-to-noise threshold of $8$, indicated by red vertical dashed lines. All the triggers from all the templates in the template bank are then clustered in time (bottom left) to identify the highest re-weighted signal-to-noise glitches in the data, indicated by the orange vertical dashes lines. The triggers found within the time slice window, with a similar \emph{time period} value---within $\pm 10\%$---of the highest re-weighted SNR trigger (bottom right) are clustered by their \emph{glitch frequency} value to find the harmonic glitches at the same time, indicated again by red dashed lines.}
    \label{4:fig:clustering_story}
\end{figure}

Our matched filtering process retains ``triggers'' (potential \scladj glitches) wherever the re-weighted signal-to-noise time series is larger than 8. We retain no more than one trigger within a window size equal to half the \emph{time period} of the template used to produce the re-weighted signal-to-noise time series, and only store triggers at local maxima. A re-weighted signal-to-noise time series with multiple peaks and identified triggers can be seen in the top right panel in figure \ref{4:fig:clustering_story}.

After matched filtering all the templates against the data, we will recover multiple triggers for any potential \scladj glitch, as we might expect to independently identify peaks in multiple templates around the true values of the glitch. We therefore collect all the triggers generated by the template bank and cluster these in time, using a window of half of the shortest duration template---$0.9$ seconds. This will result in a list of triggers corresponding to the highest re-weighted SNRs, where each trigger should correspond to a unique \scladj glitch. The bottom left panel in figure \ref{4:fig:clustering_story} shows an example of the triggers found by the search and the highest re-weighted SNR triggers found by clustering.

However, we also expect to see instances of harmonic glitches which are produced by the same scattering surface and so share the same \emph{time period} and have \emph{glitch frequency} values equal to a multiple of the lowest frequency glitch in the harmonic stack~\cite{TAccadia:2010}. We investigate each trigger in the list, searching for harmonic glitches occurring at the same time. We use the first list of triggers identified by all templates across the bank and filter by those that occur within $\pm0.05$ seconds of the \emph{centre time} of the trigger we are investigating, an example of this window can be seen in the top left panel of figure \ref{4:fig:clustering_story}. We then filter the triggers again, keeping only those with an associated \emph{time period} value within $\pm 10 \%$ of the trigger's \emph{time period}. Finally, we cluster these remaining triggers by their associated \emph{glitch frequency} using a window size of $4$Hz, a lower limit on the frequency separation of harmonic glitches, the bottom right panel in figure \ref{4:fig:clustering_story} shows the identification of harmonic glitches for the overlaid \scladj glitch in the top left panel of figure \ref{4:fig:clustering_story}.

\subsection{Hierarchical subtraction to find parameter values}

\begin{figure}
     \centering
     \begin{minipage}[t]{1.0\linewidth}
        \centering
        \includegraphics[width=0.48\linewidth]{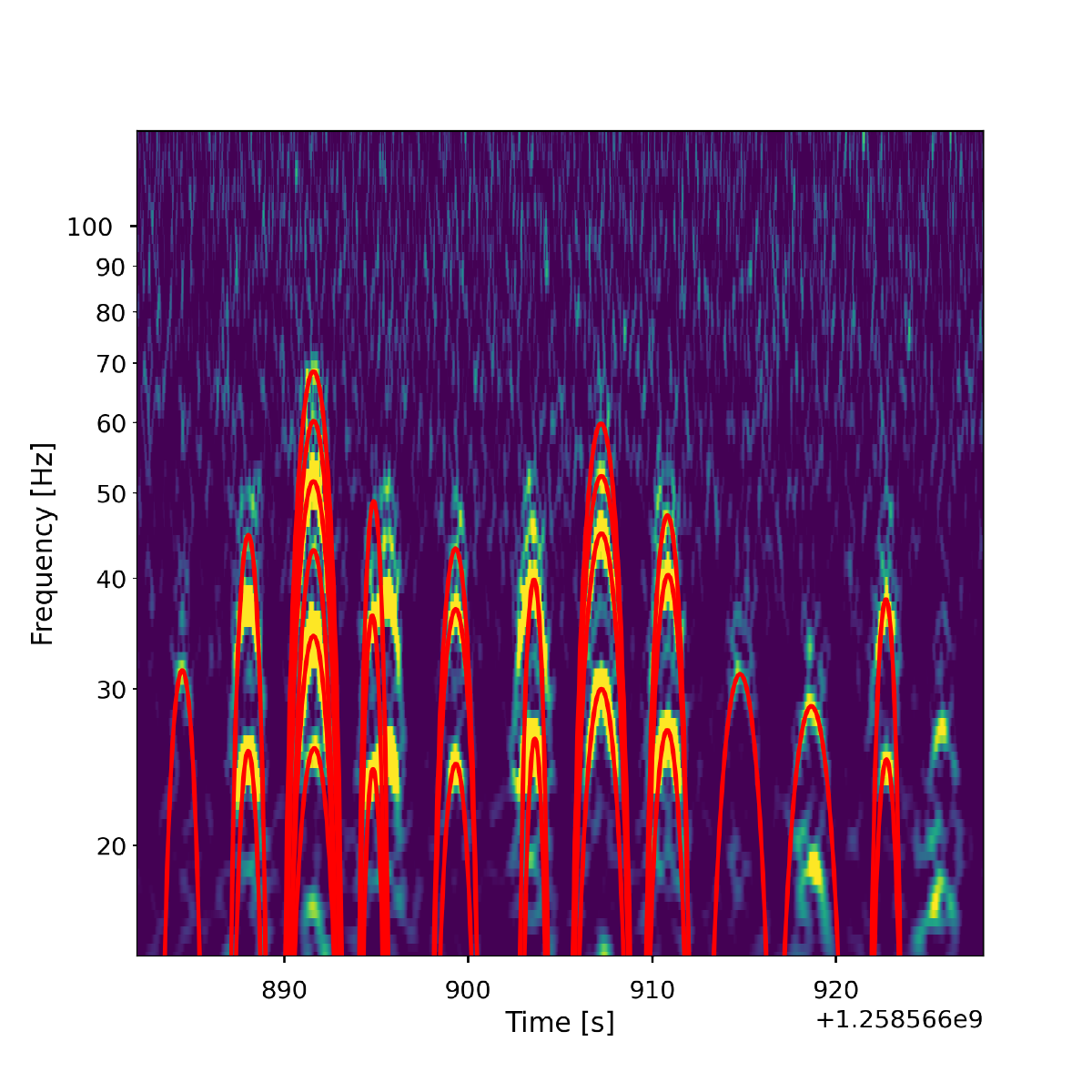}%
        \hfill%
        \includegraphics[width=0.48\linewidth]{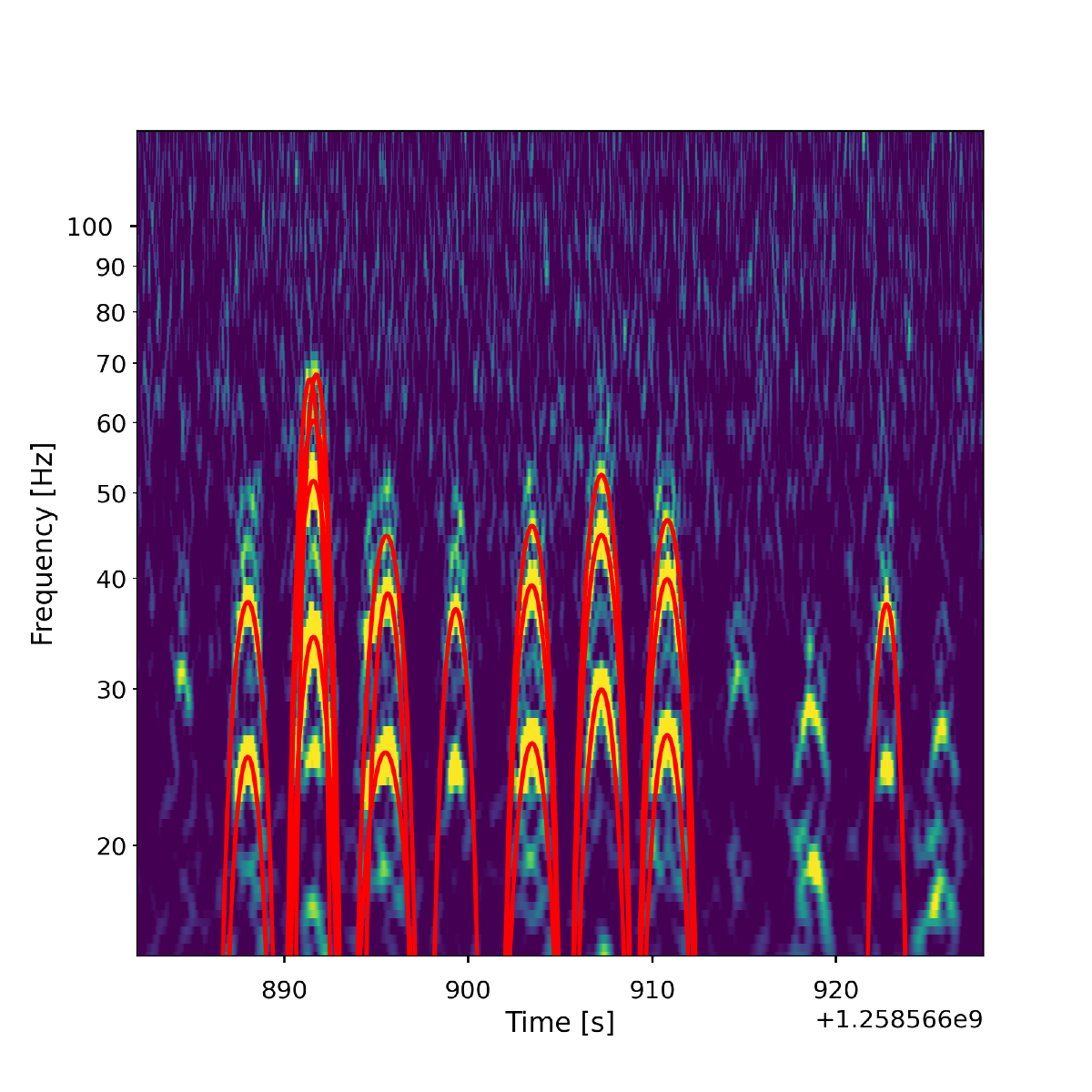}
     \end{minipage}
         \caption{LIGO-Hanford data from 2019-11-23 17:54:22--2019-11-23 17:55:12 containing \scladj glitches which have been identified by the ArchEnemy search (left), there is a misalignment in the template found for a number of glitches in this period of data and some missed glitches. Instead of finding all \scladj glitches at once we run a hierarchical subtraction to find the \scladj glitches in a chunk of time one-by-one. \Scladj glitches remaining after running the hierarchical subtraction search (right) for the same period of data, we have missed more \scladj glitches however misalignments have been removed. The highest harmonic at approximately $892$ seconds has been incorrectly split into two separate templates.}
    \label{4:fig:overlay_goods}
\end{figure}

We now have a list of identified \scladj glitches and their parameter values, however, these might not be fully accurate when there are many glitches close in time and frequency, as illustrated in Figure~\ref{4:fig:overlay_goods}. It is important that the parameters we find match well with the glitches in the data to remove as much power as possible.

To better identify the parameter values of the \scladj glitches, we perform a hierarchical procedure using information about the glitches we have found so far. Firstly, we create new segments of time which we know contain \scladj glitches, taking a window of $8$ seconds on either side of each previously identified glitch, if two glitch windows overlap they are combined into the same segment.

For each segment, we then create a reduced template bank, consisting of templates ``close'' to the \scladj glitches previously identified in the segment. We take the smallest and largest \emph{time period} and \emph{glitch frequency} glitches in the segment and bound the retained templates by these values with $\pm 0.25$ seconds on the \emph{time period} and $\pm 1$Hz on the \emph{glitch frequency}. For each data segment the reduced template bank is matched filtered with the data, the maximum re-weighted SNR value is found, and the corresponding glitch is subtracted. We then matched filter \emph{again} and remove the next largest re-weighted SNR template. This process is repeated until no templates, when matched filtered with the data, produce any re-weighted SNR values above the SNR limit of $8$. This method of hierarchical subtraction produces our final list of \scladj glitches.

A further benefit of using these shorter data segments is that we are estimating the PSD of the data using only a short period of data close to the \scladj glitches being removed. This protects us from a rapidly changing PSD in non-stationary data, which might cause Gaussian noise to be identified with larger SNR in the periods where the PSD is larger. This can be resolved by including the variation in the power spectral density as an additional statistic in the re-ranking of triggers, which has been done for CBC \gwadj searches in~\cite{PSD_var:2020}.

\begin{figure}
    \centering
    \begin{minipage}[t]{1.0\linewidth}
        \centering
        \includegraphics[width=0.45\linewidth]{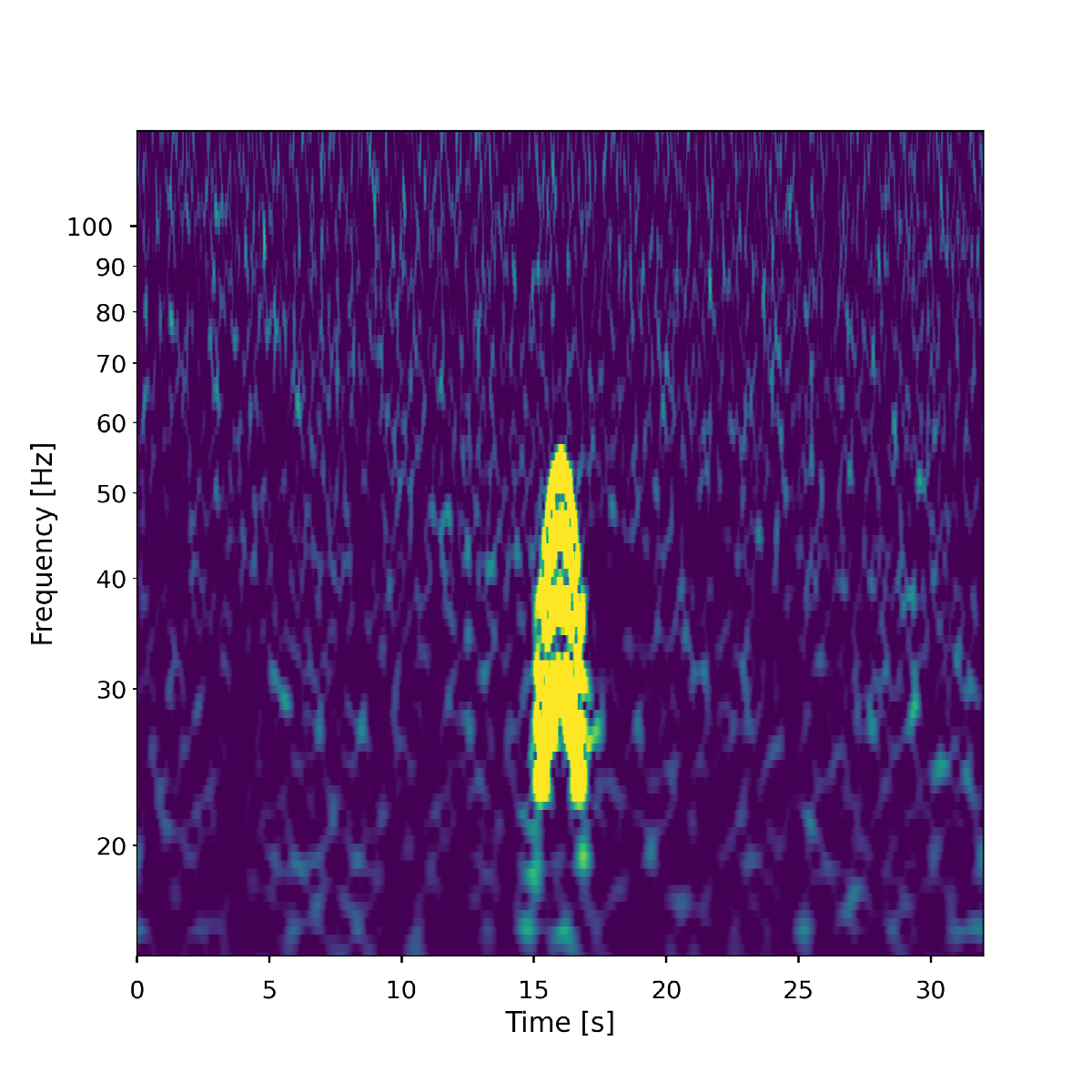}%
        \hfill%
        \includegraphics[width=0.45\linewidth]{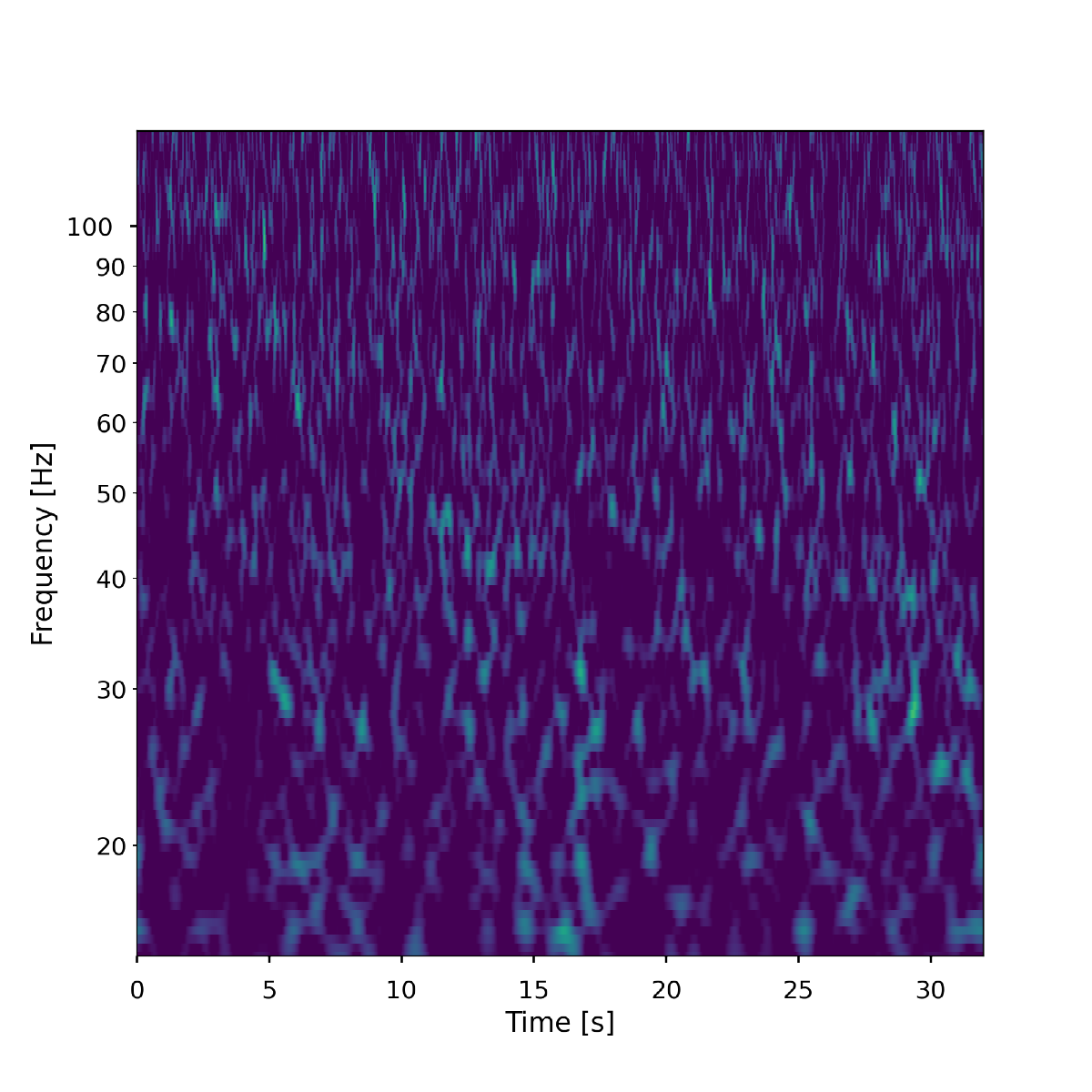}
    \end{minipage}
    \caption{Data containing an injected stack of harmonic \scladj glitches (left) and the corresponding data found when running the hierarchical subtraction search and subtracting the identified \scladj glitches from the data (right).}
    \label{4:fig:injected_glitches}
\end{figure}

We demonstrate the hierarchical subtraction step on an amount of data which contains four injected \scladj glitches in a single harmonic stack, this can be seen in figure \ref{4:fig:injected_glitches}. As shown, the \scladj glitches are found and subtracted from the data leaving behind a cleaned segment of \gwadj data with no excess noise. Figure~\ref{4:fig:overlay_goods} shows the identified \scladj glitch triggers before and after performing the hierarchical subtraction step on a stretch of data containing real \scladj glitches. We identify more triggers prior to performing hierarchical subtraction, however there are more errant mismatches between \scladj glitches and the overlaid templates. By performing the hierarchical subtraction, we more accurately identify \scladj glitches, however, we miss some glitches that were previously identified. There is still some imperfection in this process, and we do not refine the method further in this work, but highlight this as useful direction for future studies in removing \scladj glitches.

\subsection{Identified \scladj glitches}

The methodology described in previous sections is implemented in our ``ArchEnemy'' pipeline, which is capable of searching for \scladj glitches in \gwadj data using a pre-generated bank of glitch templates. We use ArchEnemy to analyse the aforementioned data, which produced a list of $2749$ \scladj glitches in data from the LIGO-Hanford observatory and $1306$ from the LIGO-Livingston observatory.

\begin{figure}
       \centering
    \includegraphics[width=0.7\linewidth]{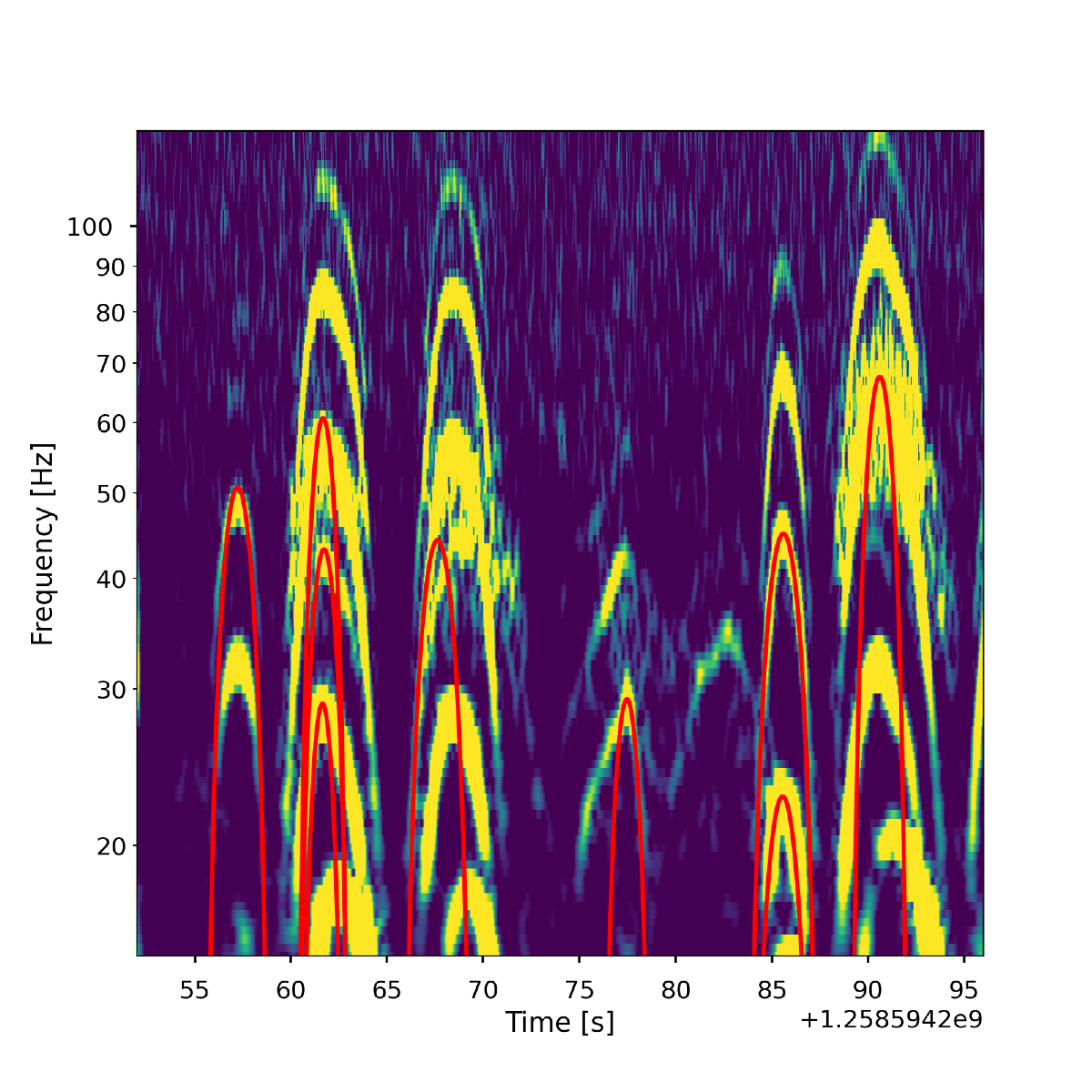}
    \caption{LIGO-Livingston data from 2019-11-24 01:30:32--2019-11-24 01:31:20 containing a very large number of \scladj glitches at multiple times and frequencies over-plotted with the \scladj glitches identified by the ArchEnemy search. Very few overlays match well onto \scladj glitches. The template bank used in this search terminates at $80 \, \text{Hz}$ and so the \scladj glitches located above this value will not be correctly identified. There are also asymmetric \scladj glitches located which will not be identified correctly by our search, which assumes symmetry in the \scladj glitch.}
    \label{4:fig:overlay_bads}
\end{figure}

The number of \scladj glitches found by the ArchEnemy pipeline can be compared to Gravity Spy for the same period of time. Gravity Spy finds $2731$ and $1396$ for LIGO-Hanford and LIGO-Livingston respectively~\cite{gravityspy:2023}. There will be a difference in the number of glitches found by ArchEnemy and Gravity Spy for at least two reasons: Gravity Spy treats an entire stack of harmonic glitches as a single \scladj glitch, whereas ArchEnemy will identify each glitch as a separate occurrence. Gravity Spy can also identify \scladj glitches which are not symmetric and fall outside our template bank, for example, the \scladj glitches shown in Figure~\ref{4:fig:overlay_bads}.

Figure~\ref{4:fig:overlay_goods} is an example of the results of the ArchEnemy pipeline and how well it has identified \scladj glitches in a period of data. A majority of the glitches have been identified with the correct parameter values and even in cases where the chosen template is not visually perfect, there is a good match between the template and the identified power in the data, particularly in the case of slightly asymmetric glitches. Figure~\ref{4:fig:overlay_bads} demonstrates a period of time when the ArchEnemy pipeline has not fitted well the \scladj glitches in the data. The glitches at this time are improperly fit by the templates due to asymmetry of the morphology of the glitches and because some glitches are outside our template bank parameter range. However, we note that this is a very extreme period of \scladj glitching and immediately after this time the detector data is no longer flagged as ``suitable for analysis''.

We have demonstrated the ArchEnemy pipeline on a stretch of data from O3 and have identified and characterised a list of \scladj glitches, which could be removed from the data. We do note that there are cases where the identification has not worked well, but we expect that subtracting our list of glitches from the data will reduce their effect on the \gwadj search. In the next section, we will demonstrate this by quantifying sensitivity with the PyCBC pipeline.

\subsection{Safety of \scl identification}
\label{4:ssec:injsafety}

The data we have searched through contains no previously identified \gwadj signals~\cite{gwtc3:2023}. However, there is a risk that the ArchEnemy search would identify real \gwadj signals as \scladj glitches. To assess this possibility, we simulate and add many \gwadj signals into the data and assess whether any signals are misidentified.

To do this, we use three separate sets of simulated \gwadj signals (or ``injection sets''), one for BBHs, another for binary neutron stars (BNSs) and a third for neutron star black hole (NSBH) systems. We use the same simulations as the LVK search of this data, detailed in the appendix of~\cite{gwtc3:2023}. Each injection set consists of $6200$ simulated signals spaced between $82$ and $120$ seconds apart. We treat these injection sets exactly the same as for the injection-less data, adding the simulations to the data, and then running ArchEnemy to produce a list of \scladj glitches for each injection set.

To determine whether we have misidentified any \gwadj injections as \scladj glitches we look for glitches we have found within the overlapping frequency band of \gwadj signals and our \scladj glitch template bank. This corresponds to approximately $15 \, \text{seconds}$ before merger time for the injections. The simulated signals occur every ${\sim}100$ seconds so we expect to see glitches within this $15$ second window, therefore, we also require that there must be more triggers identified in the \scladj glitch search \emph{with} injections when compared to the search \emph{without} injections within the window. The details of the number of \gwadj injections with overlapping \scladj triggers can be seen in Table~\ref{4:tab:coincident_triggers}.

\begin{table}[tb]
\caption{\label{4:tab:coincident_triggers}For both interferometers and all $3$ injection sets we identify the number of injections which are found to have \scladj glitches identified within $15$ seconds of merger time (``Injections with Coincident Triggers''), along with the number of \scladj glitches found within this window for these injections (``Scattered-Light Coincident Triggers''). We investigated each of these injections and recorded the number which actually had \scladj glitches identified due to the injected \gwadj signal (``Actual Overlapped Injections'').}
\footnotesize
\renewcommand{\arraystretch}{1.2}
\begin{tabular}{@{}cccccc}
\hline
&    & Injections with& Scattered-Light & Actual Overlapped \\
Interferometer & Injection Set & Coincident Triggers& Coincident Triggers&Injections \\
\hline
H1 & BBH           & 20 & 45 & 1\\
   & BNS           & 23 & 50 & 2\\
   & NSBH          & 38 & 73 & 7\\
L1 & BBH           & 13 & 21 & 2\\
   & BNS           & 18 & 30 & 2\\
   & NSBH          & 35 & 56 & 16\\
\hline
\end{tabular}

\end{table}

A \scladj glitch will be identified close to a \gwadj signal in two cases: the ArchEnemy search is misidentifying the \gwadj signal as a glitch \emph{or} the simulated signal was added close to actual glitches and a change in the data has meant a different number of glitches has been identified. The presence of real \scladj glitches means we might miss a \gwadj signal, therefore, we \emph{do} want to find and subtract glitches close to \gwadj signals, but we do not want to subtract power from the \gwadj signal itself. The \scladj glitch $\chi^{2}$ test was designed to prevent the matching of \scladj glitch templates to other causes of excess power, however, these results show it is not perfect.

\begin{figure}
  \centering
  \begin{minipage}[t]{1.0\linewidth}
    \centering
    \includegraphics[width=0.48\linewidth]{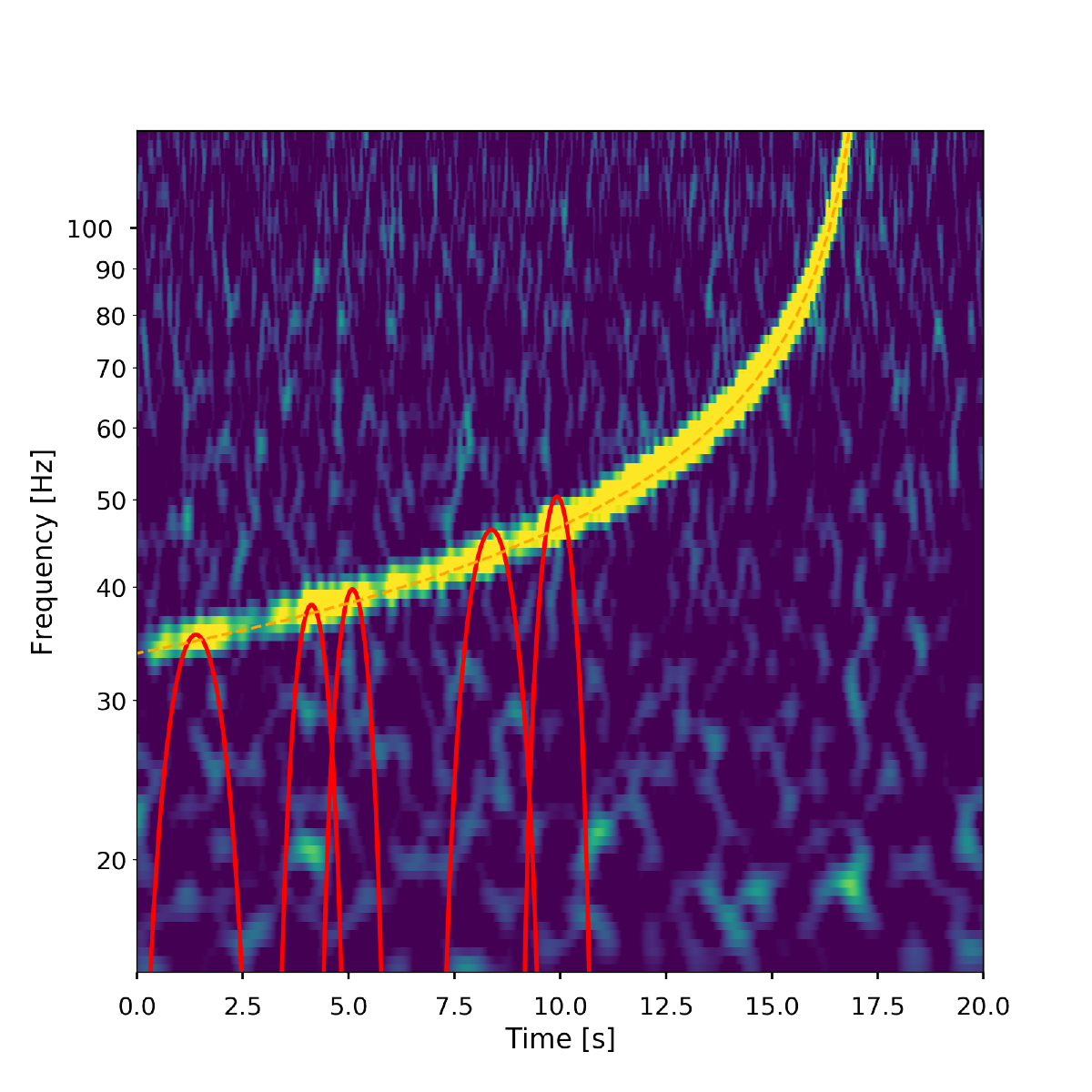}%
    \hfill%
    \includegraphics[width=0.48\linewidth]{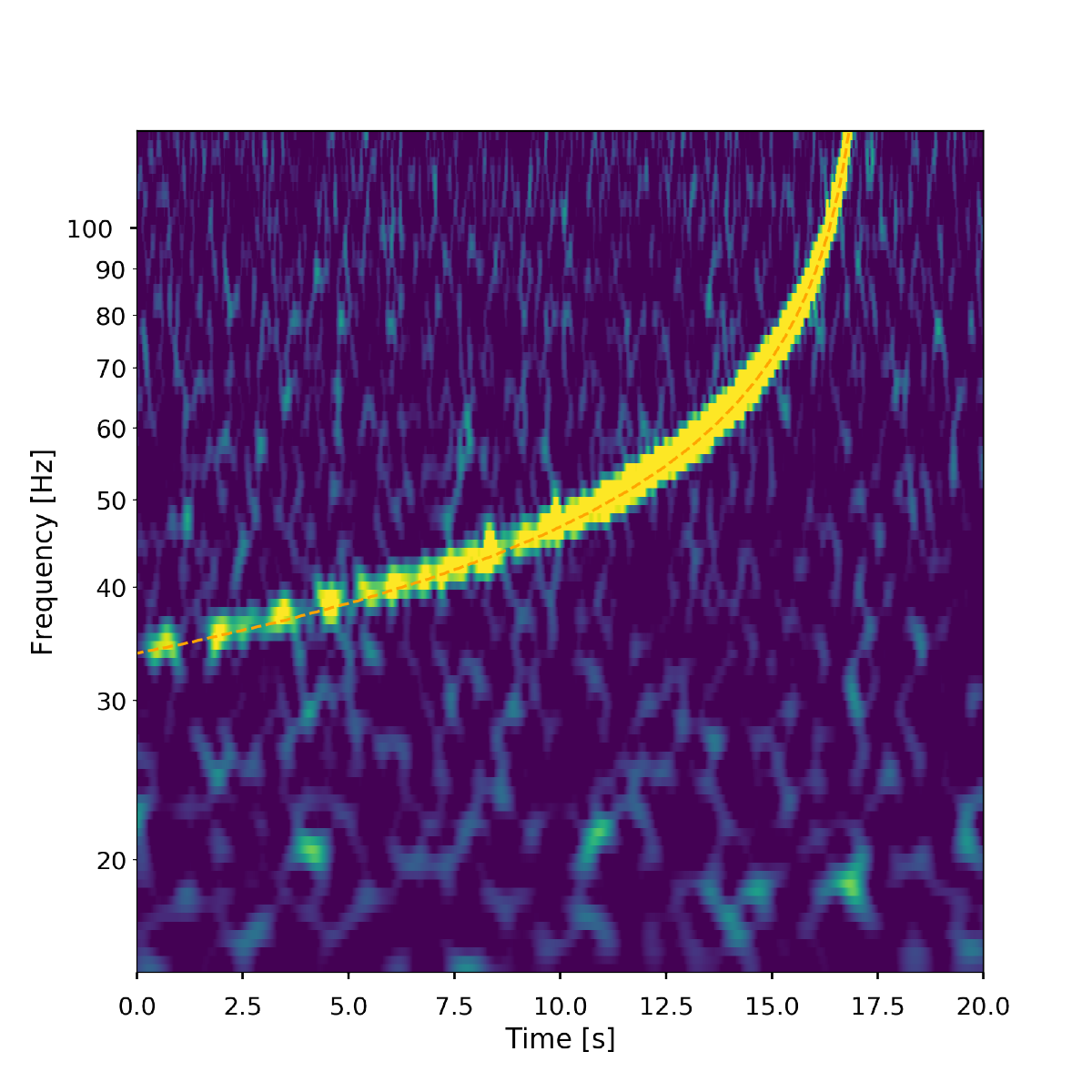}
  \end{minipage}
    \caption{An injected binary neutron star compact binary coalescence \gwadj signal, with the \scladj glitches identified by the ArchEnemy search pipeline overlaid in red (left). The same injected signal, but with the \scladj glitches removed from the data (right). It can be seen that power is removed from the signal track and also there is an amount of power added to the data above the track.}
    \label{4:fig:loud_inj}
\end{figure}

We investigate each injection with coincident \scladj triggers, seeing how many had misidentified \scladj glitches on the inspiral of the \gwadj signal, this number can be seen in the column ``Actual Overlapped Injections'' in Table~\ref{4:tab:coincident_triggers}. We have included an example of the matching of \scladj glitches onto \gwadj injections in Figure~\ref{4:fig:loud_inj}, the right panel shows the \gwadj data post glitch subtraction where it can be seen there is a portion of the power being subtracted from the signal. Although power is being removed from the signal, the \gwadj injection is still found by the search for \gws, which we will describe later. For the cases that we have investigated, we note that the behaviour shown in Figure~\ref{4:fig:loud_inj} only happens for signals that have very large SNR, and are therefore unphysically close to us. A similar effect is observed with the ``autogating'' process, described in~\cite{PyCBC:2016}, which prevents the detection of these loud signals. In contrast to the ``autogating'' though, signals like that illustrated in Figure~\ref{4:fig:loud_inj} are still identified as \gwadj signals by the PyCBC search after \scladj glitch removal.

\section{\label{4:sec:results}Assessing sensitivity gain from removing \scladj glitches}

We now assess whether removing our identified list of \scladj glitches results in a sensitivity gain when searching for CBCs. We do this by comparing the results from the offline PyCBC search on the original data, to the results of the same search but analysing data where the glitches have been removed.

\subsection{Comparing search results with and without glitch subtraction}

The PyCBC pipeline is able to assess the significance of potential CBCs in a given stretch of data, and does the same with a set of simulated signals. This significance is quoted in terms of a ``false-alarm rate'', which denotes how often we would expect to see a non-astrophysical event at least as significant as the coincident trigger being considered. In this work, we assess sensitivity at a false-alarm threshold of 2 background events every year.

The data we have searched over contained no previously found \gwadj signals~\cite{gwtc3:2023} and our search after subtracting \scladj glitches identified no new \gwadj signals. While the search hasn't found any \gws, we can still measure the improvement in the sensitivity of the detectors by comparing the number of simulated signals identified with a false-alarm rate below 2 per year for each injection set (described in Section~\ref{4:ssec:injsafety}) with and without removing glitches from the data. Table \ref{4:tab:found_injs} shows the number of injections found for all injection sets and both searches.
\begin{table}[tb]
\centering
\caption{\label{4:tab:found_injs}The number of injections found by each search with a false-alarm rate less than 2 per year alongside the number of newly-found and newly-missed injections, those found by the glitch-subtracted and not the original search and vice versa. We also show the sensitivity ratio of the glitch-subtracted search and original search for each injection set.}
\footnotesize
\renewcommand{\arraystretch}{1.2}
\begin{tabular}{@{}cccccc}
\hline
Injection & Original  & Glitch- & Sensitivity & Newly & Newly \\
Type & Search & Subtracted & ratio & Found & Missed \\
\hline
BBH & 1215 & 1222 & 1.01 & 10 & 3 \\
BNS & 1315 & 1315 & 1.00 & 5 & 5 \\
NSBH & 1260 & 1265 & 1.00 & 8 & 3 \\
\hline
\end{tabular}

\end{table}

We compare the number of injections found by both searches but also look at the \gwadj injections found by the original search and missed by the glitch-subtracted search and vice versa, this information can be seen in Table~\ref{4:tab:found_injs}. Considering signals found by the original search and missed by the glitch-subtracted search there are $3$ binary black hole injections with false-alarm rates in the original search ranging from $0.5\text{--}0.3$ per year, one of which had a glitch removed approximately $9$ seconds after the injection, there are $5$ newly-missed binary neutron star injections with false-alarm rates ranging from $0.5\text{--}0.056$ per year, two of the five binary neutron star injections had glitches removed within $60$ seconds of the injection, and there are $3$ newly-missed neutron star black hole injections with false-alarm rates ranging from $0.5\text{--}0.086$ per year, one had glitches removed within $60$ seconds of the injection. The other newly-missed injections showed no \scladj glitches within a $20$ second window for binary black hole injections and a $60$ second window for binary neutron star and neutron star black hole injections.

\begin{figure}
  \centering
  \begin{minipage}[t]{1.0\linewidth}
    \centering
    \includegraphics[width=0.48\linewidth]{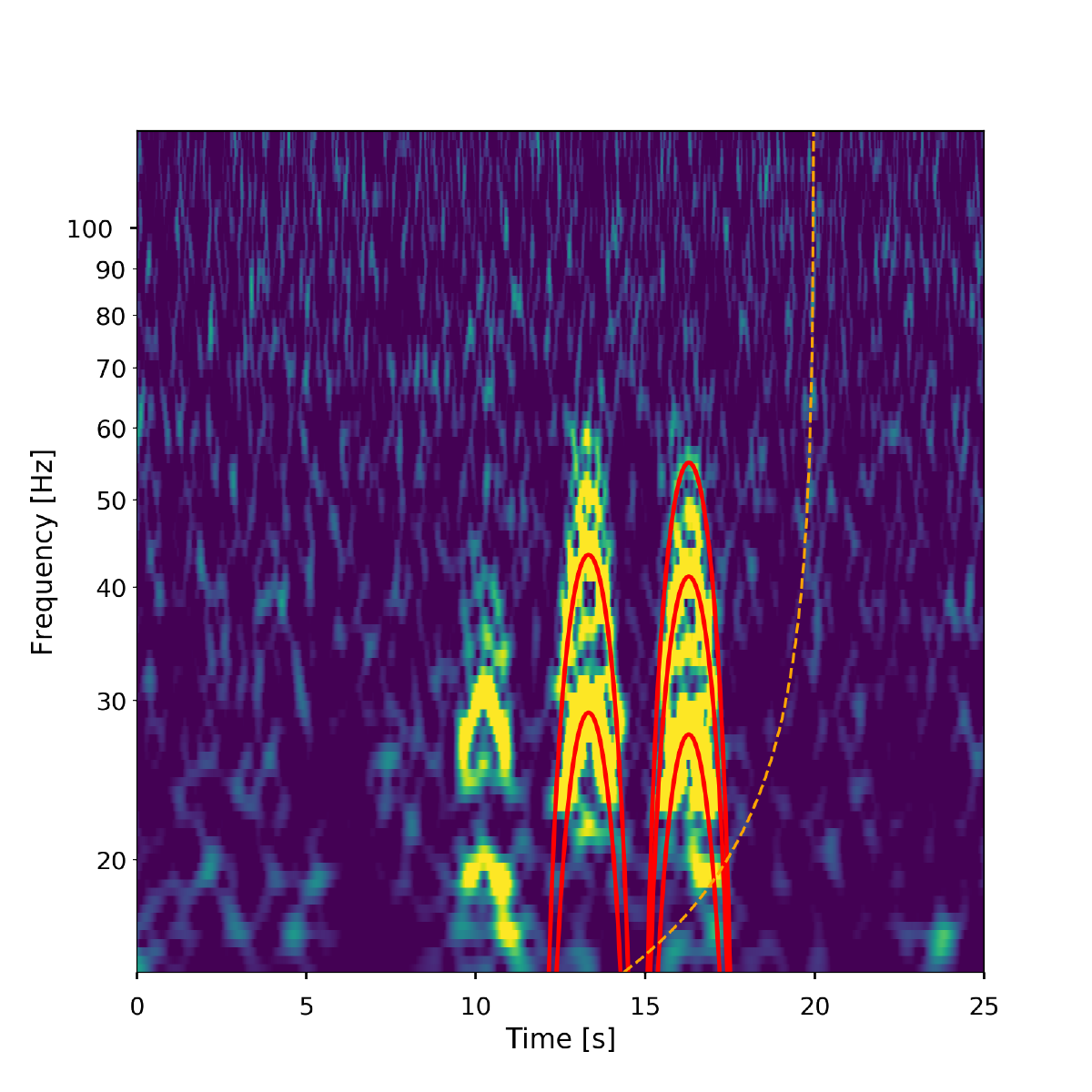}%
    \hfill%
    \includegraphics[width=0.48\linewidth]{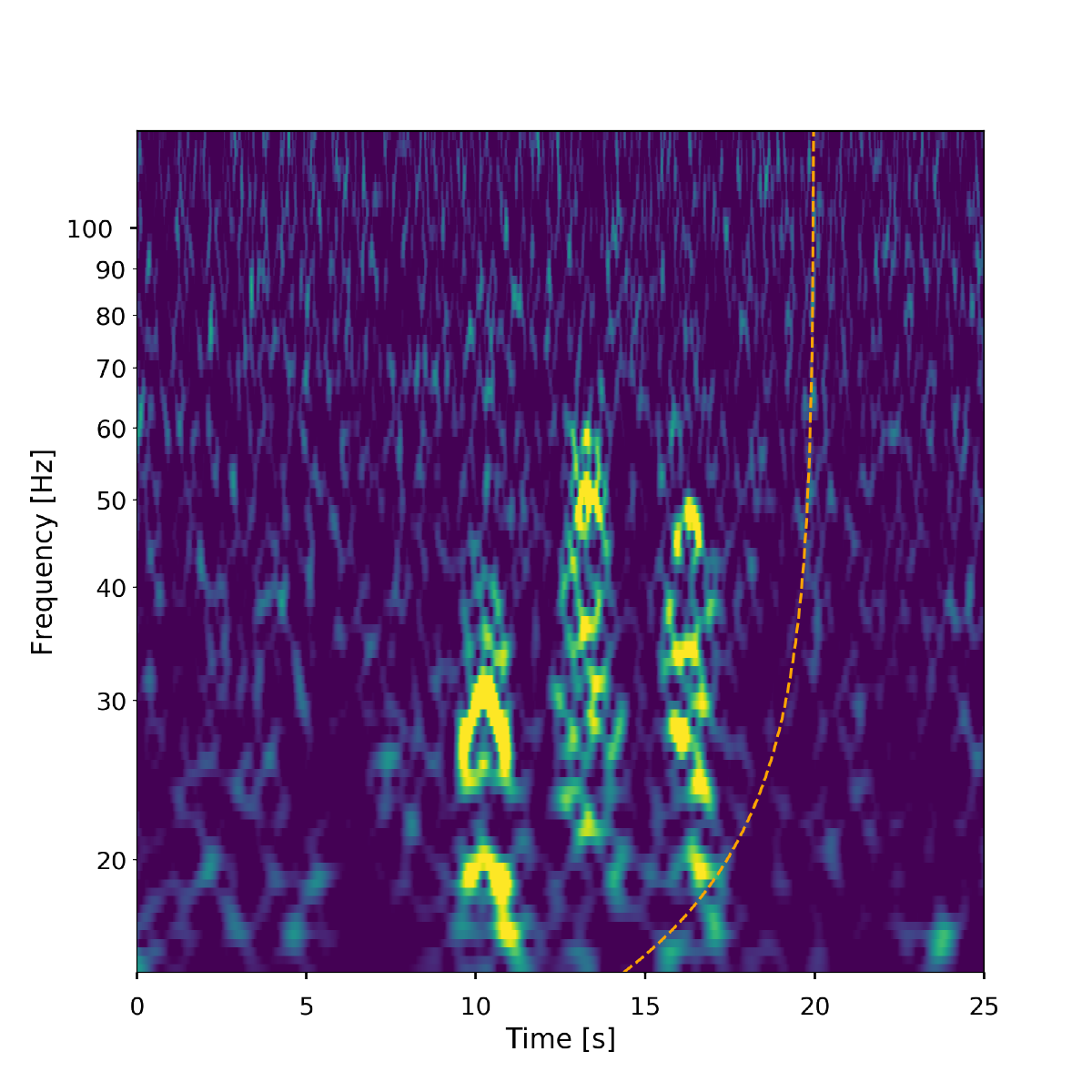}
  \end{minipage}

  \vspace{0.2cm}

  \begin{minipage}[t]{1.0\linewidth}
    \centering
    \includegraphics[width=0.48\linewidth]{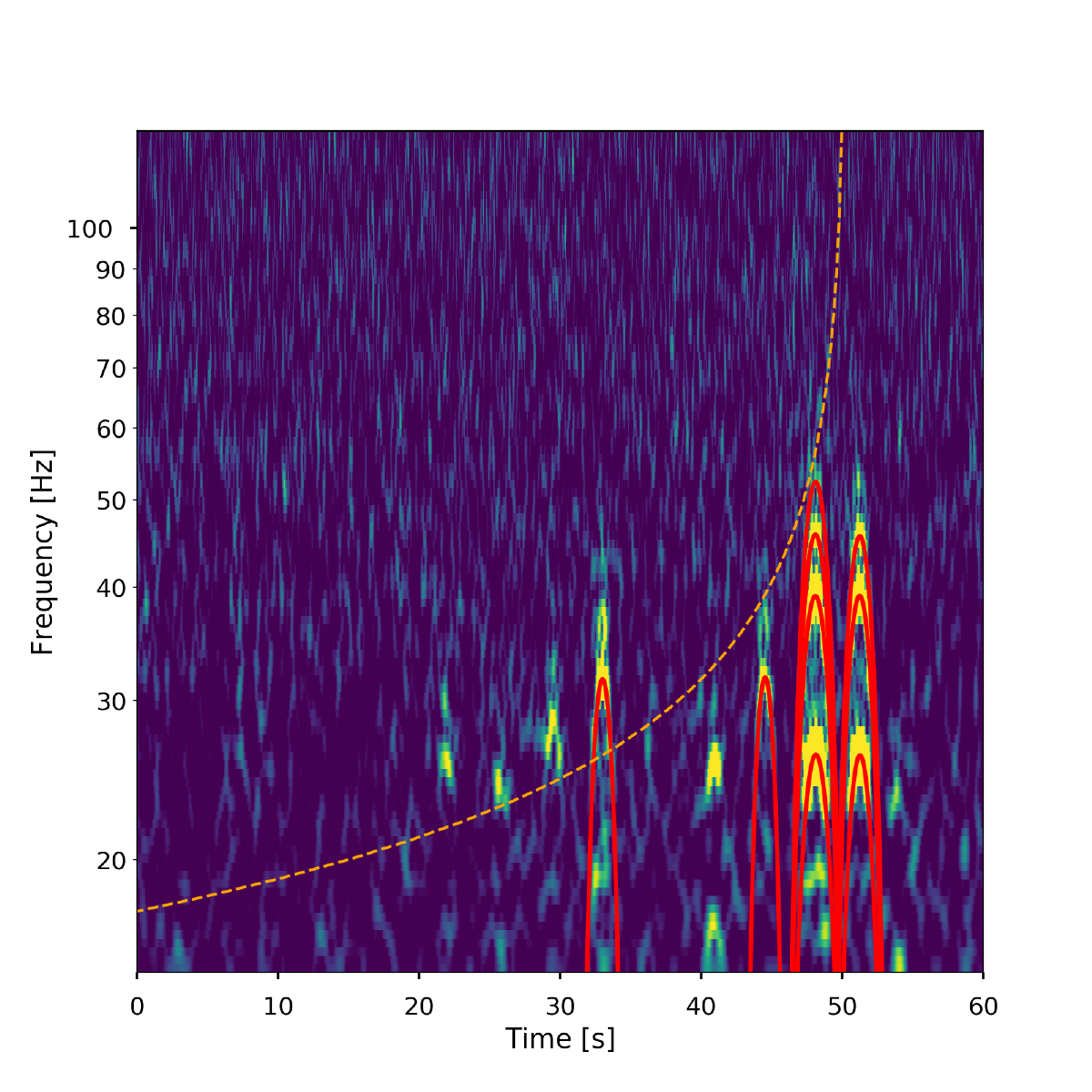}%
    \hfill%
    \includegraphics[width=0.48\linewidth]{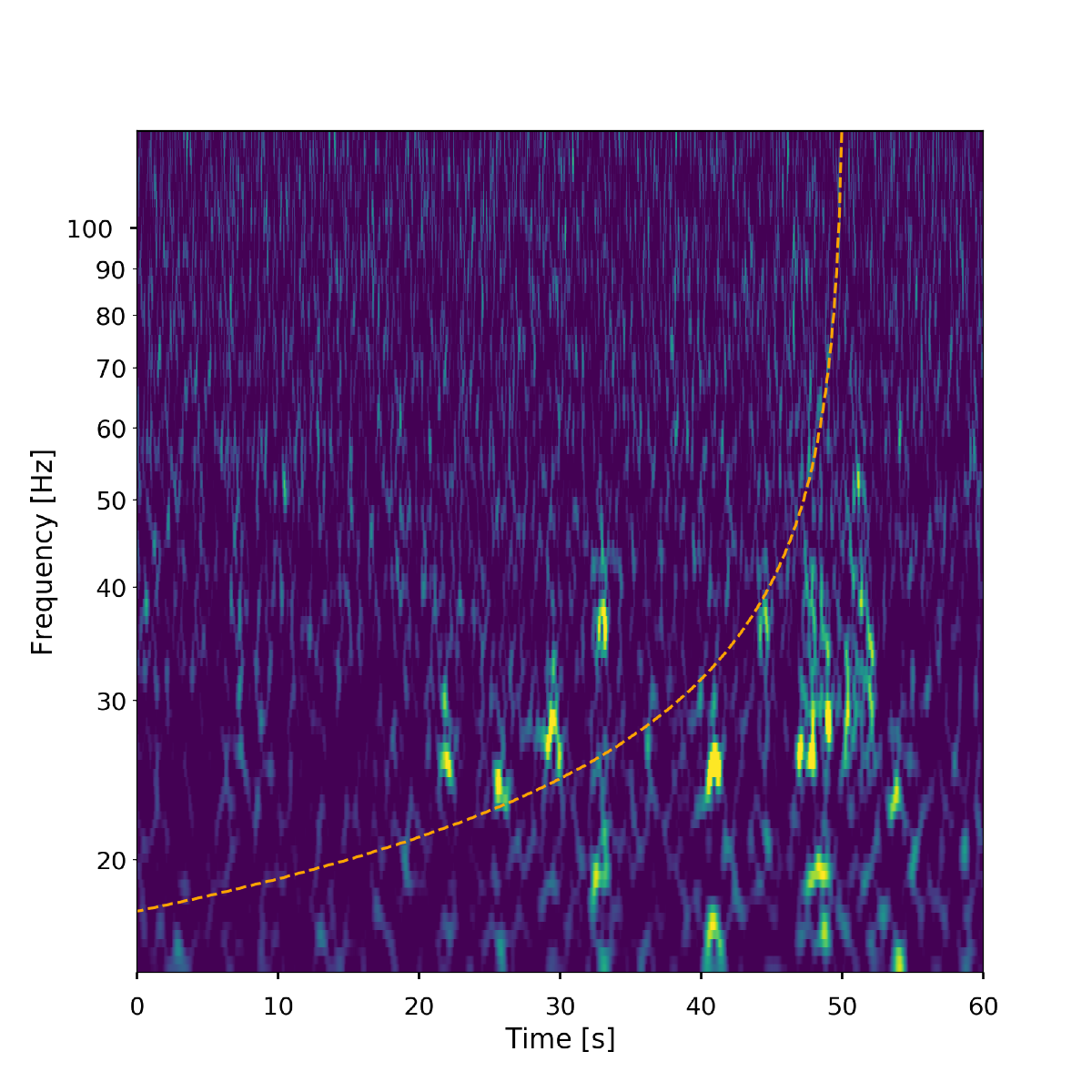}
  \end{minipage}
    \caption{Two examples of \gwadj injections found by the glitch-subtracted search for \gws which were not found by the original \gwadj search due to the presence of \scladj glitches at the same time as the \gwadj inspiral. Top left: A binary black hole injection with a false-alarm rate of 1 per $8643.73$ years is shown alongside the \scladj glitches found by the ArchEnemy search and subtracted from the data prior to performing the glitch-subtracted PyCBC search for \gws (top right). Bottom left: A neutron star black hole injection with a false-alarm rate of 1 per $9961.55$ years and the \scladj glitches found by the ArchEnemy search and subtracted from the data prior to performing the glitch-subtracted PyCBC search for \gws (bottom right).}
    \label{4:fig:ae_found}
\end{figure}

The glitch-subtracted search identifies $10$ additional binary black hole injections, the most significant of which have false-alarm rates of 1 per $190.50$, 1 per $7633.84$ and 1 per $8643.73$ years. We illustrate the last of these in Figure~\ref{4:fig:ae_found} (top). $5$ extra binary neutron star injections were found, with false-alarm rates from $0.5\text{--}0.19$ per year and $8$ neutron star black hole injections were found, where the false-alarm rate of the most significant is 1 per $9961.55$ years. This injection can also be seen in Figure~\ref{4:fig:ae_found} (bottom). We find $6$ of the $10$ binary black hole injections have \scladj glitches within a $20$ second window of the injection, $3$ of $5$ binary neutron star injections have \scladj glitches within a $60$ second window of the injection and, $6$ of the $8$ neutron star black hole injections have \scladj glitches within a $60$ second window of the injection. We provide more details about the newly-found and newly-missed injections in~\ref{4:sec:apdx_injections_table}

To quantify the sensitivity of the search, we calculate the sensitive volume in which we can observe \gwadj signals. To calculate the sensitive volume we measure the detection efficiency of different distance bins taken from the injection sets and then multiply the efficiencies by the volume enclosed by the distance bins, these volumes are then summed to find the total volume the search is sensitive to~\cite{rw_snr_eq:2012}. We are then able to calculate the ratio in sensitivities between the glitch-subtracted \gwadj search and the original \gwadj search,  revealing the improvement that subtracting \scladj glitches has made.

\begin{figure}
     \centering
     \includegraphics[width=0.7\textwidth]{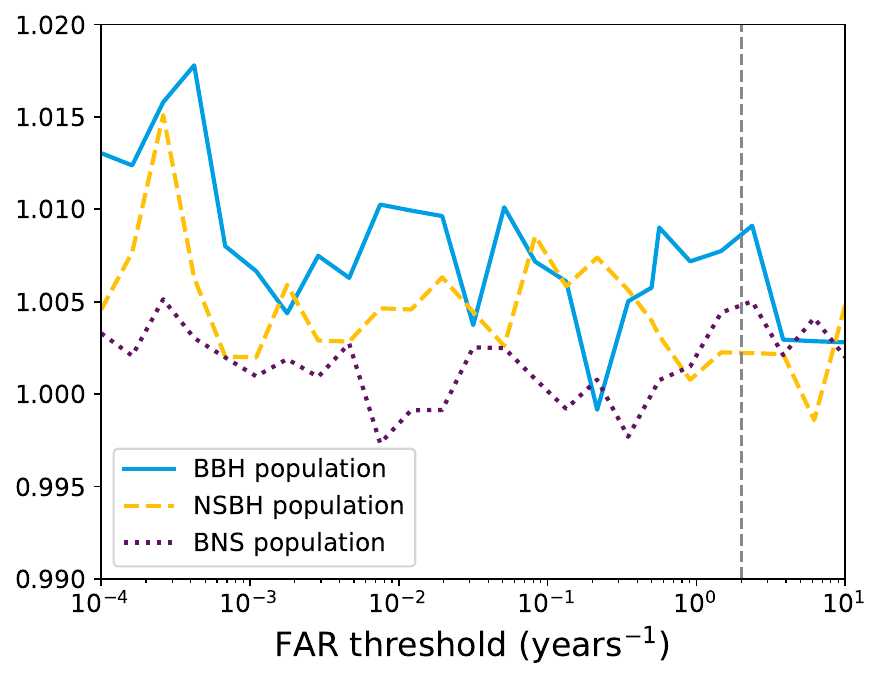}
     \caption{The ratio of the sensitive volume-time of the glitch-subtracted search and the original \gwadj search. The grey dashed line indicates a false-alarm rate of $2$ per year, which is our threshold and the point at which we measure any sensitivity improvements of the glitch-subtracted search for each of the three \gwadj injection sets. }
     \label{4:fig:allinj_vt_ratio}
\end{figure}

Figure~\ref{4:fig:allinj_vt_ratio} displays the ratio of the sensitive volume measured for the glitch-subtracted \gwadj search and the original PyCBC \gwadj search, across different false-alarm rate values, we quote our sensitivity ratios at a false-alarm rate value of 2 per year. The same set of injected signals was used for both \gwadj searches, and therefore a direct comparison of search sensitivities can be made via this ratio. Disappointingly, the measured sensitivity improvement is small in the results we obtain. For the binary black hole injections we measure a sensitivity ratio at a 2 per year false-alarm rate of $1.01$, for binary neutron stars $1.00$ and neutron-star--black-holes $1.00$.

The statistical significance of the sensitivity increase we report for the binary black hole injection set can be found by investigating the null hypothesis of seeing the same $1\%$ increase under the assumption that the subtraction of \scladj glitches does not actually increase sensitivity. When performing this analysis we find that our result is not statistically significant at the 95\% confidence interval --i.e. there is a 5.24\% chance that we would measure such an increase in sensitivity at least as large as this under the null hypothesis-- (see~\ref{4:sec:apdx_stat_sig} for details). However, the marginal sensitivity increase would not justify repeating the \scladj glitch search and glitch-subtracted \gwadj search on a larger injection set, instead more work is needed to better identify and remove \scladj glitches while remaining safe in the presence of \gwadj signals.

\subsection{\label{4:sec:apdx_injections_table}Injections tables}

Here we have three tables which contain data on the newly-found and newly-missed injections for each injection set. The tables are separated into the values for the inverse false-alarm rate and ranking statistic found for each injection in both searches then the SNR, $\chi^{2}$ and PSD variation values for each detector in both searches. The first table is the results of the binary black hole injection set, the second table is the binary neutron star injection set and the third is the neutron star black hole injection set. A horizontal dashed line separates newly-found and newly-missed injections, using a false-alarm rate threshold of 2 per year. There were 2 newly-found binary black hole injections which were not found at all by the original search, and therefore they do not appear in the first table.

There have been 34 newly-found or newly-missed \gwadj injections when subtracting \scladj glitches, it is informative to understand how the \gwadj search is influenced by the glitch subtraction to cause this outcome. The ranking statistic~\cite{PyCBC_global:2020} represents the legitimacy of a signal being astrophysical in origin and is partially computed using the re-weighted SNR, which is itself computed using the initial SNR alongside the various \gwadj discriminators and the PSD variation measurement~\cite{PSD_var:2020}. We use the trigger information saved by the \gwadj search to identify why injections that weren't found previously have been found post-glitch subtraction, and vice versa.

As an example, we take the smallest false-alarm rate (1 per $8643.73$ years), newly-found, binary black hole injection, and look at the ranking statistic, SNR, $\chi^{2}$ and, PSD variation measurements in both detectors and both searches -- these values can be found in Table~\ref{4:tab:apdx_changed_snr_bbh}. This injection was originally seen with a false-alarm rate of 100 per year, far above our threshold, a very large increase in the ranking statistic from 13.29 to 27.01 is certainly responsible for the decreased false-alarm rate. There were no changes in the values measured by the LIGO-Livingston detector between searches which is expected as no \scladj glitches were found within $512$ seconds of the injection. The LIGO-Hanford detector sees a small increase in the SNR measured, from 7.98 to 8.22, a small increase in the $\chi^{2}$ value, from 2.32 to 2.48, and a very significant decrease in the PSD variation measurement, from 3.47 to 1.50. Using equation 18 of~\cite{PSD_var:2020}, we can calculate a re-weighted SNR of $4.60$ for the original search and $5.67$ for the glitch-subtracted search, a significant increase in the SNR. Similar analyses for all the newly-found and newly-missed injections can be made using information found in tables~\ref{4:tab:apdx_changed_snr_bns} and~\ref{4:tab:apdx_changed_snr_nsbh}.

The decrease in the PSD variation is true for the three newly-found very low false-alarm rate injections, accompanied by the small changes in SNR and increase in the $\chi^{2}$ measurement. For newly-found and newly-missed injections which lie close to the 2 per year false-alarm rate threshold, there is no definitive reason as to why these injections changed state.

\newpage
\newgeometry{left=2cm,right=2cm,top=2cm,bottom=2cm} 
\begin{landscape}
\begin{table}[tb]
\centering
\caption{\label{4:tab:apdx_changed_snr_bbh}This table contains the trigger information for the newly-found and newly-missed \textbf{binary black hole injections} recorded by the original search~\cite{gwtc3:2023} and the glitch-subtracted search.}
\begin{tabular}{|c|c|c|c|c|c|c|c||c|c|c|c|c|c|c|c|}
\hline
\multicolumn{8}{|c||}{Glitch-Subtracted} & \multicolumn{8}{c|}{Original Search} \\
\hline
\multicolumn{2}{|c|}{} & \multicolumn{3}{c|}{H1} & \multicolumn{3}{c||}{L1} & \multicolumn{2}{c|}{} & \multicolumn{3}{c|}{H1} & \multicolumn{3}{c|}{L1}\\
\hline
IFAR & Ranking & SNR & $\chi^{2}$ & PSD & SNR & $\chi^{2}$ & PSD & IFAR & Ranking & SNR & $\chi^{2}$ & PSD & SNR & $\chi^{2}$ & PSD \\ &

Stat. & & & Var. & & & Var. & & Stat. & & & Var. & & & Var.\\
\hline
8643.73 & 27.01 & 8.22 & 2.48 & 1.50 & 8.05 & 2.07 & 1.00 & 0.01 & 13.29 & 7.98 & 2.32 & 3.47 & 8.05 & 2.07 & 1.00 \\
7633.84 & 26.13 & 8.18 & 2.25 & 1.22 & 7.58 & 1.46 & 1.03 & 0.37 & 16.63 & 8.19 & 1.87 & 2.17 & 7.59 & 1.45 & 1.02 \\
4.89 & 19.06 & 7.23 & 1.88 & 0.97 & 10.34 & 3.15 & 1.05 & 1.56 & 18.01 & 7.24 & 2.11 & 0.97 & 10.34 & 3.42 & 1.04 \\
3.14 & 18.65 & 5.89 & 1.54 & 1.01 & 8.86 & 2.40 & 1.09 & 1.52 & 17.99 & 5.89 & 1.54 & 1.01 & 8.86 & 2.40 & 1.09 \\
2.75 & 18.53 & 8.50 & 1.63 & 0.99 & 6.80 & 1.37 & 1.05 & 1.51 & 17.98 & 8.51 & 1.56 & 0.98 & 6.80 & 1.37 & 1.05 \\
2.39 & 18.41 & 7.83 & 2.00 & 1.01 & 8.41 & 3.41 & 1.02 & 1.06 & 17.65 & 7.81 & 2.14 & 1.01 & 8.41 & 3.41 & 1.02 \\
2.18 & 18.32 & 7.02 & 1.90 & 1.02 & 6.35 & 2.15 & 1.01 & 1.79 & 18.14 & 7.05 & 1.92 & 1.02 & 6.35 & 2.16 & 1.01 \\
2.01 & 14.11 & 7.94 & 1.89 & 1.01 & 5.67 & 1.95 & 1.16 & 1.90 & 14.14 & 7.94 & 1.90 & 1.01 & 5.07 & 0.00 & 1.02 \\
\hdashline
1.76 & 18.12 & 6.89 & 1.43 & 1.02 & 6.76 & 2.59 & 0.97 & 3.33 & 18.71 & 6.86 & 1.82 & 1.02 & 6.68 & 2.53 & 0.97 \\
1.79 & 18.14 & 6.40 & 1.69 & 1.02 & 7.56 & 2.01 & 1.01 & 2.05 & 18.27 & 6.40 & 1.69 & 1.02 & 7.56 & 2.00 & 1.01 \\
1.65 & 18.06 & 5.14 & 0.00 & 0.91 & 7.57 & 1.57 & 1.02 & 2.05 & 18.27 & 5.14 & 0.00 & 0.91 & 7.57 & 1.57 & 1.02 \\
\hline
\end{tabular}
\end{table}
\end{landscape}
\restoregeometry 

\newpage

\newgeometry{left=1cm,right=1cm,top=2cm,bottom=2cm} 
\begin{landscape}
\begin{table}[tb]
\centering
\caption{\label{4:tab:apdx_changed_snr_bns}This table contains the trigger information for the newly-found and newly-missed \textbf{binary neutron star injections} recorded by the original search~\cite{gwtc3:2023} and the glitch-subtracted search.} 
\begin{tabular}{|c|c|c|c|c|c|c|c||c|c|c|c|c|c|c|c|}
\hline
\multicolumn{8}{|c||}{Glitch-Subtracted} & \multicolumn{8}{c|}{Original Search} \\
\hline
\multicolumn{2}{|c|}{} & \multicolumn{3}{c|}{H1} & \multicolumn{3}{c||}{L1} & \multicolumn{2}{c|}{} & \multicolumn{3}{c|}{H1} & \multicolumn{3}{c|}{L1}\\
\hline
IFAR & Ranking & SNR & $\chi^{2}$ & PSD & SNR & $\chi^{2}$ & PSD & IFAR & Ranking & SNR & $\chi^{2}$ & PSD & SNR & $\chi^{2}$ & PSD \\ &
Stat. & & & Var. & & & Var. & & Stat. & & & Var. & & & Var.\\
\hline
5.12 & 19.12 & 5.86 & 1.99 & 1.01 & 7.15 & 1.94 & 1.02 & 1.49 & 17.96 & 5.74 & 2.14 & 1.01 & 7.15 & 1.94 & 1.02 \\
4.91 & 19.07 & 5.48 & 2.29 & 1.01 & 7.92 & 2.16 & 1.01 & 1.37 & 17.89 & 5.78 & 2.25 & 1.01 & 7.23 & 2.13 & 1.01 \\
4.54 & 18.99 & 4.82 & 0.00 & 1.07 & 8.68 & 2.05 & 1.06 & 1.66 & 18.07 & 4.82 & 0.00 & 1.14 & 8.68 & 2.05 & 1.06 \\
2.46 & 18.43 & 5.48 & 1.95 & 1.05 & 7.61 & 2.03 & 0.99 & 1.78 & 18.13 & 5.48 & 1.94 & 1.05 & 7.62 & 2.08 & 0.99 \\
2.34 & 18.39 & 6.25 & 2.18 & 1.01 & 6.82 & 1.96 & 1.00 & 1.45 & 17.94 & 6.09 & 2.05 & 1.01 & 6.83 & 2.01 & 1.00 \\
\hdashline
1.24 & 17.71 & 6.58 & 2.14 & 1.01 & 6.89 & 2.48 & 1.01 & 17.81 & 20.19 & 6.58 & 2.15 & 1.01 & 7.16 & 2.26 & 1.01 \\
1.14 & 17.72 & 6.00 & 2.67 & 0.98 & 8.07 & 2.66 & 1.22 & 7.14 & 19.44 & 6.17 & 2.45 & 0.97 & 8.26 & 2.80 & 1.22 \\
1.89 & 18.19 & 6.32 & 2.19 & 1.00 & 6.79 & 1.83 & 0.99 & 3.57 & 18.78 & 6.37 & 2.12 & 0.99 & 6.79 & 1.75 & 0.99 \\
0.88 & 17.46 & 6.88 & 2.14 & 0.99 & 6.19 & 2.12 & 0.99 & 2.33 & 18.39 & 6.87 & 2.07 & 0.99 & 6.19 & 1.99 & 0.99 \\
1.36 & 17.80 & 6.20 & 2.02 & 1.00 & 6.83 & 2.35 & 1.00 & 2.07 & 18.18 & 6.20 & 2.02 & 1.00 & 6.83 & 2.28 & 1.00 \\
\hline
\end{tabular}
\end{table}
\end{landscape}
\restoregeometry 

\newpage

\newgeometry{left=1cm,right=1cm,top=2cm,bottom=2cm} 
\begin{landscape}
\begin{table}[tb]
\centering
\caption{\label{4:tab:apdx_changed_snr_nsbh}This table contains the trigger information for the newly-found and newly-missed \textbf{neutron star black hole injections} recorded by the original search~\cite{gwtc3:2023} and the glitch-subtracted search.}
\begin{tabular}{|c|c|c|c|c|c|c|c||c|c|c|c|c|c|c|c|}
\hline
\multicolumn{8}{|c||}{Glitch-Subtracted} & \multicolumn{8}{c|}{Original Search} \\
\hline
\multicolumn{2}{|c|}{} & \multicolumn{3}{c|}{H1} & \multicolumn{3}{c||}{L1} & \multicolumn{2}{c|}{} & \multicolumn{3}{c|}{H1} & \multicolumn{3}{c|}{L1}\\
\hline
IFAR & Ranking & SNR & $\chi^{2}$ & PSD & SNR & $\chi^{2}$ & PSD & IFAR & Ranking & SNR & $\chi^{2}$ & PSD & SNR & $\chi^{2}$ & PSD \\ &
Stat. & & & Var. & & & Var. & & Stat. & & & Var. & & & Var.\\
\hline
9961.55 & 27.41 & 5.55 & 2.08 & 1.23 & 10.26 & 2.27 & 1.12 & 0.01 & 12.83 & 6.01 & 2.39 & 2.25 & 8.54 & 2.37 & 1.12 \\
141.35 & 22.28 & 7.76 & 2.20 & 1.02 & 6.54 & 2.35 & 1.01 & 1.21 & 17.77 & 7.83 & 1.94 & 1.02 & 7.29 & 2.50 & 1.00 \\
18.26 & 20.36 & 5.94 & 2.20 & 1.09 & 6.02 & 1.99 & 1.10 & 1.29 & 17.82 & 5.74 & 2.05 & 1.09 & 6.27 & 2.10 & 1.18 \\
17.12 & 20.29 & 7.64 & 2.00 & 1.11 & 6.66 & 2.15 & 1.05 & 1.37 & 17.89 & 7.36 & 2.35 & 1.11 & 6.66 & 2.15 & 1.05 \\
12.94 & 20.03 & 6.52 & 1.79 & 1.07 & 7.31 & 2.32 & 1.13 & 1.96 & 18.23 & 6.57 & 1.85 & 1.20 & 7.31 & 2.32 & 1.13 \\
7.12 & 19.42 & 5.13 & 0.00 & 1.05 & 8.33 & 2.08 & 1.04 & 0.03 & 14.23 & 5.30 & 2.16 & 1.05 & 7.44 & 2.43 & 1.04 \\
3.61 & 18.77 & 5.80 & 2.01 & 1.02 & 7.43 & 1.88 & 0.93 & 1.81 & 18.15 & 5.80 & 2.01 & 1.02 & 7.43 & 1.97 & 0.93 \\
2.15 & 18.22 & 7.41 & 2.04 & 1.02 & 5.57 & 2.03 & 1.03 & 0.00 & 4.34 & 5.64 & 2.18 & 0.98 & 5.04 & -0.00 & 1.01 \\
\hdashline
0.23 & 16.16 & 5.79 & 1.81 & 0.96 & 8.78 & 3.93 & 1.03 & 11.69 & 19.93 & 5.78 & 1.89 & 0.96 & 8.79 & 3.17 & 1.03 \\
1.56 & 17.92 & 6.71 & 1.91 & 0.99 & 6.73 & 2.50 & 1.05 & 2.68 & 18.41 & 6.71 & 1.90 & 0.99 & 6.72 & 2.39 & 1.05 \\
1.72 & 18.10 & 5.77 & 1.90 & 0.99 & 6.77 & 1.74 & 1.00 & 2.08 & 18.28 & 5.77 & 1.90 & 0.99 & 6.80 & 1.69 & 1.00 \\
\hline
\end{tabular}
\end{table}
\end{landscape}
\restoregeometry 

\section{\label{4:sec:apdx_stat_sig}Statistical significance}

We report a $1\%$ increase in the sensitivity for the binary black hole injection set in the glitch-subtracted \gwadj search (see Section~\ref{4:sec:results}). We wish to determine the statistical significance of this result under the null hypothesis that subtracting \scladj glitches prior to searching for \gws does not increase the sensitivity of the \gwadj search. The binary black hole injection set contains $6200$ injected signals, $10$ additional injections were found in the glitch-subtracted \gwadj search, $3$ additional injections were missed in our search, this gives us $13$ injections which have changed state.

First, we calculate the probability that an injection has been affected by the glitch removal,
\begin{equation}
    p = \frac{13}{6200} = 0.21\% ,
\end{equation}
then we calculate the standard deviation,
\begin{equation}
    \textrm{std} = \sqrt{n * p * (1 - p)} = 3.60 .
\end{equation}
Using the standard deviation we calculate the number of standard deviations our result deviates from the mean. We divide the number of positively changed (newly-found) injections by the standard deviation,
\begin{equation}
    \textrm{standard deviation} = \frac{(10 - 3)}{3.60} = 1.94 .
\end{equation}
Under the assumption of no sensitivity increase caused by the subtraction of glitches, we measure our result of +7 newly-found \gwadj injections to lie $1.94$ standard deviations from an expected value of $0$ newly-found injections. The critical value of a 95\% confidence interval, that is to say there is a 1 in 20 chance of our null hypothesis being true, is $1.96$ meaning our result is within the 95\% confidence interval. We can describe this as there being a 5.24\% chance that our result is not caused by the subtraction of glitches but is instead caused by random chance. To reduce the error in the computed sensitivity ratio a larger injection set test would be required, this would need a large time and computational power investment which we do not believe is justified in the case of such a minor increase in the sensitivity.

\section{\label{4:sec:conclusion}Conclusion}

We have demonstrated a new method for modelling \scladj glitches and identifying and characterizing these glitches in a period of \gwadj data. We have developed a \scladj glitch specific $\chi^{2}$ test which can discriminate between \scladj glitches, other types of glitches and \gwadj signals. We have searched through a representative stretch of \gwadj data known to contain \scladj glitches, found thousands of these glitches and subtracted them from the \gwadj data prior to running a search for \gws. The results of this search include a small increase in the measured sensitivity of the \gwadj search for binary black hole \gwadj signals, and modest change to sensitivity for binary neutron star and neutron star black hole \gwadj signals.

We highlight that the task of accurately identifying and parameterizing \scladj glitches in the data is not a trivial one, especially where there are repeated, and harmonic, glitches present in the data. We have developed a new $\chi^{2}$ test to reduce the number of false identifications of \scladj glitches, but we do still see cases where we have misidentified other glitches, and even some loud \gwadj signals, as caused by \scl, and cases where we do not correctly identify, or parameterize, actual \scladj glitches. Improving this identification process would be important in improving the efficacy of this process.

The possibility of using this model of \scladj glitches as a bespoke application to \gwadj signals which are known to have coincident \scladj glitches has been explored and implemented into Bilby~\cite{BILBY:2019} to perform a parameter estimation of \scladj glitches and removing these to produce glitch-free data~\cite{Udall:2023}. The inclusion of the extra term from~\cite{Was_Subtract:2021} within the model can help identify \scladj glitches more accurately. Selectively subtracting glitches based on the presence of a \gwadj signal is possible by moving the glitch subtraction process inside the \gwadj searches and including the results of other \gwadj discriminators~\cite{rw_snr_eq:2012, McIsaac_Chi:2022} to determine the legitimacy of an ArchEnemy identified glitch.

The results of the application of the ArchEnemy search pipeline, the list of \scladj glitches, can also be used in other applications. For example, it could be used in the form of a veto~\cite{O2O3_DetChar:2021}, where we use knowledge of the presence of \scladj glitches to down rank periods of time in \gwadj data. Additionally, we could use \scladj glitches previously identified by tools such as Gravity Spy~\cite{gravityspy:2023} and target known \scladj glitches with the ArchEnemy search pipeline.

As a final note, while we acknowledge that the sensitivity improvements that we have observed---${\sim} 1\%$---are very modest, the concept of removing \scladj glitches, or other identified glitch classes, from the data prior to matched filtering for CBCs is one that we encourage others to explore further. An increase in the rate of events or the rate of \scladj glitches in future observing runs will mean an increase in the number of affected events, such techniques offer a method for mitigating the effect that these glitches will have on the search, maximizing the number of observations that can be made.

\chapter[Improving the PyCBC Live Ranking Statistic]{Improving the PyCBC Live Ranking Statistic}
\label{chapter:5-pycbc-live}
\chapterquote{Have patience, all will be explained.}{Guthix}
This work will be published as part of the PyCBC Live fourth observing run methods paper, which is currently being drafted. I am responsible for the changes concerning the exponential noise model in the PyCBC Live ranking statistic as well as contributing to the development necessary to introduce these changes in the PyCBC Live infrastructure.

\section{\label{5:sec:introduction}Introduction}

Searching for \gws is performed by search pipelines. The search pipelines operate across two timelines: low-latency, rapid detection of \gwadj signals in real time, and offline, which is performed months after the data has been obtained. Low-latency search pipelines are optimised for rapid detection to disseminate information about potential \gwadj signals to the wider scientific community in as little time as possible. The latency between a \gwadj signal arriving at the detectors and being detected by search pipelines is crucial for multi-messenger events.

Several offline \gwadj search pipelines were operational during the third observing run: cWB~\cite{cWB:2020}, GstLAL~\cite{GstLAL:2020}, MBTA~\cite{MBTA:2021}, and PyCBC Offline~\cite{PyCBC_global:2020}. Table XIV in Appendix D 7a of the second half of the third observing run's catalogue paper~\cite{gwtc3:2023} highlights that the PyCBC Offline search was the most sensitive \gwadj search pipeline~\cite{PyCBC:2016, PyCBC:2017, PyCBC_package:2021}.

The low-latency \gwadj search pipelines include cWB~\cite{cWB:2020}, GstLAL~\cite{GstLAL:2020}, MBTA~\cite{MBTA:2021}, PyCBC Live~\cite{PyCBC_Live:2018}, SPIIR~\cite{SPIIR:2020}, and oLIB~\cite{oLIB:2015}. In the third observing run, the GstLAL pipeline~\cite{GstLAL:2020} uploaded the preferred event\footnote{The event with the highest SNR of those uploaded by all pipelines.} for $33$ of the $56$ low-latency super events, whereas PyCBC live uploaded the preferred event in only $15$ events~\cite{gracedb_superevents:2024}. The PyCBC live search pipeline is not the most sensitive in the low-latency regime, so we want to improve the PyCBC Live search to have the same sensitivity as its offline counterpart, and to do this we must look at the key differences between the two searches.

Offline searches for \gws can utilise post-detection information when processing their results and have a much larger limit on the amount of computational analysis that can be performed to find events. Low-latency \gwadj searches, such as PyCBC Live, receive and analyse data in regular fixed analysis segments, and all computational processing must be completed before the next analysis segment arrives. If the triggers in an analysis segment are not processed in time, lag is introduced as a backlog of triggers builds up faster than they can be processed. The maximum latency between an event arriving and being detected by PyCBC Live during the third observing run was $20$ seconds~\cite{PyCBC:2017}.

PyCBC Offline contains components and information that could not be previously used by the PyCBC Live search without violating the latency restrictions. Introducing these components into the live search will improve the sensitivity of the \gwadj search, bringing the two PyCBC searches closer to parity and enabling the detection of more \gwadj events with greater significance.

This chapter is laid out as follows: In Section~\ref{5:sec:ranking-stat} we describe how PyCBC ranks identified events to assess significance, in Section~\ref{5:sec:previous-stat} we discuss the ranking statistic used by the PyCBC searches during the third observing run and how these are constructed, in Section~\ref{5:sec:new-additions} we describe the new additions to the PyCBC Live ranking statistic we have implemented and how these have been adapted specifically for the PyCBC Live search, in Section~\ref{5:sec:injection-tests} we discuss how we evaluate the sensitivity improvement of our changes using an injection study and in Section~\ref{5:sec:sensitivity-improvements} we discuss the immediate sensitivity increase highlighted by the injection study. After this, we delve deeper into the regions of parameter space responsible for the sensitivity changes in sections~\ref{5:sec:injection-investigations} and~\ref{5:sec:investigating-regions}. We describe the test of the ranking statistic changes on the PyCBC Live mock data challenge infrastructure in Section~\ref{5:sec:mdc-test} and conclude in Section~\ref{5:sec:conclusion}.

\section{\label{5:sec:ranking-stat}The ranking statistic}

The ranking statistic is the detection statistic used to calculate the significance of a \gwadj detection. The ranking statistic values can be mapped directly to false-alarm rate (FAR), a key metric used to assess the likelihood that a detected signal is real and not a result of coincident noise triggers being found by the \gwadj search~\cite{PyCBC_global:2020}.

A ranking statistic combines numerous pieces of information about a candidate event to provide a single measure that reflects the event's significance. The ranking statistic can include information such as: single detector trigger signal-to-noise ratio (SNR), $\rho$, signal-consistency tests~\cite{Allen_Chi:2005, rw_snr_eq:2012, PyCBC_sg:2018} and, coincidence tests between detectors. It can also include more complex information like coincident phase and time difference likelihood based on source localisation and detector orientation~\cite{PyCBC:2017, PyCBC_singles:2022}. Different ranking statistics combine different pieces of information to make the final assessment of significance.

Improving the ranking statistic used in the PyCBC Live search will improve the search's ability to distinguish between real \gwadj events and false-alarms. The PyCBC Offline search's ranking statistic contains more information than the ranking statistic used by PyCBC Live during the third observing run, some of which we can adapt for the PyCBC Live search~\cite{PSD_var:2020, PyCBC:2017, PyCBC_global:2020}.

\section{\label{5:sec:previous-stat}PyCBC ranking statistics used in O3}

The PyCBC Live search will calculate the single detector ranking statistic, the SNR re-weighted by the $\chi^{2}$ tests, for each trigger for each online detector; coincidences are then formed between triggers to identify potential \gwadj events. The coincident triggers are then ranked by the coincident ranking statistic to provide the final ranking statistic value.

The PyCBC Offline search in the third observing run ranked single detector triggers by: SNR, $\rho$, calculated by the matched filter of template and data~\cite{FINDCHIRP:2012}; the traditional $\chi^{2}$, calculated by measuring the difference in the expected and actual $\rho$ for discrete frequency bins in the signal evolution~\cite{Allen_Chi:2005}; the sine-Gaussian $\chi^{2}$, which measures $\rho$ in frequency bins above the signal's maximum frequency~\cite{PyCBC_sg:2018} and; PSD variation, which estimates the effect of non-stationary noise on $\rho$, re-weighting $\rho$ prior to applying the $\chi^{2}$ tests~\cite{PSD_var:2020}. These four single detector ranking statistic components are applied to all triggers found by the search to give a new re-weighted $\rho$ value, $\hat{\rho}$~\cite{rw_snr_eq:2012}. \cite{Chatziioannou:2024} provides a detailed review of the $\chi^{2}$ tests currently being used in \gwadj searches.

After the calculation of $\hat{\rho}$ for each single detector trigger, the different detector triggers are combined and a coincidence test is applied to identify coincident triggers. The coincidence test checks whether the arrival time of the \gwadj signal at the separate detectors is possible given the light-travel time of the \gwadj signal, for example, the two LIGO detectors are separated by a straight line through the Earth $3002$ kilometres long. Given the speed of light, there is a $0.01$ second maximum physical travel time for a signal detected by one detector to appear in the other detector. On top of this time window, we add an additional $0.002$ seconds to account for timing errors between the two detectors, giving a total allowed time difference between trigger arrival times at different detectors of $0.012$ seconds.

Triggers across detectors found within the time window and above the $\hat{\rho}$ threshold (typically $4.5$) can be considered coincident and are passed to the coincident ranking statistic for ranking. The PyCBC Offline coincident ranking statistic includes: likelihood of coincident trigger time and phase differences~\cite{PyCBC:2016}, likelihood correction based on data artefacts found in correlated auxiliary channels which are insensitive to \gws~\cite{DQ_vetoes:2017, iDQ:2020}, an exponential noise model which models noise distributions of templates separately and re-ranks single detector ranking statistic based on how frequently a template has historically triggered on noise with high $\hat{\rho}$~\cite{PyCBC:2017}, and a template dependent factor which re-weights templates based on their astrophysical likelihood~\cite{PyCBC_focussed_bbh:2024}.

The PyCBC Live ranking statistic used during the third observing run was comparatively simple. The single detector ranking statistic is the same as the PyCBC Offline single detector ranking statistic, except without the PSD variation statistic. PyCBC Live only ranks coincident triggers by the likelihood of their time and phase differences~\cite{PyCBC_Live:2018}.

\subsection{\label{5:sec:old-stat-construction}Constructing the O3 PyCBC Live ranking statistic}

We describe the ranking statistic used by PyCBC Live during the third observing run in more detail to identify the potential components that can be added from the PyCBC Offline ranking statistic. The construction of the ranking statistic used by PyCBC Live during the third observing run is described in~\cite{PyCBC_Live:2018}.

As described in Section~\ref{2:sec:ranking-statistic}, the general optimal ranking statistic is defined as the difference of the log of the signal rate, $ p^{S}(\Vec{\theta})$ and noise rate,  $p^{N}(\Vec{\theta})$,
\begin{align}
    R &= \log p^{N}(\Vec{\theta}) - \log p^{S}(\Vec{\theta})
\end{align}
over the parameter space $\Vec{\theta} = \left(\hat{\rho}_{H}, \hat{\rho}_{L}, \chi^{2}_{H}, \chi^{2}_{L}, \delta\phi, \delta t, m_{1}, m_{2}, \Vec{s_{1}}, \Vec{s_{2}}\right)$. 

The construction of the ranking statistic used by PyCBC Live,
\begin{equation}
    R^{2} \propto 2 [\log p^{S}(\Vec{\theta}) - \log p^{N}(\Vec{\theta})] + constant,
    \label{5:eqn:general-detection-statistic}
\end{equation}
was chosen such that the ranking statistic of a two detector coincidence with both Gaussian and stationary noise~\footnote{The detector noise is \textbf{not} assumed to be Gaussian and stationary, but this ranking statistic is recovered if that were the case.} will have single detector noise rate, $r_{n}$, that falls off as a Gaussian,
\begin{equation}
    r_{n,det} \propto \exp \left( -\frac{(\rho_d - \mu)^2}{2 \sigma^2} \right),
    \label{5:eqn:old-noise-rate}
\end{equation}
and a combined detector noise rate that is the product of both detector noise rates,
\begin{equation}
    p^{N}(\Vec{\theta}) \propto \exp \left( -\frac{(\rho_{H}^{2} + \rho_{L}^{2})}{2} \right),
    \label{5:eqn:old-comb-noise-rate}
\end{equation}
recovering the standard quadrature sum SNR statistic,
\begin{equation}
    R^{2} = -2 \log p^{N}(\Vec{\theta}) = \rho^{2}_{H} + \rho^{2}_{L} .
\end{equation}

To expand on the quadrature sum ranking statistic, which only considers the noise rate, we can include information about the signal rate which takes the parameters $(\rho_{H}, \rho_{L}, \delta t, \delta \phi)$: the two detector SNRs, the difference in the time of arrival for the two detectors, $\delta t = t_{H} - t_{L}$, and the difference in phase of the gravitational waveforms, $\delta \phi = \phi_{H} - \phi_{L}$. These parameters, for a real signal, will depend on the source localisation and detector orientation and therefore, some combinations of parameters are more likely than others. The exact process for calculating $p^{S}(\Vec{\theta})$ is provided in detail in~\cite{PyCBC:2017}, for this chapter's purpose a look-up is made to a phase-time histogram which represents the most likely combination of the two parameters in the signal space. Including $p^{S}(\Vec{\theta})$ in the ranking statistic (Equation~\ref{5:eqn:general-detection-statistic}) yields,
\begin{keyeqn}
\begin{equation}
    R = \sqrt{\rho^{2}_{H} + \rho^{2}_{L} + 2 \log\left(p^{S}(\Vec{\theta})\right)}
    \label{5:eqn:original-statistic}
\end{equation}
\end{keyeqn}
which is the final version of the original ranking statistic used by PyCBC Live in the third observing run.

\section{\label{5:sec:new-additions}Improving the PyCBC Live ranking statistic}

The ranking statistic used by PyCBC Live in the third observing run combines the noise rate contribution, which is the quadrature sum of detector $\hat{\rho}$ in stationary, Gaussian noise, and the signal rate of the trigger. We know our detector noise is \textbf{not} stationary and Gaussian~\cite{LIGO_data_quality:2015} and therefore by including an accurate noise model in the ranking statistic we can improve the sensitivity of the PyCBC Live search.

There are two components of the PyCBC Offline ranking statistic that directly address these two deficiencies in our ranking statistic: PSD variation, which corrects the single detector $\rho$ for non-stationary noise present in the data and; the exponential noise model, which models the noise distribution of each template by fitting an exponential to the noise falloff.

These changes dramatically improve our noise model for the data, providing a more accurate noise rate for use in calculating the ranking statistic and assigning significance to our \gwadj events. A better noise model allows the ranking statistic to better be able to distinguish between signal and noise events. There are other components of the PyCBC Offline ranking statistic that are in development in parallel to our work to be used in the PyCBC Live ranking statistic in the future: a kernel density estimation method to estimate probability density functions of signal rate for templates in the template bank, applying a weighting to the signal rate of a coincidence trigger depending on the signal search parameters and the template used to find the signal~\cite{PyCBC_focussed_bbh:2024}; \texttt{iDQ} uses data quality information from auxiliary channels to provide more information to the noise rate for triggers in PyCBC Live~\cite{iDQ:2020}. All of these changes will improve PyCBC Live's significance estimation in future observing runs.

\subsection{\label{5:sec:methodology}Adapting PyCBC Live to compute generic offline detection statistic in low-latency}

We must discuss the architecture of the PyCBC Live~\cite{PyCBC_Live:2018} search before we can discuss how the changes to the ranking statistic have been made. Data from the \gwadj observatories is distributed to computer clusters, such as the LIGO Data Analysis System (LDAS) at Caltech in the United States of America~\cite{ldas_caltech:2024}, which is made available to the PyCBC Live search pipeline on a number of cluster nodes. PyCBC Live has exclusive access to $151$ cluster nodes for the explicit purpose of running the PyCBC Live full-bandwidth search pipeline, $150$ of these nodes are `worker' nodes and $1$ (rank 0) is the `control' node.

The template bank is split equally across these worker nodes which are independent and produce their own set of single detector triggers. The triggers from all worker nodes are transferred to the control node which combines them into coincident events, calculates ranking statistic and false-alarm rate. PyCBC Live has strong limitations placed on it by the low-latency nature of the search pipeline, the computational processing of a data segment must be completed prior to the end of the next data segment lest data will build up waiting to be processed. These latency requirements limit additions that can be made to the search pipeline because they might either take too long to compute or will require the data to wait for extra information before the processing can be completed.


PSD variation requires the matched filtering of a generic CBC filter to calculate the variation time series. This matched filter is calculated by the control node because the PSD variation values needs to be distributed to all worker nodes for single detector ranking statistic calculation. This calculation introduces a negligible amount of latency on top of the ${\sim}700$,$000$ matched filters already being performed every data stride. Section~\ref{5:sec:psd-var} details the implementation of PSD variation in live.

The template fits require a large number of triggers for each template in order for the noise model to be representative of the true noise distribution. In PyCBC Offline the template fits are created using the triggers produced by the templates in the whole segment of data. The template fit statistics are then calculated and the triggers are then re-ranked depending on the resulting template fit statistics of the templates. We would not be able to do the same for PyCBC Live because we cannot create the template fit statistics every search segment, these files can take up to an hour to produce.


To incorporate the exponential noise model into the PyCBC Live ranking statistic, we begin by assuming that the noise distribution remains approximately constant over a period of at least one week. This assumption simplifies the task of modelling the noise, allowing us to use data from recent PyCBC Live triggers and capture the same statistics for each template. Specifically, we generate the template fits based on the previous week’s worth of PyCBC Live triggers, and use these fits to produce the necessary statistic files. These statistic files are then employed to re-rank new PyCBC Live triggers.

This approach offers a significant advantage in terms of efficiency, as it allows the creation of template fit statistic files to occur in parallel with the PyCBC Live search, avoiding any additional latency in processing \gwadj events. Furthermore, this system is flexible enough to accommodate regular updates to the statistic files. While we currently update the statistic files on a weekly basis, a more dynamic approach could involve a rolling window, where the noise model is refreshed more frequently---for example, by using a sliding week-long window that is updated every day. Such a rolling update would allow for a more refined, real-time adjustment to changing noise conditions. These can be tuned in the future.

\subsection{\label{5:sec:psd-var}PSD variation}

The power spectral density (PSD) describes how much power is present in the data at each frequency, essentially how the data power is distributed in the frequency domain. The power in each frequency band can change over time due to instrumental noise such as seismic noise, thermal noise and quantum noise and cannot be mitigated except through detector improvements. We call this noise non-stationary noise. The noise profile can change over the period of a search segment, meaning the PSD will be inaccurate at different times. Identifying triggers in the PyCBC searches is done by matched filtering the data with a signal template, described in Section~\ref{2:sec:matched-filter}, which depends on the PSD in the noise-weighted inner product.

A PSD which doesn't accurately describe the noise profile of the data will lead to a misestimation in the calculation of $\rho$. Measuring the actual PSD of the data is particularly challenging due to the presence of non-stationary and non-Gaussian noise therefore, the offline search empirically measures the PSD of the \gwadj data being analysed using Welch's method, however, instead of using a mean average of the overlapping PSDs a median average it used to remove the effect of short duration glitches in the data.

In PyCBC Offline, the estimated PSD is used for very long search segments ($512$ seconds) and while the effects of short duration glitches have been mitigated with the median average method we still suffer from the effects of non-stationary noise. 

The mismatch between the estimated PSD, $S_{E}$, and actual PSD, $S_{A}$, of the data will have the effect of underestimating and overestimating the noise at different times in our search segment. We are able to track the sources of the non-stationary noise in some cases, for example, when the source of the non-stationary noise is a period of increased seismic activity this will be identified by the seismometer auxiliary channels and can then be subtracted from the \gwadj data. This isn't true for all noise however, some non-stationary noise has no immediately identifiable source and we are unable to make the subtraction.

Therefore, we must track the difference between $S_{E}$ and $S_{A}$ using the PSD variation statistic~\cite{PSD_var:2020}. The PSD variation statistic models the relationship between $S_{E}$ and $S_{A}$ as,
\begin{equation}
    S_{A} = \nu_{s} S_{E}, 
\end{equation}
where $\nu_{s}$ is treated as a frequency independent parameter. The PSD variation is defined as the time series which tracks $\nu_{s}$, therefore, by computing $\nu_{s}(t)$ we can calculate the difference between $S_{A}$ and $S_{E}$ for all times in our search data.

In the offline search $\nu_{s}(t)$ is calculated using an approximate expression for a typical CBC template, $|h(f)| \propto f^{\frac{-7}{6}}$, and $S_{E}$ to construct a filter,
\begin{equation}
    F = \frac{|h(f)|}{S_{E}} ,
    \label{5:eqn:psd-var-filter}
\end{equation}
which is band-passed between $20$Hz and $480$Hz, smoothed with a Hann window and combined with the data to produce an equation for the $\nu_{s}(t)$,
\begin{equation}
    \nu_{s}(t) \equiv N \langle \rho \rangle(t), 
\end{equation}
where $N$ is a constant such that the expectation value of $\nu_{s}$ in Gaussian noise is $1$ and $\langle\rho\rangle$ is the variance of $\rho$. This produces $\nu_{s}(t)$ that can then be applied to the offline search as part of the single detector ranking statistic where triggers are assigned $\nu_{s}$ at trigger time, $t$, which re-weights $\rho$ prior to any signal consistency tests,
\begin{equation}
    \rho_{scaled} = \frac{\rho}{\sqrt{\nu_{s}}} .
    \label{5:eqn:psd-var-snr-reweighting}
\end{equation}


The live search maintains a data ring buffer of $512$ seconds, rolling the newest eight seconds of data on as the older eight seconds are rolled off. The live search can be operating for potentially weeks at a time, during which the detector noise can change, this means the live search requires a dynamic PSD which can update during runtime. The initial PSD is estimated using Welch's method with a median average and is first created when enough data has been accumulated after starting the search. Every analysis segment (eight seconds) a new PSD, $S_{N}$, is estimated and compared to the current search PSD, $S_{C}$. The comparison is made by calculating the distance a binary neutron star system with equal $1.4 \, \text{M}_{\odot}$ masses would need to be from the detectors to be observed with $\rho \!=\! 8.0$, this is called the BNS distance and it depends exclusively on the PSD. The BNS distance is calculated both $S_{C}$ and $S_{N}$ and if the BNS distance of $S_{N}$ differs from the BNS distance of $S_{C}$ by $\pm1\%$ then $S_{N}$ replaces $S_{C}$. If the BNS distance of $S_{N}$ is within the threshold then $S_{N}$ is discarded and the search keeps using $S_{C}$.

The live search does not need to calculate $\nu_{s}(t)$ for the entire data ring buffer, new triggers are only ever found in the latest analysis segment and therefore this is the only period of time for which $\nu_{s}(t)$ needs to be calculated and $\nu_{s}$ values distributed to the new triggers. To calculate the PSD variation values we need to track $S_{E}$ for each detector and create a new filter (Equation~\ref{5:eqn:psd-var-filter}) for each detector which is updated every time $S_{E}$ is updated. The filters are then convolved with the latest eight seconds in the data ring buffer, the mean square of $\nu_{s}(t)$ is calculated every $0.25$ seconds to find outliers caused by short duration glitches. These outliers are then replaced with an average of the adjacent elements in $\nu_{s}(t)$ and the time series is then further averaged every second to produce the final $\nu_{s}(t)$. For each trigger, $\nu_{s}$ is then extracted by interpolating between the two nearest whole seconds in the time series and is then used to scale $\rho$ before $\hat{\rho}$ is calculated, as shown in Equation~\ref{5:eqn:psd-var-snr-reweighting}.

\subsection{\label{5:subsec:template-fits}Modelling the noise distribution in live}

\begin{figure}
    \centering
    \includegraphics[width=1.0\linewidth]{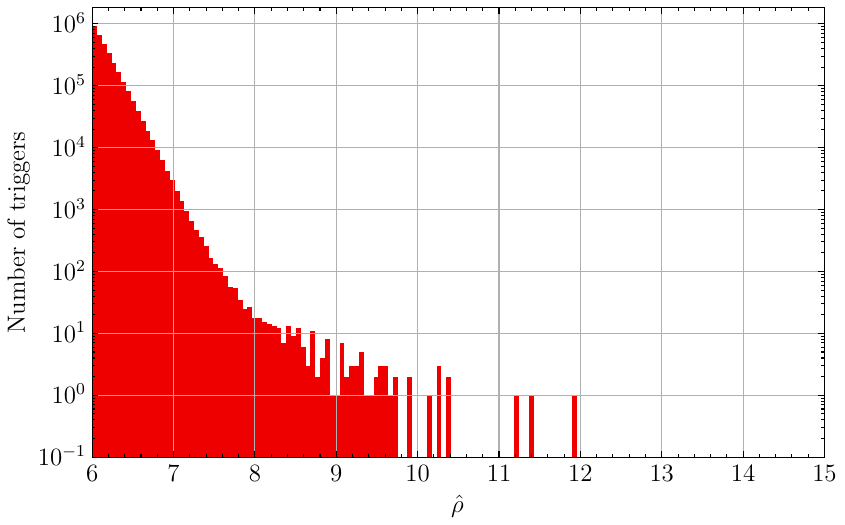}
    \caption{The new-SNR, $\hat{\rho}$, distribution of LIGO-Hanford triggers for a chunk of data searched through using PyCBC Offline in the third observing run. The higher SNR triggers can be attributed to non-Gaussian noise transients. $\hat{\rho}$ has been calculated using the Allen-$\chi^{2}$~\cite{Allen_Chi:2005} and sine-Gaussian $\chi^{2}$~\cite{PyCBC_sg:2018} to re-weight SNR.}
    \label{5:fig:H1_long_tails}
\end{figure}

Non-Gaussian noise artefacts (glitches) can produce high SNR triggers, these are partially mitigated with the single detector ranking statistic signal consistency tests, but there is still a long tail in the noise distribution (Figure~\ref{5:fig:H1_long_tails}) that can be attributed to the effects of glitches.
This noise falloff has historically been assumed to be Gaussian in the PyCBC Live search and all templates are treated equally. There are a number of very common glitches that occur frequently in the data that can match more closely to some templates in the template bank and not others. An example of this is blip glitches~\cite{blips:2019} which resemble the \gwadj signals of high mass binary black hole mergers, therefore the templates which describe these signals will trigger more regularly on these glitches with higher $\hat{\rho}$ compared to other templates in the bank which do not look similar to any glitch class. Not only will the noise falloff no longer be Gaussian, each template is treated equally, where some trigger far more frequently with higher $\hat{\rho}$.

Single detector triggers found by PyCBC searches are only saved if the $\rho > 4.5$. All triggers above the threshold from the noise distribution we want to model. The noise falloff is modelled in PyCBC by taking all the triggers produced by a template and fitting an exponential function to the slope of the trigger distribution and is quantified by the exponential fit factor, $\alpha$, and the count of triggers found with $\hat{\rho} > 6.0$, $\mu$. This process is commonly called `template fitting' and the output parameters are called the `template fits'.

PyCBC Offline searches through pre-defined chunks of IGWN data~\cite{gwtc3:2023}, which are all around one calendar week long. The search will matched filter the template bank and the data in one of these chunks, and all triggers above the previously mentioned $\rho$ threshold for each template are collected and saved. These triggers are then used to calculate $\alpha$ and $\mu$ for each template and these parameters are saved to data files which contain all the template fits for the entire template bank for that chunk of data.
%
%
To understand how the template fits contributes to the ranking statistic, we must first construct the ranking statistic which includes the exponential noise model. We previously defined a general detection statistic in Equation~\ref{5:eqn:general-detection-statistic} which was constructed to obtain the quadrature sum statistic for a detector with stationary, Gaussian noise. This ranking statistic models the noise falloff in the distribution as a Gaussian. The new exponential fit to the noise falloff doesn't make this assumption so we return to,
\begin{equation}
    \Lambda(\Vec{\theta}) = \mu_{s} \frac{p^{S}(\Vec{\theta})}{p^{N}(\Vec{\theta})} ,
    \label{5:eqn:optimal-detection-statistic}
\end{equation}
where $\Lambda$ is the likelihood ratio (Equation~\ref{2:eq:likelihood_ratio}), $\mu_{s}$ is the astrophysical signal rate and is assumed to be constant, $p^{S}(\Vec{\theta})$ is the signal event rate density (signal rate) and $p^{N}(\Vec{\theta})$ is the noise event rate density (noise rate). Due to expected values of signal and noise rate spanning many orders of magnitude we take the logarithm of Equation~\ref{5:eqn:optimal-detection-statistic},
\begin{equation}
    R = \log \Lambda = \log p^{S}(\Vec{\theta}) - \log p^{N}(\Vec{\theta}).
    \label{5:eqn:signal-minus-noise-rate}
\end{equation}
The noise rate is a combination of the single detector noise rates,
\begin{equation}
    p^{N}(\Vec{\theta}) = A_{N} \prod_{d} r_{d}(\hat{\rho}_{d}),
\label{5:eqn:comb-noise-rate}
\end{equation}
where $A_{N}$ is the `allowed area' for coincident noise events from two detectors, $A_{N\{12\}} \!=\! 2\tau_{12}$, where $\tau_{12}$ is the \gwadj travel time window between detectors $1$ \& $2$ with a small allowance for timing error (currently $0.03$ seconds) and $r_{d}$ is the single detector noise event rate density for each detector $d$. The natural logarithm of this noise rate is taken to get the final combined log noise rate,
\begin{equation}
    \log(p^{N}(\Vec{\theta})) = \log(A_{N}) +  \sum_{d} \log(r_{d}(\hat{\rho}_{d})) .
\label{5:eqn:comb-log-noise-rate}
\end{equation}
The single detector noise rate is calculated with the product of the count of triggers above the $\hat{\rho} > 6.0$, $\mu$, and the probability of $\hat{\rho}$ given the template (including $\alpha$ and $\mu$) and noise,
\begin{equation}
    r_{d}(\hat{\rho}_{d}; {\Vec{\theta}}; N) = \mu(\Vec{\theta}) p(\hat{\rho} | \Vec{\theta}, N) ,
\label{5:eqn:single-noise-rate}
\end{equation}
where
\begin{equation}
    p(\hat{\rho} | \Vec{\theta}, N) = \alpha(\Vec{\theta}) \exp\left(-\alpha(\Vec{\theta})\cdot\left(\hat{\rho} - \hat{\rho}_{thresh}\right)\right)
\label{5:eqn:p-definition}
\end{equation}
thereby giving the single detector noise rate as,
\begin{equation}
    r_{d}(\hat{\rho}; {\Vec{\theta}}, N) = \mu(\Vec{\theta}) \alpha(\Vec{\theta}) \exp\left(-\alpha(\Vec{\theta}) \cdot (\hat{\rho} - \hat{\rho}_{thresh})\right)
\label{5:eqn:single-noise-rate-full}
\end{equation}
and when the natural logarithm is taken we arrive at the complete single detector log noise rate,
\begin{equation}
    \log r_{d}(\hat{\rho}) = \log\mu(\Vec{\theta}) +  \log\alpha(\Vec{\theta}) - \alpha(\Vec{\theta}) \cdot(\hat{\rho} - \hat{\rho}_{thresh})
\label{5:eqn:single-log-noise-rate}
\end{equation}
for each detector, $d$.

\section{\label{5:sec:injection-tests}Measuring improvements in search sensitivity}

To quantifiably demonstrate that the new additions to the ranking statistic have improved the live search we can investigate the differences in sensitivity when the PSD variation and exponential noise model are included. We quantify the sensitivity of the search by calculating the sensitive volume in which we can observe \gwadj signals. The sensitive volume is calculated by measuring the detection efficiency of different distance bins that are taken from an injection set and then multiplying by the volume enclosed by the distance bins. The volume of each bin is then summed to find the total sensitive volume of the search~\cite{rw_snr_eq:2012}.

We can search through a period of data which contains an injection set with two instances of the PyCBC Live search, one containing the new ranking statistic additions and one representing the original ranking statistic. The ratio of the new statistic sensitive volume to the original statistic sensitive volume will indicate the change in sensitivity when including the new ranking statistic components.

We have chosen to test the ranking statistic changes to PyCBC Live using a smaller template bank than the current PyCBC Live template bank, which contains over $730$,$000$ templates~\cite{PyCBC_Live:2018}. PyCBC-BBH is a search focussed on binary black hole (BBH) signals performed in the third observing run~\cite{gwtc3:2023, PyCBC_focussed_bbh:2024} and used a template bank with only $15$,$436$ templates with the drawback that we are only considering BBH signals. This smaller template bank uses fewer computational resources and allows a full week of data to be searched through with PyCBC Live in as little as a day. The injection set used in the third observing run is shown in Table~\ref{5:tab:injection-set}~\cite{gwtc3:2023} with rows for the specific cuts on injections for both the PyCBC-broad (full parameter space search) and PyCBC-BBH search.
\begin{table}[ht]
    \centering
    \begin{tabular}{c c ccc }
        \multicolumn{2}{c}{ } & Mass & Mass &  Spin \\
        \multicolumn{2}{c}{ } & distribution & range ($M_{\odot}$) & range  \\
        \hline
        \multirow{6}{*}{{Injections}} & \multirow{2}{*}{BBH} &  $\left.p(m_{1}) \propto m_{1}{}^{-2.35}\right.$ \rule{0pt}{1.05\normalbaselineskip} & $2 < m_{1} < 100$ & \multirow{2}{*}{$\left|\chi_{1,2}\right| < 0.998$} \\
         & & $p(m_{2}|m_{1}) \propto m_{2}$ & $2 < m_{2} < 100$ & \\[0.05\normalbaselineskip]
         & \multirow{2}{*}{NSBH} & $\left.p(m_{1}) \propto m_{1}^{-2.35}\right.$ & $2.5 < m_{1} < 60$ & $\left|\chi_1\right| < 0.998$ \\
         & & Uniform & $1 < m_{2} < 2.5$ & $\left|\chi_2\right| < 0.4$  \\[0.05\normalbaselineskip]
         & \multirow{2}{*}{BNS} & \multirow{2}{*}{Uniform} & $1 < m_{1} < 2.5$ & \multirow{2}{*}{$\left|\chi_{1,2}\right| < 0.4$} \\
         & & & $1 < m_{2} < 2.5$ & \\
        \hline
        \multirow{3}{*}{PyCBC-broad} & BBH & & $\mathcal{M} > 4.353 $ & \\[0.05\normalbaselineskip]
        & NSBH & & \hspace{-1.25cm} $ 2.176 < \mathcal{M} < 4.353 $ & \\[0.05\normalbaselineskip]
        & BNS & & $ \mathcal{M} < 2.176 $ & \\
        \hline
        {PyCBC-BBH} & {BBH} & & $\mathcal{M} > 4.353 $ \\
    \end{tabular}
    \caption{Recreated from~\cite{gwtc3:2023}. The intrinsic parameter distributions used to generate the injection set used for computing the sensitive volume change when including the PSD variation statistic and exponential noise model. Here PyCBC-broad is the PyCBC offline search across the full template bank and parameter space, including templates representing binary black hole, neutron star black hole and binary neutron star signals. This is comparable to the template bank used by the PyCBC Live full-bandwidth search. PyCBC-BBH considers only templates where $\mathcal{M} > 4.353 \, \text{M}_{\odot}$, which includes only binary black hole signals. This is the same template bank that is used for the analysis in this chapter.}
    \label{5:tab:injection-set}
\end{table}
%
%

As previously mentioned, the template fits are measured from the triggers found in the previous week to the week currently being searched through. Therefore, to perform our injection set test we need to do an initial search with PyCBC Live using the original ranking statistic to generate a week's worth of triggers and create template fit statistics. The second week of data is then created containing a large number of injected \gwadj signals that is searched through with PyCBC Live using both the original ranking statistic and the new ranking statistic. The two weeks of O3b \gwadj data used lasted from 06 January 2020 23:59:42 to 20 January 2020 23:59:42 and, according to the \gwadj catalogue for the third observing run~\cite{gwtc3:2023}, there are no known \gwadj signals present. After both searches have completed analysing the second week of data we can estimate the sensitivity improvement. We can choose a threshold false-alarm rate (FAR) and count the number of injections in both searches that have been found with a FAR lower than that number, if the counted number of injections is higher when including the new ranking statistic additions we have seen an increase in the sensitivity of the search. We can do this for a range of FAR values in $[10^{4}, 10^{-4}]$ to build up a picture of the sensitivity ratio curve.

\section{\label{5:sec:sensitivity-improvements}Improvements in sensitivity}

By adding the PSD variation and the exponential noise model to the ranking statistic we expect to find \gwadj injections with a greater significance than before. This significance is determined by the FAR of the injection, which has a direct mapping to the ranking statistic value, an event found with a FAR of $1$ per year would indicate that an event with the exact same parameters is expected to be seen in the data caused by non-astrophysical noise-not a real event-once per year of analysis time. We commonly use the inverse false-alarm rate (IFAR) with units of $years$ where a large IFAR indicates a more confident detection. An IFAR threshold is chosen to separate events which are real from those that are false alarms. A balance must be made to avoid contaminating the detection catalogue with a large number of potential false alarms, while not missing any real events. In this analysis we are considering only two detectors, allowing potentially more false-alarms from coincident noise when compared to a three detector configuration and therefore we have chosen a more conservative IFAR threshold of $1$ year to assess sensitivity improvements. The current threshold for LIGO/Virgo alert generation is $1$ per $2$ months~\cite{PyCBC_Live:2018} so low-latency \gwadj searches will identify more \gwadj events, but these detections have additional post detection checks from data quality checks from detector characterisation~\cite{O2O3_DetChar:2021} and parameter estimation~\cite{gwtc3:2023} which can also rule out false-alarms and to prevent catalogue contamination.


To measure the sensitivity improvement of the new ranking statistic we count the number of injections found by both searches over a range of IFAR values in $[10^{-4}, 10^{4}]$ and take the ratio of the counts. The injection set is pre-weighted to ensure each injection represents an equal volume, therefore instead of distance binning, as described in Section~\ref{5:sec:injection-tests}, we can simply count the number of injections seen by each search.
%
\begin{figure}
  \centering
  \begin{minipage}[t]{1.0\linewidth}

    \includegraphics[width=1.0\textwidth]{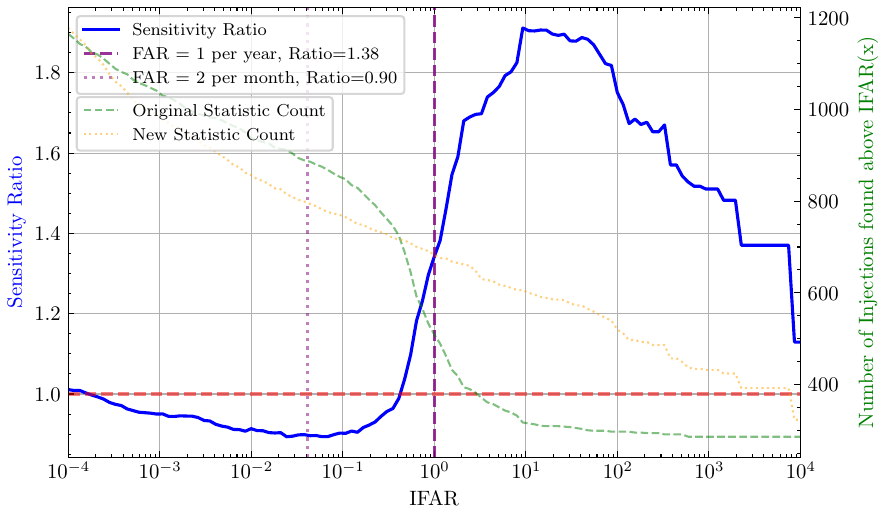}
    \caption{The sensitive volume ratio between a PyCBC Live search whose ranking statistic contains PSD variation and an exponential noise model and a PyCBC Live search using the original ranking statistic used by PyCBC Live during the third observing run as a function of inverse false-alarm rate (IFAR). The blue curve represents the sensitivity ratio computed by comparing the number of injections detected with an IFAR exceeding the corresponding x-axis value. An increase in sensitivity is indicated by values greater than 1, the horizontal red-dashed line. Two vertical dashed lines at fixed false-alarm rates (FAR) highlight specific thresholds: FAR = 1 per year (dark purple, dashed) with a sensitivity ratio of 1.38, and FAR = 2 per month (light purple, dotted) with a sensitivity ratio of 0.90. The green and orange curves represent the cumulative number of injections detected in the old and new searches, respectively, as a function of IFAR.}
    \label{5:fig:fits-psdvar-sensitivity}

  \end{minipage}
\end{figure}
%
Figure~\ref{5:fig:fits-psdvar-sensitivity} shows the sensitivity ratio comparing the new statistic search to the original statistic search. We see $35\%$ more injections with an IFAR greater than our threshold of $1$ year which is a very large increase in sensitivity, however at a FAR of $2$ per month we see a sensitivity decrease of $10\%$. To understand the shape of the sensitivity curve we can look at the new additions to the ranking statistic and understand further how they are changing the significance of individual injections.

\section{\label{5:sec:injection-investigations}Changes in injection significance}


\begin{figure}
      \centering
    \includegraphics[width=1.0\textwidth]{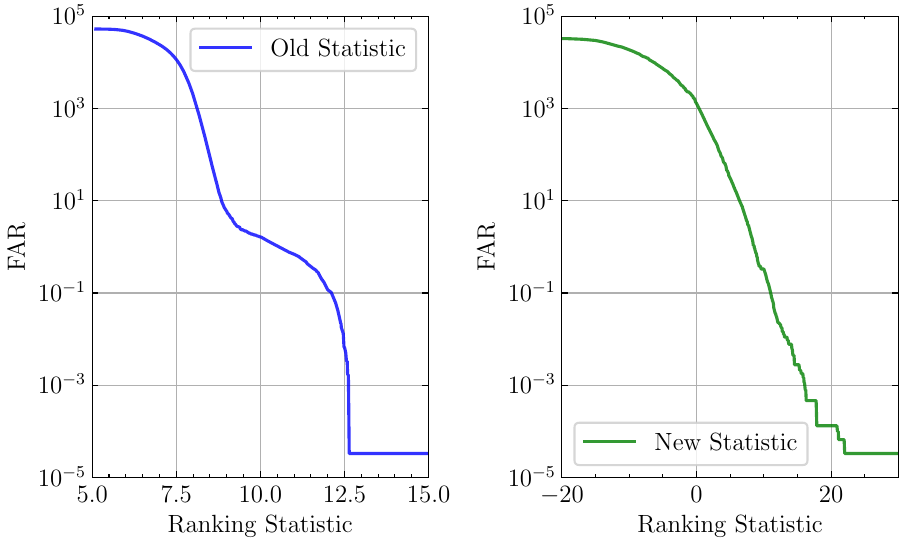}
    \caption{The mapping between false-alarm rate (FAR) and ranking statistic for the original PyCBC Live ranking statistic used in the third observing run (left, blue) and the new PyCBC Live ranking statistic including PSD variation and an exponential noise model (right, green). Ranking statistic values cannot be compared between ranking statistics, but injection FARs can be. Higher ranking statistic values indicate better, lower FARs. The original ranking statistic has a narrower range of allowed ranking statistic values $\in [5, 12.5]$ along with a distinct curvature whereas the new statistic ranking statistic can take a broader range of ranking statistic values $\in [-20, 22]$ and has a relatively smooth drop off with increasing ranking statistic values. The `shoulder' in the original ranking statistic was an unwanted feature which has now been removed.}
    \label{5:fig:fits-psdvar-far-stat}
\end{figure}
The ranking statistic assigned to each event can be mapped directly to FAR, shown in Figure~\ref{5:fig:fits-psdvar-far-stat}. Changing the ranking statistic calculation will lead to incomparable values between searches, but the individual injection IFAR values can be compared. The PyCBC Live search will matched filter the template bank and the data and where the resulting $\rho$ and $\hat{\rho}$ is above pre-defined thresholds (both $4.5$ in the third observing run) the trigger is kept. The coincidence ranking statistic value is calculated and then a ranking statistic value is assigned to the event. The initial $\rho$ values are the same between the two searches because these depend only on template and data but $\hat{\rho}$ depends on the single detector ranking statistic which has the addition of PSD variation in our new statistic, therefore it is possible for the PSD variation to prevent the saving of a trigger due to a trigger's $\hat{\rho}$ being re-weighted below the threshold.

\begin{figure}
      \centering
    \includegraphics[width=1.0\textwidth]{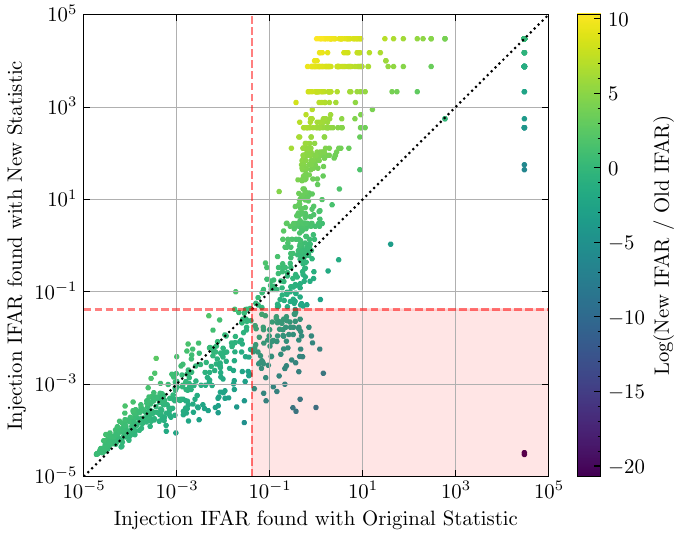}
    \caption{The inverse false-alarm rate (IFAR) values recorded for each injection that was seen by both the PyCBC Live search that includes the PSD variation and exponential noise model in the ranking statistic and the PyCBC Live search using the same ranking statistic as used in the third observing run. Each point represents an injection from the injection set, with the x-axis showing the IFAR recorded using the original ranking statistic and the y-axis showing the IFAR recorded using the new ranking statistic. The colour bar indicates the logarithmic difference in IFAR between the two searches, with positive values (yellow) indicating better performance of the new statistic and negative values (purple) indicating a worse performance with the new statistic. The dashed line at y=x represents equal IFAR values for both searches. The vertical and horizontal red dashed lines denote a false-alarm rate threshold of $2$ per month. The highlighted region in the bottom right of the plot indicates excess injections that have been seen by the original ranking statistic above the $2$ per month threshold and by the new ranking statistic below the $2$ per month threshold.}
    \label{5:fig:ifar-ifar-fits-psdvar-shaded}
\end{figure}

We can plot the IFAR value of each injection that has been found in both searches, seen in
Figure~\ref{5:fig:ifar-ifar-fits-psdvar-shaded}.
Within Figure~\ref{5:fig:ifar-ifar-fits-psdvar-shaded} we have plotted a line on the diagonal $y=x$, injections above this line ($y > x$) have been found with a larger IFAR with the new ranking statistic and injections below ($y < x$) have been found with a lower IFAR with the new statistic. Figure~\ref{5:fig:fits-psdvar-sensitivity} highlighted the decrease in sensitivity at a FAR of $2$ per month, this decrease is due to an excess number of injections in the red-shaded box in the bottom-right of Figure~\ref{5:fig:ifar-ifar-fits-psdvar-shaded}, this region represents injections which have been found by the original ranking statistic below a FAR of $2$ per month but with a FAR above $2$ per month when found by the new ranking statistic. We now proceed to explore these results, investigating why the injections found in the red-shaded box have been down-ranked and other interesting features.

\subsection{\label{5:sec:ignoring-psdvar}Effects of PSD variation on sensitivity}

The injections shown in Figure~\ref{5:fig:ifar-ifar-fits-psdvar-shaded} that lie below the $y = x$ diagonal line have been found with a lower significance when including the PSD variation and the exponential noise model in the ranking statistic. We can separate the new additions and investigate the individual contributions to the changes in sensitivity.

\begin{figure}
       \centering
    \includegraphics[width=1\textwidth]{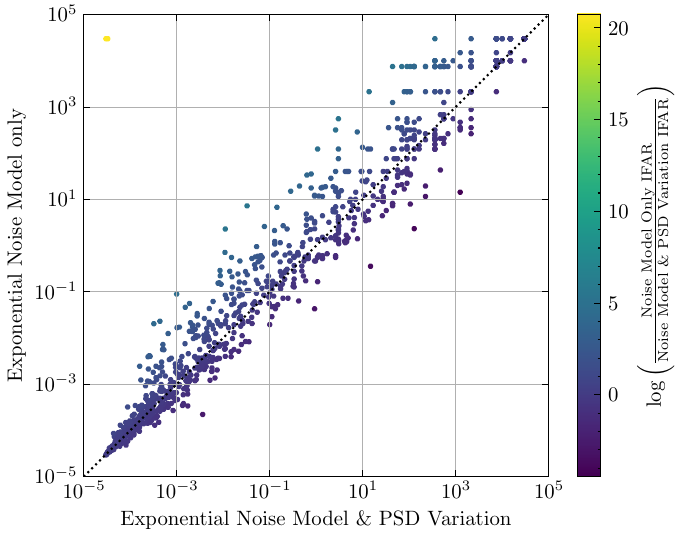}
    \caption{The inverse false-alarm rate (IFAR) values recorded for injections that were seen by both the PyCBC Live search that includes new additions of the PSD variation statistic and exponential noise model in the ranking statistic and the PyCBC Live search which has new additions of only the exponential noise model. Each point represents an injection from the injection set, with the x-axis showing the IFAR recorded by the PyCBC Live search including the exponential noise model and PSD variation ranking statistic and the y-axis showing the IFAR recorded by the PyCBC Live search which has added only the exponential noise model. The colour bar indicates the logarithmic difference in IFAR between the two searches, with positive values (yellow) indicating better performance when \textbf{not} including PSD variation and negative values (purple) indicating a better performance with the PSD variation statistic. The dashed line at y=x represents equal IFAR values for both searches.}
    \label{5:fig:ifar-ifar-fits-vs-fits-psdvar}
\end{figure}

The PSD variation statistic measures the non-stationary noise in the data and applies a weighting to the trigger $\rho$ value calculated by the matched filter of template and data, prior to the calculation of $\hat{\rho}$. To understand the effect of the PSD variation statistic on the injection significance we can compare the IFAR of injections found by the new statistic which includes both the PSD variation statistic and the exponential noise model and the IFAR of the same injections but with only the addition of the exponential noise model. This comparison is shown in Figure~\ref{5:fig:ifar-ifar-fits-vs-fits-psdvar}
In this comparison, $550$ injections were found with a higher significance when \textbf{not} including PSD variation, $399$ were found with a higher IFAR when including PSD variation and $326$ were found with the same IFAR. We can also observe a greater dispersion of injections found above the diagonal line when compared to those found below it, meaning injections found above the diagonal line are found with a greater significance difference than those found below the line. Alongside this general trend we can see a few injections in the top-left of the plot which have had a very large increase in significance when we do not include PSD variation in the statistic, these injections correspond to the heavily down-ranked injections in the bottom-right of Figure~\ref{5:fig:ifar-ifar-fits-psdvar-shaded}.

\begin{figure}
       \centering
    \includegraphics[width=1.0\textwidth]{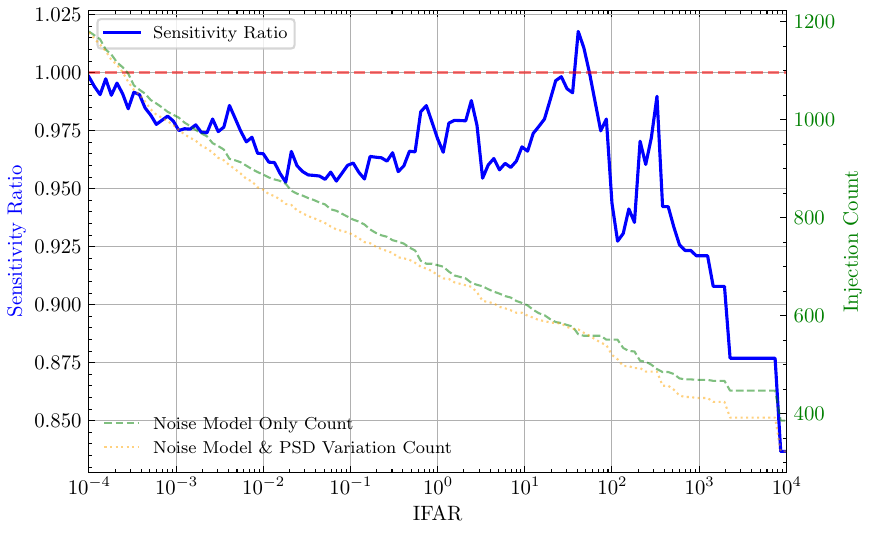}
    \caption{A comparison of sensitive volume ratio between two PyCBC Live searches: one which includes the addition of the PSD variation statistic and an exponential noise model in the ranking statistic and one which includes only the addition of the exponential noise model. The blue curve represents the sensitivity ratio computed by comparing the number of injections detected with an inverse false-alarm rate (IFAR) exceeding the corresponding x-axis value. An increase in sensitivity is indicated by values greater than 1, the horizontal red-dashed line. The green dashed curve represents the cumulative number of injections detected as a function of IFAR by the PyCBC Live search with added PSD variation and the exponential noise model. The orange dashed curve represents the cumulative number of injections detected as a function of IFAR by the PyCBC Live search with only the exponential noise model added.}
    \label{5:fig:vt-ratio-f-fo}
\end{figure}

We can compare the sensitivity change when including the PSD variation statistic by plotting the sensitivity curve in Figure~\ref{5:fig:vt-ratio-f-fo}.
From this, it is clear to see there is a small decrease in sensitivity when including the PSD variation statistic in the new ranking statistic. We note that the current implementation of PSD variation in PyCBC Offline has problems, specifically, PSD variation is eliminating extremely significant low-mass injections. Investigation into the PSD variation issues is being conducted and future iterations of PSD variation may improve its performance in PyCBC Live.

We have previously described how the PyCBC Live search re-calculates the PSD every analysis segment and replaces the currently used PSD when the BNS distance of the new PSD differs from the current PSD's by more than $1\%$. This reduces the need to include the PSD variation statistic to tackle non-stationary noise in the data. In light of the sensitivity decrease when including PSD variation, we make the choice to not use PSD variation in the ranking statistic going forward.

\begin{figure}
       \centering
    \includegraphics[width=1.0\textwidth]{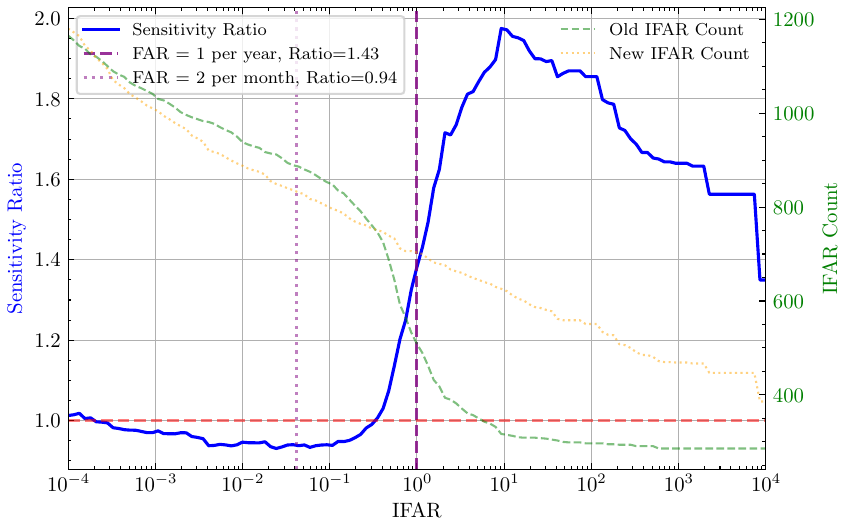}
    \caption{A comparison of sensitive volume ratio between a PyCBC Live search whose ranking statistic contains an exponential noise model and a PyCBC Live search using the original ranking statistic used by PyCBC Live during the third observing run as a function of inverse false-alarm rate (IFAR). The blue curve represents the sensitivity ratio computed by comparing the number of injections detected with an IFAR exceeding the corresponding x-axis value. An increase in sensitivity is indicated by values greater than 1, the horizontal red-dashed line. Two vertical dashed lines at fixed false-alarm rates (FAR) highlight specific thresholds: FAR = 1 per year (dark purple, dashed) with a sensitivity ratio of 1.43, and FAR = 2 per month (light purple, dotted) with a sensitivity ratio of 0.94. The green and orange curves represent the cumulative number of injections detected in the old and new searches, respectively, as a function of IFAR.}
    \label{5:fig:vt-ratio-fits-only}
\end{figure}

When adding only the exponential noise model to the ranking statistic we can re-analyse the sensitivity improvements to produce Figure~\ref{5:fig:vt-ratio-fits-only}.
This shows a greater increase across the IFAR range, with sensitivity increase at a FAR of $1$ per year of $43\%$ and a decrease of $6\%$ at a FAR of $2$ per month.

\begin{figure}
       \centering
    \includegraphics[width=1.0\textwidth]{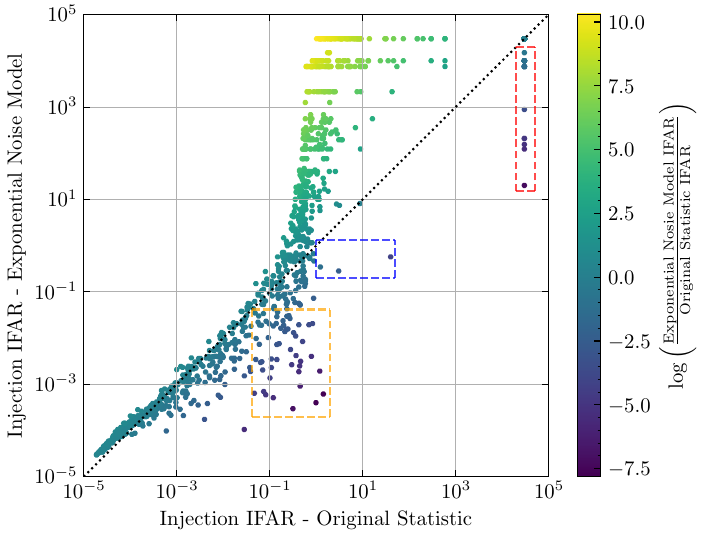}
    \caption{The inverse false-alarm rate (IFAR) values recorded for each injection seen by both the PyCBC Live search that includes the exponential noise model in the ranking statistic and the PyCBC Live search using the same ranking statistic used in the third observing run. Each point represents an injection from the injection set, with the x-axis showing the IFAR recorded using the original ranking statistic and the y-axis showing the IFAR recorded using the exponential noise model ranking statistic. The colour bar indicates the logarithmic difference in IFAR between the two searches, with positive values (yellow) indicating better performance when including the exponential noise model in the statistic and negative values (purple) indicating a better performance with the original statistic. The dashed line at y=x represents equal IFAR values for both searches. Three regions containing injections that have been down-ranked and will be investigated have been highlighted in dashed boxes.}
    \label{5:fig:ifar-ifar-fits-only-regions}
\end{figure}

We can plot the IFAR vs IFAR plot for the search including only the exponential noise model in the ranking statistic compared to the original search, seen in Figure~\ref{5:fig:ifar-ifar-fits-only-regions}, and when we compared to Figure~\ref{5:fig:ifar-ifar-fits-psdvar-shaded} we can see that the distribution of injections at low IFARs has decreased in dispersion around the diagonal line.
We have highlighted three regions of interest which contains injections that have decreased in significance that we will investigate further and understand the cause of their down-ranking.

\subsection{\label{5:sec:poor-temp-fits}Templates which trigger frequently with large SNRs}



The new exponential noise model in the ranking statistic allows us to model individual template noise distributions using that template's historical triggers. This allows us to down-rank templates which frequently trigger on non-Gaussian noise with a high $\hat{\rho}$. As describe in Equation~\ref{5:eqn:comb-noise-rate}, the exponential noise model is characterised by an exponential fit to the slope of the trigger distribution and is quantified by the exponential fit factor, $\alpha$, and the count of triggers above $\hat{\rho}_{thresh}$, $\mu$. Therefore, using these two parameters we must be able to differentiate between templates which trigger frequently on noise with large $\hat{\rho}$ and those that rarely trigger on noise. All templates in the bank will trigger on both Gaussian noise and glitches in our data, and after applying all signal consistency tests these triggers might still might have $\hat{\rho} > 8.0$.

\begin{figure}
    \centering
    \includegraphics[width=1.0\textwidth]{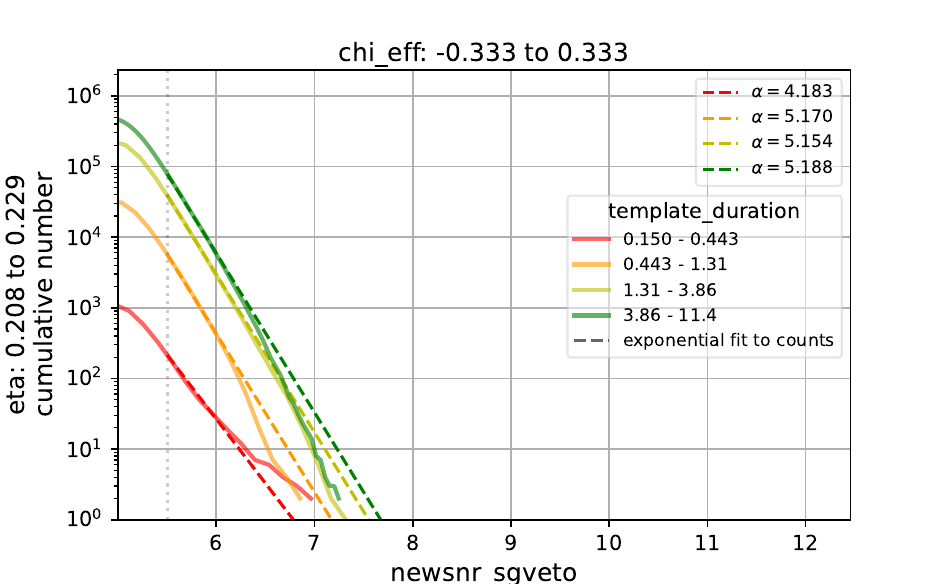}
    \caption{The noise distribution and template fits for a bin which contains triggers from templates which have symmetric mass ratio, $\eta$, between $0.208 - 0.229$ and effective spin, $\chi_{eff}$, between $-0.333 - 0.333$. The template triggers are further split into 4 bins depending on the duration of the template responsible for the trigger (in seconds). The solid coloured lines represent the cumulative counts of triggers that have been found in a template duration bin with $\rho$ that has been re-weighted by the traditional $\chi^{2}$~\cite{Allen_Chi:2005} and the sine-Gaussian $\chi^{2}$~\cite{PyCBC_sg:2018} to obtain new SNR, $\hat{\rho}$. The dashed coloured lines represent the exponential fit that has been made to the cumulative counts for each template duration bin where the exponential fit factor value, $\alpha$, is displayed in the top right most legend.}
    \label{5:fig:template-fits}
\end{figure}

Figure~\ref{5:fig:template-fits} shows an example of the exponential fitting to a group of template's noise distributions to obtain the exponential fit factors. This group of templates have effective spin $\chi_{eff} \in [-0.333, 0.333]$, symmetric mass ratio $\eta \in [0.208, 0.229]$ and the different template duration bins and corresponding exponential fitting factors have been plotted over the top.
This figure shows the difference in exponential fit slopes for the varying template bins and it can be seen that the bins which trigger more frequently with a higher new SNR have shallower slopes (lower $\alpha$) compared to the steeper slopes (higher $\alpha$) in other bins. Equation~\ref{5:eqn:single-log-noise-rate} shows that higher $\alpha$ values correspond to lower noise rates and lower $\alpha$ and higher $\mu$ values produce higher noise rates.

\begin{figure}
    \centering
    \includegraphics[width=1\textwidth]{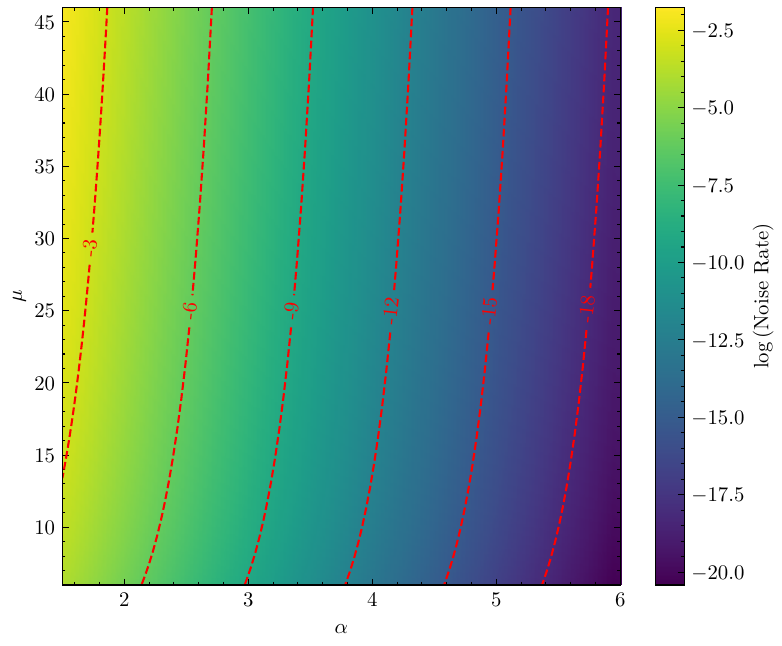}
    \caption{The single detector ($\log$) noise rate for varying combinations of exponential fit factor, $\alpha$, and trigger rates above a single detector statistic value ($\hat{\rho}_{thresh} = 6.0$), $\mu$. Constant contours of single detector noise rate are shown as dashed red lines, with the single detector noise rate value labelled within the contour line. This figure has been made using Equation~\ref{5:eqn:single-log-noise-rate} where $\hat{\rho} = 10.0$.}
    \label{5:fig:log-noise-static-snr}
\end{figure}

\begin{figure}
    \centering
    \includegraphics[width=1\textwidth]{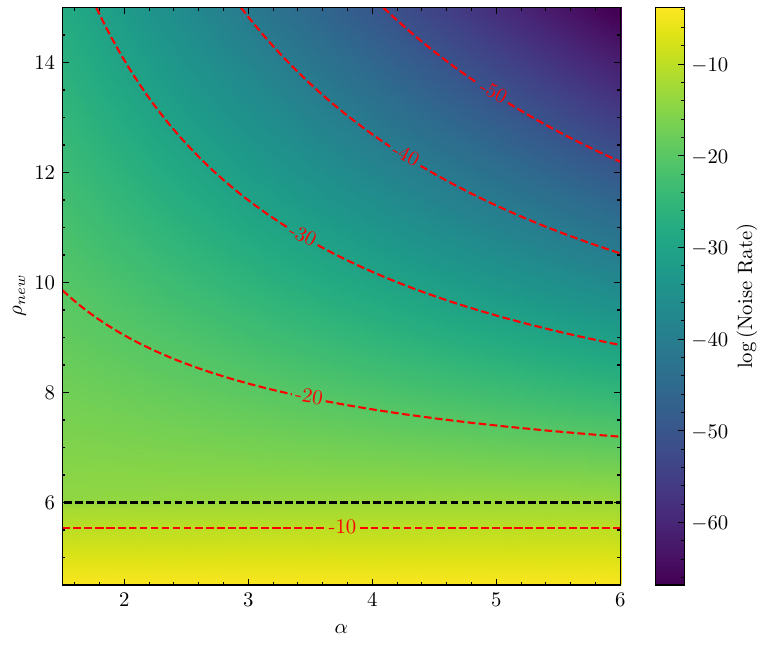}
    \caption{The single detector ($\log$) noise rate for varying combinations of exponential fit factor, $\alpha$, and single detector statistic value, $\hat{\rho}$. Constant contours of single detector noise rate are shown as dashed red lines, with the single detector noise rate value labelled within the contour line. This figure has been made using Equation~\ref{5:eqn:single-log-noise-rate} where $\mu$ has been set equal to the benchmark noise rate $-14.6$.}
    \label{5:fig:log-noise-static-rate}
\end{figure}

Templates which trigger more frequently and with a large $\hat{\rho}$ on noise will be down ranked by the ranking statistic, but real \gwadj signals that match this template will also be down ranked if they have low single detector $\hat{\rho}$. We can plot Equation~\ref{5:eqn:single-log-noise-rate}, which calculates the single detector noise rate, in Figure~\ref{5:fig:log-noise-static-snr} to show the dependence on $\alpha$ and $\mu$. In our search $\alpha$ can take values in $[1.5, 6.0]$ and $\mu \in [5, 45]$ and we have used a static $\hat{\rho}$ of $10.0$ in the equation. It can be seen that a higher $\alpha$ value has a greater effect on the noise rate and the contribution from $\mu$ has a maximum value of $3.8$, $ + \log\mu$ in Equation~\ref{5:eqn:single-log-noise-rate}.
Figure~\ref{5:fig:log-noise-static-rate} shows the noise rate's dependence on $\alpha$ and $\hat{\rho}$ while $\mu$ has been set to the benchmark noise rate of $-14.6$, which is the representative noise rate measured empirically during the second observing run. 
A black dashed line has been plotted at $\hat{\rho} = 6.0$ due to triggers below this $\hat{\rho}$ threshold all being assigned $\alpha = 6.0$, therefore the noise rate below this threshold is constant for all $\alpha$ values on the plot. We can clearly see a strong correlation between $\hat{\rho}$ and noise rate with a heavy dependency on $\alpha$, to the point where high $\hat{\rho}$ triggers with low $\alpha$--while still significant--have higher noise rates than triggers with high $\alpha$.

\begin{figure}
    \centering
    \includegraphics[width=\textwidth]{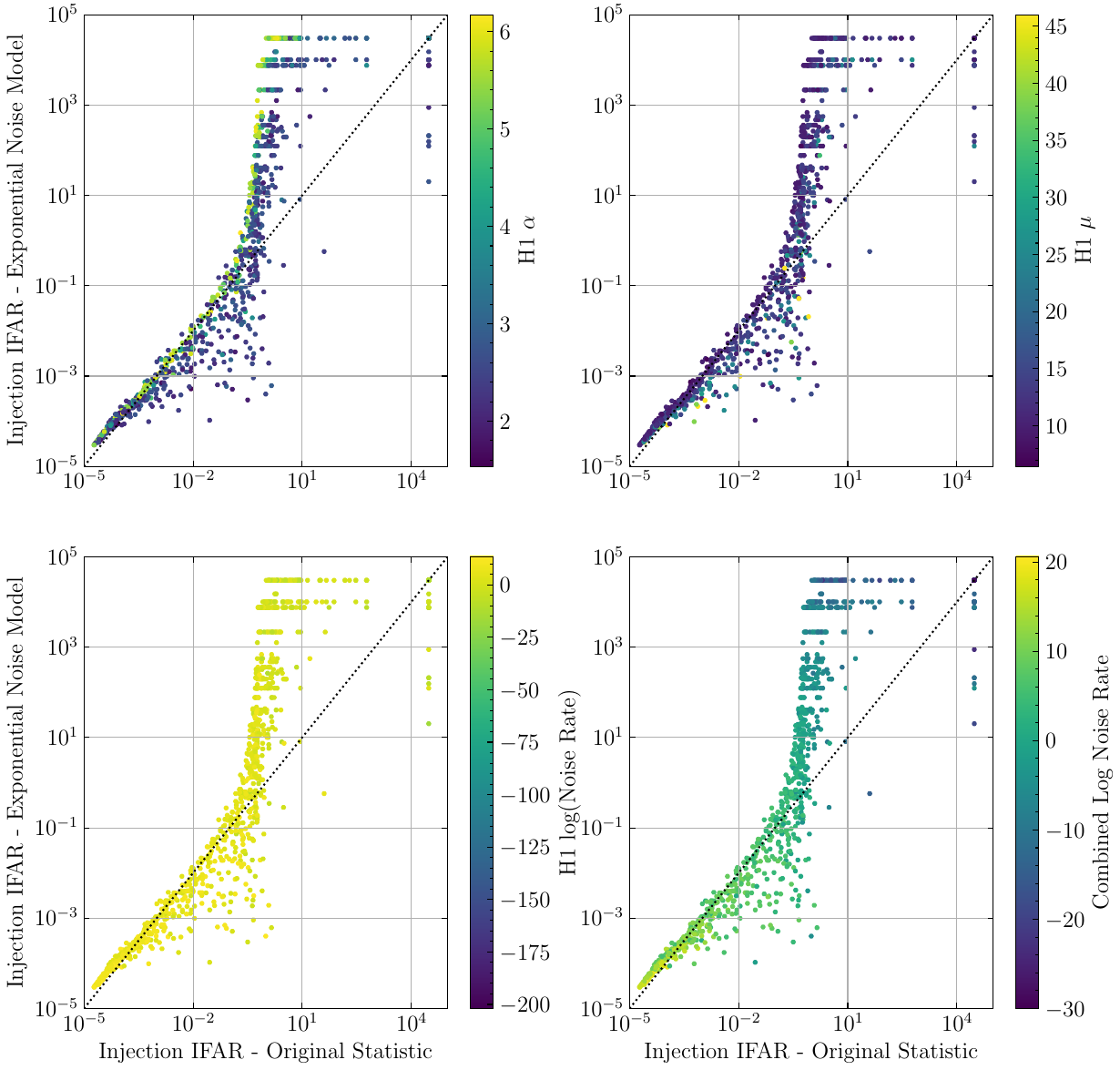}
    \caption{The inverse false-alarm rate (IFAR) values recorded for each injection seen by both the PyCBC Live search that includes the exponential noise model in the ranking statistic and the PyCBC Live search using the same ranking statistic used in the third observing run. Each point represents an injection from the injection set, with the x-axis showing the IFAR recorded using the original ranking statistic and the y-axis showing the IFAR recorded using the exponential noise model ranking statistic. The top left plot is coloured by the LIGO-Hanford trigger template's exponential fit factor, $\alpha$. The top right plot is coloured by the LIGO-Hanford trigger template's count of triggers found above $\hat{\rho}_{threshold} = 6.0$, $\mu$. The bottom left plot is coloured by the LIGO-Hanford single detector ($\log$) noise rate, calculated using Equation~\ref{5:eqn:single-log-noise-rate}. The bottom right plot is coloured by the combined ($\log$) noise rate, Equation~\ref{5:eqn:comb-log-noise-rate}.}
    \label{5:fig:ifar-ifar-subplots}
\end{figure}

We can plot the IFAR vs IFAR distribution for the injection set for a single detector's $\alpha$, $\mu$ and log noise rate to see if there is any clear dependence on the template fit parameters, seen in figures~\ref{5:fig:ifar-ifar-subplots}.
It is not clear from these figures that $\alpha$ (top-left) and $\mu$ (top-right) are having a large effect on the log noise shown in the bottom-left figure. There is a curve of high $\alpha$ points above the $y=x$ line in the top-left figure but no clear distinction between higher and lower $\alpha$ values corresponding to higher and lower IFAR values. As mentioned before and reinforced by the top-right plot, $\mu$ has a small effect on the noise rate and low $\mu$ values are more common than high ones.

Figure~\ref{5:fig:ifar-ifar-subplots} shows the combined detector noise rates (Equation~\ref{5:eqn:comb-log-noise-rate}), where we can see a clearer distinction between high and low noise rates and how this is affecting the IFAR values. However, looking at this plot we can see that combined noise rate does not completely explain the improved significance for a number of injections, we can see clearly there are some low noise rate injections below the $y=x$ line which should have caused an increase in the significance, but we have seen a decrease. Therefore, we must investigate other changes to the ranking statistic.

\subsection{\label{5:sec:comparing-statistic-construction}Ranking statistic construction}

The original ranking statistic and our new ranking statistic both take contributions from the signal event rate density, $p^{S}(\Vec{\theta})$, and the noise event rate density, $p^{N}(\Vec{\theta})$. The construction of these ranking statistics has been detailed in sections~\ref{5:sec:old-stat-construction} and~\ref{5:subsec:template-fits} where it can be seen they include the contributions from $p^{S}(\Vec{\theta})$ differently. For comparison, the original statistic is constructed as,
\begin{keyeqntitled}{Original Ranking Statistic}
\begin{equation}
    R = \sqrt{\hat{\rho}^{2}_{H} + \hat{\rho}^{2}_{L} + 2\log\left(p^{S}(\Vec{\theta})\right)}
    \label{5:eqn:original-statistic-repeat}
\end{equation}
\end{keyeqntitled}
and the statistic including the exponential noise model is,
\begin{keyeqntitled}{New Ranking Statistic}
\begin{equation}
    R = \log p^{S}(\Vec{\theta}) - \log\left(A_{N}\right) + \sum_{d} \log\left(r_{d}(\hat{\rho})\right),
    \label{5:eqn:new-statistic}
\end{equation}
\end{keyeqntitled}
where $p^{N}(\Vec{\theta})$ has been substituted with the combined noise rate defined in Equation~\ref{5:eqn:comb-log-noise-rate}.

The general detection statistic (Equation~\ref{5:eqn:general-detection-statistic}) was constructed so that in stationary, Gaussian noise the ranking statistic recovers the quadrature sum ranking statistic. When including the exponential noise model we are explicitly modelling the detector noise and therefore have chosen to use the optimal detection statistic (Equation~\ref{5:eqn:optimal-detection-statistic}) which is simply the ratio of signal and noise event rate densities.

Equations~\ref{5:eqn:original-statistic-repeat} and~\ref{5:eqn:new-statistic} show that the original statistic quadratically adds signal rate and noise rate whereas the new noise distribution model ranking statistic will add signal and noise rates linearly, changing the ratio of contributions from signal rate and noise rate in the ranking statistic calculation.

\begin{figure}
  \centering
  \begin{minipage}[t]{1.0\linewidth}
    \includegraphics[width=1\textwidth]{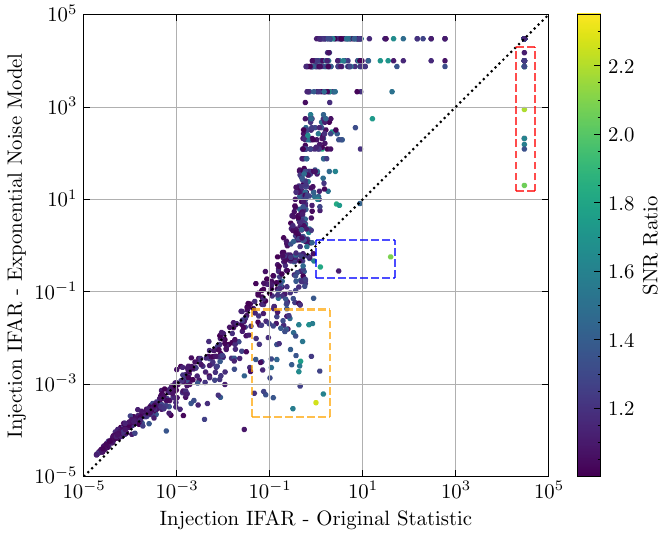}
  \end{minipage}
  \caption{The inverse false-alarm rate (IFAR) values recorded for each injection seen by both the PyCBC Live search that includes the exponential noise model in the ranking statistic and the PyCBC Live search using the same ranking statistic used in the third observing run. Each point represents an injection from the injection set, with the x-axis showing the IFAR recorded using the original ranking statistic and the y-axis showing the IFAR recorded using the exponential noise model ranking statistic. The injections are coloured by the SNR ratio of the two detector $\hat{\rho}$ values, set so that the SNR ratio is always greater than $1$. The regions highlighted by dashed boxes contain down-ranked injections and are investigated in subsequent sections.}
  \label{5:fig:ifar-ifar-snr-ratio}
\end{figure}

One particular way the different noise rate models manifests is in the ratio of the two detector $\hat{\rho}$ values. Modelling the noise distribution falloff as Gaussian causes the ranking statistic to attempt to maximise the squared sum of the $\hat{\rho}$ values found by each detector with only the signal-consistency tests to re-weight $\rho$. Our new model adds additional weights in the form of the template fits so that if a different trigger is seen with slightly lower $\hat{\rho}$ in both detectors, but the template has a lower noise rate (by having higher $\alpha$ and lower $\mu$), then it will be preferred by the noise rate model. Figure~\ref{5:fig:ifar-ifar-snr-ratio} shows the injections coloured by the SNR ratio in the new statistic search.
It can be seen that there are a number of points in regions of interest that have been down-ranked by the new search specifically because they have a high SNR ratio in the old statistic.

\subsection{\label{5:sec:diff-triggers}Different triggers selected by each search}

The ranking statistic is responsible for determining the most significant triggers in an analysis segment in the live search. As stated in Equation~\ref{5:eqn:signal-minus-noise-rate} the ranking statistic is composed of two primary contributions, the signal rate and the noise rate. As described in the previous section (\ref{5:sec:comparing-statistic-construction}), both the signal and the noise rate are template dependent and therefore when the construction of the ranking statistic is changed a different template might become responsible for the preferred trigger compared with the template chosen by the original statistic search. We investigate some injections which have now been found with a different trigger when compared to the original ranking statistic search to understand how changing the ranking statistic individually changes the injections.


Each injection in the initial injection set has an assigned index number. When a trigger is found by the search the identification number of the template responsible is saved with the trigger. Using these two values we can directly compare the template ID of each injection index to find injections which have been found with a different template ID and therefore a different trigger in both searches. Table~\ref{5:tab:200-new-stat} shows an injection, with injection index $\!=\! 200$, that has been found with a different trigger in both searches.
\begin{table}[ht]
    \centering
    \rowcolors{2}{white}{lightgray}
    \begin{tabular}{lcc}
        \toprule
        \textbf{Injection Index = 200} & \textbf{Original Trigger} & \textbf{New Trigger} \\
        \midrule
        $\log p^{N}(\Vec{\theta})$  & -15.53 & -18.79 \\
        $\log p^{S}(\Vec{\theta})$ & -15.77 & -15.67 \\
        $R_{new}$ & -0.24 & 3.12 \\
        \bottomrule
    \end{tabular}
    \caption{The ($\log$) signal rate, $\log p^{S}(\Vec{\theta})$, ($\log$) noise rate, $\log p^{N}(\Vec{\theta})$, ranking statistic value, $R_{new}$ and inverse false-alarm rate (IFAR), calculated using the ranking statistic which includes an exponential noise model, for the preferred trigger found by the PyCBC Live search when using the original ranking statistic used during the third observing run (Original Trigger) and the different preferred trigger found by the PyCBC Live search using the exponential noise model ranking statistic (New Trigger) for the injection with index = 200. The ranking statistic value for both triggers is calculated from the signal rate and noise rate of the injections using Equation~\ref{5:eqn:new-statistic}.}

    \label{5:tab:200-new-stat}
\end{table}
This table displays the values for the signal rate, $\log p^{S}(\Vec{\theta})$, the noise rate, $\log p^{N}(\Vec{\theta})$ and the ranking statistic and IFAR for both triggers calculated using the new statistic. The new trigger has a ranking statistic value of $3.12$ which makes it the preferred trigger over the trigger found using the original statistic which has a ranking statistic value of $-0.24$. This increase in ranking statistic is caused almost entirely by the lower noise rate of the new trigger and while it also has a larger signal rate this is a marginal increase. We then compare this with Table~\ref{5:tab:200-old-stat} which shows the components needed to calculate the original ranking statistic value and why the original trigger was chosen over the trigger preferred by the new statistic in the original search.
\begin{table}[ht]
    \centering
    \rowcolors{2}{white}{lightgray}
    \begin{tabular}{lcc}
        \toprule
        \textbf{Injection Index = 200} & \textbf{Original Trigger} & \textbf{New Trigger} \\
        \midrule
        $\rho_{new, H1}$  & 15.06 & 13.50 \\
        $\rho_{new, L1}$   & 6.67 & 6.58 \\
        $\rho_{new, H1}^2 + \rho_{new, L1}^2$   & 271.29 & 225.25 \\
        $\log p^{S}(\Vec{\theta})$ & -15.77 & -15.67 \\
        $R_{old}$ & 15.48 & 13.93 \\
        \bottomrule
    \end{tabular}
    \caption{The H1 new SNR, $\hat{\rho}_{H1}$, L1 new SNR, $\hat{\rho}_{L1}$, squared sum of the H1 and L1 $\hat{\rho}$s, ($\log$) signal rate, $\log p^{S}(\Vec{\theta})$, ranking statistic value, $R_{old}$, and inverse false-alarm rate (IFAR), calculated using the original ranking statistic used by PyCBC Live during the third observing run (Equation~\ref{5:eqn:original-statistic}), for the preferred trigger found by the PyCBC Live search when using the original ranking statistic used during the third observing run and the different preferred trigger found by the PyCBC Live search using the exponential noise model ranking statistic for the injection with index = 200.}
    \label{5:tab:200-old-stat}
\end{table}
We can see that the new trigger has lower $\hat{\rho}$ in both detectors and a slightly higher signal rate. Therefore, the ranking statistic value of the original trigger is greater than that of the trigger preferred by the new ranking statistic.

We can do another analysis on the injection with index $= 445$ which was also found with different triggers in both searches. Table~\ref{5:tab:445-new-stat} shows the noise and signal rates for the injection where the noise rates are very similar but the new trigger's increase ranking statistic value is primarily due to an increase in signal rate with the new trigger's template.
\begin{table}[ht]
    \centering
    \rowcolors{2}{white}{lightgray}
    \begin{tabular}{lcc}
        \toprule
        \textbf{Injection Index = 445} & \textbf{Original Trigger} & \textbf{New Trigger} \\
        \midrule
        $\log\left(\text{Noise Rate}\right)$  & -12.96 & -12.69 \\
        $\log\left(\text{Signal Rate}\right)$ & -4.40 & -2.74 \\
        $R_{new}$ & 8.56 & 9.95 \\
        \bottomrule
    \end{tabular}
    \caption{The ($\log$) signal rate, $\log p^{S}(\Vec{\theta})$, ($\log$) noise rate, $\log p^{N}(\Vec{\theta})$, ranking statistic value, $R_{new}$ and inverse false-alarm rate (IFAR), calculated using the ranking statistic which includes an exponential noise model, for the preferred trigger found by the PyCBC Live search when using the original ranking statistic used during the third observing run (Original Trigger) and the different preferred trigger found by the PyCBC Live search using the exponential noise model ranking statistic (New Trigger) for the injection with index = 445. The ranking statistic value for both triggers is calculated from the signal rate and noise rate of the injections using Equation~\ref{5:eqn:new-statistic}.}
    \label{5:tab:445-new-stat}
\end{table}
Similarly, when we look at the original statistic calculation in Table~\ref{5:tab:445-old-stat} we can see very similar original ranking statistic values where the signal rate increase with the new trigger is not great enough to overcome the increased $\hat{\rho}$ of the original triggers.
\begin{table}[ht]
    \centering
    \rowcolors{2}{white}{lightgray}
    \begin{tabular}{lcc}
        \toprule
        \textbf{Injection Index = 445} & \textbf{Original Trigger} & \textbf{New Trigger} \\
        \midrule
        $\rho_{H1,new}$  & 10.44 & 10.78 \\
        $\rho_{L1,new}$   & 8.14 & 7.16 \\
        $\rho_{H1,new}^2 + \rho_{L1,new}^2$   & 175.25 & 167.47 \\
        $\log\left(\text{Signal Rate}\right)$ & -4.40 & -2.74 \\
        $R_{old}$ & 12.90 & 12.73 \\
        \bottomrule
    \end{tabular}
    \caption{The H1 new SNR, $\hat{\rho}_{H1}$, L1 new SNR, $\hat{\rho}_{L1}$, squared sum of the H1 and L1 $\hat{\rho}$, ($\log$) signal rate, $\log p^{S}(\Vec{\theta})$, ranking statistic value, $R_{old}$, and inverse false-alarm rate (IFAR), calculated using the original ranking statistic used by PyCBC Live during the third observing run (Equation~\ref{5:eqn:original-statistic}), for the preferred trigger found by the PyCBC Live search when using the original ranking statistic used during the third observing run (Original Trigger) and the different preferred trigger found by the PyCBC Live search using the exponential noise model ranking statistic (New Trigger) for the injection with index = 445.}
    \label{5:tab:445-old-stat}
\end{table}

The addition of the exponential noise model in the ranking statistic can explain the increase in sensitivity for some injections, but others have an increase in sensitivity due to the different constructions of the ranking statistics. Different weighting has been placed on the signal and noise rates between the two searches and therefore a different trigger, with a better significance, can be chosen with a similar noise rate (the container for the exponential noise model) to the original trigger, but the template associated with that trigger has a better signal rate.

\section{\label{5:sec:investigating-regions}Investigating down-ranked regions}

In sections~\ref{5:sec:ignoring-psdvar} we investigated the effect of including the PSD variation in the ranking statistic and concluded that it will not enter the ranking statistic going forward. However, when looking at the sensitivity ratio for the new ranking statistic, including only the exponential noise model, against the original statistic (Figure~\ref{5:fig:vt-ratio-fits-only}) we can see that the decrease in sensitivity around at an FAR of $2$ per month is still around $6\%$.

We have discussed how changing the ranking statistic can change the significance of individual injections in sections~\ref{5:sec:poor-temp-fits}, \ref{5:sec:comparing-statistic-construction} and, \ref{5:sec:diff-triggers} and looking at the IFAR vs IFAR plot in Figure~\ref{5:fig:ifar-ifar-fits-only-regions} we can highlight three regions of interest which contain down-ranked injections that we must investigate and understand the reasoning behind the down-ranking. The highlighted yellow region is of particular interest as being responsible for the decrease in sensitivity at a FAR of $2$ per month.

\subsection{\label{5:sec:bottom-left-region}Injections down-ranked around FAR = 2 per month}

\begin{figure}
    \centering
    \includegraphics[width=1\textwidth]{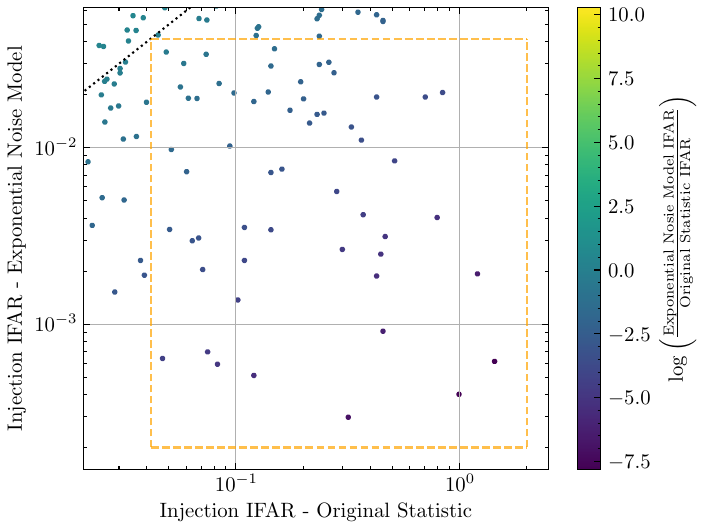}
    \caption{The inverse false-alarm rate (IFAR) values recorded for each injection seen by both the PyCBC Live search that includes the exponential noise model in the ranking statistic and the PyCBC Live search using the same ranking statistic used in the third observing run. Each point represents an injection from the injection set, with the x-axis showing the IFAR recorded using the original ranking statistic and the y-axis showing the IFAR recorded using the exponential noise model ranking statistic. The colour bar indicates the logarithmic difference in IFAR between the two searches, with positive values (yellow) indicating better performance when including the exponential noise model in the statistic and negative values (purple) indicating a better performance with the original statistic. The dashed line at y=x represents equal IFAR values for both searches. This figure is zooms in on the yellow dashed box shown in Figure~\ref{5:fig:ifar-ifar-fits-only-regions}.}
    \label{5:fig:bottom-left-region}
\end{figure}

The sensitive volume ratio between the new ranking statistic search and the original ranking statistic search (Figure~\ref{5:fig:vt-ratio-fits-only}) highlights a drop in the sensitive ratio at an approximate FAR of $2$ per month. The region responsible for this drop is highlighted on Figure~\ref{5:fig:ifar-ifar-fits-only-regions} in a yellow dashed box and a zoomed in version of this figure can be seen in Figure~\ref{5:fig:bottom-left-region}.
The original IFAR values of the injections found in this region range in $[0.02, 2.5]\,\text{years}$ and the new IFAR values range in $[0.0015, 0.0625]\,\text{years}$, all these injections are found below the $y = x$ diagonal line, indicating a lower IFAR found for the injections with the new ranking statistic.
\begin{figure}
    \centering
    \includegraphics[width=1\textwidth]{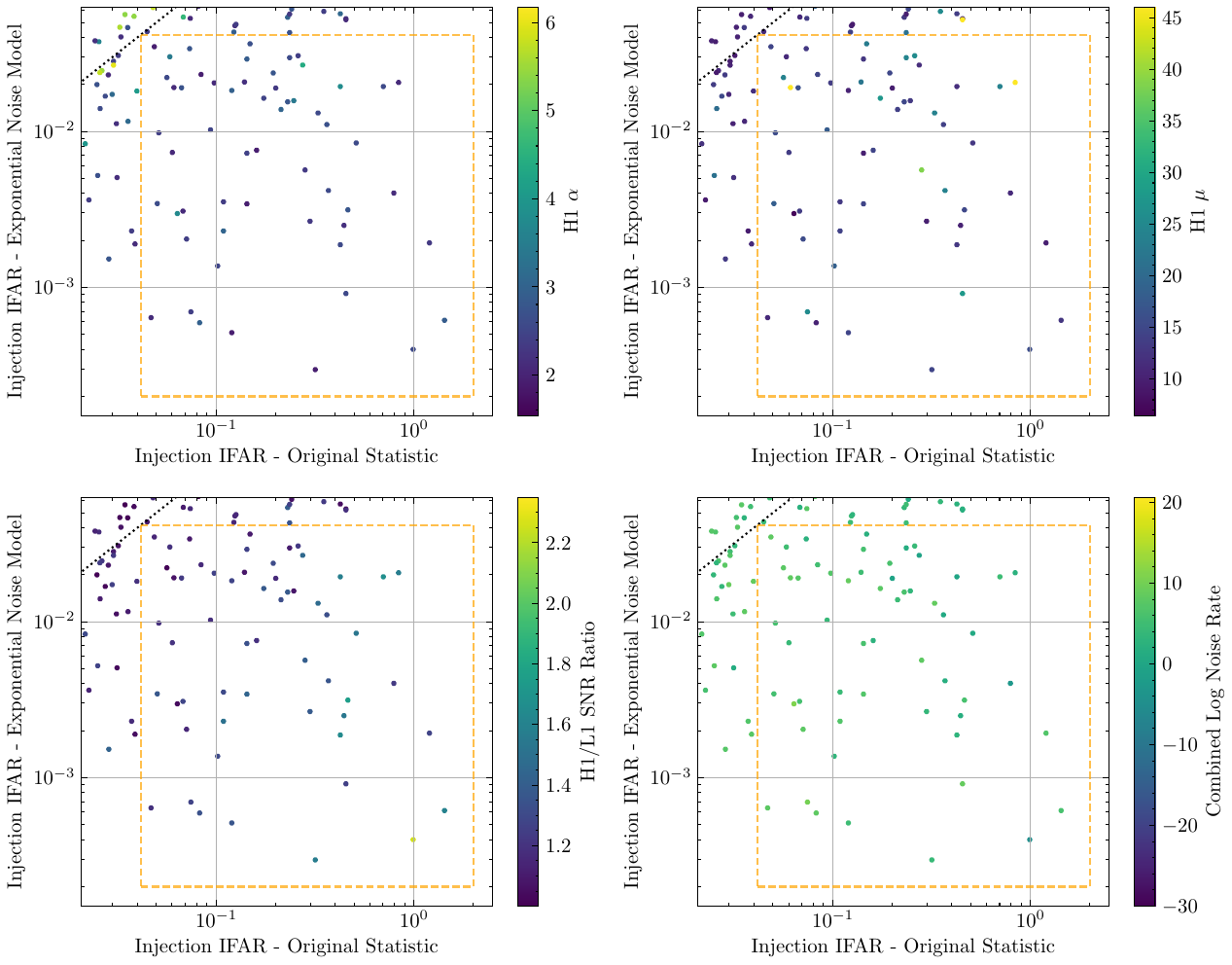}
    \caption{The inverse false-alarm rate (IFAR) values recorded for each injection seen by both the PyCBC Live search that includes the exponential noise model in the ranking statistic and the PyCBC Live search using the same ranking statistic used in the third observing run. Each point represents an injection from the injection set, with the x-axis showing the IFAR recorded using the original ranking statistic and the y-axis showing the IFAR recorded using the exponential noise model ranking statistic. The top left plot is coloured by the LIGO-Hanford trigger template's exponential fit factor, $\alpha$. The top right plot is coloured by the LIGO-Hanford trigger template's count of triggers found above $\hat{\rho}_{threshold} = 6.0$, $\mu$. The bottom left plot is coloured by the SNR ratio of the two detector $\hat{\rho}$, chosen such that the SNR ratio is always greater than $1$. The bottom right plot is coloured by the combined ($\log$) noise rate~\ref{5:eqn:comb-log-noise-rate}.}
    \label{5:fig:bottom-left-subplots}
\end{figure}
Figure~\ref{5:fig:bottom-left-subplots} highlights four different parameters we can investigate to understand why these injections have been down-ranked. Injections that are found further from the diagonal $y = x$ line have been down-ranked more significantly when using the exponential noise model in the ranking statistic. The 2 figures on the top row of Figure~\ref{5:fig:bottom-left-subplots} show H1 $\alpha$ and H1 $\mu$ of the injections which represent the contribution to the ranking statistic from the exponential noise model. The majority of the injections have been found with a high noise rate, indicated by low $\alpha$ and high $\mu$ values, but there is no correlation between these values and the distance from the diagonal line.

The bottom right figure shows the combined log noise event rate density for the injections and again, there is no clear correlation between distance from the diagonal and noise rate value, the majority of injections have a high noise rate but some templates which have been heavily down-ranked have low noise rates relative to other injections in this region.

The bottom left figure shows the ratio of the two detector $\hat{\rho}$ values identified in the search using the new statistic, calculated such that the ratio is always greater than $1$. This figure shows that injections that have been most significantly down-ranked when using the new statistic have higher SNR ratios, this is caused by the noise rate contributions to the ranking statistic described in Section~\ref{5:sec:comparing-statistic-construction} and how the original statistic maximises the squared sum of the detector $\hat{\rho}$ values in the noise rate calculation and the new statistic takes a linear sum of $\hat{\rho}$ with a weighting based on historical noise rate of the trigger template. Therefore, in the new statistic search a low $\hat{\rho}$ in one detector can no longer be compensated for by a much larger $\hat{\rho}$ in the other detector, leading to a decrease in the significance of the injections. This is a consequence of the ranking statistic formulation. To solve this, the SNR ratios could be included in the noise rate calculations but, we have not analysed the potential negative effects this might cause.

\subsection{\label{5:sec:top-right-region}Down-ranked highly significant injections}

The region indicated in the top right of Figure~\ref{5:fig:ifar-ifar-fits-only-regions} by a dashed red box contains injections that were found with the maximum IFAR value with the original ranking statistic but have now been down-ranked by including the exponential noise rate model in the ranking statistic. These injections are still found with a high significance in the new statistic. This region contains $24$ injections, which we can split into two groups for investigation: $12$ injections found with the same trigger in both searches and $12$ injections found with different triggers in both searches. 

When an injection is found with the same trigger across both searches we can dismiss any changes in $\hat{\rho}$ being the cause for an IFAR drop. We can analyse these injections to understand whether these injections were down-ranked due to high noise rate templates being responsible for the triggers, or if the construction of the ranking statistic has changed their significance.

For each injection found with the same trigger in both searches, Table~\ref{5:tab:top-right-same-trigs-fits} shows the individual detector $\hat{\rho}$, the template fit parameters $\alpha$ and $\mu$, the single detector noise event rate density, r$_n$, the combined noise event rate density,$p^{N}(\Vec{\theta})$ , the signal event rate density, $p^{S}(\Vec{\theta})$, and finally the ranking statistic value that these injections were found with using the new ranking statistic.
\begin{table}[ht]
    \centering
    \small
    \setlength{\tabcolsep}{4pt}
    \rowcolors{2}{white}{lightgray}
    \begin{tabular}{lccccccccccc}
        \toprule
        & \multicolumn{4}{c}{\textbf{H1}} & \multicolumn{4}{c}{\textbf{L1}} \\
        \cmidrule(lr){2-5} \cmidrule(lr){6-9}
        \textbf{Index} & \textbf{$\hat{\rho}$} & \textbf{$\alpha$} & \textbf{$\mu$} & \textbf{$\log r_n$} & \textbf{$\hat{\rho}$} & \textbf{$\alpha$} & \textbf{$\mu$} & \textbf{$\log r_n$} & \textbf{$\log p^{N}(\Vec{\theta})$} & \textbf{$\log p^{S}(\Vec{\theta})$} & \textbf{Rank. Stat.} \\
        \midrule
        8015 & 8.53 & 2.60 & 24.5 & -2.42 & 9.35 & 2.75 & 22.5 & -5.07 & -11.22 & 0.55 & 11.77 \\
        8026 & 9.31 & 2.37 & 9.4 & -4.74 & 8.88 & 2.52 & 14.4 & -3.67 & -12.14 & -0.91 & 11.23 \\
        8110 & 8.99 & 2.37 & 9.4 & -3.97 & 10.30 & 2.52 & 14.4 & -7.26 & -14.96 & -4.78 & 10.18 \\
        9222 & 10.09 & 2.78 & 15.0 & -7.63 & 9.11 & 2.63 & 23.0 & -4.07 & -15.44 & -3.04 & 12.40 \\
        9365 & 7.34 & 2.57 & 24.0 & 0.68 & 10.88 & 2.28 & 29.0 & -6.94 & -9.99 & 0.14 & 10.13 \\
        9777 & 8.33 & 2.57 & 33.75 & -1.51 & 9.30 & 2.78 & 34.25 & -4.63 & -9.86 & 1.91 & 11.77 \\
        10105 & 9.23 & 2.01 & 13.67 & -3.18 & 8.89 & 2.95 & 20.0 & -4.46 & -11.37 & -0.31 & 11.06 \\
        10251 & 9.52 & 2.37 & 9.4 & -5.23 & 10.26 & 2.52 & 14.4 & -7.17 & -16.14 & -10.22 & 5.92 \\
        10979 & 8.31 & 2.01 & 13.67 & -1.33 & 10.23 & 2.95 & 20.0 & -8.41 & -13.47 & -2.02 & 11.45 \\
        11386 & 12.34 & 2.37 & 9.4 & -11.91 & 6.47 & 2.52 & 14.4 & 2.41 & -13.22 & -5.02 & 8.20 \\
        11608 & 9.93 & 1.93 & 14.5 & -4.24 & 7.68 & 2.83 & 23.75 & -0.56 & -8.52 & 1.07 & 9.59 \\
        13845 & 8.40 & 2.37 & 9.4 & -2.57 & 10.37 & 2.52 & 14.4 & -7.44 & -13.74 & -1.53 & 12.21 \\
        \bottomrule
    \end{tabular}
    \caption{Properties of preferred triggers detected by both the LIGO-Hanford (H1) and LIGO-Livingston (L1) detectors for injections indicated by index number in the injection set. The table shows the new signal-to-noise ratio ($\hat{\rho}$), the exponential fit factor ($\alpha$), the number of triggers above the threshold ($\mu$), and the single detector ($\log$) noise rate ($\log r_n$) for the preferred trigger found by both detectors for each injection. The final columns provide the combined ($\log$) noise rate ($\log p^{N}(\Vec{\theta})$), the ($\log$) signal rate ($\log p^{S}(\Vec{\theta})$), and the ranking statistic value calculated using the ranking statistic including the exponential noise model.}
    \label{5:tab:top-right-same-trigs-fits}
\end{table}
%


To demonstrate how the addition of the exponential noise model can cause the down-ranking of an injection, we pick injection $11608$ which had the most significant drop, from maximum IFAR in the old statistic to an IFAR of $209.55$ years when found with the new statistic. We display the ranking statistic components for both searches in tables~\ref{5:tab:11608-new-stat} and~\ref{5:tab:11608-old-stat}.
\begin{table}[ht]
    \centering
    \setlength{\tabcolsep}{4pt}
    \begin{minipage}{0.48\textwidth}
        \centering
        \rowcolors{2}{white}{lightgray}
        \begin{adjustbox}{minipage=\linewidth-0cm, margin=0.5cm 0cm 0cm 0cm}
        \begin{tabular}{lcc}
            \toprule
            \textbf{Index = 11608} & \textbf{New Trigger} \\
            \midrule
            $\log p^{N}$  & -8.52 \\
            $\log p^{S}$ & 1.07 \\
            $R_{new}$ & 9.59 \\
            IFAR & 209.55 \\
            \bottomrule
        \end{tabular}
        \end{adjustbox}
        \caption{}
        \label{5:tab:11608-new-stat}
    \end{minipage}
    \hfill
    \begin{minipage}{0.48\textwidth}
        \centering
        \rowcolors{2}{white}{lightgray}
        \begin{tabular}{lcc}
            \toprule
            \textbf{Index = 11608} & \textbf{Original Trigger}\\
            \midrule
            $\rho_{H1,new}^2 + \rho_{L1,new}^2$   & 157.59 \\
            $\log p^{S}$ & 1.07 \\
            $R_{new}$ & 12.60 \\
            IFAR & 30174.69 \\
            \bottomrule
        \end{tabular}
        \caption{}
        \label{5:tab:11608-old-stat}
    \end{minipage}
\end{table}
It is important to remember that we cannot directly compare ranking statistic values between the searches, the only values we can compare are the signal rate which is the same for both searches due to this injection being found with the same trigger in both, and the IFAR.

Looking at Table~\ref{5:tab:top-right-same-trigs-fits} for injection $11608$ it has low $\alpha$ values of $1.93$ and $2.83$ and therefore is being heavily down-ranked by the template fits, these amount to the highest noise rate in the table of $-8.52$. This injection has a signal rate of $1.07$ which is low, the original statistic initially assigned such a large significance to this injection due to the high quadrature sum of the new SNRs heavily outweighing the low signal rate in the ranking statistic calculation (Equation~\ref{5:eqn:original-statistic}) whereas the new statistic placed less significance on the $\hat{\rho}$ from both detectors for this injection, with an equal contribution from both the signal and noise rate (Equation~\ref{5:eqn:new-statistic}). While the exponential noise model has down-ranked the significance of this injection, it has still been found with a very high IFAR.

When an injection has been found with a different trigger by the new ranking statistic and has still been down-ranked with a lower significance it indicates that the original trigger found by the original statistic would have performed even worse than the new trigger. In Section~\ref{5:sec:diff-triggers} we have discussed why different triggers are chosen by the new statistic and we can further demonstrate these using the injections in this region that have been found with a different trigger in the new ranking statistic search. We create the same table for the injections found with different template IDs, Table~\ref{5:tab:top-right-diff-temp-fits}.
\begin{table}[ht]
    \centering
    \small
    \setlength{\tabcolsep}{4pt}
    \rowcolors{2}{white}{lightgray}
    \begin{tabular}{lccccccccccc}
        \toprule
        & \multicolumn{4}{c}{\textbf{H1}} & \multicolumn{4}{c}{\textbf{L1}} \\
        \cmidrule(lr){2-5} \cmidrule(lr){6-9}
        \textbf{Index} & \textbf{$\hat{\rho}$} & \textbf{$\alpha$} & \textbf{$\mu$} & \textbf{$\log r_n$} & \textbf{$\hat{\rho}$} & \textbf{$\alpha$} & \textbf{$\mu$} & \textbf{$\log r_n$} & \textbf{$\log p^{N}(\Vec{\theta})$} & \textbf{$\log p^{S}(\Vec{\theta})$} & \textbf{Rank. Stat.} \\
        \midrule
        8466 & 13.50 & 2.78 & 13.00 & -17.27 & 6.58 & 3.01 & 17.50 & 2.21 & -18.79 & -15.67 & 3.12 \\
        8776 & 6.06 & 2.39 & 25.33 & 3.96 & 12.48 & 2.62 & 26.67 & -12.74 & -12.51 & -0.22 & 12.29 \\
        9040 & 6.20 & 1.93 & 14.50 & 2.95 & 11.01 & 2.83 & 23.75 & -9.99 & -10.77 & -1.30 & 9.47 \\
        9338 & 10.78 & 2.68 & 8.73 & -9.66 & 7.16 & 2.08 & 10.96 & 0.70 & -12.69 & -2.74 & 9.95 \\
        9819 & 7.00 & 1.93 & 14.50 & 1.41 & 11.32 & 2.83 & 23.75 & -10.87 & -13.19 & -1.25 & 11.94 \\
        9859 & 10.39 & 2.58 & 12.86 & -7.83 & 6.61 & 3.69 & 14.57 & 1.74 & -9.81 & -4.62 & 5.19 \\
        10071 & 6.65 & 2.09 & 13.43 & 1.97 & 14.53 & 3.03 & 20.71 & -21.69 & -23.45 & -16.34 & 7.11 \\
        10101 & 10.99 & 2.52 & 18.84 & -8.72 & 7.65 & 3.87 & 17.00 & -2.21 & -14.66 & -6.69 & 7.97 \\
        10134 & 6.55 & 2.60 & 24.50 & 2.72 & 8.76 & 2.75 & 22.50 & -3.46 & -4.47 & 0.09 & 4.56 \\
        10297 & 10.59 & 2.58 & 12.86 & -8.35 & 7.96 & 3.69 & 14.57 & -3.27 & -5.35 & -5.53 & -0.18 \\
        10788 & 9.28 & 2.58 & 12.86 & -4.95 & 8.38 & 3.69 & 14.57 & -4.79 & -13.47 & -0.64 & 12.83 \\
        11505 & 10.78 & 2.78 & 13.00 & -9.71 & 6.67 & 3.01 & 17.50 & 1.95 & -11.48 & -5.97 & 5.51 \\
        \bottomrule
    \end{tabular}
    \caption{Properties of preferred triggers detected by both the LIGO-Hanford (H1) and LIGO-Livingston (L1) detectors for injections indicated by index number in the injection set. The table shows the new signal-to-noise ratio ($\hat{\rho}$), the exponential fit factor ($\alpha$), the number of triggers above the threshold ($\mu$), and the single detector ($\log$) noise rate ($\log r_n$) for the preferred trigger found by both detectors for each injection. The final columns provide the combined ($\log$) noise rate ($\log p^{N}$), the ($\log$) signal rate ($\log p^{S}$), and the ranking statistic calculated using the ranking statistic including the exponential noise model.}
    \label{5:tab:top-right-diff-temp-fits}
\end{table}
These injections have very poor fit parameter values with low $\alpha$ values (especially in H1) and high $\mu$ values. One thing that is very clear, especially when compared to the same trigger table, is that these injections were all found with substantial SNR ratios. These injections and their SNR ratios can be seen in Figure~\ref{5:fig:ifar-ifar-snr-ratio} and this is a clear demonstration of how high SNR ratio triggers are favoured in the old statistic and are not in the new statistic. Overall, most injections have been found with high combined detector noise rates and the signal rates aren't large enough to compensate. This region demonstrates that while injections may be down-ranked using the new ranking statistic we are still achieving an overall increase in sensitivity in this IFAR region (Figure~\ref{5:fig:vt-ratio-fits-only}) and we understand the causes of the down-ranking of these highly significant injections.

\subsection{\label{5:sec:middle-region}Down-ranked marginal injections}

The injections found in the blue dashed box in the middle of Figure~\ref{5:fig:ifar-ifar-fits-only-regions} are injections which have been found with IFAR above ${\sim}1$ year with the original ranking statistic but have been down-ranked by the new ranking statistic. An IFAR of $1$ year is a conservative threshold that distinguishes \gwadj events as being real or fake, the sensitive ratio plot (Figure~\ref{5:fig:vt-ratio-fits-only}) highlights the sensitivity increase at this IFAR threshold as being $43\%$. These are the few injections in this region that have been down-ranked and so we want to understand what has caused this down-ranking. This region only contains $3$ injections and therefore we can investigate each injection individually to understand the cause of their down-ranking.

We can look at the numerical values associated with each injection, this is shown in Table~\ref{5:tab:bottom-right-rank-stat}.
\begin{landscape}
\begin{table}[tb]
    \centering
    \small
    \setlength{\tabcolsep}{5pt}
    \rowcolors{2}{white}{lightgray}
    \begin{tabular}{lcccccccccccccc}
        \toprule
        & \multicolumn{4}{c}{\textbf{H1}} & \multicolumn{4}{c}{\textbf{L1}} & \multicolumn{3}{c}{\textbf{New}} & \multicolumn{3}{c}{\textbf{Old}} \\
        \cmidrule(lr){2-5} \cmidrule(lr){6-9} \cmidrule(lr){10-12} \cmidrule(lr){13-15}
        \textbf{Index} & \textbf{$\hat{\rho}$} & \textbf{$\alpha$} & \textbf{$\mu$} & \textbf{$\log r_n$} & \textbf{$\hat{\rho}$} & \textbf{$\alpha$} & \textbf{$\mu$} & \textbf{$\log r_n$} & \textbf{$\log p^{N}$} & \textbf{$\log p^{S}$} & \textbf{IFAR} & \textbf{$\hat{\rho}_{H1}^{2} + \hat{\rho}_{L1}^{2}$} & \textbf{$\log p^{S}$} & \textbf{IFAR} \\
        \midrule
        8313  & 5.87 & 2.39 & 15.50 & 5.33 & 12.21 & 3.28 & 19.75 & -16.20 & -14.60 & -15.17 & 0.57 & 183.43 & -15.17 & 40.18 \\
        
        10256 & 9.57 & 2.71 & 16.16 & -5.92 & 8.04 & 3.77 & 19.30 & 8.95 & -0.69 & -1.65 & 0.35 & 119.74 & -1.65 & 1.24 \\
        
        10995 & 8.73 & 2.32 & 10.00 & -2.05 & 10.07 & 1.44 & 18.21 & -1.26 & -7.04 & -8.38 & 0.28 & 151.36 & -8.38 & 3.06 \\
        \bottomrule
    \end{tabular}
    \caption{Properties of preferred triggers detected by both the LIGO-Hanford (H1) and LIGO-Livingston (L1) detectors for injections indicated by index number in the injection set. The table shows the new signal-to-noise ratio ($\hat{\rho}$), the exponential fit factor ($\alpha$), the number of triggers above the threshold ($\mu$), and the single detector ($\log$) noise rate ($\log r_n$) for the preferred trigger found by both detectors for each injection. The `New' columns provide the combined ($\log$) noise rate ($\log p^{N}$), the ($\log$) signal rate ($\log p^{S}$), and the inverse false-alarm rate (IFAR) calculated using the ranking statistic including the exponential noise model. The `Old' columns provide the squared sum of the two detector $\hat{\rho}$s, the ($\log$) signal rate ($\log p^{S}$), and the IFAR values calculated using the ranking statistic used by the PyCBC Live search during the third observing run.}
    \label{5:tab:bottom-right-rank-stat}
\end{table}
\end{landscape}
All three of these injections have been found with the same trigger in both searches and demonstrate a few different cases for injection down-ranking. Injection $8313$ was found with the same trigger in both searches and while the new statistic's combined noise rate is very low it isn't enough to overcome the very low signal rate whereas the original statistic placed more weight on the very high L1 $\hat{\rho}$ which has compensated for the relatively low H1 $\hat{\rho}$. This is another case where the SNR ratio has a big impact on the significance of recovered events in the original statistic search. Injection $10256$ was originally found with an IFAR of $1.24$ and has been down-ranked to an IFAR of $0.35$. The L1 noise rate is very high, contributing to a large combined noise rate which, when matched with the signal rate, has not produced a very significant ranking statistic value. Finally, injection $10955$ is seen with decent $\hat{\rho}$ in both detectors, but the $\alpha$ values are poor, especially in L1. This produces a high combined log noise rate which the signal rate couldn't overcome.

\section{\label{5:sec:mdc-test}Testing using the full template bank}

To test the implementation of the new additions to the ranking statistic we ran the PyCBC Live search in an `offline' mode where the PyCBC Live search operates how it will for the real low-latency search but all the data is available before the search begins and there is no need to wait for the next analysis segment. This allows the PyCBC Live search to process the data as fast as possible and when using the PyCBC-BBH template bank we could process one week of \gwadj data in as little as a day.

The PyCBC-BBH template bank contains ${\sim}15$,$000$ templates, whereas the template bank used by the PyCBC Live search during the fourth observing run has over $700$,$000$ templates. Individual users are limited to the amount of memory that can be requested when to perform generating the template bank and so the changes could not be tested using the full template bank during development. To fully test these changes, we must use the full template bank and a live dataset which simulates exactly how the real search will operate.

The \gwadj mock data challenge (MDC) replays data from the third observing run with a known set of injections and streams it to the same data infrastructure which streams live data from the \gwadj detectors. This allows the full simulation of the low-latency \gwadj search and unlocks the ability to test changes to the \gwadj low-latency search pipelines even when the \gwadj detectors are not operating, either between observing runs, during maintenance downtime or when the detectors are not working correctly.

We were able to test our changes to the PyCBC Live ranking statistic on the PyCBC Live MDC computational resources, allowing the new ranking statistic to be run a low-latency environment with live-streamed data to test for any latency introduced into the search by the new ranking statistic components. The computational resources assigned to the PyCBC Live MDC search are equal to those being used in the PyCBC Live search during the fourth observing run and therefore we do not encounter the previously mentioned memory allocation limitations and related issues.

\begin{figure}
    \centering
    \includegraphics[width=1.0\linewidth]{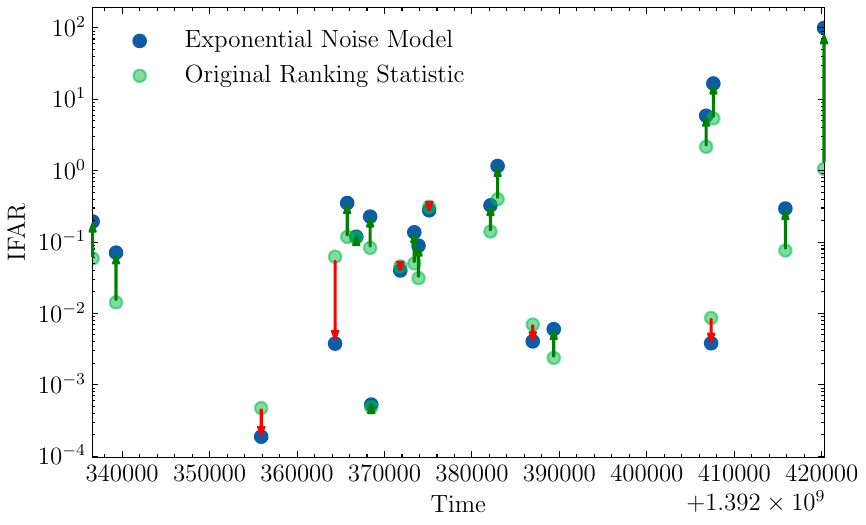}
    \caption{\Gwadj candidates jointly seen by the PyCBC Live full bandwidth Mock Data Challenge (MDC) searches using the original PyCBC Live third observing run ranking statistic (Original Ranking Statistic, green circles) and the PyCBC Live ranking statistic including the exponential noise model (Exponential Noise Model, blue circles). Events which have been seen with an increased inverse false-alarm rate with the exponential noise model ranking statistic have a green arrow between the original search's circle and the exponential noise model's circle, events seen with a decreased inverse false-alarm rate have a red arrow.}
    \label{5:fig:mdc_results}
\end{figure}

To test the ranking statistic changes, we created static template fits for all templates in the full template bank for a week of data and ran the search on the proceeding weeks during the test. The template fits were not updated during this test and the purpose of the test was not to measure sensitivity improvements but to test for bugs in implementation and if the additional ranking statistic components introduce any lag to the search. During this test we identified a number of bugs which were introducing major lag to the search, after fixing the bugs there is no latency difference between the configuration used during the third observing run and the PyCBC Live search including the new ranking statistic components. Figure~\ref{5:fig:mdc_results} shows the IFAR for \gwadj candidates identified by both the original ranking statistic and the exponential noise model ranking statistic for a single day during the MDC test.

\section{\label{5:sec:conclusion}Conclusion}

We have demonstrated that two ranking statistic components taken from the PyCBC Offline search's ranking statistic can be adapted and used in the PyCBC Live search's ranking statistic to improve the overall sensitivity of the PyCBC Live search for \gws. We have implemented the new ranking statistic additions, PSD variation and an exponential noise model, into the PyCBC Live ranking statistic and searched through $1$ week of \gwadj data containing thousands of \gwadj injections with a PyCBC Live search using the original ranking statistic used by PyCBC Live during the third observing run and another PyCBC Live search using a ranking statistic that includes PSD variation and the exponential noise model and compared the false-alarm rates of the injections found by both searches to measure the increase in sensitivity from including the new additions to the ranking statistic. 

We measure an overall increase in sensitivity when both PSD variation and the exponential noise model are included in the ranking statistic however, we measure a greater increase in sensitivity when we do \textbf{not} include PSD variation in the ranking statistic. We also investigate a decrease in sensitivity around a false-alarm rate of $2$ per month by identifying the changes the exponential noise model has made to the ranking statistic to cause the down-ranking of the injections originally found with this false-alarm rate.

We have tested the changes to the PyCBC Live search using the smaller PyCBC-BBH template bank and the complete PyCBC Live template bank being currently used in the fourth observing run and we highlight the difficulty of increasing the complexity of the low-latency search pipeline ranking statistic where changes have the potential to introduce latency to the search pipeline, delaying the dissemination of search results and how the new additions to ranking statistic have not increased the analysis latency of the PyCBC Live search.

The new additions to the PyCBC Live ranking statistic are currently being tested alongside PyCBC Live infrastructure changes to allow template fit statistics to update weekly based on the previous week's triggers to adapt to changing template noise distributions. The new PyCBC Live ranking statistic will also include data quality information from \texttt{iDQ} and will be run on the second half of the fourth observing run.

\chapter[Evaluating the PyCBC Live Early Warning Search]{Evaluating the PyCBC Live Early Warning Search}
\label{chapter:6-earlywarning}
\chapterquote{Stars are better off without us.}{Josephus Miller}
The work that forms this chapter was performed as my final project during the PhD to investigate and improve the PyCBC Live Early Warning search. This work has highlighted a few problems with the search and has identified the solutions to these problems, if more time were available the next steps would be to implement these solutions into PyCBC Live. These solutions will be implemented by other researches at the University of Portsmouth.
The PyCBC Live early warning search identifies \gwadj signals prior to merger, giving warning to the international astronomy community of an incoming \gwadj event. The early warning search is an adaptation to the PyCBC Live full bandwidth search---discussed heavily in chapter~\ref{chapter:5-pycbc-live}---and we have identified problems relating to search for pre-merger signals which require changes to be made to the early warning infrastructure to improve the sensitivity of the pipeline.

In Section~\ref{6:sec:multi-messenger-astronomy} we discuss the joint detection of astrophysical events with both \gw and electromagnetic observatories, in Section~\ref{6:sec:gw170817} we highlight the only multi-messenger event seen thus far and the timeline of observations. The PyCBC Live early warning search is described in Section~\ref{6:sec:early-warning-search} with the template bank construction in Section~\ref{6:sec:early-warning-template-bank} and how sky localisation for events is obtained in Section~\ref{6:sec:event-localisation}. Section~\ref{6:sec:gw170817-in-ew} uses GW170817 as an example to demonstrate the capabilities of the early warning search and in Section~\ref{6:sec:injection-tests} we describe a full injection set test performed using the early warning search to identify any problems and the results of this search. Finally, in sections~\ref{6:sec:false-problems} and~\ref{6:sec:missing-cands} we discuss the problems found when testing the early warning search, the cause of the problems and the potential solutions to these problems.

\section{\label{6:sec:multi-messenger-astronomy}Multi-messenger astronomy}

The PyCBC \gwadj search pipeline searches for \gwadj signals from the merger of two compact objects~\cite{PyCBC:2016}. Black holes do not emit electromagnetic radiation and therefore the merger of two black holes can currently only be directly observed with \gwadj detectors~\cite{Ghez:2000}. Neutron stars, on the other hand, are electromagnetically bright themselves and the merger of two neutron stars might produce a kilonova~\cite{Kilonovae:2017}. Kilonovae emit a broad range of electromagnetic radiation across the spectrum~\cite{kilonova_lightcurve:2017} which can be observed with many of the ground and space-based electromagnetic observatories we have on Earth. The combination of \gw and electromagnetic emissions allows us to understand more about these events than either observation could provide alone~\cite{multi_mess_astro:2019}. Astrophysical events which have been observed with more than one type of signal are called multi-messenger events, this can be from any of the three: electromagnetic signals, \gws or neutrinos. We will use the term `multi-messenger event' to describe an event which has been seen with a \gwadj signal and an electromagnetic counterpart.

One such quantity that can be measured from observing kilonovae with both \gws and electromagnetic radiation is the Hubble constant~\cite{Schutz:1986}, the measured rate of expansion of the Universe~\cite{hubble:1929}. The Hubble constant is currently determined across two distance scales using purely electromagnetic observations: firstly, the large scale cosmological measurements using the cosmic microwave background~\cite{WMAP_H0:2003} and baryon acoustic oscillations~\cite{BAO_H0:2009} and secondly, the local Universe scale using the astrophysical standard candle Cepheid variable stars~\cite{Cepheids_H0:2001} and type Ia supernovae~\cite{TypeIa_H0:1998}. The Hubble constant values obtained at the different scales do not agree~\cite{H0_tension:2020} so by combining the observation of binary neutron star signals using both \gw and electromagnetic observatories we can provide an independent measurement of the Hubble constant to break this tension.

At nearby distances ($d \lesssim 50 \, \text{Mpc}$) the Hubble constant can be calculated using Hubble's law~\cite{hubble:1929},
\begin{equation}
    v = H_{0} d,
    \label{6:eq:basic_hubbles_law}
\end{equation}
where $v$ is the recessional velocity of a source, $H_{0}$ is the Hubble constant and $d$ is the proper distance to the source. For nearby objects ($z \lesssim 0.1$): $v$ can be approximated as $v \approx cz$, where $z$ is the redshift of the object and $c$ is the speed of light; luminosity distance is approximately equal to the proper distance $d \approx D_{L}$. On small distance scales the expansion of the universe doesn't greatly affect these measurements. We obtain an equation for the Hubble constant at short distances
\begin{keyeqntitled}{Hubble's Law}
\begin{equation}
    H_{0} = c \frac{z}{D_{L}}.
    \label{6:eq:hubbles-law}
\end{equation}
\end{keyeqntitled}

The expansion rate of the Universe historically is not constant. For objects at larger distances we have to account for redshift in the luminosity distance. We begin by defining the luminosity distance in terms of the comoving distance,
\begin{equation}
    D_{L}(z) = (1 + z)D_{c}(z),
\end{equation}
which accounts for the expanding universe in the term $(1 + z)$. The comoving distance is given by
\begin{equation}
    D_{c}(z) = c \int^{z}_{0} \frac{dz^{\prime}}{H(z^{\prime})},
\end{equation}
where the Hubble constant, $H(z^{\prime})$, is now the Hubble parameter at redshift $z^{\prime}$ and depends on the model of cosmology.
We can use the comoving distance to get an expression for the luminosity distance,
\begin{equation}
    D_{L}(z) = (1 + z) c \int^{z}_{0} \frac{dz^{\prime}}{H(z^{\prime})}.
\end{equation}
$H_{0}$ will depend on the cosmological parameters representing the density of matter, radiation and dark energy. To measure the Hubble constant we need to measure the redshift and luminosity distance of multiple sources.

The luminosity distance of a \gwadj event is inversely proportional to the amplitude of the \gwadj strain~\cite{Schutz:1986},%
\begin{equation}
    h(t) \propto \frac{1}{D_{L}} , 
\end{equation}
which is obtained via \gwadj searches and parameter estimation. The \gwadj strain amplitude does have a degeneracy with the inclination angle, $\iota$, of the binary neutron star system which is described as the angle between the line of sight to the observer and the orbital angular momentum of the source~\cite{inclin_degen_2:2019}.

The \gwadj strain dependence on inclination angle manifests differently in the two \gwadj polarisations and can be simply described as~\cite{inclin_degen:2018},
\begin{align}
    h_{+} &\propto \left(1+\cos^{2}\iota\right), \\
    h_{\times} &\propto \cos\iota .
    \label{6:eqn:inclin_polarisations}
\end{align}
It can be seen from Equation~\ref{6:eqn:inclin_polarisations} that an edge-on system ($\iota = 90^{\circ}$) will produce a signal entirely in the $h_{+}$ polarisation but a face-on system will produce a mixed signal with contributions from both $h_{+}$ and $h_{\times}$. The inclination angle degeneracy could be broken if we were able to measure $h_{+}$ and $h_{\times}$ independently, but the individual \gwadj detectors are sensitive only to a linear combination of the polarisations~\cite{aLIGO:2015}. Multiple \gwadj detectors with different orientations will be sensitive to different linear combinations of $h_{+}$ and $h_{\times}$ and are therefore required to break this degeneracy and fully reconstruct the source orientation~\cite{inclin_degen_2:2019}.

The redshift of a \gwadj event is embedded in the \gwadj signal via the chirp rate of the system. The chirp rate is defined as the rate of change of the \gwadj frequency and is given to leading order as~\cite{Jaranowski:2009}
\begin{keyeqntitled}{Gravitational Wave Chirp Rate}
\begin{equation}
    \frac{df}{dt} = \frac{96}{5} \pi^{8/3} \left(\frac{G\mathcal{M}}{c^{3}}\right)^{5/3} f^{11/3}.
    \label{6:eq:chirp_rate}
\end{equation}
\end{keyeqntitled}
The rate of change of the \gwadj frequency is easy to measure and will give us the chirp mass of the system,
\begin{equation}
    \mathcal{M} = \frac{(m_1 m_2)^{\frac{3}{5}}}{(m_1 + m_2)^{\frac{1}{5}}}.
    \label{6:eq:mchirp}
\end{equation}
which depends on the two masses of the system, $m_{1}$ and $m_{2}$. The observed $\mathcal{M}$ is redshift dependent,
\begin{equation}
    \mathcal{M}_{obs} = \mathcal{M}_{source}(1 + z), 
    \label{6:eq:mchirp_obs}
\end{equation}
and $\mathcal{M}_{obs}$ is scaled by the redshift where events with larger redshifts will appear with larger observed chirp masses than their source chirp mass, $\mathcal{M}_{source}$. We would require a measurement of the Hubble constant to determine the redshift using the luminosity distance and we reach a circular argument.

We can look at other ways to measure the redshift of the event without using the \gwadj signal. The primary candidate is an electromagnetic emission from the event, for example, a kilonova from a binary neutron star system. Using the electromagnetic emission we can locate the host galaxy of the source and find the approximate redshift of the binary neutron star system. We will then have a luminosity distance, measured by the \gwadj signal, and a redshift, measured by the electromagnetic observation, and using Equation~\ref{6:eq:hubbles-law} we can calculate the Hubble constant. This combination of event information from two separate sources of information for the same event means we can use electromagnetically bright \gwadj sources as another standard siren for measuring the Hubble constant, highlighting the importance of observing multi-messenger events.

\section{\label{6:sec:gw170817}GW170817: The only multi-messenger event so far}

We have described multi-messenger events as those that have been seen with a \gw and an electromagnetic signal, in reality the event will broadcast information across the electromagnetic spectrum at varying timescales so we might see a single \gwadj signal but multiple electromagnetic signals. There are observatories around the globe and in space which span this entire spectrum and can observe counterparts to our \gwadj signal in all the different frequency ranges: from high frequency gamma rays to long wavelength radio waves. 

\begin{figure}
    \centering
    \includegraphics[width=0.8\linewidth]{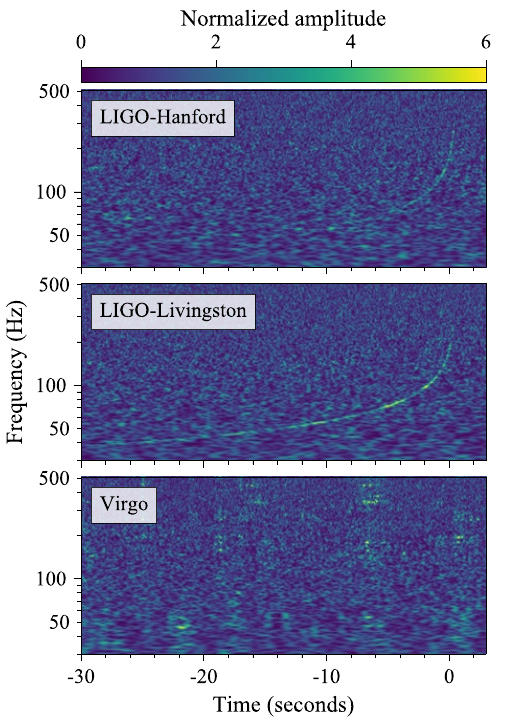}
    \caption{The three detector (LIGO-Hanford (top), LIGO-Livingston (middle) and Virgo (bottom)) time-frequency images (Q-scan~\cite{qscan:2004} showing data containing the first binary neutron star \gwadj event, GW170817~\cite{GW170817:2017}, with times shown relative to 17th August 2017 12:41:04 UTC. The detector amplitudes are individually normalised for that detector's amplitude spectral density. The glitch observed in the LIGO-Livingston detector has been subtracted, the technique and results when doing so are presented in section IV of~\cite{GW170817:2017} and this figure is taken directly from figure 1 in~\cite{GW170817:2017}.}
    \label{6:fig:gw170817_qscan}
\end{figure}

GW170817~\cite{GW170817:2017} is the first and only multi-messenger event we have observed, the Q-scans depicting the \gwadj signal found by the three detectors online at the time of the event can be seen in Figure~\ref{6:fig:gw170817_qscan} and the GCN circular~\cite{gcn_circulars:2024} (rapid astronomical bulletins submitted by and distributed to the global physics community) can be found at: \href{https://gcn.gsfc.nasa.gov/other/G298048.gcn3}{https://gcn.gsfc.nasa.gov/other/G298048.gcn3}. On the 17th August 2017 at precisely 12:41:04 UTC, the \gws from the merger of a binary neutron star system were seen by LVK detectors. Only $1.7$ seconds later, a gamma-ray burst (GRB 170817A~\cite{gw170817_joint:2017}) was seen by Fermi~\cite{Fermi:2022, Fermi_GW170817:2017} and INTEGRAL~\cite{INTEGRAL:2003, INTEGRAL_GW170817:2017}. The kilonova emissions were seen next, visual light by the Hubble Space Telescope (HST)~\cite{HST:2000, HST_GW170817:2021} at 11 hours, infrared by HST, VISTA~\cite{VISTA:2015, VISTA_GW170817:2017} at 12 hours and UV by Swift's UVOT~\cite{Swift:2004, Swift_GW170817:2017} at 15.3 hours post-merger. The final electromagnetic frequencies to be seen were X-rays by Chandra~\cite{Chandra_GW170817:2017} at 9 days post-merger and radio waves by VLA~\cite{VLA:2019, VLA_GW170817:2017} at 16 days post-merger.

Analysing GW170817's \gwadj signal and electromagnetic emissions reveal a binary neutron star system located in the Hydra constellation belonging to the galaxy NGC 4993~\cite{NGC4993:1998}. Using these signals the Hubble constant can be inferred as $H_0 \!=\! 70.0^{+12.0}_{-8.0} \, \text{km} \, \text{s}^{-1} \,\text{Mpc}^{-1}$~\cite{GW170817_H0:2017} which is consistent with both the measured cosmological value for $H_0$ ($67.74 \pm0.46 \, \text{km} \, \text{s}^{-1} \,\text{Mpc}^{-1}$~\cite{Planck_H0:2015}) and the astrophysical value for $H_0$ ($73.24 \pm1.74 \, \text{km} \, \text{s}^{-1} \,\text{Mpc}^{-1}$~\cite{Riess_H0:2016}) but, it is not accurate enough to break the tension between both values. To constrain the value of the Hubble constant further, we need to observe many more electromagnetically bright events~\cite{Palmese:2021}. Using GW170817 we were able to determine other physical properties such as the speed of gravity~\cite{Harry_speed_of_gravity:2022, Baker_speed_of_gravity:2022}.

Using the electromagnetically bright multi-messenger events to constrain the measurement of the Hubble constant is straight forward. We are also able to use `dark siren' events, which have no electromagnetic counterpart and so rely entirely on the \gwadj emission to measure the redshift. To identify the redshift of dark sirens must use galaxy catalogues and \gwadj sky maps to locate the host galaxy of the \gwadj signal. This is not discussed further in this chapter, for more information on dark sirens please look to~\cite{DES:2019, Dalang_dark_sirens:2023}

\section{\label{6:sec:early-warning-search}Detecting \gwadj events pre-merger}

\begin{figure}
    \centering
    \includegraphics[width=1.0\linewidth]{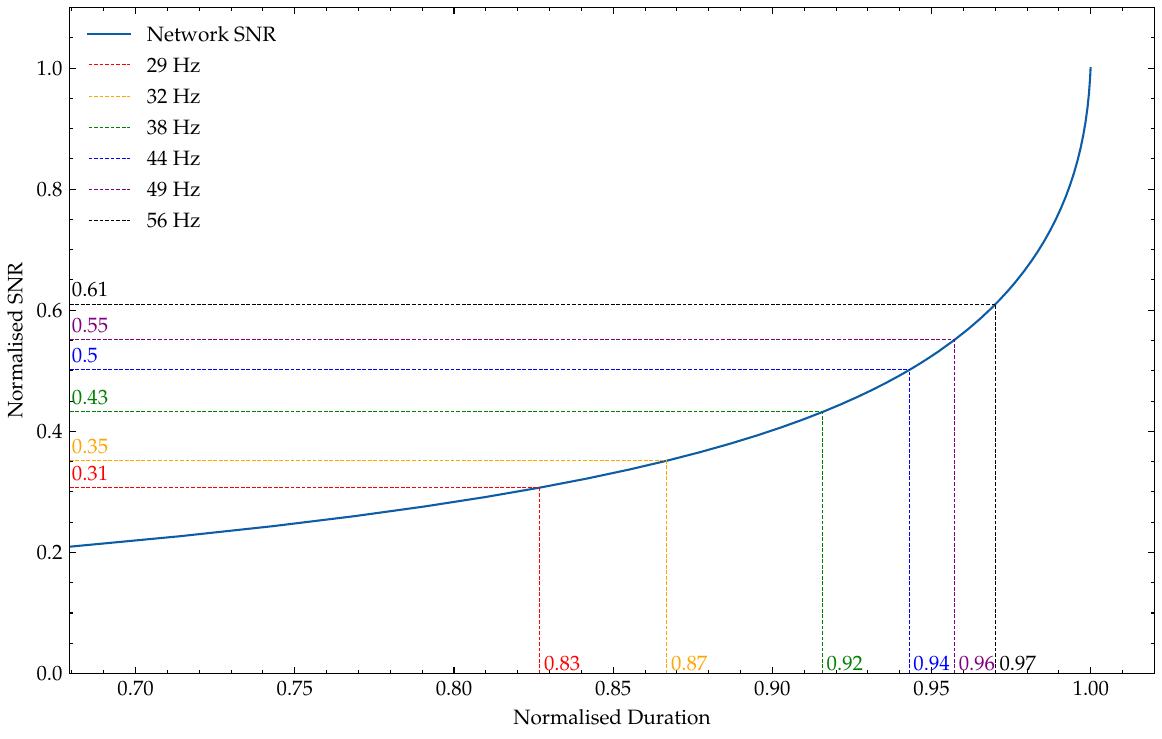}
    \caption{The accumulation of signal-to-noise ratio over the duration of a $1.4 \, \text{M}_{\odot}{\text{--}}1.4 \, \text{M}_{\odot}$ binary neutron star system. SNR and duration have been normalised such that the SNR and duration at an initial signal frequency of $15 \, \text{Hz}$ and a final signal frequency of $1024 \, \text{Hz}$ is equal to $1$.}
    \label{6:fig:snr_accumulation}
\end{figure}

\Gwadj signals from low-mass sources merge at frequencies above the sensitive frequency band of the LIGO \gwadj detectors~\cite{aLIGO_design_curve:2018}. The lower sensitive frequency of the LIGO detectors is ${\sim}20 \, \text{Hz}$ meaning the \gwadj signal can exist in the sensitive band for potentially hundreds of seconds before it is identified by the \gwadj search pipelines. Figure~\ref{6:fig:snr_accumulation} shows the accumulation of SNR over a $1.4 \, \text{M}_{\odot}{\text{--}}1.4 \, \text{M}_{\odot}$ binary neutron star signal; at $29 \, \text{Hz}$, 
$31 \, \text{\%}$ of the SNR is accumulated at $83 \, \text{Hz}$ of the total template duration.

This presents an opportunity to detect the \gwadj signal before it merges, in the inspiral regime of the system. The visible electromagnetic counterpart to the \gwadj signal is produced post-merger, therefore, by disseminating information about an ongoing \gwadj event prior to the merger we are able to give enough warning to electromagnetic observatories to slew their telescopes to the approximate region on the sky in which we will indicate a \gwadj event will occur. Providing this warning pre-merger is commonly referred to as `early warning'. This section details the methodologies used in the PyCBC Live early warning search to achieve the detection of electromagnetically bright \gwadj signals pre-merger.

Even with an accurate sky map and dedicated electromagnetic telescopes (for example, GOTO is used for \gwadj event follow-up~\cite{GOTO:2020}), the gamma-ray burst counterparts can arrive less than two seconds after merger, not giving us enough time to identify the signal, produce a sky map and slew any telescopes to the correct sky location.

\subsection{\label{6:sec:pycbc-ew-search}The PyCBC Live early warning search}

To identify potential \gwadj signals prior to the merger of the compact objects, the PyCBC Live search pipeline has created an optimised early warning search for pre-merger gravitational waveforms. In this section, we describe the design choices which allow the early warning search to operate in the early warning regime.

We use two terms when referring to search detection times: latency refers to the time taken by the search to identify a \gwadj template and produce a \gwadj event; warning refers to the time prior to the \gwadj signal's merger that the event was identified. The early warning attempts to maximise the warning time, which is improved by decreasing the latency between detection and event production.

\begin{landscape}
\begin{figure}
    \centering
    \includegraphics[width=\columnwidth]{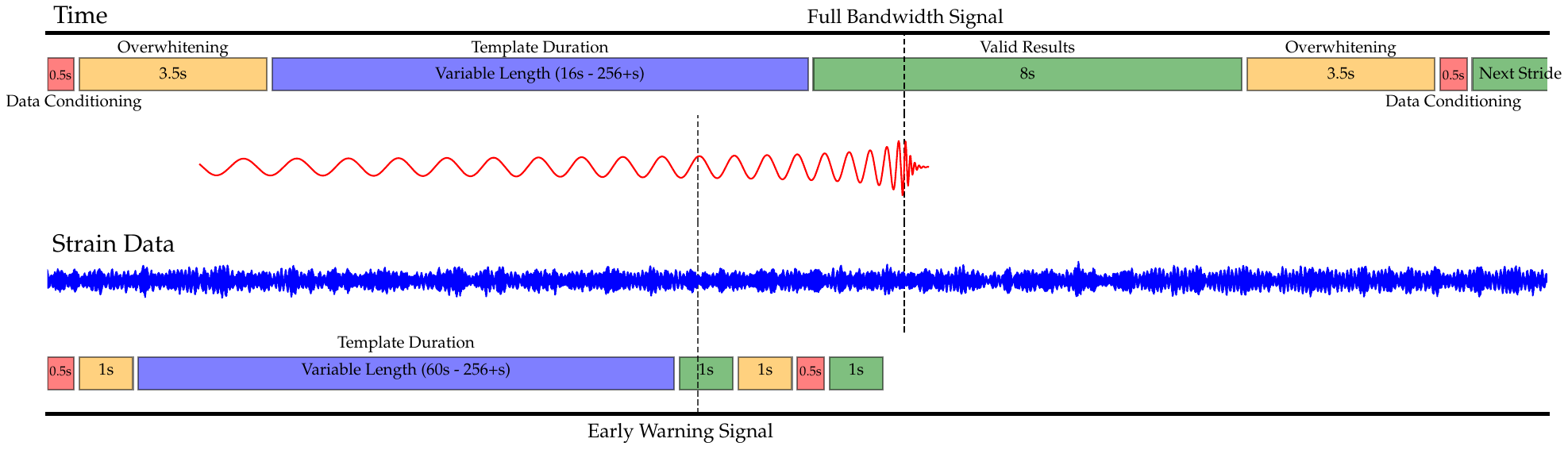}
    \caption{An illustration of the detection latency between template end and event upload to GraceDB~\cite{ligo_gracedb:2024}. The top half of the plot pertains to the PyCBC Live Full Bandwidth search and indicates that the search has a maximum latency of $20$ seconds, the bottom half describes the PyCBC Live Early Warning search which has a maximum latency of $3.5$ seconds. From the end of the \gwadj template the search must wait for the current analysis stride to end, another period of data must then be waited on (overwhitening and data conditioning) and then the search analyses the previous analysis stride during the next analysis stride.}
    \label{6:fig:latency_plot}
\end{figure}
\end{landscape}

We can divide the latency of detection into three contributing components: firstly, the analysis stride of the early warning search is one second, therefore a signal will have to wait until the entire second has arrived before it can be processed; secondly, when highpassing and overwhitening\footnote{Dividing the data by the PSD~\cite{FINDCHIRP:2012}.} our data we introduce corruption to the edge samples, we pad the data by an extra $1.5 \, \text{seconds}$ to account for this, which the signal has to wait for; thirdly, we need to analyse the data to detect the signal, we impose a strict analysis time requirement of less than the analysis stride, so data does not build up in the processing queue. In total, we have a minimum latency of $2.5 \, \text{seconds}$ (if the signal appears in the final few samples of the analysis stride) and a maximum latency of $3.5 \, \text{seconds}$ (if the signal appears in the first few samples of the analysis stride). Figure~\ref{6:fig:latency_plot} shows minimum and maximum latencies of both the full bandwidth and early warning search.
By comparison, the full bandwidth search has a minimum and maximum latency of $12$ to $20$ seconds respectively, with no possibility to observe the event pre-merger.

To ensure the processing time for the triggers is completed in less than one second aspects of the search have been simplified: we search over only two interferometers (LIGO-Hanford and LIGO-Livingston) and we use a simple new SNR single detector ranking statistic and phase-time-amplitude histogram coincident ranking statistic. Alongside this, the early warning search has a far smaller template bank of only $9180$ templates compared to the ${\sim}700$,$000$ templates in the full bandwidth template bank.

\subsection{\label{6:sec:early-warning-template-bank}Frequency truncated template bank}

The template bank used by the PyCBC Live early warning search is composed of frequency truncated templates. These are \gwadj templates which start at a lower frequency and end in the inspiral phase at a upper frequency cutoff. The bank contains six different frequency cutoffs at: $29$, $32$, $38$, $44$, $49$ and $56 \, \text{Hz}$. These frequencies correspond to approximately $60$, $46$, $29$, $20$, $15$ and $10$ seconds before merger for a $1.4 \, \text{M$_\odot$}$\text{--}$1.4\, \text{M$_\odot$}$ binary neutron star signal.

The PyCBC Live early warning template bank is constructed by generating six separate template banks (one for each frequency cutoff) using a geometric placement algorithm~\cite{Harry_Lundgren:2012} with a minimal match between templates of $0.97$. The template parameters of the bank are chosen to represent potentially electromagnetically bright signals, these are produced by the merger of low mass binary neutron star or neutron star black hole systems. The masses of the two compact objects is limited between $1\text{--}3 \, \text{M$_\odot$}$ and to reduce the size and complexity of the template bank we use zero-spin on both components. The PSD used to generate the template bank is representative of the Advanced LIGO design curve that includes estimates for the main fundamental noises of the interferometer, in particular: seismic noise, thermal noise and quantum noise~\cite{aLIGO_design_curve:2018} with a binary neutron star range of $140 \, \text{Mpc}$~\cite{ligo_prospects:2016}. The six early warning template banks combined contain $9180$ templates.
\begin{figure}
    \centering
    \includegraphics[width=\textwidth]{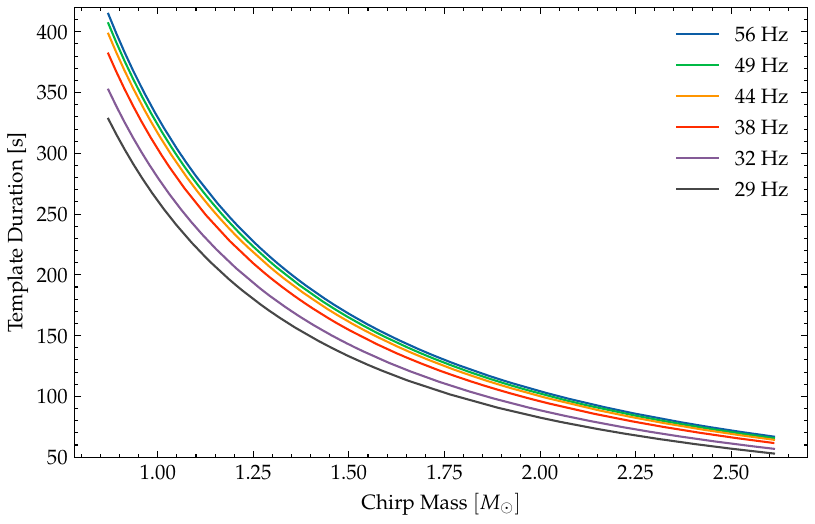}
    \caption{Template duration versus chirp mass, $\mathcal{M}$, for different frequency cutoffs in the early warning template bank. Each curve corresponding to a different frequency cutoff, $f_{final}$, ($29\text{--}56  \, \text{Hz}$). The chirp mass is calculated from the component masses of the binary system, and the template duration is calculated between the lower frequency, $17 \, \text{Hz}$ and the frequency cutoff, $f_{final} \, \text{Hz}$.}
    \label{6:fig:tb_duration_mchirp}
\end{figure}
The duration of templates at each final frequency cutoff can be seen in Figure~\ref{6:fig:tb_duration_mchirp} as a function of the chirp mass, $\mathcal{M}$, of the template. It can be seen that higher $\mathcal{M}$ templates have shorter duration and templates with a lower frequency cutoff also have shorter duration when compared to templates with the same $\mathcal{M}$ but a higher frequency cutoff.

\subsection{\label{6:sec:event-localisation}Determining event sky locations}


We call a \gwadj candidate obtained by the early warning search an `event'. These events have information about the \gwadj parameters and make a prediction of the time of coalescence of the \gwadj signal. Identifying the sky location of the event is critical for multi-messenger astronomy. With a sky location, the \gwadj observatories are able to tell electromagnetic observatories where to slew their telescopes to see the event prior to merger, to have the best chance of seeing electromagnetic counterparts as possible.

\begin{figure}
    \centering
    \includegraphics[width=1.0\linewidth]{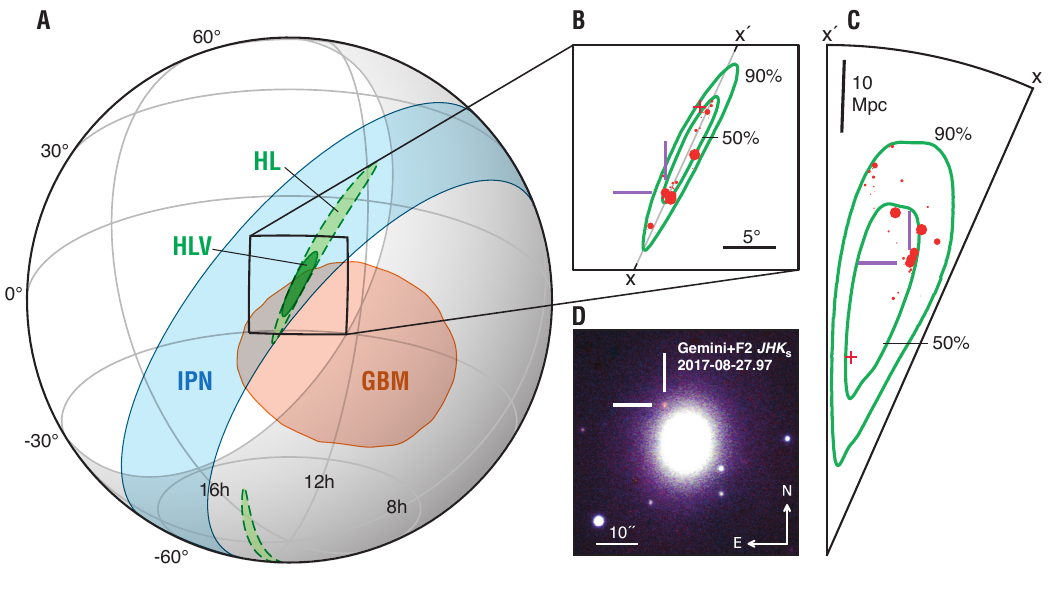}
    \caption{The localisation of GW170817 and associated electromagnetic counterpart. The rapid LIGO localisation is indicated by the green dashed contour, and the LIGO \& Virgo localisation by solid green. Fermi~\cite{Fermi:2022} is shown in orange, and the Interplanetary Network triangulation from Fermi and INTEGRAL~\cite{INTEGRAL:2003} in blue. This figure is taken directly from~\cite{gw170817_skymap:2017}}
    \label{6:fig:gw170817_skymap}
\end{figure}

We describe the sky location with three parameters: right ascension, declination, and distance. Distance is measured from the \gwadj signal, as previously described and we  measure the right ascension and declination using two key pieces of information: the time difference between time of arrivals at different detectors which allows us to calculate a time-of-arrival triangulation; and amplitude differences between the signal strength measured by each detector to determine the \gwadj polarisation, the individual detector quadrupolar sensitivity pattern of the antenna response will favour particular sky locations where the detector is more sensitive to \gws coming from certain angles. The greater number of \gwadj detectors in the network lead to more accurate sky location triangulation. Figure~\ref{6:fig:gw170817_skymap} shows the sky map created for GW170817~\cite{gw170817_skymap:2017} with the source location identified by the LIGO-Hanford and LIGO-Livingston detectors shown in the green dashed contour (representing the 90\% confidence interval).

\begin{figure}
    \centering
    \includegraphics[width=\textwidth]{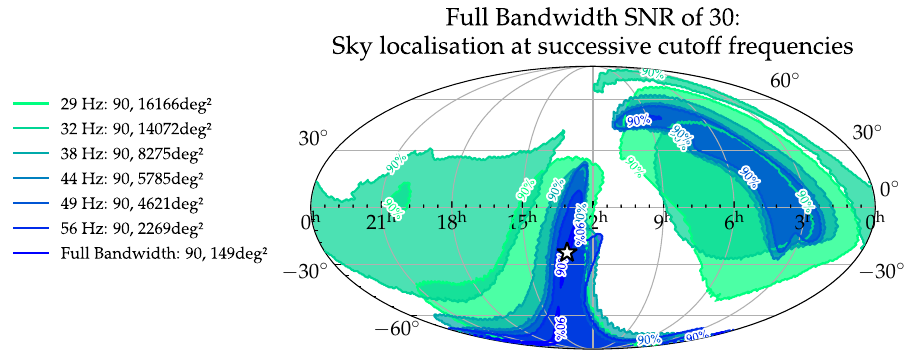}
    \caption{The \gwadj sky localisation probability map (sky map) corresponding to a $1.4 \, \text{M$_\odot$}\text{--}1.4 \, \text{M$_\odot$}$ binary neutron star signal with a full-bandwidth (upper frequency cutoff $\!=\! 1024 \, \text{Hz}$) signal-to-noise ratio of $30$. The successive $90\%$ confidence interval sky map contours for frequency cutoffs from $29\text{--}56 \, \text{Hz}$ are shown, where it can be seen that a higher frequency cutoff leads to a more accurate sky map.}
    \label{6:fig:ew_30SNR_multiple}
\end{figure}

A greater number of detectors observing the event will produce more accurate sky locations; adding a third detector to the network (like Virgo~\cite{aVirgo:2015}) is required for a point sky location. BAYESTAR~\cite{BAYESTAR:2016} is a tool for providing accurate sky maps from \gwadj events in as little as a few seconds. As a \gwadj signal is observed by templates with higher and higher frequency cutoffs, the sky map will become more accurate due to the increase in accumulated SNR. To demonstrate the improving sky map localisation as the signal progresses through our template bank we can produce the sky maps found by the early warning search when looking for a $1.4 \, \text{M$_\odot$}\text{--}1.4 \, \text{M$_\odot$}$ binary neutron star injection at a random point on the sky. The \gwadj injection is placed at a distance such that the full bandwidth SNR of the signal is $30$. Figure~\ref{6:fig:ew_30SNR_multiple} shows the sky map displaying the 90\% confidence interval contours for every frequency cutoff in the template bank as well as the full bandwidth sky map, the sky location of the injection can be seen as a star on the sky map. In the legend of the figure, you can see the decreasing sky localisation area as frequency cutoff increases.
\begin{table}[ht]
    \centering
    \setlength{\tabcolsep}{4pt}
    \rowcolors{4}{white}{lightgray}
    \begin{tabular}{cccc}
        \toprule
        \textbf{Full SNR} & \textbf{10} & \textbf{20} & \textbf{30 (Fig.~\ref{6:fig:ew_30SNR_multiple})} \\
        \midrule
        \textbf{Frequency [Hz]} & \multicolumn{3}{c}{\textbf{Localisation area [deg$^{2}$] (90\% credible area) }} \\
        \cmidrule(lr){2-4}
        29 & Not Found & 26,583 & 16,166 \\
        32 & Not Found & 22,464 & 14,072 \\
        38 & Not Found & 12,303 & 8,275 \\
        44 & 18,584 & 8,408 & 5,785 \\
        49 & 14,068 & 6,758 & 4,621 \\
        56 & 9,636 & 4,679 & 2,269 \\
        1024 & 1,908 & 300 & 149 \\
        \bottomrule
    \end{tabular}
    \caption{The sky localisation areas found by BAYESTAR~\cite{BAYESTAR:2016} for three different \gwadj injections injected at the same sky location but with different distances so as their full bandwidth signal-to-noise ratio to be equal to the `Full SNR'. The sky localisation 90\% confidence interval is shown for the six frequency cutoffs found in the PyCBC Live early warning template bank~\cite{PyCBC_earlywarning:2020} as well as the sky localisation area for the full bandwidth ($1024 \, \text{Hz}$) event. Events which weren't found above the PyCBC Live early warning search signal-to-noise ratio threshold for a frequency cutoff have no sky localisation and have been given a value of `Not Found' in the table.}
    \label{6:tab:skymap_early_warning}
\end{table}

We can do the same for an injection placed at a distance such that the full bandwidth SNR is $10$ and $20$ and display the sky localisation areas at the different frequency cutoffs in Table~\ref{6:tab:skymap_early_warning}. The $10$ SNR injection was not seen by the early warning search for frequency cutoffs $29$ and $32 \, \text{Hz}$, this is due to an event not being found above the search's SNR threshold. It is clear that a larger frequency cutoff leads to a more accurate sky localisation.

\section{\label{6:sec:gw170817-in-ew}Detecting GW170817 in early warning}

GRB 170817A was seen by Fermi $1.7$ seconds after GW170817 was seen by LVK. Fermi observes the complete sky over a ${\sim}90 \, \text{minute}$ period~\cite{Fermi:2022} and can enter a pointing mode to re-point toward high profile events, such as those reported by \gwadj detectors. With the GRB counterpart being the earliest confirmation of a multi-messenger event, it is even more important that \gwadj observatories provide as much warning for an event to allow other telescopes to be in position for detecting electromagnetic counterparts. The \gwadj signal from GW170817 could have been detected by a $29 \, \text{Hz}$ template $67$ seconds pre-merger, that is $67$ seconds warning we can provide to electromagnetic observatories.

\begin{figure}
    \centering
    \includegraphics[width=1.0\linewidth]{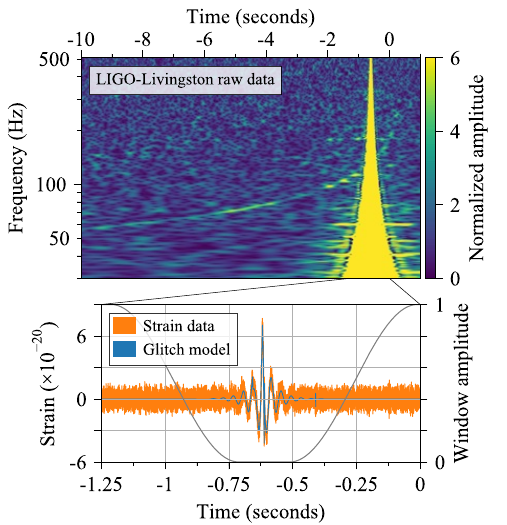}
    \caption{The Q-scan~\cite{qscan:2004} of the LIGO-Livingston data containing GW170817~\cite{GW170817:2017}, with a very loud wide bandwidth glitch present (top). The glitch was initially subtracted from the data using a windowing function to reduce the amplitude of the data containing the glitch to $0$, this windowing function can be seen as a grey line in the bottom figure. The glitch was also modelled using Bayeswave~\cite{BayesWave:2015} and subtracted from the data to preserve the \gwadj signal power found beneath the glitch, this glitch model (blue line) can be seen overlapping the strain data (orange line) in the bottom figure. This figure has been taken from~\cite{GW170817:2017}.}
    \label{6:fig:gw170817_glitch}
\end{figure}

GW170817 was seen by \gwadj searches and the initial GCN circular was distributed $27$ minutes after merger time. The LIGO-Livingston data contained a very loud glitch which had to be removed before a sky map could be made, leading to a sky map latency of $5$ hours and $14$ minutes. The Q-scan containing the glitch can be seen in Figure~\ref{6:fig:gw170817_glitch}~\cite{GW170817:2017}. The original event was uploaded to GraceDB (the Gravitational Wave Candidate Event Database~\cite{ligo_gracedb:2024}) and can be seen at on this web page: \href{https://gracedb.ligo.org/events/G298048}{https://gracedb.ligo.org/events/G298048}.
In the fourth observing run we now have well tested and robust search pipelines for \gws and as previously stated, we expect a maximum latency of $20$ seconds post-merger~\cite{PyCBC_Live:2018}, glitches can be auto-gated and BAYESTAR~\cite{BAYESTAR:2016} can produce sky maps in a few seconds.

We can demonstrate the early warning search's efficacy on GW170817. Table~\ref{6:tab:gw170817_early_warning} displays the template bank cutoff frequencies alongside the network SNR of the early warning search searching for GW170817 in the original data and a GW170817-like injection in simulated data with the current detector noise curve, also shown is the time pre-merger (warning) at which GW170817 will reach the specified frequency. The single detector SNR threshold is $4.5$, giving a network SNR threshold of $6.36$. GW170817 was much louder in the LIGO-Livingston interferometer, LIGO-Hanford's single detector SNR did not increase above the threshold until the $56 \, \text{Hz}$ frequency cutoff, at which the network SNR also crossed the threshold---$12 \, \text{seconds}$ prior to the merger. The GW170817-like injection into the O4 noise curve has very large SNRs in all the frequencies, and the $29 \, \text{Hz}$ template was observed with a $67 \, \text{second}$ warning.
\begin{table}[ht]
    \centering
    \setlength{\tabcolsep}{4pt}
    \rowcolors{4}{white}{lightgray}
    \begin{tabular}{cccc}
        \toprule
        \multicolumn{1}{c}{\textbf{Frequency}} & \multicolumn{2}{c}{\textbf{Network SNR}} & \multicolumn{1}{c}{\textbf{Warning}} \\
        \cmidrule(lr){1-4}
        \textbf{Cutoff [Hz]} & \textbf{Original Data} & \textbf{O4 Injection} & \textbf{Time [s]} \\
        \midrule
        29 & 6.19 & 13.26 & 67 \\
        32 & 5.76 & 18.58 & 52 \\
        38 & 6.11 & 21.96 & 33 \\
        44 & 6.25 & 25.87 & 22 \\
        49 & 6.19 & 24.42 & 17 \\
        56 & 7.04 & 35.24 & 12 \\
        \bottomrule
    \end{tabular}
    \caption{The signal-to-noise ratio values for the \gwadj events with corresponding frequency cutoff found by the PyCBC Live early warning search when searching over the data which contains GW170817~\cite{GW170817:2017} (Original Data) and data which contains a GW170817-like injection into representative data from the fourth observing run (O4 Injection). Also shown is the time before merger in which the event was found (Warning Time).}
    \label{6:tab:gw170817_early_warning}
\end{table}
By implementing an early warning search in PyCBC Live we are capable of identifying \gwadj signals before merger. This will aid in localising potentially electromagnetically bright \gwadj signals and inform telescopes of all frequency ranges to capture as much of the electromagnetic counterpart as possible, enabling greater quality multi-messenger astronomy.

\section{\label{6:sec:injection-tests}Testing the early warning search}

The early warning search has been operating throughout the fourth observing run, and is yet to see any early warning events. A Mock Data Challenge, which cyclically re-runs $40$ days of third observing run data, is being used to test the capabilities of the early warning search and has suggested that the search is not running optimally.

To thoroughly evaluate this search pipeline's performance, we want to carry out the first large scale injection study of early warning-like \gwadj signals to directly test the efficacy of the search. By performing this injection study, we will identify any problems and improvements that can be made to enable the greatest possibility of detecting multi-messenger events via early warning in current and future observing runs.

In this section, we describe the injection study of the early warning search, detailing the parameters of the injection set used and the results of the study; the total number of injections found with at least one candidate event and the total number of candidate events found for all injections.

\subsection{\label{6:sec:injection-set}Searching for thousands of \gwadj injections}

To test the capabilities of the early warning search an injection campaign was performed, similar to the injection campaigns performed in chapters~\ref{chapter:4-archenemy} and~\ref{chapter:5-pycbc-live}. The injection parameter value ranges match the early warning template bank, therefore, the injection set used in this test contains exclusively signals that should be seen by the early warning search. The injection set contains $32$,$599$ unique injections placed every ${\sim}100$ seconds, encompassing $3$,$456$,$000$ seconds or exactly $40$ days worth of data. The signal parameters and value ranges can be seen in Table~\ref{6:tab:ew_inj_params}.
\begin{table}[ht]
    \centering
    \setlength{\tabcolsep}{4pt}
    \rowcolors{2}{white}{lightgray}
    \begin{tabular}{ccc}
        \toprule
        \multicolumn{3}{c}{\textbf{Variable Parameters}} \\
        \cmidrule(lr){1-3}
        \textbf{Parameter} & \textbf{Value Range} & \textbf{Prior Distribution} \\
        \midrule
        Primary Mass, $m_1$ & $1.0\text{--}3.0$ [$\text{M$_{\odot}$}$] & uniform \\
        Secondary Mass, $m_2$ & $1.0\text{--}3.0$ [$\text{M$_{\odot}$}$] & uniform \\
        Primary Spin z-component, $spin1z$ & $-0.05\text{--}0.05$ & uniform \\
        Secondary Spin z-component, $spin2z$ & $-0.05\text{--}0.05$ & uniform \\
        Phase, $\phi_{c}$ & $0\text{--}2\pi$ & uniform angle \\
        Inclination, $\iota$ & $0\text{--}\pi$ & $\sin$ angle \\
        Polarisation, $\psi$ & $0\text{--}2\pi$ & uniform angle \\
        Right Ascension, $\alpha$ & $0\text{--}2\pi$ & uniform sky \\
        Declination, $\delta$ & $-\frac{\pi}{2}\text{--}\frac{\pi}{2}$ & uniform sky \\
        \bottomrule
        \multicolumn{3}{c}{\textbf{Static Parameters}} \\
        \cmidrule(lr){1-3}
        \textbf{Parameter} & \textbf{Value} & \textbf{} \\
        \midrule
        Waveform Approximant & IMRPhenomXAS~\cite{IMRPhenomXAS:2020} & \\
        Lower Frequency, $f_{lower}$ & 17 [$\text{Hz}$] & \\
        Reference Frequency, $f_{ref}$ & 17 [$\text{Hz}$] & \\
        \bottomrule
    \end{tabular}
    \caption{The parameters used to create the injection set used to test the PyCBC Live early warning search. Variable parameters will be assigned a random value within in the value range, distributed across all injections according to the prior distribution. Static parameters are the same for all injections in the injection set. Created using PyCBC~\cite{PyCBC_package:2021}.}
    \label{6:tab:ew_inj_params}
\end{table}

We first simulate the injections to create the injection set. The injections are then distributed uniformly in SNR between $12\text{--}60$. We then inject the injections into simulated coloured noise therefore, no non-Gaussian artefacts exist in the data.

\subsection{\label{6:sec:results}Results of the early warning injection test}

We expect injections can be seen by up to six templates corresponding to the six different frequency cutoffs. With $32$,$599$ injections in the injection set we could expect a maximum of $195$,$594$ candidate events in the results. We can expect some deviation to this number due to false events caused by random noise and the lower frequency cutoff templates not finding events for low SNR injections---some injections are just too quiet to be seen.

\begin{figure}
    \centering
    \includegraphics[width=1.0\textwidth]{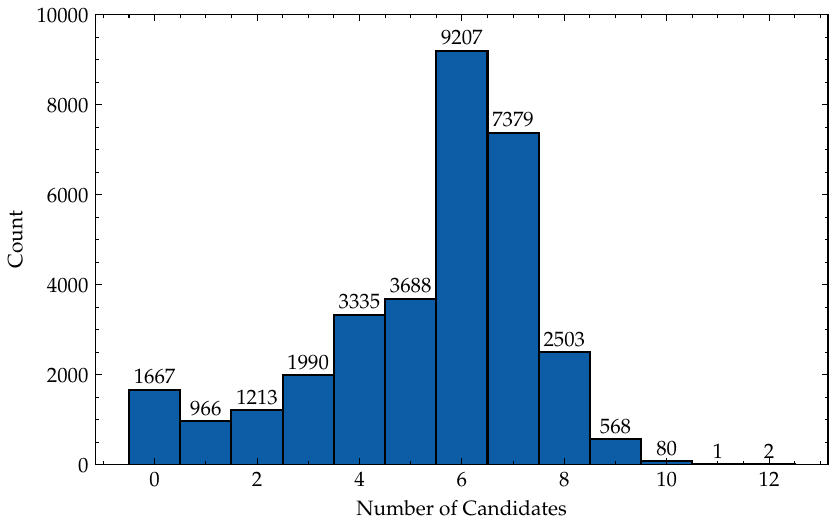}
    \caption{A histogram showing the number of injections (Count) found with varying numbers of candidate events (Number of Candidates) by the PyCBC Live early warning search. The exact number of counts is displayed on top of the histogram bars.}
    \label{6:fig:cand_hist_dupes}
\end{figure}

We create and run the early warning search on the $40$ days of data containing the injections for both LIGO-Hanford and LIGO-Livingston to find two-detector coincidences. This is consistent with the PyCBC Live early warning search which currently doesn't use Virgo data to identify candidate events. The total number of candidate events found is $175$,$219$, we can correlate a candidate event's time of coalescence with the known injection times to eliminate $1$,$211$ candidate events which were not found within a short $2 \, \text{second}$ window of an injection time of coalescence. The total number of candidate events found for all injections is $174$,$008$, Figure~\ref{6:fig:cand_hist_dupes} shows a histogram of the number of candidate events found per injection.

Upon inspection of Figure~\ref{6:fig:cand_hist_dupes} we immediately identify an issue with the early warning search; the histogram shows us a large number of injections which have been found with greater than six candidate events. This is unexpected and we will explore the cause of this in the following sections, as well as other identified issues.

\section{\label{6:sec:false-problems}False events}

With these results we can evaluate the performance of the early warning search and identify the problems that we initially set out to find. For each identified problem, we describe its source, estimate the number of injections affected, and detail potential solutions to be implemented into the early warning search to prevent the problem from occurring in future observing runs. We will begin with the problem identified by Figure~\ref{6:fig:cand_hist_dupes}, where the early warning search is identifying too many candidate events for some injections.

\subsection{\label{6:sec:duplicate-frequency-cands}Finding duplicate frequency cutoff candidates}


The template bank used by the early warning search contains templates for six different frequency cutoffs. The time before merger corresponding to each frequency cutoff is injection specific and it is expected that when the early warning search reaches that time, it would identify a candidate event found by a template with the correct frequency cutoff. This then places a maximum number of events found for each injection using the template bank at six events.

We have observed injections that have been seen with more than six candidate events. These candidates must have frequency cutoffs equal to one of the six found in our template bank and therefore some frequency cutoffs have been found multiple times for an injection. This should not be possible as the time before merger for the frequency cutoff would correspond to a stride in the early warning search and only the best coincident event can be chosen by the early warning search as a candidate.

\subsubsection{\label{6:sec:cands-across-bounds}Candidates straddling search boundaries}
%
%
\begin{figure}
       \centering
    \includegraphics[width=0.75\textwidth]{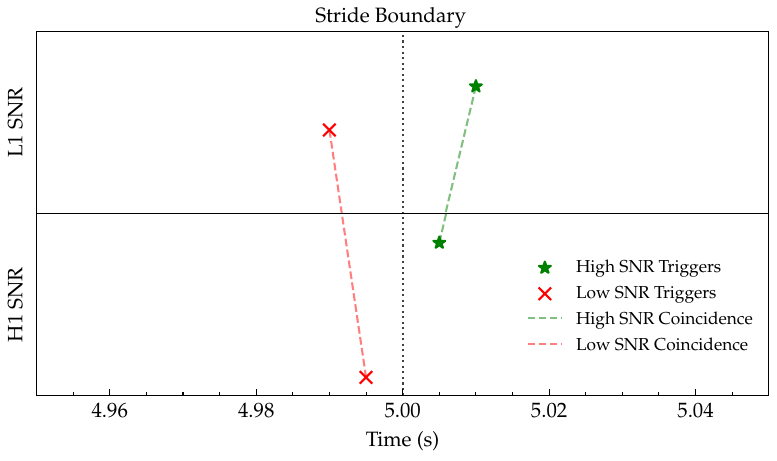}
    \caption{An illustration of two candidate events being found with the same frequency cutoff for a single injection. The red coincidence has lower single detector and coincidence SNR, but both are above SNR threshold due to the frequency cutoff template ending very close to the analysis stride boundary, this event is uploaded as a candidate event. The green coincidence is the `proper' coincidence found after the frequency cutoff template has ended and it also uploaded after the first one, leading to two events with the same frequency cutoff being uploaded for one \gwadj signal.}
    \label{6:fig:candidates_across_boundaries}
\end{figure}

The early warning search operates on a single second analysis stride, meaning every single second the template bank is matched filtered with the data to discover any \gwadj events. It is possible that the pre-merger time corresponding to a frequency cutoff will lie very near to the boundary between search strides. In this case, the search will record an event found in the first stride and will then again record another event in the second stride with the same frequency cutoff.

Figure~\ref{6:fig:candidates_across_boundaries} is a demonstration of this possibility. Either event can have a higher SNR when compared to the other, but the two scenarios can be treated differently to prevent some duplicate events being uploaded in the future. If the event found in the first stride has a higher SNR than the second stride we can prevent an upload of the second event (provided the template frequency cutoff is the same)---this is already done by the PyCBC Live Single Detector search~\cite{PyCBC_singles:2022}---but if the event found in the second stride has a higher SNR than the first then we will have to upload two events. We cannot delay uploading the first event.

\subsubsection{\label{6:sec:trigs-across-bounds}Triggers across search boundaries}
%
%
\begin{figure}
       \centering
    \includegraphics[width=0.8\textwidth]{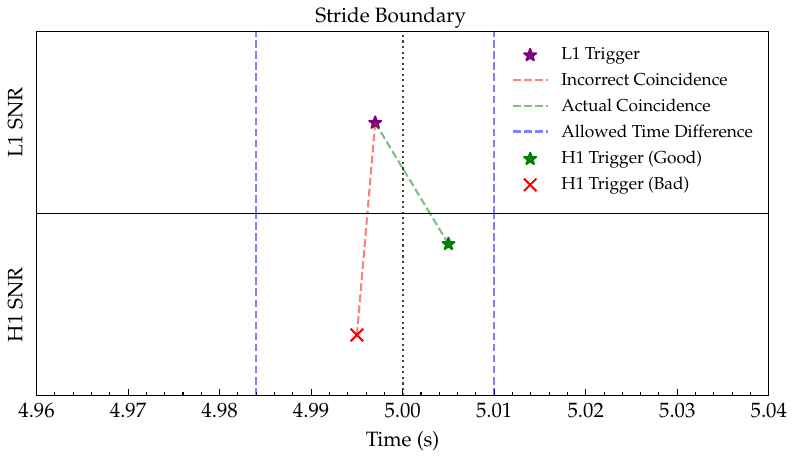}
    \caption{An illustration of two coincident candidate events being uploaded for the same frequency cutoff. The L1 trigger for the coincidence arrives in the analysis stride and the H1 trigger arrives in the next analysis stride. The early warning search has found a low significance H1 trigger in the first analysis stride and made a low significance coincidence. When the second analysis stride is analysed the correct coincidence will be found and the same L1 trigger is uploaded again with another coincident trigger at the same frequency cutoff.}
    \label{6:fig:triggers_across_boundaries}
\end{figure}
The stride boundary can have another effect, which causes duplicate frequency events to be uploaded. Coincident triggers between detector 1 and detector 2 have an allowed time difference in each detector to account for the physical light-travel time of the \gw and a smaller amount of computational timing errors. It is possible for the two single detector triggers to appear across a stride boundary.

The trigger will first be seen in detector 1, which will be seen as a highly significant single detector trigger by the early warning search, where it will attempt to match with a coincident trigger in the other detector. As previously mentioned, this could be Gaussian noise which happens to match well enough for a low-significance coincidence to be made, or it could be the final point in a rising SNR time series. When detector 2 records the trigger in the next stride the early warning search will create the actual significant coincident event using the same detector 1 trigger twice. Therefore, we will get two candidate events at the same final frequency cutoff with one being less significant than the other. A demonstration of this case is seen in Figure~\ref{6:fig:triggers_across_boundaries}.

This is an unavoidable failure mode of the early warning search, we require the light travel time in the ranking statistic to form coincidences between single detector triggers and this will naturally fall across the stride boundary for some injections.

\begin{figure}
    \centering
    \includegraphics[width=1.0\textwidth]{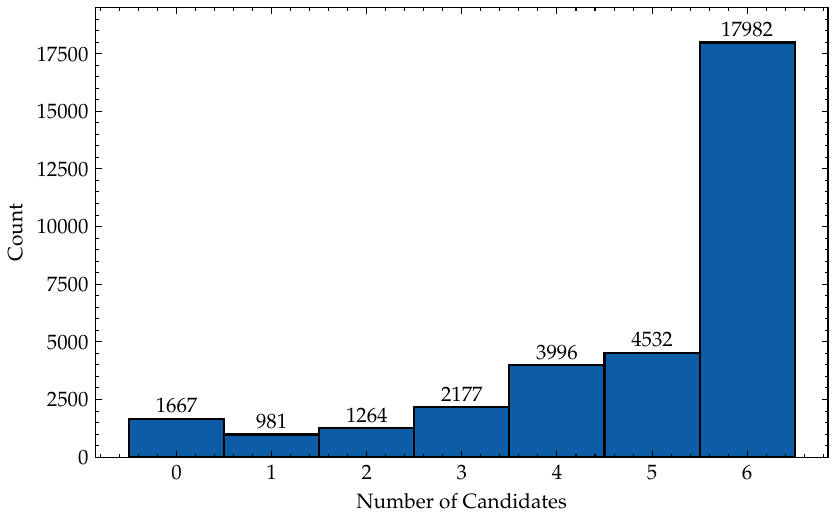}
    \caption{A histogram showing the number of injections (Count) found with varying numbers of candidate events (Number of Candidates) by the PyCBC Live early warning search. The exact number of counts is displayed on top of the histogram bars.}
    \label{6:fig:cand_hist_dupes_removed}
\end{figure}

We find that $13$,$098$ injections ($40.17\%$ of the total) were found with at least one duplicate frequency and $17$,$432$ duplicate frequencies can be removed from our results leaving $156$,$576$ unique candidate events remaining. Figure~\ref{6:fig:cand_hist_dupes_removed} recreates Figure~\ref{6:fig:cand_hist_dupes} where the duplicate candidate events have not been included.

Candidate events are all treated independently and while the additional events uploaded to GraceDB may be confusing, \gwadj signals on GraceDB commonly have multiple events uploaded per pipeline with multiple pipelines. Therefore, the event with the highest SNR will be preferred until a higher SNR event is identified.

\subsection{\label{6:sec:false-alarms}Low significance candidate events}

We have discussed false alarm candidate events that do not have times of coalescence consistent with any known injection time of coalescence and have removed $1$,$211$ candidates from our final results. There is a possibility of a false alarm candidate event to have a time of coalescence that falls nearby to an injection. These are easy to identify because those that appear at a lower frequency will not have the corresponding candidates from the next frequencies in the template bank. This is not something the early warning search can take into account in low-latency, but it is something that can be used post-detection to confirm the legitimacy of an event.
\begin{figure}
       \centering
    \includegraphics[width=\textwidth]{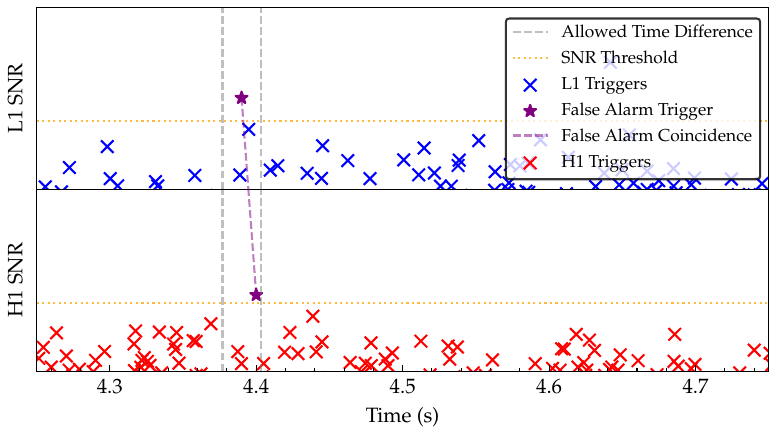}
    \caption{An illustration of a coincident candidate being found by the alignment of Gaussian noise triggers in two detectors. The crosses represent the single detector triggers identified by the \gwadj search, triggers found above the SNR threshold can be considered a single detector trigger. Coincident triggers have to lie inside the allowed coincidence window, shows by the grey dashed lines around the L1 trigger.}
    \label{6:fig:low_significance_candidates}
\end{figure}
The injection set used in this search was made using a PSD representative of the advanced LIGO noise curve~\cite{aLIGO_design_curve:2018} and therefore is coloured Gaussian noise, not containing any non-Gaussian noise transients. This means that when we identify a false-alarm candidate it has been caused by Gaussian noise triggers aligning in between the two detectors and finding a coincidence, a demonstration of this can be seen in Figure~\ref{6:fig:low_significance_candidates}.

We can identify further false-alarm candidate events in two ways: look for injections with single candidates that have frequency cutoffs not equal to the highest cutoff in the template bank ($56 \, \text{Hz}$); and by comparing the injection template and the candidate event template. We observed $99$ injections with single candidates and a further $39$ candidate events with a mismatch between template $\mathcal{M}$ and injection $\mathcal{M}$ of ${\pm}1\%$.

\begin{figure}
    \centering
    \includegraphics[width=1.0\linewidth]{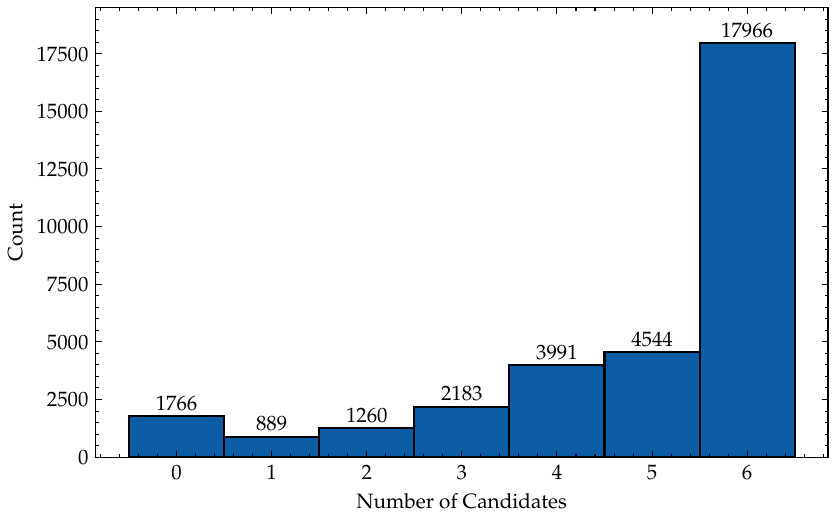}
    \caption{A histogram showing the number of injections (Count) found with varying numbers of candidate events (Number of Candidates) by the PyCBC Live early warning search. The exact number of counts is displayed on top of the histogram bars.}
    \label{6:fig:cand_hist_all_removed}
\end{figure}

Of the original $174$,$088$ candidate events identified and we have been able to eliminate $17$,$570$ candidate events that were caused by boundary issues or false-alarms, leaving $156$,$438$ candidate events identified by the early warning search during the injection study. Figure~\ref{6:fig:cand_hist_all_removed} shows the distribution of the number of candidate events found per injection, where $13$,$225$ injections have additional unexpected candidate events.

\section{\label{6:sec:outside-coinc-window}Candidate events we did not see}

\begin{figure}
       \centering
    \includegraphics[width=\textwidth]{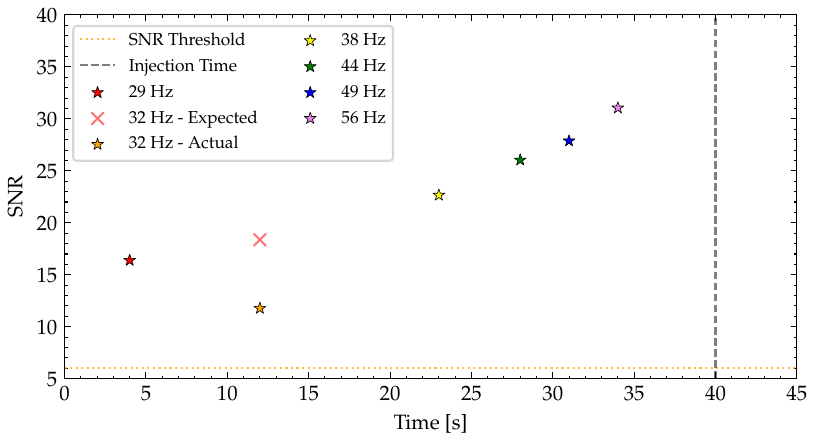}
    \caption{An illustration of the expected and SNRs for the case where one of the event's coincident triggers have fallen outside the coincident timing window. In the first case we see a missing frequency in the SNR evolution (indicated by the red cross), in the second case we see a non-monotonic increase in SNR where the search has found a low significance event to `fill' the gap (shown by the orange star).}
    \label{6:fig:non-monotonic-snr}
\end{figure}

We have identified injections that have seen extra candidate events from what we expect and now we can look at injections where we expected to see more events than we did. We count the number of candidate events for each injection and compare to the expected number of candidate events considering the lowest frequency cutoff the injection was found with. For example, if an injection has been seen at $29 \, \text{Hz}$ then we would expect six candidate events in total, whereas for an injection seen first at $49 \, \text{Hz}$ we would only expect two candidate events ($49 \, \text{Hz}$ and $56 \, \text{Hz}$). A further inspection of these candidate event SNRs can reveal another issue with the observations, we expect a monotonic increase in the SNR of each subsequent frequency cutoff in the template bank. If the SNR between consecutive candidates does not exhibit this SNR increase then we know this injection has encountered a problem we must investigate. Figure~\ref{6:fig:non-monotonic-snr} shows an illustration of these injections, where the 
$32 \, \text{Hz}$ has either not been seen at all or has been seen with a much lower SNR than what is expected.

\subsection{\label{6:sec:light-travel-time}Coincident triggers outside the light travel time}

We investigate the frequency of the missing candidate events for injections that have a gap in frequency evolution and obtain the following counts per frequency cutoff: $29 \, \text{Hz} \!=\! 0$ (We do not distinguished between missed and not found for the first frequency cutoff), $32 \, \text{Hz} \!=\! 288$, $38 \, \text{Hz} \!=\! 162$, $44 \, \text{Hz} \!=\! 132$, $49 \, \text{Hz} \!=\! 168$, $56 \, \text{Hz} \!=\! 140$. Some injections have multiple missing frequencies, and in total we find $873$ injections ($2.68\%$) are missing at least one expected candidate event in its frequency evolution. This is a lower limit on the true number of injections with missing candidate events, we are unable to count injection where: the $56 \, \text{Hz}$ template was the only one above the SNR threshold and was missed; the missed frequency is the lowest frequency in the evolution, for example $29 \, \text{Hz}$, then we are just assuming that the $29 \, \text{Hz}$ did not have enough SNR to be seen and wasn't `missed' in this context.

The early warning search will output all the triggers found when matched filtering the template bank and the data every second. Candidate events are detected in the analysis segment containing the end time of the template, which represents the frequency cutoff for the injection. Therefore, we can look at the triggers found by the search to determine why a candidate event was not identified. The injections with missing frequencies all record single detector triggers above the SNR threshold and with the best matching template for both detectors but, we have identified that the time difference between these two triggers is greater than our allowed coincidence timing window---a component of the coincident ranking statistic to demand physical time differences between triggers.

\begin{figure}
    \centering
    \includegraphics[width=\textwidth]{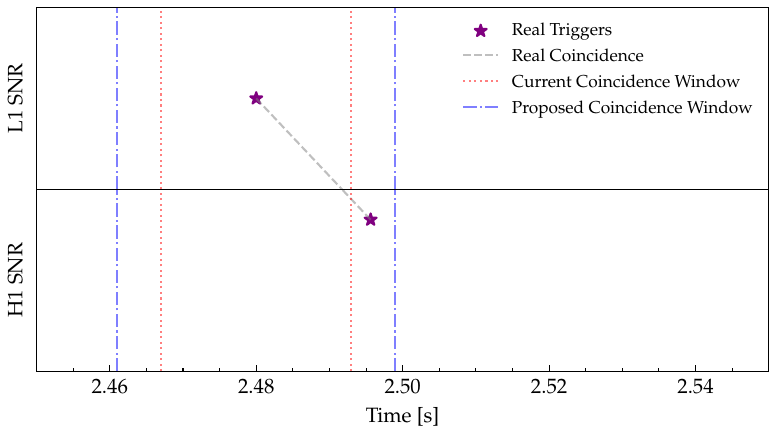}
    \caption{An illustration of two single detector triggers being found outside the allowed coincidence time difference window. The current coincidence window shows the triggers falling outside the window and therefore the search does not form a coincidence. With the new proposed coincidence time window the coincidence would be found.}
    \label{6:fig:outside_coinc_window}
\end{figure}

The two LIGO detectors are separated by a straight line through the Earth $3002$ kilometres long. Given the speed of light there is a $0.01 \, \text{second}$ window for the \gwadj signal to travel from one detector to the other. We allow a further $0.003 \, \text{seconds}$ in addition to the light travel time to account for any detection response, timing, or waveform uncertainties. This give a total allowed coincidence timing window of $0.013 \, \text{seconds}$. An illustration of this problem can be seen in Figure~\ref{6:fig:outside_coinc_window}. How is it possible for our `coincident' triggers from our injection set to lie outside the allowed coincidence timing window?

\subsubsection{\label{6:sec:timing_error}Timing accuracy error}

As described in~\cite{Fairhurst:2010}, the accuracy of the candidate time in a detector can be determined by,
\begin{keyeqntitled}{Timing Accuracy}
\begin{equation}
    \sigma_{t} = \frac{1}{2\pi\rho\sigma_{f}}
    \label{6:eqn:timing_acurracy}
\end{equation}
\end{keyeqntitled}
where the timing accuracy, $\sigma_{t}$, is inversely proportional to the SNR, $\rho$, and the effective bandwidth of the source, $\sigma_{f}$ which is defined as
\begin{equation}
    \sigma_{f}^2 = \left(\int^{\infty}_{0} df \frac{f^{2}|\hat{h}(f)|^{2}}{S(f)}\right) - \left( \int^{\infty}_{0} df \frac{f|\hat{h}(f)|^{2}}{S(f)}\right)^{2} .
    \label{6:eqn:eff_bandiwdth}
\end{equation}
where the waveform $\hat{h}$ is normalised such that $\int \frac{|\hat{h}(f)|^{2}}{S(f)}df = 1$. We use the same PSD to normalise that has been used throughout the injection study. As shown in Table 1. of~\cite{Fairhurst:2010}, the timing error for a binary neutron star system in an advanced LIGO configuration with a BNS range of $160 \, \text{Mpc}$ is $0.46 \, \text{ms}$. This is negligible when compared to our light-travel time of $10 \, \text{ms}$. We can approximately calculate the timing accuracy of the early warning templates using equations~\ref{6:eqn:timing_acurracy} and~\ref{6:eqn:eff_bandiwdth}, where the \gwadj strain term, $|h(f)|$, is assumed to be equal to the inspiral frequency evolution of a binary neutron star signal, $f^{-\frac{7}{6}}$. The estimated effective bandwidths and timing accuracy errors can be seen in Table~\ref{6:tab:timing_errors}.
\begin{table}[ht]
    \centering
    \setlength{\tabcolsep}{4pt}
    \rowcolors{3}{white}{lightgray}
    \begin{tabular}{ccc}
        \toprule
        \textbf{Frequency [Hz]} & $\sigma_{f}$ [Hz] & $\sigma_{t}$ [ms] \\
        \midrule
        29 & 3.38 & 5.89 \\
        32 & 4.19 & 4.75 \\
        38 & 5.79 & 3.44 \\
        44 & 7.37 & 2.70 \\
        49 & 8.68 & 2.29 \\
        56 & 10.51 & 1.89 \\
        1024 & 90.05 & 0.22 \\
        \bottomrule
    \end{tabular}
    \caption{The effective bandwidth, $\sigma_{f}$, and timing accuracy error, $\sigma_{t}$, for the frequency cutoffs (Frequency) found in the early warning template bank as well as the full bandwidth frequency ($1024 \, \text{Hz}$).}
    \label{6:tab:timing_errors}
\end{table}
Using an SNR of $8$ (and a lower frequency of $17$ Hz) yields $1$ sigma timing errors from $5.89$ to $1.89$ milliseconds for the early warning frequency cutoffs, where higher frequencies have smaller timing errors. The current allowed error is $3 \, \text{ms}$ therefore we can expect to miss candidates from all the frequency cutoff templates when the timing error places coincident triggers outside the current allowed coincidence time window. This can be remedied with a widening of the window which could be done on a template dependent basis, with a timing accuracy error dependent on the frequency cutoff.

The PyCBC Live early warning search uses a sample rate of $256 \, \text{Hz}$ which contributes to a worse timing accuracy. The time between samples is $3.9 \, \text{ms}$ and a trigger found by another detector $3$ samples later will be within the allowed window of $13 \, \text{ms}$ but being $4$ samples later equals a time difference of $15.6  \, \text{ms}$; falling outside the window. We can perform a subsample interpolation between samples to recover the correct trigger time within the allowed window.

\subsubsection{\label{6:sec:ew_phasetd}Early warning tuned Phase-Time-Amplitude histogram}
\begin{figure}
    \centering
    \includegraphics[width=1.0\linewidth]{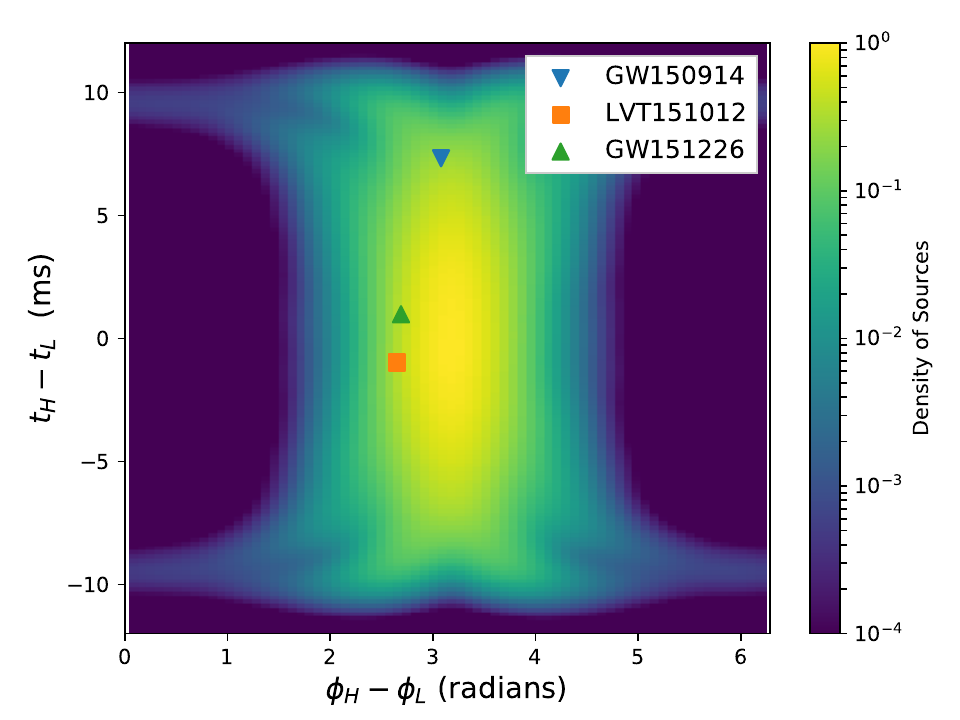}
    \caption{Phase-Time-Amplitude histogram for the PyCBC ranking statistic, showing the density of \gwadj candidate sources as a function of time delay on the y-axis, $t_H - t_L$ in milliseconds, and phase difference on the x-axis, $\phi_H - \phi_L$ in radians, between detectors. The first three \gwadj candidates have been highlighted on the plot~\cite{gwtc1:2019}. The colour scale indicates the density of sources in logarithmic scale. The histogram can be used to determine how likely a signal is based on the differences in phase, time, and amplitude recorded by each detector. Taken from~\cite{PyCBC:2017}.}
    \label{6:fig:phase-time-histogram}
\end{figure}
Increasing the allowed coincidence time window allows the early warning search to form coincident triggers which share the same template parameters and are physically possible (when accounting for timing accuracy error). However, we will face another problem where the coincident ranking statistic used by the early warning search will consider these events to have a very low signal rate. The coincident ranking statistic used by the early warning search is \texttt{phasetd} (phase-time delay) and is used to assess the coincidence likelihood between two triggers from different detectors.

For a two detector coincidence, the likelihood value is found by sampling from a pre-created \texttt{phasetd} 3-dimensional histogram in amplitude, time, and phase space. The file is created by simulating \gwadj signals from an isotropic distribution of sources, where each signal is assigned a random sky location (right ascension and declination), inclination and polarisation. The detector responses to the signal are calculated, from which we get the amplitudes, time, and phases for each injection in each detector. For a pair of detectors, the time difference, phase difference and relative signal amplitude are then binned into discrete bins which represent different combinations of amplitude, time, and phase difference. An example of the phase-time histogram (marginalised over amplitude) can be seen in Figure~\ref{6:fig:phase-time-histogram}~\cite{PyCBC:2017} where the density of sources at time differences greater or less than $13 \, \text{ms}$ is extremely low. Therefore, when we sample from the histogram with large time difference, we will receive very low value of the signal rate. The timing accuracy error will need to be taken account explicitly in the time difference on both trigger times.

Another improvement that can be made in the generation of the \texttt{phasetd} statistic files is including the sample rate of the search as a parameter. For a Monte Carlo simulation, the trigger time differences are as accurate as a \texttt{float64} parameter can be. Whereas the timing difference between triggers in the PyCBC Live search will only take discrete values of the time difference dependent on the sample rate, for example, the early warning search has a sample rate of $256 \, \text{Hz}$ and therefore the time difference between samples is approximately $3.9$ milliseconds. Therefore, our \texttt{phasetd} file will only ever been sampled from using time difference that are $n$-multiples of this time between samples. We can create \texttt{phasetd} files with bins in the time difference parameter space which represent the sample rate of the search and therefore can be sampled more accurately.

\subsubsection{\label{6:sec:missing-cands}Missing candidates}

\begin{figure}
    \centering
    \includegraphics[width=1.0\linewidth]{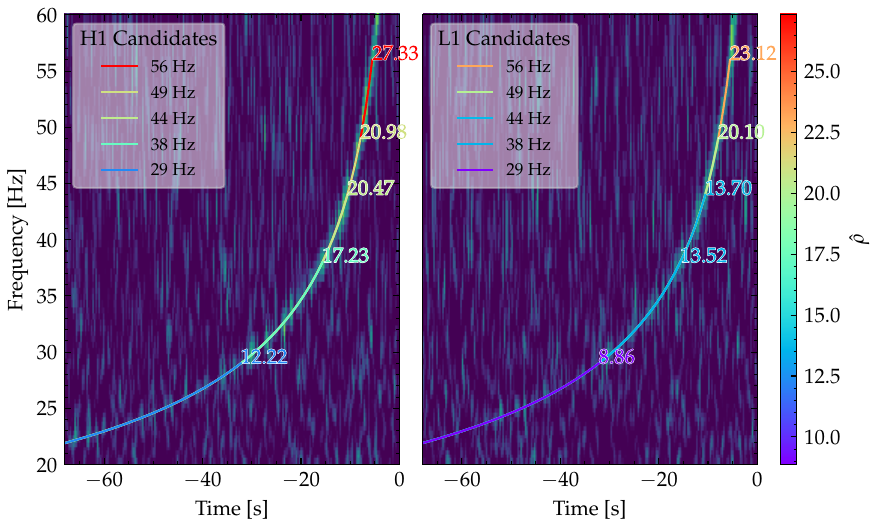}
    \caption{An example of a \gwadj injection seen by the PyCBC Live early warning search with a missing template with frequency cutoff $32 \, \text{Hz}$.}
    \label{6:fig:missing-freq-eg}
\end{figure}

We know now the search will produce triggers that fall outside the allowed coincidence timing window and fail to form a coincident event. Figure~\ref{6:fig:missing-freq-eg} shows an example of an injection which has been found with all but $1$ final frequency cutoff template. 
To verify the time difference problem, we investigated the expected network SNRs for this injection at each final frequency cutoff in both a noiseless and expected noise regime (not performed using the PyCBC Live search). The values for these can be seen in Table~\ref{6:tab:noise_snrs}.
\begin{table}[ht]
    \centering
    \setlength{\tabcolsep}{4pt}
    \rowcolors{2}{white}{lightgray}
    \begin{tabular}{ccc}
        \toprule
        \textbf{Frequency Cutoff [Hz]} & \textbf{Noiseless SNR} & \textbf{Noisy SNR} \\
        \midrule
        29 & 16.43 & 15.38 \\
        32 & 19.16 & 17.95 \\
        38 & 24.00 & 21.50 \\
        44 & 28.11 & 25.26 \\
        49 & 31.04 & 27.94 \\
        56 & 34.48 & 29.02 \\
        \bottomrule
    \end{tabular}
    \caption{The expected signal-to-noise ratios in both noiseless and noisy data for the \gwadj injection shown in Figure~\ref{6:fig:missing-freq-eg}. The noiseless SNR has been calculated by taking the un-normalised matched-filter of the template with itself~\cite{Brown_Thesis:2004}. The noisy SNR has been calculated by matched-filtering the template with the injection that has been injected into simulated \gwadj data.}
    \label{6:tab:noise_snrs}
\end{table}
We can clearly see that we are expecting to see a large network SNR for the $32 \, \text{Hz}$ template, which isn't recovered by the search. Next we are able to re-run the early warning search using a template which perfectly describes the injection parameters and in this case we still fail to recover the candidate event but, we can look into the trigger files to identify the single trigger found by both searches with the same template above the SNR threshold. We find a H1 trigger with SNR $14.55$ and an L1 trigger with SNR $12.38$ (giving a network SNR of $19.11$) and when the time difference between triggers is calculated we get $15.625 \, \text{ms}$, greater than the currently allowed $13 \, \text{ms}$. 

We then perform another search with a sample rate of $2048 \,\text{Hz}$, which doesn't find a coincidence but has a smaller time difference of $14.6 \, \text{ms}$ (still outside the window) proving that the time accuracy error is playing a part and it isn't simply a sampling inaccuracy. When performing a final search in which we expand the allowed coincidence timing window by an additional $10 \, \text{ms}$ (this can be tuned in the future) we successfully recover all $6$ events for this injection.


\subsubsection{\label{6:sec:non-mono-snr}Non-monotonic SNR evolution}

\begin{figure}
    \centering
    \includegraphics[width=1.0\linewidth]{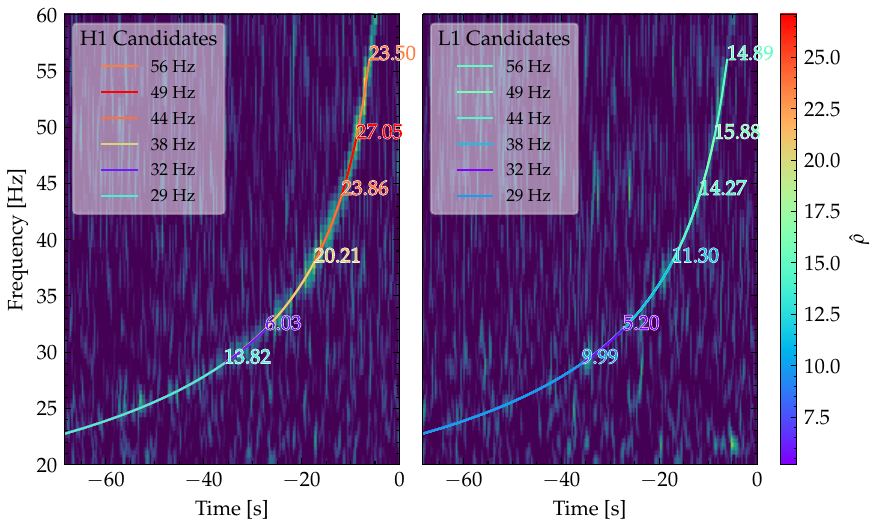}
    \caption{An example of a \gwadj injection that has been found with a non-monotonically increasing signal-to-noise ratio between successive frequency cutoffs. The $32 \, \text{Hz}$ frequency cutoff \gwadj template has been seen with a lower SNR in the H1 and L1 single detector searches than the $29 \, \text{Hz}$ frequency cutoff template. This is caused by the actual $32 \, \text{Hz}$ template coincident triggers being found outside the allowed coincidence timing window.}
    \label{6:fig:non_mono_eg}
\end{figure}

Another manifestation of the timing accuracy error is the possibility for the missing frequency to be found but with a SNR not following a monotonic SNR increase across the entire injection event timeline. A demonstration of this can be seen in Figure~\ref{6:fig:non-monotonic-snr} and an example of this happening for an injection in our early warning search can be seen in Figure~\ref{6:fig:non_mono_eg}.

It can be seen that the $32 \, \text{Hz}$ has been found by the search for this injection, but with a much lower SNR than expected and not following the monotonic SNR increase that we would expect. When investigating the triggers found we can see that the search has managed to find a low-significance trigger within the coincidence time window and has managed to form a coincidence with low SNR, similar to the low significance candidates described in Section~\ref{6:sec:false-alarms} but where one trigger is actually real. Looking into the single detector trigger files we can once again find the coincident trigger pair that should've been made if the coincidence timing window was larger and when running the search over this injection with the larger coincidence time window we find all $6$ events for this injection with the expected SNR.

We find that $997$ injections had non-monotonically increasing SNR between all successive frequency cutoffs. However, we expect a number of injections to have non-monotonically increasing SNRs due to Gaussian noise in the detector influencing the SNR values recovered for some templates. Therefore, we only count injections in which the drop between successive SNRs is greater than $5\%$ of the previous frequency cutoff template's SNR. When applying this limit, we obtain only $120$ injections with non-monotonically increasing SNR ($0.37\%$).


\section{\label{6:sec:conclusion}Conclusion}

We injected $32$,$599$ simulated \gwadj signals into $40$ days of simulated \gwadj data and searched through this data using the PyCBC Live early warning search. The PyCBC Live early warning search found $175$,$219$ candidate events, $174$,$008$ of these candidate events can be associated with an injection in our injection set and after removing extra unexpected candidate events from duplicate frequency events and false-alarms we obtain $156$,$438$ candidate events for all injections.

The more significant problem identified by this injection study is the missed events that we should've seen but were didn't due to the inadequate allowed coincidence timing window between coincident triggers. We identify $873$ injections that were missing at least one expected candidate event and $120$ more which missed the coincident event but found a far less significant coincidence with a significant deviation from the expected monotonic SNR increase. In total, we missed $1$,$766$ injections completely however, of these we have been unable to estimate those missed due to the coincidence timing window.

To prevent \gwadj signals from being missed by the PyCBC Live early warning search a configuration change is needed to increase this window alongside an improved phase-time-amplitude histogram which accounts for coincident \gwadj triggers at these extended time differences. 

In conclusion, the early warning search is more than capable of observing \gwadj signals tens of seconds prior to the merger of the \gwadj event. We are able to disseminate information about this event to the international scientific community however, a statistic change is required in the current and future observing runs to enable the identification of \gwadj signals with time differences between detectors outside the current allowed coincidence timing window. While the early warning search is more than capable of observing these events, electromagnetically bright events have proven to be exceedingly rare and therefore we do not expect to observe an early warning event in the near future. Additionally, the early warning search is robust in its capability to detect up to five individual candidate events per \gwadj event therefore, if we miss a single one of these due to timing differences then there is a still a high likelihood of observing the \gwadj event in early warning via the other four frequency cutoffs.

\chapter[Optimising Signal-to-Noise Ratio in PyCBC Live]{Optimising Signal-to-Noise Ratio in PyCBC Live}
\label{chapter:7-snr-optimiser}
\chapterquote{Are we there yet?}{Donkey}
The work that forms this chapter was performed as my first initial project upon starting my PhD and small additions were made over the course of the PhD. The PyCBC Live SNR optimiser has not been detailed in any PyCBC methods publication to date and will form a section in the upcoming fourth observing run PyCBC Live methods paper where these changes will be published. I was responsible for tuning the current PyCBC Live SNR optimiser, developing and introducing a new optimiser to PyCBC Live, and implementing astrophysically-motivated bounds on the spin template parameters.
The PyCBC Live search pipeline identifies \gwadj signals in real-time from the \gwadj data provided by the current international \gwadj network. Sky maps are produced of the found \gwadj events, which indicate the sky position of the source of the signal. The sky localisation is more accurate for an event with a greater signal-to-noise ratio (SNR), therefore, as part of the pipeline we rapidly optimise event SNR using an optimisation algorithm to recover the greatest SNR and more accurate sky localisation possible.

There are many optimisation algorithms with many tuneable hyperparameters which determine operating parameters of the algorithm, the appropriate algorithms and hyperparameter values need to be chosen to ensure the optimisation performs optimally. The optimal methods for optimising SNR in PyCBC Live were researched and improved multiple times across the course of the PhD. These changes are being written up in the PyCBC Live fourth observing run methods paper.

\section{\label{7:sec:introduction}Introduction}


As described in chapters~\ref{chapter:5-pycbc-live} and~\ref{chapter:6-earlywarning} the PyCBC Live search pipeline (among others) is responsible for detecting and producing accurate sky maps of \gwadj signals in low-latency. Search pipeline events are uploaded to the Gravitational-Wave Candidate Event Database~\cite{ligo_gracedb:2024} (GraceDB) from which a \textit{preferred event} is chosen to send a GCN Circular~\cite{gcn_circulars:2024} (a rapid astronomical bulletin submitted by and distributed to astronomy community members worldwide). The preferred event is chosen based on a listed selection criterion. These rules are taken verbatim from~\cite{gracedb_superevent_selection}:

When multiple online searches report events at the same time, the preferred event is decided by applying the following rules, in order:
\begin{itemize}
    \item A publishable event, meeting the public alert threshold, is given preference over one that does not meet the threshold.
    \item An event from CBC searches is preferred over an event from unmodeled Burst searches (see Searches for details on search pipelines).
    \item In the case of multiple CBC events, three-interferometer events are preferred over two-interferometer events, and two-interferometer events are preferred over single-interferometer events.
    \item In the case of multiple CBC events with the same number of participating interferometers, the event with the highest SNR is preferred. The SNR is used to select the preferred event among CBC candidates because higher SNR implies better sky location and parameter estimates from low-latency searches. In the case of multiple burst events, the event with the lowest FAR is preferred.
\end{itemize}
The PyCBC Live search is a CBC search and when analysing the same number of interferometers as other CBC searches there is a tie-breaker for preferred event dependent on the SNR of the event.

It is clear from these set of rules that to become the preferred event with the most accurate sky map, we must upload the event with the highest SNR. This is where the PyCBC Live SNR optimiser becomes relevant. We construct our template bank with a minimum match of $0.97$, allowing up to a $3\%$ SNR loss for any event. If we want to maximise the SNR of our event, we must perform a finer search over the template bank parameter space around the initially found template to recover that potentially lost $3\%$.

The SNR optimiser uses optimisation algorithms which tune input parameters (template parameters) to maximise (or minimise) an output (network SNR). There are a number of different optimisation algorithms operating under different mathematical optimisation techniques, all of which have hyperparameters which can be tuned to the problem space to optimise optimiser performance---both output value and computational time.

We detail the investigation into an optimal optimisation algorithm for PyCBC Live along with the tuning of the hyperparameters for this optimisation algorithm, how these algorithms were tested and the results of using the SNR optimiser on a number of PyCBC Live events.


In Section~\ref{7:sec:optimising_snr_in_low_latency} we discuss how the SNR optimiser fits into the PyCBC Live search pipeline and when it is run, alongside this we detail the generic form of the optimisation algorithms and the current optimiser being used in PyCBC Live. In Section~\ref{7:sec:original_bounds} we list the parameter value bounds for the optimised parameters and how these were calculated, in Section~\ref{7:sec:original_de} we detail the \textit{differential evolution} optimisation algorithm and in Section~\ref{7:sec:de_hyperparameter_tuning} how the hyperparameter of this algorithm were tuned and the improvements from tuning the hyperparameter. In Section~\ref{7:sec:exploring_alt_opts} we explore other optimisers and describe Particle Swarm Optimisation, the newly implemented optimisation algorithm. We detail some additional improvements to the SNR optimiser in Section~\ref{7:sec:additional-improvements} and in Section~\ref{7:sec:results} we analyse the average SNR increase of the SNR optimiser over the events uploaded by PyCBC in 2024. We conclude in Section~\ref{7:sec:conclusion} and detail potential future improvements to the SNR optimiser.

\section{\label{7:sec:optimising_snr_in_low_latency}Optimising SNR in low-latency}

The PyCBC Live search will matched filter the entire template bank with the data ring buffer every analysis stride (see chapter~\ref{chapter:5-pycbc-live} for greater detail). The matched filter will produce SNR triggers when any template has been found with a matched filter above an SNR of $4.5$. The PyCBC Live search will then pass these triggers to the ranking statistic, taking into account any signal-consistency test values or other ranking statistic components, to find the most significant coincident set of triggers. After the uploading of an event to GraceDB, the PyCBC Live search will spawn a subprocess and the PyCBC programme \texttt{pycbc\_optimize\_snr} is initialised. This programme takes the \gwadj data and is given the event template parameters to run a finer search over the parameter space, specifically optimising for the network SNR. This finer search is performed not using a template bank approach but with an optimisation function, a process that finds the `best' possible solution by maximising for a particular objective given input parameters that can be tuned. For PyCBC Live the objective being maximised is the network SNR and the tuneable input parameters are the template bank parameters: primary component mass $m_{1}$, secondary component mass $m_{2}$, primary spin z-component $s_{1z}$ and, secondary spin z-component $s_{2z}$. Boundaries are applied to the input parameters, which are calculated using the initial event parameter values.

\begin{figure}
    \centering
    \includegraphics[width=1.0\linewidth]{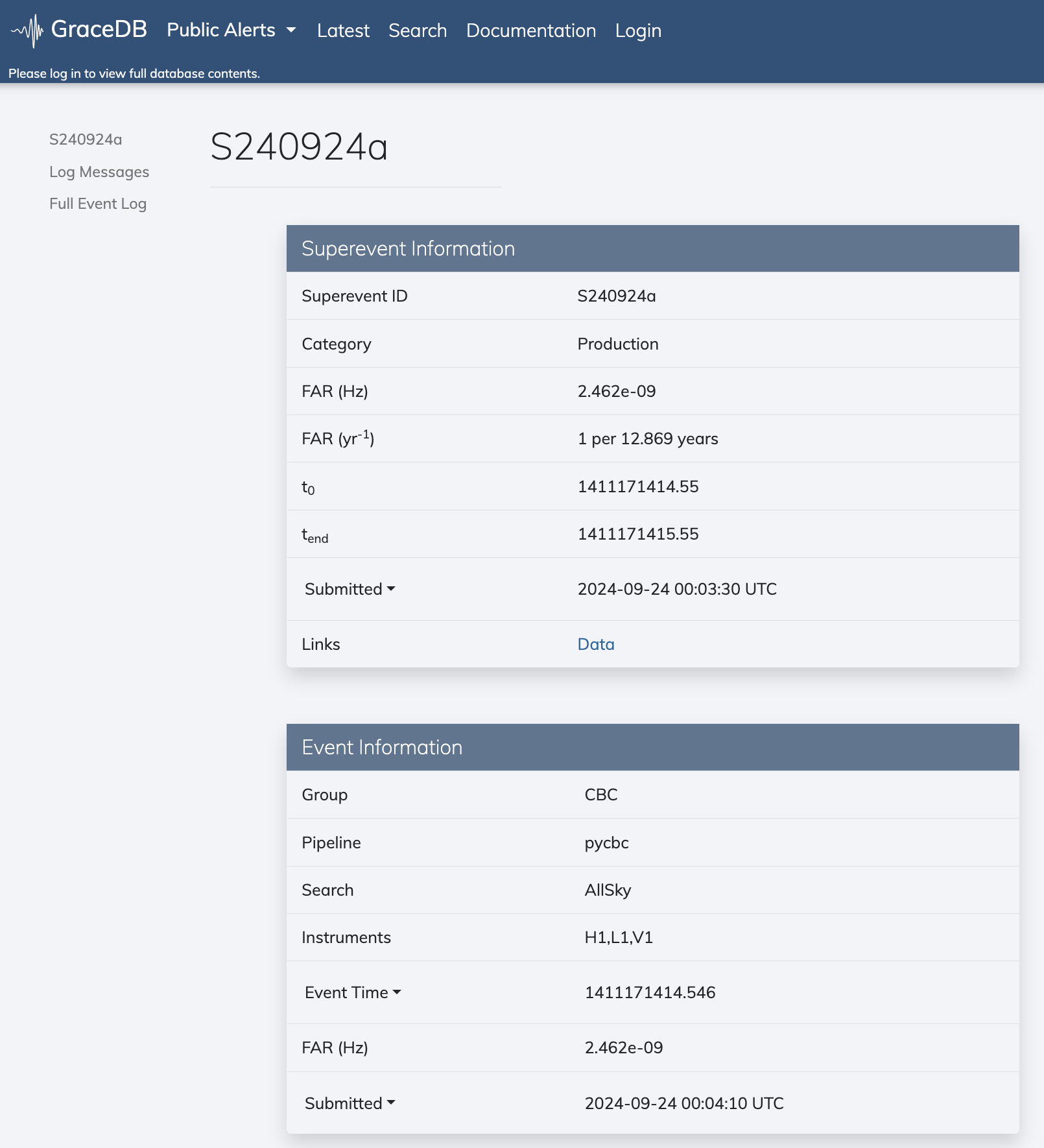}
    \caption{The public webpage for the superevent \gwadj signal observed on the 24$^{\text{th}}$ September 2024, S240924a~\cite{superevent_S240924a}. The preferred search pipeline for this event can be seen in the table titled `Event Information', where it can be seen that a PyCBC Live is the preferred event.}
    \label{7:fig:gracedb_pref_event}
\end{figure}

\begin{figure}
    \centering
    \includegraphics[width=1.0\linewidth]{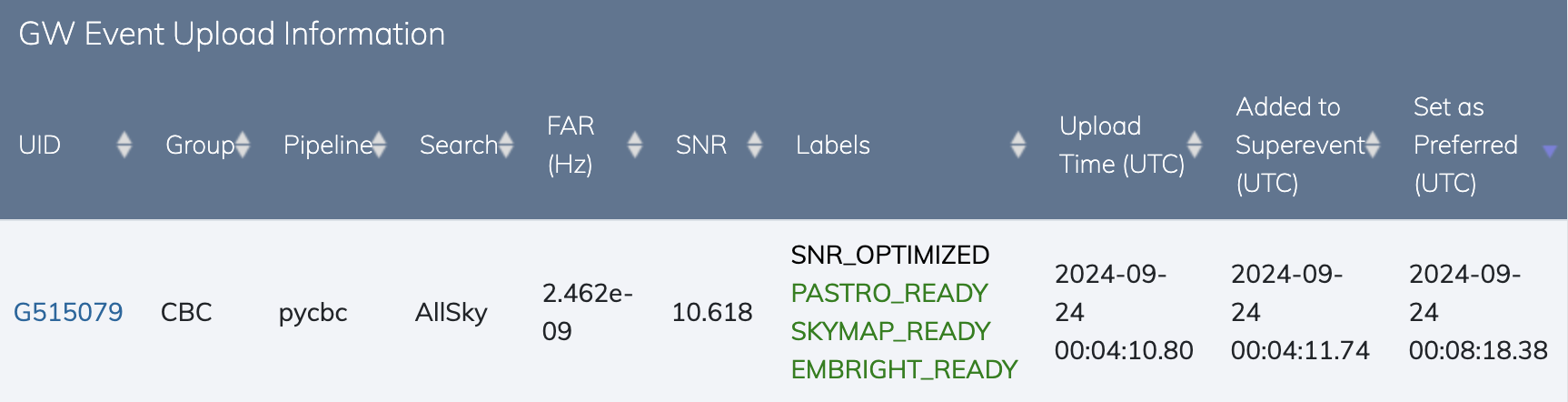}
    \caption{The preferred event for the super event S240924a~\cite{superevent_S240924a} identified by the PyCBC Live search pipeline (among others). This PyCBC Live event was created by the SNR optimiser.}
    \label{7:fig:gracedb_snr_optimizer}
\end{figure}

We run this SNR optimisation to produce better sky maps for the GraceDB `super event' (the \gwadj signal each pipeline event has observed). An example for a recent \gwadj signal (S240924a~\cite{superevent_S240924a}) can be seen in Figure~\ref{7:fig:gracedb_pref_event}.
When all pipelines pass the public alert threshold and use the same number of interferometers, the tie for preferred event is broken by the SNR of the event. For S240924a, we can see the PyCBC event (which was the preferred event) was an SNR optimised event, indicated by the label `SNR\_OPTIMIZED' in Figure~\ref{7:fig:gracedb_snr_optimizer}.

The original PyCBC SNR optimiser was introduced during the second half of the third observing run to improve the sky maps that are generated by PyCBC Live using the SciPy~\cite{SciPy:2020} optimiser \texttt{differential evolution}~\cite{DE:1997}. Examples noted in the original GitHub pull request~\cite{pycbc_pull_request_2659} see increases in network SNR from $9.818$ to $10.149$, an increase of $3.6\%$.

Optimisation algorithms can generally be categorised into two main types based on their approach to searching the parameter space: population-based algorithms and single-solution algorithms. Population-based algorithms initialise a set of candidate solutions (a population) and explore the parameters space iteratively using interactions between the candidates to locate the optimal solution. Single-solution algorithms use a single candidate point that is iteratively improved based on some criteria. Only population-based algorithms have been considered for the PyCBC Live SNR optimiser for their computational efficiency in parallelised computing environments where the network SNR of each point can be computed in parallel using a large number of computing cores. These population-based algorithms all follow the same general steps in optimisation.
\begin{enumerate}
    \item Initialise a population of size $N$ with random parameter values
    \item Evaluate the network SNR of the point
    \item Adjust each point in the population using the base optimisation algorithm
    \item Re-evaluate the network SNR of each point
    \item Keep points with greater network SNR
    \item Repeat until the number of max iterations has been reached \textbf{OR} the population has converged on a point
    \item Output the optimised template solution
\end{enumerate}
These instructions can vary slightly for optimisation algorithms and we will explore these differences as well as the point adjusting algorithms in the following sections.

\subsection{\label{7:sec:original_bounds}Parameter bounds}

The optimiser will be passed the template parameters of the event that triggers the SNR optimisation. From these parameters, the \textit{bounds} are defined. These bounds represent the minimum or maximum value that the optimiser can explore for a specific parameter, preventing it from searching areas we do not want it to search.

The mass parameter bounds are created by converting $m_{1}$ and $m_{2}$ into chirp mass, $\mathcal{M}$, and symmetric mass ratio, $\eta$; this combination leads to a more physically meaningful difference in waveform when changed. The $\mathcal{M}$ bounds are calculated,
\begin{align}
    \mathcal{M}_{\min} &=
    \begin{cases}
        \mathcal{M} \cdot \left(1 - \frac{\mathcal{M}}{50.0}\right) & \text{if } \mathcal{M} \cdot \left(1 - \frac{\mathcal{M}}{50.0}\right) \geq 1 \\
        1 & \text{if } \mathcal{M} \cdot \left(1 - \frac{\mathcal{M}}{50.0}\right) < 1
    \end{cases} \\
    \mathcal{M}_{\max} &=
    \begin{cases}
        \mathcal{M} \cdot \left(1 + \frac{\mathcal{M}}{50.0}\right) & \text{if } \mathcal{M} \cdot \left(1 + \frac{\mathcal{M}}{50.0}\right) \leq 80 \\
        80 & \text{if } \mathcal{M} \cdot \left(1 + \frac{\mathcal{M}}{50.0}\right) > 80.
    \end{cases}
\end{align}
The remaining parameter bounds are set to
\begin{align}
    0.01 &\ge \eta \ge 0.2499, \\
    -0.9 &\ge s_{1z} \ge 0.9, \\
    -0.9 &\ge s_{2z} \ge 0.9, \\
\end{align}
independent of initial event parameters.

Upon creating an instance of the optimiser, a population of candidate solutions is randomly initialised. The first hyperparameter we consider is population size, $N$, determining the number of points in the population. Each point is a 4-dimensional vector, $\textbf{x}_{i}$, given a random value for each parameter inside the bounds, $\textbf{x}_{i} \!=\! (\mathcal{M}_{i}, \eta_{i}, s_{1z, i}, s_{2z, i}$), this is our initial population, $\textbf{x}_{N}$.

\subsection{\label{7:sec:original_de}The original SNR optimiser}

The differential evolution optimiser is a stochastic, population-based algorithm that iteratively 'mutates' the population to improve individual SNRs. We can visualise the algorithm as a process similar to natural evolution, where, in each iteration, individuals undergo genetic variation through mutation. This combines features from other individuals to create new trial candidates. Just as nature favours the survival of the fittest, the algorithm evaluates the performance of these trial candidates and retains those that exhibit greater SNR, gradually steering the population towards the optimal solution.

When using differential evolution, we have four key hyperparameters:
\begin{itemize}
    \item \textbf{Population size, \( N \)}: Determines how many individuals are in the population.
    \item \textbf{Maximum number of iterations}: Sets how long the algorithm runs, i.e., the maximum number of generations.
    \item \textbf{Mutation factor, \( F \)}: Controls the magnitude of mutation and influences how much individuals are altered by the mutation vectors.
    \item \textbf{Recombination constant, \( R \)}: Also known as the crossover probability, it determines the likelihood that a trial candidate takes parameters from the mutation vector instead of the original individual.
\end{itemize}
Another hyperparameter, which we do not tune, is the number of processes that are spawned to run the algorithm in parallel. This is set to ``as many as possible'' when running with PyCBC Live.

The algorithm begins by spawning a population of individuals with initial parameter values within the defined bounds. The network SNR for each individual is calculated, and the individual with the highest SNR, \( \mathbf{x}_{best} \), is identified. Next, a set of mutation vectors are created,
\begin{equation}
    \mathbf{v}_i = \mathbf{x}_{best} + F \cdot (\mathbf{x}_{r1} - \mathbf{x}_{r2}),
\end{equation}
where \( \mathbf{x}_{r1} \) and \( \mathbf{x}_{r2} \) are distinct random individuals from the population. A trial candidate, \( u_i \), is created by combining the mutation vector \( \mathbf{v}_i \) with the target individual \( \mathbf{x}_i \), according to the rule
\begin{equation}
    u_{ij} =
    \begin{cases}
    v_{i,j} & \text{if } r_j < R \\
    x_{i,j} & \text{otherwise}.
    \end{cases}
\end{equation}
Here, \( u_{ij} \) is the \( j \)-th parameter of the trial candidate \( u_i \), \( v_{ij} \) is the \( j \)-th parameter of the mutation vector \( v_i \), and \( x_{ij} \) is the \( j \)-th parameter of the target individual \( x_i \). \( r_j \) is generated uniformly in the interval \([0, 1)\). If \( r_j < R \), the mutation vector value is used; otherwise, the original target individual's value is retained. One exception is that the final parameter (4th) \textbf{always} takes its value from the mutation vector, ensuring that no individual remains completely unchanged in the new population.

Once all trial candidates have been generated, their network SNRs are calculated and compare to the target individual SNR. If a trial candidate has a higher network SNR than the original target individual, it replaces the target. If not, the trial candidate is discarded. The process of creating mutation vectors and generating trial candidates is repeated until the maximum number of iterations is reached, or another stopping criterion is satisfied.

\section{\label{7:sec:additional-improvements}Improvements to the optimiser between observing runs}

The PyCBC programme \texttt{pycbc\_optimize\_snr} has been under irregular but constant development to improve the functionality. There are a number of improvements to the programme which have improved the computing time and maximum SNR found. These have been implemented between the third observing run, when the SNR optimiser was created, and the fourth observing run where the SNR optimiser is currently running on live data.

\subsection{\label{7:sec:de_hyperparameter_tuning}Tuning hyperparameters}

Tuning the four hyperparameters for the differential evolution optimiser allows for the best performance of the SNR optimiser in the live search. We optimised the following hyperparameters:
\begin{itemize}
    \item Population size
    \item Maximum number of iterations
    \item Mutation factor
    \item Recombination factor
\end{itemize}
where a balance must be struck between SNR and computing time. To maximise network SNR we would use larger population sizes, more iterations, higher mutation factors and a lower recombination rate, but this will widen the effective search radius and increase the amount of time taken to find an optimised value.

As shown previously, the original values for the hyperparameters were able to recover the $3\%$ missing SNR from any template mismatch therefore, we chose to tune hyperparameters to reduce the time to run the programme. The computing time is directly related to both the population size and maximum number of iterations, which had initial values $N \!=\! 200$ and $\texttt{maxiter} \!=\! 100$. Using a simple grid search over these two hyperparameters for a number of events, we were able to maintain the SNR recovery while reducing the computing time by ${\sim}25\%$ on average using $N \!=\! 100$ and $\texttt{maxiter} \!=\! 50$. These were the first set of improvements for the PyCBC Live SNR optimiser.

\subsection{Initial point in the population}

The SNR optimiser can miss the maximum SNR of the parameter space if it is not searched with a fine enough set of particles. Not only can the maximum point be missed, but the maximum can be found below the SNR recovered by the initial PyCBC Live event. To prevent this, we can set a floor on the optimiser SNR by including our initial guess into the initial swarm (or population). Including this initial template into the swarm will set the best position, allowing for a faster convergence on the maximum SNR.

\subsection{Astrophysical spin bounds}

The bounds described in Section~\ref{7:sec:original_bounds} are very general and cover a wide parameters space. The PyCBC Live search is capable of detecting \gwadj signals from binary black hole, binary neutron star and, neutron star — black holes. We assume a physical limit on  neutron star spin~\cite{Harry_Lundgren:2012} therefore we can impose more astrophysically informed spin bounds on objects with masses consistent with neutron stars
\begin{align}
    \text{spin bounds} &=
    \begin{cases}
        [-0.4, 0.4] & \text{if } \text{max mass} < 3 \, M_{\odot}, \\
        [-0.9, 0.9] & \text{if } \text{max mass} > 3 \, M_{\odot},
    \end{cases}
\end{align}
it is important to impose the condition dependent on the maximum mass the optimiser is allowed to search over. If the optimiser is allowed to assign masses to particles greater than the upper mass limit on neutron stars then we have to let the spin vary by up to the black hole spin bounds because the optimal template could be in this region even if the initial template (from which the bounds are calculated) isn't.

\subsection{Chirp time boundaries}

Another improvement made to the parameter bounds is determining the $\mathcal{M}$ bounds using a constant chirp time window. We define the chirp time as the time between the start of the signal (at some define lower frequency, $f_{L}$) and the peak frequency prior to the merger
\begin{equation}
    \tau_0 = A_{0}\left(\mathcal{M}\right)^{-5/3}.
\end{equation}
where
\begin{equation}
    A_{0} = \frac{5}{256 \left(\pi f_{L}\right)^{8/3}}.
\end{equation}
The chirp time to Newtonian order, $\tau_{0}$, depends only on the two masses, $m_{1}$ and $m_{2}$~\cite{Cokelaer:2007}. By using a chirp time window we can search the $\mathcal{M}$ parameter space for signals that look far more physically similar than if we were just changing $\mathcal{M}$ itself.

A higher mass \gwadj event will merge at lower frequencies and therefore have a shorter chirp time. To calculate the $\mathcal{M}$ bounds we first calculate the chirp time of the initial event, then we subtract the constant chirp time window ($2$ seconds in PyCBC Live) and convert back to $\mathcal{M}$ to find the maximum $\mathcal{M}$ bound and add the window and convert to find the minimum $\mathcal{M}$ bound.

We also improve the $\eta$ bounds by imposing a minimum mass of any component object $\text{min}(m) \!=\! 0.99$ and calculating the maximum mass of $m_{1}$ using the maximum $\mathcal{M}$. Using this, we can calculate the \textit{minimum} value of $\eta$, again by using the minimum mass of any component and the maximum mass of $m_{1}$. Then using the same maximum $\eta$ previously defined, we are able to define the upper mass bound on $m_{2}$.

These changes to the parameter bound calculation give us the boundaries for $\mathcal{M}$ and $\eta$ as,
\begin{align}
    \left(\frac{A_{0}}{\tau_{0} + 2.0}\right)^{\frac{3}{5}} &\ge \mathcal{M} \ge \left(\frac{A_{0}}{\tau_{0} - 2.0}\right)^{\frac{3}{5}}, \\[10pt]
    \frac{\text{max}(m_{1}) \cdot 0.99}{(\text{max}(m_{1}) + 0.99)^{2}} &\ge \eta \ge 0.2499,
\end{align}
which allow for less time to be wasted in computing regions of the parameter space which we know will not contain the maximum point.

\section{\label{7:sec:results}O3 optimisation results}

To evaluate the performance of the SNR optimiser in the PyCBC Live search we analyse the events uploaded by the PyCBC Live search to GraceDB. There is a \gwadj mock-data challenge (MDC) that runs continuously to enable live search pipelines to test pipelines and changes to the pipelines prior to enabling them on the real \gwadj data. The MDC replays $40$ days of third observing run \gwadj data with simulated injections continuously, providing a much larger rate of signals to test changes on.

We query GraceDB for all PyCBC `grace events'\footnote{Grace event refers to an individual event uploaded by a pipeline. The initial PyCBC upload and the SNR optimiser upload are separate grace events.} uploaded by the PyCBC full bandwidth MDC search between the dates \texttt{2024-05-30} and \texttt{2024-07-16} and retrieve $10270$ grace events, of which $5091$ were uploaded by the SNR optimiser. Each grace event has an associated `super event' to which multiple pipelines (and multiple grace events from each pipeline) are assigned, of the SNR optimised grace events, $5008$ were assigned a preferred grace event uploaded by PyCBC Live. Finally, of these PyCBC preferred super events, $4600$ had SNR optimised grace events as the preferred events. In $91.85\%$ of cases, when the SNR optimiser has uploaded a grace event, it is preferred over the original PyCBC Live uploaded grace event.

\begin{figure}
    \centering
    \includegraphics[width=1.0\linewidth]{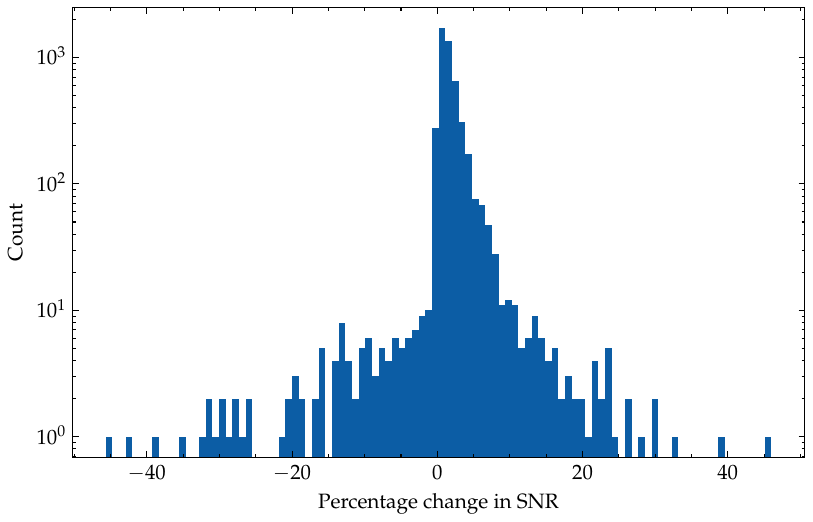}
    \caption{The percentage change in SNR between the optimised SNR and original SNR for GraceDB super events found between \texttt{2024-05-30} and \texttt{2024-07-16} that were found with PyCBC Live and have an SNR optimised grace event in the \gwadj data third observing run replay mock-data challenge.}
    \label{7:fig:snr-diffs}
\end{figure}

\begin{figure}
    \centering
    \includegraphics[width=1.0\linewidth]{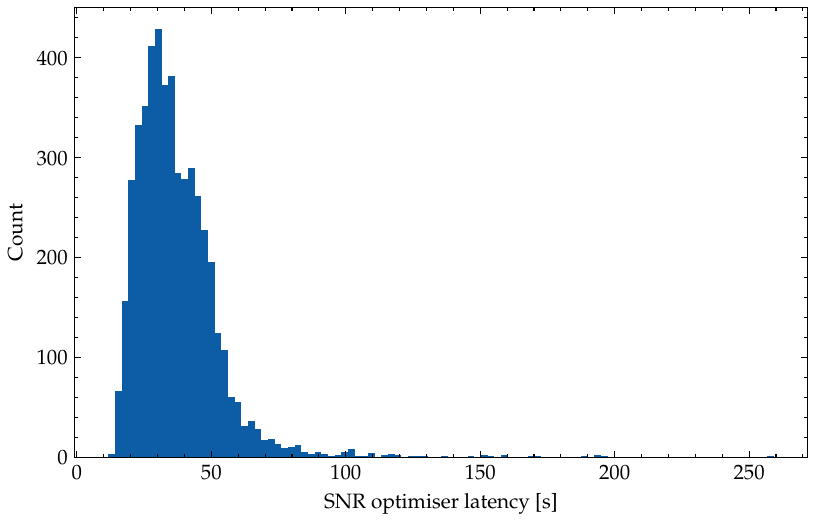}
    \caption{The time after original grace event upload for the SNR optimiser to upload an SNR optimised grace event for GraceDB super events found between \texttt{2024-05-30} and \texttt{2024-07-16} that were found with PyCBC Live and have an SNR optimised grace event in the \gwadj data third observing run replay mock-data challenge.}
    \label{7:fig:latencies}
\end{figure}

\begin{figure}
    \centering
    \includegraphics[width=1.0\linewidth]{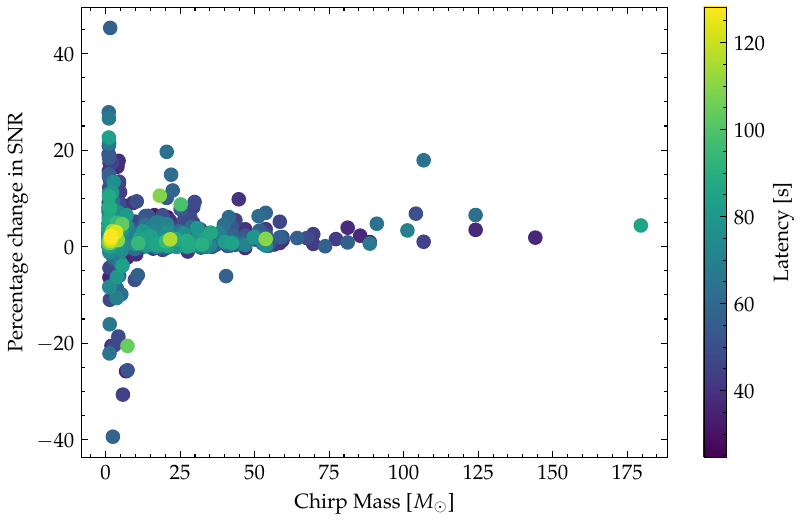}
    \caption{The percentage change in SNR against the chirp mass for each \gwadj event identified by the MDC search with an SNR optimised event, the injections are coloured by the latency between initial event upload and optimised event upload which is directly related to the time required for the optimiser to run.}
    \label{7:fig:mchirp_latency_diff}
\end{figure}

Figure~\ref{7:fig:snr-diffs} shows the change in SNR between original upload and SNR optimised upload for all these SNR optimised events and Figure~\ref{7:fig:latencies} shows the latency between original upload and SNR optimised upload.
The average change in SNR optimised events is a $+1.64\%$ increase in SNR. We can see many events which have lost significant amounts of SNR when optimised, these are caused by the bounds of the SNR optimisation parameter space not overlapping with the template bank parameter space or the super event being caused by an exceptional event. As an example, the event with the highest loss in SNR had a negative change in SNR of $-45.67\%$, the original grace event parameters were: $m_{1}\!=\!68.91\,\text{M}_{\odot}$, $m_{2}\!=\!1.08\,\text{M}_{\odot}$, $s_{1z}\!=\!0.994$, $s_{2z}\!=\!0.0$, this event is very significantly precessing and has a very large mass ratio making the parameter space difficult to optimise over. Even though the original candidate was within the parameter space, the differential evolution optimiser enforces a mutation at every iteration, so the initial best point can be lost and never recovered with the number of iterations being performed. Complicated parameter spaces are difficult for the optimiser to explore in the number of iterations. When the SNR optimiser uploads a grace event, it will achieve an average increase in SNR of $1.64\%$ with an average latency between uploads of $36.97 \, \text{seconds}$. Figure~\ref{7:fig:mchirp_latency_diff} shows the SNR change in optimised events against the event's chirp mass coloured by the latency, it can be clearly seen that lower chirp mass events have longer optimisation times.

\section{\label{7:sec:exploring_alt_opts}Exploring alternate optimisation algorithms}

After improving the differential evolution optimisation algorithm, we began to investigate other optimisation algorithms that could be used to greater success for the PyCBC Live problem space. The SciPy package has a module dedicated to optimisers, \texttt{SciPy.optimize} which offers $2$ alternative optimisers suitable for this optimisation task: brute force optimisation, which involves performing a grid search over the input parameters; and simplicial homology global optimisation (SHGO), an optimisation technique which divides the parameter space into smaller regions and evaluates points in these regions to identify the most promising areas where the best solution might be found~\cite{shgo:2018}.

We dismiss the brute force optimiser immediately and do not consider the SHGO optimiser due to requirements of the parameter space being smooth or non-noisy that we are unsure are fulfilled~\cite{shgo:2018}. We therefore looked outside SciPy and identified the particle swarm optimisation (PSO) algorithm~\cite{pso:1995} as a candidate for the PyCBC Live SNR optimiser. This algorithm has been used for \gwadj searches as shown by~\cite{pso_search_1:2018} and more recently~\cite{pso_search_2:2023}. This made it a good candidate for optimising network SNR in PyCBC Live.

\subsection{\label{7:sec:pso}Particle Swarm Optimisation}

We can describe the particle swarm optimisation (PSO) algorithm similar to how we previously described the differential evolution algorithm. PSO has the hyperparameters:
\begin{itemize}
    \item Number of particles, $N$
    \item Maximum number of iterations, $T$
    \item Inertia weight: weight next iteration velocity based on the previous iteration velocity, $w$
    \item Personal learning rate, $c_{1}$
    \item Social learning rate, $c_{2}$
\end{itemize}
the number of particles and maximum number of iterations are identical to the differential evolution hyperparameters. The personal and social learning rates can be considered as acceleration coefficients of the particle.

Once again, we are maximising the network SNR of the PyCBC Live event, adjusting the template parameters ($m_{1}$, $m_{2}$, $s_{1z}$ and $s_{2z}$) to find the optimal values to give the maximum network SNR. PSO will initialise a ``swarm'' of particles, each with a unique combination of template parameters which can be considered the position of the particle in the parameter space, $\mathbf{x}_{i}$, the particle will also be given a random velocity, $\mathbf{v}_{i}$. The network SNR of each particle is calculated and the personal best of every particle is set, $\textbf{p}_{i}$ (as this is the only position the particle has occupied). From the swarm, the global best position is determined, $\textbf{g}$.

Next, we update the velocity of each particle
\begin{keyeqn}
\begin{equation}
    v_{i(t+1)} = w \cdot \mathbf{v}_i(t) + c_1 \cdot r_1 \cdot (\mathbf{p}_i - \mathbf{x}_i(t)) + c_2 \cdot r_2 \cdot (\mathbf{g} - \mathbf{x}_i(t)),
    \label{7:eq:pso_velocity_equation}
\end{equation}
\end{keyeqn}
where $w$ is the inertial velocity of the particle, $c_{1}$ is the personal learning rate, $c_{2}$ is the social learning rate, $r_{1}$ and $r_{2}$ are random numbers uniformly distributed in $\left[0, 1\right]$, $\textbf{p}_{i}$ is the personal best position of the particle $i$ and $\textbf{g}$ is the global best position of the swarm. We can break Equation~\ref{7:eq:pso_velocity_equation} down into three distinct contributions to the velocity of the particle at iteration $t + 1$: $w \cdot \textbf{v}_{i}(t)$, how much weight is placed on the previous velocity of the particle, adjusted using the inertia weight hyperparameter; $c_{1} \cdot r_{1} \cdot \left(\textbf{p}_{i} - \textbf{x}_{i}(t)\right)$, how strongly the particle pulls toward its personal best, adjusted using the personal learning weight hyperparameter; $c_{2} \cdot r_{2} \cdot \left(\textbf{g} - \textbf{x}_{i}(t)\right)$, how strongly the particle pulls toward the global best, adjusted using the social learning weight hyperparameter.

After the velocity of each particle is calculated, the particle positions are updated
\begin{equation}
    \mathbf{x}_i(t+1) = \mathbf{x}_i(t) + \mathbf{v}_i(t+1),
\end{equation}
and the network SNR of each particle is calculated again. The particle personal bests and the global best are re-evaluated and updated if they have changed, and the iterative process repeats. Upon reaching the number of iterations or if there is convergence in the network SNR value, the algorithm is stopped and the optimal template parameter values are returned which maximise the network SNR.

\subsection{\label{7:sec:tuning-pso}Tuning and evaluating the PSO optimiser}

The PSO optimiser has not been used in the PyCBC Live search on either the MDC or real data, but we have implemented it into PyCBC Live to allow the possibility of it being used in the future. We have tested PSO on the Chapter~\ref{chapter:6-earlywarning} injection set (described in Table~\ref{6:tab:ew_inj_params}) which focuses on the low-mass parameter space, the region which the optimiser has struggled to search over.

Using \texttt{optuna}~\cite{optuna:2019}, a hyperparameter tuner, we tune the five hyperparameters of the PSO optimiser to achieve the greatest SNR for a single \gwadj injection from the injection set. We get rounded hyperparameter values: $\text{iterations} \!=\! 50$, $\text{particles} \!=\! 150$, $\text{inertia weight} \!=\! 0.2$, $\text{personal learning rate} \!=\! 0.75$, $\text{social learning rate} \!=\! 1.0$. Using these hyperparameter values we search through $21$,$600 \, \text{seconds}$ of \gwadj data containing \gwadj injections every ${\sim}100 \, \text{seconds}$, with both the current differential evolution optimiser and the PSO optimiser. The differential evolution optimiser achieves an average percent increase in SNR of $0.512 \%$ in an average time of $30.489 \, \text{seconds}$ whereas the PSO optimiser achieves an average increase in SNR of $0.543\%$ in an average time of $32.058 \, \text{seconds}$. The PSO optimiser achieves a greater average SNR increase with a slightly longer average optimisation time. We perform the same test and cut the number of iterations to $25$, leading to an average optimisation time of  $14.638 \, \text{seconds}$ while maintaining an average SNR increase of $0.543\%$.

These tests indicate that the PSO optimiser has the potential to replace the differential evolution optimiser in PyCBC Live and with greater hyperparameter tuning we can produce an SNR optimiser which explores the low-mass parameter space effectively.

\section{\label{7:sec:conclusion}Conclusion and future improvements}

The SNR optimiser will increase the SNR obtained by the PyCBC Live search pipeline for a \gwadj event in the vast majority of cases, with an average $1.64\%$ increase. This will lead to more accurate sky maps and improved parameter estimation for the \gwadj events. However, as identified by Figure~\ref{7:fig:snr-diffs}, we have a large number of events ($8.15\%$) where the SNR optimiser has \textbf{not} increased the found SNR. We have identified that this is true when the parameter space is more complex, in the case of signals with high mass ratio, non-aligned spins or long duration templates. We have a latency requirement on our optimiser and so are unable to necessarily allow the deep exploration of these parameter spaces.


We have identified the particle swarm optimisation algorithm as a candidate for further investigation to improve the PyCBC Live SNR optimiser and we have shown that hyperparameter tuning can lead to improvements in optimised SNR as well as decrease optimisation time therefore requiring further investigation into hyperparameter tuning to improve both the current and future SNR optimiser configurations.

The implementation of the SNR optimiser in PyCBC Live has improved the prospects of \gwadj detection in low-latency and will enable higher quality science in the post-detection regime. Providing more accurate sky maps for low-latency events will allow better multi-messenger follow-up of these events. The SNR optimiser has many areas of potential improvement that if more time were available, we would seek to investigate and implement. In this subsection, I will discuss some of the improvements that could be made to the SNR optimiser.

\subsubsection{Template bank initial population positions}

The two optimisers that are currently implemented---differential evolution and particle swarm optimisation---rely on spawning an initial random population of points to explore the template parameter space. We have prior knowledge about the parameter space from our templates bank and can place this initial population of points at the points of our template bank, the points will be drawn toward the best matching point from the template bank and hopefully explore the area between template bank points more deeply. If the parameter space bounds are large for an event this will place the population approximately at the correct density for the parameter space as defined by the stochastic template bank placement algorithm (described in Section~\ref{2:sec:template-bank}), more densely covered areas of the template bank will be assigned more initial points to explore that space.

\subsubsection{Template dependent optimisers and hyperparameters}

We have adjusted the parameter space bounds according to the initial template parameters, the mass bounds are dependent on the template chirp time and the spin bounds are dependent on the component object (black hole or neutron star). We could apply further options to both optimisers and the hyperparameters of the optimisers, depending on the template. If high-mass BBH signals require fewer iterations we can adjust that hyperparameter, if low-mass signals require a finer exploration of the parameter space we can change the mutation factor and recombination factor or the inertia and learning rates. We might be able to find optimisers that respond better to specific regions of the parameter space, similar to how the template banks are composed using different techniques for different regions.

\subsubsection{Machine learning optimisers and hyperparameter tuning}

Since starting this PhD in 2020 the rise of machine learning has been relentless, techniques that were considered novel and cutting-edge have fallen by the wayside and newer, more efficient and effective techniques have replaced them. The SNR optimiser can benefit heavily from this catalogue of machine learning based events that could be highly optimised for an SNR optimisation problem. We have a large number of historical events to train such a model on to allow enhanced parameter space exploration, which can allow either a faster optimisation or a deeper optimisation, especially for low-mass signals.

Another area in which machine learning can help is with hyperparameter tuning. Optimising the SNR optimiser hyperparameters will return instant reward for the current configuration of the optimiser, the hyperparameters chosen initially or after our initial optimisation may not still be the correct choice in the fourth observing run and in the future. We generate a template bank for each observing run so we should optimise the hyperparameters too.

\chapter[Working in Industry]{Working in Industry}
\label{chapter:8-industry}
\chapterquote{I can't wait for you to finish your PhD so I can retire.}{Wiktoria Kedziora}
As a DISCnet-funded PhD student, I was required to undertake two 3-month placements in industry. These placements were instrumental to my success as a PhD student: the skills, knowledge, and change in environment drove me to learn more in those six months than in some years of my PhD. The placements also helped inform my career decisions and clarified my post-PhD direction.

In this chapter, I will briefly describe the details of my employment and the projects I undertook during these placements. Due to non-disclosure agreements, confidentiality, and the passing of time, I am unable to provide extensive details or specific figures from my work. Section~\ref{8:sec:shell} covers my time at Shell Research UK, while Section~\ref{8:sec:MCA} discusses my work at the Maritime and Coastguard Agency, a UK Government civil service agency.

\section{\label{8:sec:shell}Shell Research UK}

I was hired as a Data Scientist Intern at Shell and placed within the Predictive Maintenance team. The team’s primary focus is to use advanced data science techniques to predict maintenance needs for the many machines at Shell sites globally, helping to prevent unplanned disruptions caused by failing components.

Predictive maintenance aims to anticipate equipment failures by analysing operational data. A machine may refer to anything from a simple pump to a complex refinery system. Components in these machines are fitted with sensors that monitor parameters like temperature, pressure, and vibration. These readings form a high-dimensional dataset reflecting the machine's operational health. For example, a long-term rise in temperature may indicate wear or degradation, signalling an impending failure.

We employed a reconstruction error approach to identify failure patterns. This method involves training machine learning models to understand the normal behaviour of a machine. Reconstruction error measures how closely the model's predictions match actual sensor readings. Significant deviations between predicted and observed values---i.e., a high reconstruction error---can indicate abnormal behaviour and possible component failure.

The machine learning algorithms used in this approach fall under unsupervised learning, which does not rely on labelled data for classification. During my time at Shell, I developed and implemented Principal Component Analysis (PCA) as part of the predictive maintenance pipeline for monitoring machine components.

\section{\label{8:sec:MCA}Maritime and Coastguard Agency}

I worked as a Data Scientist Intern at the Maritime and Coastguard Agency (MCA), which is responsible for maritime safety, environmental protection, and coordinating search and rescue operations under the UK Department for Transport.

I was tasked with on-boarding a new Automatic Identification System (AIS) dataset, containing billions of data points tracking every vessel in UK coastal waters over the past four years. This dataset contained numerous errors and anomalies that needed to be processed efficiently and in parallel to ensure the data was usable by other teams within the agency.

At the MCA, I extensively used PySpark---a Python API for Apache Spark that supports large-scale data processing in distributed environments---and Databricks, a cloud platform that provides these environments. With these tools, I developed scalable pipelines to process billions of geospatial data points, impute missing values, calculate new information, and identify errors in the dataset. The unique nature of working with geospatial data allowed me to focus on identifying vessels moving anomalously over land, analysing historical traffic trends in protected areas, and preventing vessel collisions at sea.

\chapter[Conclusion]{Conclusion}
\label{chapter:conclusion}
\chapterquote{Oh, finally!}{Donkey}
We have explored the current methods that have been used to directly detect hundreds of gravitational waves from compact binary coalescences using the most sensitive observatories ever constructed. In this thesis, the research output has focused on improving the search for gravitational wave signals by both developing new tools and applying powerful existing tools in novel ways.

First, we created a novel \scladj noise artefact model and searched for these noise artefacts, removing them and re-searching for signals to recover previously missed gravitational wave injections. This method was the first of its kind, establishing a glitch search pipeline that employed techniques similar to gravitational wave search pipelines, but tailored to account for properties unique to specific glitch classes. We demonstrated that the approach used in the ArchEnemy pipeline did not significantly improve search sensitivity, but it did show that modelling and removing glitches could allow for the recovery of previously missed gravitational wave injections. In the future, the feasibility of glitch search pipelines could be expanded to include more classes of glitches, enabling a pre-search data cleaning step before gravitational wave detection.

We implemented an exponential noise model in the PyCBC Live search for gravitational waves in low-latency, enhancing the detection ranking statistic used to evaluate gravitational wave signals. By using a noise model that more accurately describes the characteristics of PyCBC Live’s noise, we demonstrated significant gains in detection sensitivity. While PyCBC Live lacks some ranking statistic components that have been developed for offline gravitational wave searches, future work could convert and optimise these components for PyCBC Live to further improve both the detection ranking statistic and the significance estimation of gravitational wave signals.

Then, we analysed, evaluated, and proposed improvements for the PyCBC Live Early Warning search to maximise the number of gravitational wave events that can be detected in the early warning regime. Running the PyCBC Live Early Warning search over a 40-day mock data set of gravitational wave injections revealed significant deficiencies in the current ranking statistic. Based on this analysis, we suggested increasing the slope of the coincidence timing window allowance and reconfiguring the \texttt{phasetd} phase-time-amplitude histograms with specific tuning for early warning. This would allow coincidences to be made when single-detector trigger pairs fall outside the light-travel time. These adjustments could lead to more gravitational wave signals being detected in early warning, and we hope to demonstrate their effectiveness with future binary neutron star events.

We discussed the SNR optimiser used in the PyCBC Live search, as well as improvements to the SNR optimiser algorithm and hyperparameter optimisation performed during this PhD, highlighting the potential for future development in this area. Finally, I discussed my industrial placements during the course of the PhD, where I gained a wealth of knowledge and skills that I brought to my research.

The current epoch of gravitational wave astronomy is ripe for further improvements in search techniques, independent of detector upgrades. New methods can increase gravitational wave search sensitivity in parallel with detector advancements, and these techniques will be applicable to the next generation of detectors, which will explore an even larger gravitational wave parameter space. By enhancing our searches, we can observe more events, and in the case of early warning, provide as much notice as possible, thereby deepening our understanding of the Universe.


{
  \titleformat{name=\chapter,numberless}[block]
    {\normalfont}
    {}
    {0pt}
    {%
      \noindent
      \begin{tabular}{ @{} b{0.70\textwidth} @{\hspace{0.5cm}} !{\tikz[overlay]\draw[accentgold, line width=1.5pt] (0,-10pt) -- (0, \paperheight);} @{\hspace{0.5cm}} l @{} }
         \raggedleft\chaptertitlefont #1
         &
         \hspace{0.2em}\chapternumberfont\resizebox{!}{0.7ex}{Ref}
      \end{tabular}%
    }

  \cleardoublepage
  \addcontentsline{toc}{chapter}{Bibliography} 
  \bibliographystyle{unsrtnat}
  \bibliography{main}

\begin{thebibliography}{205}
\providecommand{\natexlab}[1]{#1}
\providecommand{\url}[1]{\texttt{#1}}
\expandafter\ifx\csname urlstyle\endcsname\relax
  \providecommand{\doi}[1]{doi: #1}\else
  \providecommand{\doi}{doi: \begingroup \urlstyle{rm}\Url}\fi

\bibitem[Einstein(1914{\natexlab{a}})]{Einstein_1:1914}
Albert Einstein.
\newblock {The Formal Foundation of the General Theory of Relativity}.
\newblock \emph{Sitzungsber. Preuss. Akad. Wiss. Berlin (Math. Phys. )},
  1914:\penalty0 1030--1085, 1914{\natexlab{a}}.

\bibitem[Einstein(1914{\natexlab{b}})]{Einstein_2:1914}
Albert Einstein.
\newblock {On the Relativity Problem}.
\newblock \emph{Scientia}, 15:\penalty0 337--348, 1914{\natexlab{b}}.

\bibitem[Einstein(1915)]{Einstein_3:1915}
Albert Einstein.
\newblock {The Field Equations of Gravitation}.
\newblock \emph{Sitzungsber. Preuss. Akad. Wiss. Berlin (Math. Phys. )},
  1915:\penalty0 844--847, 1915.

\bibitem[Einstein(1916)]{Einstein_4:1916}
Albert Einstein.
\newblock {The foundation of the general theory of relativity.}
\newblock \emph{Annalen Phys.}, 49\penalty0 (7):\penalty0 769--822, 1916.
\newblock \doi{10.1002/andp.19163540702}.

\bibitem[Einstein(1917)]{Einstein_5:1917}
Albert Einstein.
\newblock {Cosmological Considerations in the General Theory of Relativity}.
\newblock \emph{Sitzungsber. Preuss. Akad. Wiss. Berlin (Math. Phys. )},
  1917:\penalty0 142--152, 1917.

\bibitem[Einstein and Rosen(1936)]{Einstein_6:1936}
Albert Einstein and N.~Rosen.
\newblock {Two-Body Problem in General Relativity Theory}.
\newblock \emph{Phys. Rev.}, 49:\penalty0 404--405, 1936.
\newblock \doi{10.1103/PhysRev.49.404.2}.

\bibitem[Einstein and Rosen(1937)]{Einstein_7:1937}
Albert Einstein and N.~Rosen.
\newblock {On Gravitational waves}.
\newblock \emph{J. Franklin Inst.}, 223:\penalty0 43--54, 1937.
\newblock \doi{10.1016/S0016-0032(37)90583-0}.

\bibitem[Einstein et~al.(1938)Einstein, Infeld, and Hoffmann]{Einstein_8:1938}
Albert Einstein, L.~Infeld, and B.~Hoffmann.
\newblock {The Gravitational equations and the problem of motion}.
\newblock \emph{Annals Math.}, 39:\penalty0 65--100, 1938.
\newblock \doi{10.2307/1968714}.

\bibitem[Einstein(1939)]{Einstein_9:1939}
Albert Einstein.
\newblock {On a stationary system with spherical symmetry consisting of many
  gravitating masses}.
\newblock \emph{Annals Math.}, 40:\penalty0 922--936, 1939.
\newblock \doi{10.2307/1968902}.

\bibitem[Einstein(1948)]{Einstein_10:1948}
Albert Einstein.
\newblock {A Generalized Theory of Gravitation}.
\newblock \emph{Rev. Mod. Phys.}, 20:\penalty0 35--39, 1948.
\newblock \doi{10.1103/RevModPhys.20.35}.

\bibitem[Abbott et~al.(2016{\natexlab{a}})]{GW150914:2016}
B.~P. Abbott et~al.
\newblock {Observation of Gravitational Waves from a Binary Black Hole Merger}.
\newblock \emph{Phys. Rev. Lett.}, 116\penalty0 (6):\penalty0 061102,
  2016{\natexlab{a}}.
\newblock \doi{10.1103/PhysRevLett.116.061102}.

\bibitem[Acernese et~al.(2015)]{aVirgo:2015}
F.~Acernese et~al.
\newblock {Advanced Virgo: a second-generation interferometric gravitational
  wave detector}.
\newblock \emph{Class. Quant. Grav.}, 32\penalty0 (2):\penalty0 024001, 2015.
\newblock \doi{10.1088/0264-9381/32/2/024001}.

\bibitem[Abbott et~al.(2016{\natexlab{b}})]{GW150914_TGR:2016}
B.~P. Abbott et~al.
\newblock {Tests of general relativity with GW150914}.
\newblock \emph{Phys. Rev. Lett.}, 116\penalty0 (22):\penalty0 221101,
  2016{\natexlab{b}}.
\newblock \doi{10.1103/PhysRevLett.116.221101}.
\newblock [Erratum: Phys.Rev.Lett. 121, 129902 (2018)].

\bibitem[Akutsu et~al.(2021)]{KAGRA:2021}
T.~Akutsu et~al.
\newblock {Overview of KAGRA: Detector design and construction history}, 2021.

\bibitem[Abbott et~al.(2019{\natexlab{a}})]{gwtc1:2019}
B.~P. Abbott et~al.
\newblock {GWTC-1: A Gravitational-Wave Transient Catalog of Compact Binary
  Mergers Observed by LIGO and Virgo during the First and Second Observing
  Runs}.
\newblock \emph{Phys. Rev. X}, 9\penalty0 (3):\penalty0 031040,
  2019{\natexlab{a}}.
\newblock \doi{10.1103/PhysRevX.9.031040}.

\bibitem[Abbott et~al.(2021{\natexlab{a}})]{gwtc2:2021}
R.~Abbott et~al.
\newblock {GWTC-2: Compact Binary Coalescences Observed by LIGO and Virgo
  During the First Half of the Third Observing Run}.
\newblock \emph{Phys. Rev. X}, 11:\penalty0 021053, 2021{\natexlab{a}}.
\newblock \doi{10.1103/PhysRevX.11.021053}.

\bibitem[Abbott et~al.(2024)]{gwtc21:2024}
R.~Abbott et~al.
\newblock {GWTC-2.1: Deep extended catalog of compact binary coalescences
  observed by LIGO and Virgo during the first half of the third observing run}.
\newblock \emph{Phys. Rev. D}, 109\penalty0 (2):\penalty0 022001, 2024.
\newblock \doi{10.1103/PhysRevD.109.022001}.

\bibitem[Abbott et~al.(2023)]{gwtc3:2023}
R.~Abbott et~al.
\newblock {GWTC-3: Compact Binary Coalescences Observed by LIGO and Virgo
  during the Second Part of the Third Observing Run}.
\newblock \emph{Phys. Rev. X}, 13\penalty0 (4):\penalty0 041039, 2023.
\newblock \doi{10.1103/PhysRevX.13.041039}.

\bibitem[Nitz et~al.(2019)Nitz, Capano, Nielsen, Reyes, White, Brown, and
  Krishnan]{1OGC:2018}
Alexander~H. Nitz, Collin Capano, Alex~B. Nielsen, Steven Reyes, Rebecca White,
  Duncan~A. Brown, and Badri Krishnan.
\newblock {1-OGC: The first open gravitational-wave catalog of binary mergers
  from analysis of public Advanced LIGO data}.
\newblock \emph{Astrophys. J.}, 872\penalty0 (2):\penalty0 195, 2019.
\newblock \doi{10.3847/1538-4357/ab0108}.

\bibitem[Nitz et~al.(2020{\natexlab{a}})Nitz, Dent, Davies, Kumar, Capano,
  Harry, Mozzon, Nuttall, Lundgren, and T\'apai]{2OGC:2020}
Alexander~H. Nitz, Thomas Dent, Gareth~S. Davies, Sumit Kumar, Collin~D.
  Capano, Ian Harry, Simone Mozzon, Laura Nuttall, Andrew Lundgren, and
  M\'arton T\'apai.
\newblock {2-OGC: Open Gravitational-wave Catalog of binary mergers from
  analysis of public Advanced LIGO and Virgo data}.
\newblock \emph{Astrophys. J.}, 891:\penalty0 123, 3 2020{\natexlab{a}}.
\newblock \doi{10.3847/1538-4357/ab733f}.

\bibitem[Nitz et~al.(2021{\natexlab{a}})Nitz, Capano, Kumar, Wang, Kastha,
  Sch\"afer, Dhurkunde, and Cabero]{3OGC:2021}
Alexander~H. Nitz, Collin~D. Capano, Sumit Kumar, Yi-Fan Wang, Shilpa Kastha,
  Marlin Sch\"afer, Rahul Dhurkunde, and Miriam Cabero.
\newblock {3-OGC: Catalog of Gravitational Waves from Compact-binary Mergers}.
\newblock \emph{Astrophys. J.}, 922\penalty0 (1):\penalty0 76,
  2021{\natexlab{a}}.
\newblock \doi{10.3847/1538-4357/ac1c03}.

\bibitem[Nitz et~al.(2023)Nitz, Kumar, Wang, Kastha, Wu, Sch\"afer, Dhurkunde,
  and Capano]{4OGC:2021}
Alexander~H. Nitz, Sumit Kumar, Yi-Fan Wang, Shilpa Kastha, Shichao Wu, Marlin
  Sch\"afer, Rahul Dhurkunde, and Collin~D. Capano.
\newblock {4-OGC: Catalog of Gravitational Waves from Compact Binary Mergers}.
\newblock \emph{Astrophys. J.}, 946\penalty0 (2):\penalty0 59, 2023.
\newblock \doi{10.3847/1538-4357/aca591}.

\bibitem[Venumadhav et~al.(2019)Venumadhav, Zackay, Roulet, Dai, and
  Zaldarriaga]{Princeton_1:2019}
Tejaswi Venumadhav, Barak Zackay, Javier Roulet, Liang Dai, and Matias
  Zaldarriaga.
\newblock {New search pipeline for compact binary mergers: Results for binary
  black holes in the first observing run of Advanced LIGO}.
\newblock \emph{Phys. Rev. D}, 100\penalty0 (2):\penalty0 023011, 2019.
\newblock \doi{10.1103/PhysRevD.100.023011}.

\bibitem[Venumadhav et~al.(2020)Venumadhav, Zackay, Roulet, Dai, and
  Zaldarriaga]{Princeton_2:2019}
Tejaswi Venumadhav, Barak Zackay, Javier Roulet, Liang Dai, and Matias
  Zaldarriaga.
\newblock {New binary black hole mergers in the second observing run of
  Advanced LIGO and Advanced Virgo}.
\newblock \emph{Phys. Rev. D}, 101\penalty0 (8):\penalty0 083030, 2020.
\newblock \doi{10.1103/PhysRevD.101.083030}.

\bibitem[Olsen et~al.(2022)Olsen, Venumadhav, Mushkin, Roulet, Zackay, and
  Zaldarriaga]{Princeton_3a:2022}
Seth Olsen, Tejaswi Venumadhav, Jonathan Mushkin, Javier Roulet, Barak Zackay,
  and Matias Zaldarriaga.
\newblock {New binary black hole mergers in the LIGO-Virgo O3a data}.
\newblock \emph{Phys. Rev. D}, 106\penalty0 (4):\penalty0 043009, 2022.
\newblock \doi{10.1103/PhysRevD.106.043009}.

\bibitem[Mehta et~al.(2023)Mehta, Olsen, Wadekar, Roulet, Venumadhav, Mushkin,
  Zackay, and Zaldarriaga]{Princeton_3b:2023}
Ajit~Kumar Mehta, Seth Olsen, Digvijay Wadekar, Javier Roulet, Tejaswi
  Venumadhav, Jonathan Mushkin, Barak Zackay, and Matias Zaldarriaga.
\newblock {New binary black hole mergers in the LIGO-Virgo O3b data}, 11 2023.

\bibitem[{LIGO Scientific Collaboration and Virgo
  Collaboration}(2024{\natexlab{a}})]{gracedb_superevents:2024}
{LIGO Scientific Collaboration and Virgo Collaboration}.
\newblock {Gravitational Wave Candidate Event Database O4 Significant Detection
  Candidates}.
\newblock \url{https://gracedb.ligo.org/}, 2024{\natexlab{a}}.
\newblock URL \url{https://gracedb.ligo.org/superevents/public/O4/}.
\newblock Accessed: 2024-09-17.

\bibitem[Abbott et~al.(2017{\natexlab{a}})]{GW170817:2017}
B.~P. Abbott et~al.
\newblock {GW170817: Observation of Gravitational Waves from a Binary Neutron
  Star Inspiral}.
\newblock \emph{Phys. Rev. Lett.}, 119\penalty0 (16):\penalty0 161101,
  2017{\natexlab{a}}.
\newblock \doi{10.1103/PhysRevLett.119.161101}.

\bibitem[Abbott et~al.(2020{\natexlab{a}})]{GW190425:2020}
B.~P. Abbott et~al.
\newblock {GW190425: Observation of a Compact Binary Coalescence with Total
  Mass {${\sim}3.4 M_{\odot}$}}.
\newblock \emph{Astrophys. J. Lett.}, 892\penalty0 (1):\penalty0 L3,
  2020{\natexlab{a}}.
\newblock \doi{10.3847/2041-8213/ab75f5}.

\bibitem[Abbott et~al.(2021{\natexlab{b}})]{nsbh:2021}
R.~Abbott et~al.
\newblock {Observation of Gravitational Waves from Two Neutron
  Star\textendash{}Black Hole Coalescences}.
\newblock \emph{Astrophys. J. Lett.}, 915\penalty0 (1):\penalty0 L5,
  2021{\natexlab{b}}.
\newblock \doi{10.3847/2041-8213/ac082e}.

\bibitem[Moore(2012)]{Moore_book:2012}
Thomas~A. Moore.
\newblock \emph{A General Relativity Workbook}.
\newblock University Science Books, 2012.
\newblock ISBN 978-1-891389-82-5.

\bibitem[Maggiore(2007)]{Maggiore_book:2007}
Michele Maggiore.
\newblock \emph{{Gravitational Waves: Volume 1: Theory and Experiments}}.
\newblock Oxford University Press, 10 2007.
\newblock ISBN 9780198570745.
\newblock \doi{10.1093/acprof:oso/9780198570745.001.0001}.
\newblock URL \url{https://doi.org/10.1093/acprof:oso/9780198570745.001.0001}.

\bibitem[Schutz(2009)]{Schutz_book:2009}
Bernard Schutz.
\newblock \emph{A First Course in General Relativity}.
\newblock Cambridge University Press, 2 edition, 2009.

\bibitem[Creighton and Anderson(2011)]{Creighton_book:2009}
Jolien Creighton and Warren Anderson.
\newblock Gravitational-wave physics and astronomy: An introduction to theory,
  experiment and data analysis.
\newblock \emph{Gravitational-Wave Physics and Astronomy: An Introduction to
  Theory, Experiment and Data Analysis}, 09 2011.
\newblock \doi{10.1002/9783527636037}.

\bibitem[Usman et~al.(2016)]{PyCBC:2016}
Samantha~A. Usman et~al.
\newblock {The PyCBC search for gravitational waves from compact binary
  coalescence}.
\newblock \emph{Class. Quant. Grav.}, 33\penalty0 (21):\penalty0 215004, 2016.
\newblock \doi{10.1088/0264-9381/33/21/215004}.

\bibitem[Hammond et~al.(2014)Hammond, Hild, and Pitkin]{gw_polarisation_plots}
Giles Hammond, Stefan Hild, and Matthew Pitkin.
\newblock Advanced technologies for future laser-interferometric gravitational
  wave detectors.
\newblock \emph{Journal of Modern Optics}, 61, 02 2014.
\newblock \doi{10.1080/09500340.2014.920934}.

\bibitem[Dhurkunde(2024)]{Dhurkunde:2024}
Rahul Dhurkunde.
\newblock \emph{{Unveiling the cosmos: advanced gravitational wave searches for
  eccentric and precessing binary mergers and their astrophysical
  implications}}.
\newblock PhD thesis, Leibniz U., Hannover, 2024.

\bibitem[McIsaac(2023)]{McIsaac_Thesis:2023}
Connor McIsaac.
\newblock \emph{Improving the Reach of Gravitational-Wave Astronomy}.
\newblock Doctoral thesis, University of Portsmouth, February 2023.
\newblock URL
  \url{https://researchportal.port.ac.uk/en/studentTheses/improving-the-reach-of-gravitational-wave-astronomy}.

\bibitem[Aasi et~al.(2015)]{aLIGO:2015}
J.~Aasi et~al.
\newblock {Advanced LIGO}.
\newblock \emph{Class. Quant. Grav.}, 32:\penalty0 074001, 2015.
\newblock \doi{10.1088/0264-9381/32/7/074001}.

\bibitem[Willke et~al.(2002)]{GEO600:2002}
B~Willke et~al.
\newblock The geo 600 gravitational wave detector.
\newblock \emph{Classical and Quantum Gravity}, 19\penalty0 (7):\penalty0 1377,
  3 2002.
\newblock \doi{10.1088/0264-9381/19/7/321}.
\newblock URL \url{https://dx.doi.org/10.1088/0264-9381/19/7/321}.

\bibitem[Kondrashov et~al.(2008)Kondrashov, Simakov, Khalili, and
  Danilishin]{IFO_diagram:2008}
I.~S. Kondrashov, D.~A. Simakov, F.~Ya. Khalili, and S.~L. Danilishin.
\newblock {Optimizing the regimes of Advanced LIGO gravitational wave detector
  for multiple source types}.
\newblock \emph{Phys. Rev. D}, 78:\penalty0 062004, 2008.
\newblock \doi{10.1103/PhysRevD.78.062004}.

\bibitem[Brown(2004{\natexlab{a}})]{Brown_Thesis:2004}
Duncan~A. Brown.
\newblock \emph{{Searching for gravitational radiation from binary black hole
  MACHOs in the galactic halo}}.
\newblock Other thesis, UWM, 12 2004{\natexlab{a}}.

\bibitem[Meers(1988)]{Meers:1988}
B.~J. Meers.
\newblock {Recycling in Laser Interferometric Gravitational Wave Detectors}.
\newblock \emph{Phys. Rev. D}, 38:\penalty0 2317--2326, 1988.
\newblock \doi{10.1103/PhysRevD.38.2317}.

\bibitem[Blanchet et~al.(1996)Blanchet, Iyer, Will, and Wiseman]{2PN_1:1996}
Luc Blanchet, Bala~R. Iyer, Clifford~M. Will, and Alan~G. Wiseman.
\newblock {Gravitational wave forms from inspiralling compact binaries to
  second postNewtonian order}.
\newblock \emph{Class. Quant. Grav.}, 13:\penalty0 575--584, 1996.
\newblock \doi{10.1088/0264-9381/13/4/002}.

\bibitem[Will and Wiseman(1996)]{2PN_2:1996}
Clifford~M. Will and Alan~G. Wiseman.
\newblock {Gravitational radiation from compact binary systems: Gravitational
  wave forms and energy loss to second postNewtonian order}.
\newblock \emph{Phys. Rev. D}, 54:\penalty0 4813--4848, 1996.
\newblock \doi{10.1103/PhysRevD.54.4813}.

\bibitem[Blanchet et~al.(1995)Blanchet, Damour, Iyer, Will, and
  Wiseman]{2PN_3:1995}
Luc Blanchet, Thibault Damour, Bala~R. Iyer, Clifford~M. Will, and Alan~G.
  Wiseman.
\newblock {Gravitational radiation damping of compact binary systems to second
  postNewtonian order}.
\newblock \emph{Phys. Rev. Lett.}, 74:\penalty0 3515--3518, 1995.
\newblock \doi{10.1103/PhysRevLett.74.3515}.

\bibitem[Buonanno et~al.(2009)Buonanno, Iyer, Ochsner, Pan, and
  Sathyaprakash]{PN_models:2009}
Alessandra Buonanno, Bala Iyer, Evan Ochsner, Yi~Pan, and B.~S. Sathyaprakash.
\newblock {Comparison of post-Newtonian templates for compact binary inspiral
  signals in gravitational-wave detectors}.
\newblock \emph{Phys. Rev. D}, 80:\penalty0 084043, 2009.
\newblock \doi{10.1103/PhysRevD.80.084043}.

\bibitem[Abbott et~al.(2019{\natexlab{b}})]{GW170817_TGR:2019}
B.~P. Abbott et~al.
\newblock {Tests of General Relativity with GW170817}.
\newblock \emph{Phys. Rev. Lett.}, 123\penalty0 (1):\penalty0 011102,
  2019{\natexlab{b}}.
\newblock \doi{10.1103/PhysRevLett.123.011102}.

\bibitem[Abbott et~al.(2021{\natexlab{c}})]{O3_TGR:2021}
R.~Abbott et~al.
\newblock {Tests of General Relativity with GWTC-3}, 12 2021{\natexlab{c}}.

\bibitem[Antelis and Moreno(2017)]{IMR_plot:2016}
Javier~M. Antelis and Claudia Moreno.
\newblock {Obtaining gravitational waves from inspiral binary systems using
  LIGO data}.
\newblock \emph{Eur. Phys. J. Plus}, 132\penalty0 (1):\penalty0 10, 2017.
\newblock \doi{10.1140/epjp/i2017-11283-5}.
\newblock [Erratum: Eur.Phys.J.Plus 132, 103 (2017)].

\bibitem[Buonanno and Damour(1999)]{EOB_1:1998}
A.~Buonanno and T.~Damour.
\newblock {Effective one-body approach to general relativistic two-body
  dynamics}.
\newblock \emph{Phys. Rev. D}, 59:\penalty0 084006, 1999.
\newblock \doi{10.1103/PhysRevD.59.084006}.

\bibitem[Buonanno and Damour(2000)]{EOB_2:2000}
Alessandra Buonanno and Thibault Damour.
\newblock {Transition from inspiral to plunge in binary black hole
  coalescences}.
\newblock \emph{Phys. Rev. D}, 62:\penalty0 064015, 2000.
\newblock \doi{10.1103/PhysRevD.62.064015}.

\bibitem[Damour et~al.(2000)Damour, Jaranowski, and Schaefer]{EOB_3:2000}
Thibault Damour, Piotr Jaranowski, and Gerhard Schaefer.
\newblock {On the determination of the last stable orbit for circular general
  relativistic binaries at the third postNewtonian approximation}.
\newblock \emph{Phys. Rev. D}, 62:\penalty0 084011, 2000.
\newblock \doi{10.1103/PhysRevD.62.084011}.

\bibitem[Damour(2001)]{EOB_4:2001}
Thibault Damour.
\newblock {Coalescence of two spinning black holes: an effective one-body
  approach}.
\newblock \emph{Phys. Rev. D}, 64:\penalty0 124013, 2001.
\newblock \doi{10.1103/PhysRevD.64.124013}.

\bibitem[Damour et~al.(2008)Damour, Nagar, Hannam, Husa, and
  Bruegmann]{EOB_5:2008}
Thibault Damour, Alessandro Nagar, Mark Hannam, Sascha Husa, and Bernd
  Bruegmann.
\newblock {Accurate Effective-One-Body waveforms of inspiralling and coalescing
  black-hole binaries}.
\newblock \emph{Phys. Rev. D}, 78:\penalty0 044039, 2008.
\newblock \doi{10.1103/PhysRevD.78.044039}.

\bibitem[Buonanno et~al.(2007)Buonanno, Pan, Baker, Centrella, Kelly,
  McWilliams, and van Meter]{EOB_6:2007}
Alessandra Buonanno, Yi~Pan, John~G. Baker, Joan Centrella, Bernard~J. Kelly,
  Sean~T. McWilliams, and James~R. van Meter.
\newblock {Toward faithful templates for non-spinning binary black holes using
  the effective-one-body approach}.
\newblock \emph{Phys. Rev. D}, 76:\penalty0 104049, 2007.
\newblock \doi{10.1103/PhysRevD.76.104049}.

\bibitem[Pompili et~al.(2023)]{SEOBNRv5:2023tna}
Lorenzo Pompili et~al.
\newblock {Laying the foundation of the effective-one-body waveform models
  SEOBNRv5: Improved accuracy and efficiency for spinning nonprecessing binary
  black holes}.
\newblock \emph{Phys. Rev. D}, 108\penalty0 (12):\penalty0 124035, 2023.
\newblock \doi{10.1103/PhysRevD.108.124035}.

\bibitem[Ramos-Buades et~al.(2023)Ramos-Buades, Buonanno, Estell\'es, Khalil,
  Mihaylov, Ossokine, Pompili, and Shiferaw]{SEOBNRv5_PHM-Buades:2023ehm}
Antoni Ramos-Buades, Alessandra Buonanno, H\'ector Estell\'es, Mohammed Khalil,
  Deyan~P. Mihaylov, Serguei Ossokine, Lorenzo Pompili, and Mahlet Shiferaw.
\newblock {Next generation of accurate and efficient multipolar precessing-spin
  effective-one-body waveforms for binary black holes}.
\newblock \emph{Phys. Rev. D}, 108\penalty0 (12):\penalty0 124037, 2023.
\newblock \doi{10.1103/PhysRevD.108.124037}.

\bibitem[Ajith et~al.(2007)]{IMR_1:2007}
Parameswaran Ajith et~al.
\newblock {Phenomenological template family for black-hole coalescence
  waveforms}.
\newblock \emph{Class. Quant. Grav.}, 24:\penalty0 S689--S700, 2007.
\newblock \doi{10.1088/0264-9381/24/19/S31}.

\bibitem[Estell\'es et~al.(2022)Estell\'es, Husa, Colleoni, Keitel,
  Mateu-Lucena, Garc\'\i{}a-Quir\'os, Ramos-Buades, and Borchers]{IMR_2:2020}
H\'ector Estell\'es, Sascha Husa, Marta Colleoni, David Keitel, Maite
  Mateu-Lucena, Cecilio Garc\'\i{}a-Quir\'os, Antoni Ramos-Buades, and Angela
  Borchers.
\newblock {Time-domain phenomenological model of gravitational-wave subdominant
  harmonics for quasicircular nonprecessing binary black hole coalescences}.
\newblock \emph{Phys. Rev. D}, 105\penalty0 (8):\penalty0 084039, 2022.
\newblock \doi{10.1103/PhysRevD.105.084039}.

\bibitem[Ajith et~al.(2011)]{IMRPhenomD:2009}
P.~Ajith et~al.
\newblock {Inspiral-merger-ringdown waveforms for black-hole binaries with
  non-precessing spins}.
\newblock \emph{Phys. Rev. Lett.}, 106:\penalty0 241101, 2011.
\newblock \doi{10.1103/PhysRevLett.106.241101}.

\bibitem[Pratten et~al.(2021)]{IMRPhenomXPHM:2020}
Geraint Pratten et~al.
\newblock {Computationally efficient models for the dominant and subdominant
  harmonic modes of precessing binary black holes}.
\newblock \emph{Phys. Rev. D}, 103\penalty0 (10):\penalty0 104056, 2021.
\newblock \doi{10.1103/PhysRevD.103.104056}.

\bibitem[Boyle et~al.(2019)]{Surr_1:2019}
Michael Boyle et~al.
\newblock {The SXS Collaboration catalog of binary black hole simulations}.
\newblock \emph{Class. Quant. Grav.}, 36\penalty0 (19):\penalty0 195006, 2019.
\newblock \doi{10.1088/1361-6382/ab34e2}.

\bibitem[Healy and Lousto(2022)]{Surr_2:2022}
James Healy and Carlos~O. Lousto.
\newblock {Fourth RIT binary black hole simulations catalog: Extension to
  eccentric orbits}.
\newblock \emph{Phys. Rev. D}, 105\penalty0 (12):\penalty0 124010, 2022.
\newblock \doi{10.1103/PhysRevD.105.124010}.

\bibitem[Varma et~al.(2019)Varma, Field, Scheel, Blackman, Gerosa, Stein,
  Kidder, and Pfeiffer]{NRSur7dq4:2019}
Vijay Varma, Scott~E. Field, Mark~A. Scheel, Jonathan Blackman, Davide Gerosa,
  Leo~C. Stein, Lawrence~E. Kidder, and Harald~P. Pfeiffer.
\newblock {Surrogate models for precessing binary black hole simulations with
  unequal masses}.
\newblock \emph{Phys. Rev. Research.}, 1:\penalty0 033015, 2019.
\newblock \doi{10.1103/PhysRevResearch.1.033015}.

\bibitem[Collaboration(2021)]{pipelines}
LIGO~Scientific Collaboration.
\newblock Online pipelines — ligo/virgo public alerts user guide 16
  documentation, 2021.
\newblock URL
  \url{https://emfollow.docs.ligo.org/userguide/analysis/searches.html}.

\bibitem[Nitz et~al.(2017)Nitz, Dent, Dal~Canton, Fairhurst, and
  Brown]{PyCBC:2017}
Alexander~H. Nitz, Thomas Dent, Tito Dal~Canton, Stephen Fairhurst, and
  Duncan~A. Brown.
\newblock {Detecting binary compact-object mergers with gravitational waves:
  Understanding and Improving the sensitivity of the PyCBC search}.
\newblock \emph{Astrophys. J.}, 849\penalty0 (2):\penalty0 118, 2017.
\newblock \doi{10.3847/1538-4357/aa8f50}.

\bibitem[Cannon et~al.(2020)]{GstLAL:2020}
Kipp Cannon et~al.
\newblock {GstLAL: A software framework for gravitational wave discovery}.
\newblock \emph{SoftwareX}, 13, 10 2020.
\newblock \doi{10.1016/j.softx.2021.100680}.

\bibitem[Chu et~al.(2022)]{SPIIR:2020}
Qi~Chu et~al.
\newblock {SPIIR online coherent pipeline to search for gravitational waves
  from compact binary coalescences}.
\newblock \emph{Phys. Rev. D}, 105\penalty0 (2):\penalty0 024023, 2022.
\newblock \doi{10.1103/PhysRevD.105.024023}.

\bibitem[Aubin et~al.(2021)]{MBTA:2021}
F.~Aubin et~al.
\newblock {The MBTA pipeline for detecting compact binary coalescences in the
  third LIGO\textendash{}Virgo observing run}.
\newblock \emph{Class. Quant. Grav.}, 38\penalty0 (9):\penalty0 095004, 2021.
\newblock \doi{10.1088/1361-6382/abe913}.

\bibitem[Drago et~al.(2020)]{cWB:2020}
M.~Drago et~al.
\newblock {Coherent WaveBurst, a pipeline for unmodeled gravitational-wave data
  analysis}.
\newblock \emph{SoftwareX}, 6 2020.
\newblock \doi{10.1016/j.softx.2021.100678}.

\bibitem[Lynch et~al.(2017)Lynch, Vitale, Essick, Katsavounidis, and
  Robinet]{oLIB:2015}
Ryan Lynch, Salvatore Vitale, Reed Essick, Erik Katsavounidis, and Florent
  Robinet.
\newblock {Information-theoretic approach to the gravitational-wave burst
  detection problem}.
\newblock \emph{Phys. Rev. D}, 95\penalty0 (10):\penalty0 104046, 2017.
\newblock \doi{10.1103/PhysRevD.95.104046}.

\bibitem[Skliris et~al.(2020)Skliris, Norman, and Sutton]{MLy:2020qax}
Vasileios Skliris, Michael R.~K. Norman, and Patrick~J. Sutton.
\newblock {Real-Time Detection of Unmodelled Gravitational-Wave Transients
  Using Convolutional Neural Networks}, 9 2020.

\bibitem[Nitz et~al.(2021{\natexlab{b}})Nitz, Harry, Brown, Biwer, Willis,
  Canton, Capano, Dent, et~al.]{PyCBC_package:2021}
Alex Nitz, Ian Harry, Duncan Brown, Christopher~M. Biwer, Josh Willis, Tito~Dal
  Canton, Collin Capano, Thomas Dent, et~al.
\newblock gwastro/pycbc: Python package for gravitational wave astronomy,
  August 2021{\natexlab{b}}.
\newblock URL \url{https://doi.org/10.5281/zenodo.5347736}.

\bibitem[Biswas et~al.(2012)]{Biswas:2012}
Rahul Biswas et~al.
\newblock {Likelihood-ratio ranking of gravitational-wave candidates in a
  non-Gaussian background}.
\newblock \emph{Phys. Rev. D}, 85:\penalty0 122008, 2012.
\newblock \doi{10.1103/PhysRevD.85.122008}.

\bibitem[Allen et~al.(2012)Allen, Anderson, Brady, Brown, and
  Creighton]{FINDCHIRP:2012}
Bruce Allen, Warren~G. Anderson, Patrick~R. Brady, Duncan~A. Brown, and Jolien
  D.~E. Creighton.
\newblock {FINDCHIRP: An Algorithm for detection of gravitational waves from
  inspiraling compact binaries}.
\newblock \emph{Phys. Rev. D}, 85:\penalty0 122006, 2012.
\newblock \doi{10.1103/PhysRevD.85.122006}.

\bibitem[Babak et~al.(2013)]{IHOPE:2012zx}
S.~Babak et~al.
\newblock {Searching for gravitational waves from binary coalescence}.
\newblock \emph{Phys. Rev. D}, 87\penalty0 (2):\penalty0 024033, 2013.
\newblock \doi{10.1103/PhysRevD.87.024033}.

\bibitem[Droz et~al.(1999)Droz, Knapp, Poisson, and Owen]{Droz:1999}
Serge Droz, Daniel~J. Knapp, Eric Poisson, and Benjamin~J. Owen.
\newblock {Gravitational waves from inspiraling compact binaries: Validity of
  the stationary phase approximation to the Fourier transform}.
\newblock \emph{Phys. Rev. D}, 59:\penalty0 124016, 1999.
\newblock \doi{10.1103/PhysRevD.59.124016}.

\bibitem[Brown et~al.(2012)Brown, Harry, Lundgren, and
  Nitz]{Harry_Lundgren:2012}
Duncan~A. Brown, Ian Harry, Andrew Lundgren, and Alexander~H. Nitz.
\newblock {Detecting binary neutron star systems with spin in advanced
  gravitational-wave detectors}.
\newblock \emph{Phys. Rev. D}, 86:\penalty0 084017, 2012.
\newblock \doi{10.1103/PhysRevD.86.084017}.

\bibitem[Owen and Sathyaprakash(1999)]{Owen_Sathya:1999}
Benjamin~J. Owen and B.~S. Sathyaprakash.
\newblock {Matched filtering of gravitational waves from inspiraling compact
  binaries: Computational cost and template placement}.
\newblock \emph{Phys. Rev. D}, 60:\penalty0 022002, 1999.
\newblock \doi{10.1103/PhysRevD.60.022002}.

\bibitem[Sathyaprakash and Dhurandhar(1991)]{geom_bank_1:1991}
B.~S. Sathyaprakash and S.~V. Dhurandhar.
\newblock {Choice of filters for the detection of gravitational waves from
  coalescing binaries}.
\newblock \emph{Phys. Rev. D}, 44:\penalty0 3819--3834, 1991.
\newblock \doi{10.1103/PhysRevD.44.3819}.

\bibitem[Dhurandhar and Sathyaprakash(1994)]{geom_bank_2:1992}
S.~V. Dhurandhar and B.~S. Sathyaprakash.
\newblock {Choice of filters for the detection of gravitational waves from
  coalescing binaries. 2. Detection in colored noise}.
\newblock \emph{Phys. Rev. D}, 49:\penalty0 1707--1722, 1994.
\newblock \doi{10.1103/PhysRevD.49.1707}.

\bibitem[Balasubramanian et~al.(1996)Balasubramanian, Sathyaprakash, and
  Dhurandhar]{geom_bank_3:1995}
R.~Balasubramanian, B.~S. Sathyaprakash, and S.~V. Dhurandhar.
\newblock {Gravitational waves from coalescing binaries: Detection strategies
  and Monte Carlo estimation of parameters}.
\newblock \emph{Phys. Rev. D}, 53:\penalty0 3033--3055, 1996.
\newblock \doi{10.1103/PhysRevD.53.3033}.
\newblock [Erratum: Phys.Rev.D 54, 1860 (1996)].

\bibitem[Owen(1996)]{geom_bank_4:1995}
Benjamin~J. Owen.
\newblock {Search templates for gravitational waves from inspiraling binaries:
  Choice of template spacing}.
\newblock \emph{Phys. Rev. D}, 53:\penalty0 6749--6761, 1996.
\newblock \doi{10.1103/PhysRevD.53.6749}.

\bibitem[Harry et~al.(2009)Harry, Allen, and Sathyaprakash]{Harry_sbank:2009}
Ian~W. Harry, Bruce Allen, and B.~S. Sathyaprakash.
\newblock {A Stochastic template placement algorithm for gravitational wave
  data analysis}.
\newblock \emph{Phys. Rev. D}, 80:\penalty0 104014, 2009.
\newblock \doi{10.1103/PhysRevD.80.104014}.

\bibitem[Babak(2008)]{Stochastic_tb:2008}
Stanislav Babak.
\newblock {Building a stochastic template bank for detecting massive black hole
  binaries}.
\newblock \emph{Class. Quant. Grav.}, 25:\penalty0 195011, 2008.
\newblock \doi{10.1088/0264-9381/25/19/195011}.

\bibitem[Nuttall et~al.(2015)]{LIGO_data_quality:2015}
L.~Nuttall et~al.
\newblock {Improving the Data Quality of Advanced LIGO Based on Early
  Engineering Run Results}.
\newblock \emph{Class. Quant. Grav.}, 32\penalty0 (24):\penalty0 245005, 2015.
\newblock \doi{10.1088/0264-9381/32/24/245005}.

\bibitem[Allen(2005)]{Allen_Chi:2005}
Bruce Allen.
\newblock {${\chi}^{2}$ time-frequency discriminator for gravitational wave
  detection}.
\newblock \emph{Phys. Rev. D}, 71:\penalty0 062001, 2005.
\newblock \doi{10.1103/PhysRevD.71.062001}.

\bibitem[Nitz(2018)]{PyCBC_sg:2018}
Alexander~Harvey Nitz.
\newblock {Distinguishing short duration noise transients in LIGO data to
  improve the PyCBC search for gravitational waves from high mass binary black
  hole mergers}.
\newblock \emph{Class. Quant. Grav.}, 35\penalty0 (3):\penalty0 035016, 2018.
\newblock \doi{10.1088/1361-6382/aaa13d}.

\bibitem[McIsaac and Harry(2022)]{McIsaac_Chi:2022}
Connor McIsaac and Ian Harry.
\newblock {Using machine learning to autotune chi-squared tests for
  gravitational wave searches}.
\newblock \emph{Phys. Rev. D}, 105\penalty0 (10):\penalty0 104056, 2022.
\newblock \doi{10.1103/PhysRevD.105.104056}.

\bibitem[Davies et~al.(2020)Davies, Dent, T\'apai, Harry, McIsaac, and
  Nitz]{PyCBC_global:2020}
Gareth~S. Davies, Thomas Dent, M\'arton T\'apai, Ian Harry, Connor McIsaac, and
  Alexander~H. Nitz.
\newblock {Extending the PyCBC search for gravitational waves from compact
  binary mergers to a global network}.
\newblock \emph{Phys. Rev. D}, 102\penalty0 (2):\penalty0 022004, 2020.
\newblock \doi{10.1103/PhysRevD.102.022004}.

\bibitem[Abadie et~al.(2012)]{rw_snr_eq:2012}
J.~Abadie et~al.
\newblock {Search for Gravitational Waves from Low Mass Compact Binary
  Coalescence in LIGO's Sixth Science Run and Virgo's Science Runs 2 and 3}.
\newblock \emph{Phys. Rev. D}, 85:\penalty0 082002, 2012.
\newblock \doi{10.1103/PhysRevD.85.082002}.

\bibitem[Harry et~al.(2014)Harry, Nitz, Brown, Lundgren, Ochsner, and
  Keppel]{Harry_precession:2013}
Ian~W. Harry, Alexander~H. Nitz, Duncan~A. Brown, Andrew~P. Lundgren, Evan
  Ochsner, and Drew Keppel.
\newblock {Investigating the effect of precession on searches for
  neutron-star-black-hole binaries with Advanced LIGO}.
\newblock \emph{Phys. Rev. D}, 89\penalty0 (2):\penalty0 024010, 2014.
\newblock \doi{10.1103/PhysRevD.89.024010}.

\bibitem[Ajith et~al.(2014)Ajith, Fotopoulos, Privitera, Neunzert, and
  Weinstein]{Ajith:2012}
P.~Ajith, N.~Fotopoulos, S.~Privitera, A.~Neunzert, and A.~J. Weinstein.
\newblock {Effectual template bank for the detection of gravitational waves
  from inspiralling compact binaries with generic spins}.
\newblock \emph{Phys. Rev. D}, 89\penalty0 (8):\penalty0 084041, 2014.
\newblock \doi{10.1103/PhysRevD.89.084041}.

\bibitem[Privitera et~al.(2014)Privitera, Mohapatra, Ajith, Cannon, Fotopoulos,
  Frei, Hanna, Weinstein, and Whelan]{Privitera:2013}
Stephen Privitera, Satyanarayan R.~P. Mohapatra, Parameswaran Ajith, Kipp
  Cannon, Nickolas Fotopoulos, Melissa~A. Frei, Chad Hanna, Alan~J. Weinstein,
  and John~T. Whelan.
\newblock {Improving the sensitivity of a search for coalescing binary black
  holes with nonprecessing spins in gravitational wave data}.
\newblock \emph{Phys. Rev. D}, 89\penalty0 (2):\penalty0 024003, 2014.
\newblock \doi{10.1103/PhysRevD.89.024003}.

\bibitem[Capano et~al.(2016)Capano, Harry, Privitera, and
  Buonanno]{pycbc_template_bank:2016}
Collin Capano, Ian Harry, Stephen Privitera, and Alessandra Buonanno.
\newblock {Implementing a search for gravitational waves from binary black
  holes with nonprecessing spin}.
\newblock \emph{Phys. Rev. D}, 93\penalty0 (12):\penalty0 124007, 2016.
\newblock \doi{10.1103/PhysRevD.93.124007}.

\bibitem[Aasi et~al.(2013)]{S5:2012}
J.~Aasi et~al.
\newblock {Einstein@Home all-sky search for periodic gravitational waves in
  LIGO S5 data}.
\newblock \emph{Phys. Rev. D}, 87\penalty0 (4):\penalty0 042001, 2013.
\newblock \doi{10.1103/PhysRevD.87.042001}.

\bibitem[Robinson et~al.(2008)Robinson, Sathyaprakash, and
  Sengupta]{Robinson:2008}
C.~A.~K. Robinson, B.~S. Sathyaprakash, and Anand~S. Sengupta.
\newblock {A Geometric algorithm for efficient coincident detection of
  gravitational waves}.
\newblock \emph{Phys. Rev. D}, 78:\penalty0 062002, 2008.
\newblock \doi{10.1103/PhysRevD.78.062002}.

\bibitem[Brown(2004{\natexlab{b}})]{Brown:2003}
Duncan~A. Brown.
\newblock {Testing the LIGO inspiral analysis with hardware injections}.
\newblock \emph{Class. Quant. Grav.}, 21:\penalty0 S797--S800,
  2004{\natexlab{b}}.
\newblock \doi{10.1088/0264-9381/21/5/060}.

\bibitem[Biwer et~al.(2017)]{Biwer:2016}
C.~Biwer et~al.
\newblock {Validating gravitational-wave detections: The Advanced LIGO hardware
  injection system}.
\newblock \emph{Phys. Rev. D}, 95\penalty0 (6):\penalty0 062002, 2017.
\newblock \doi{10.1103/PhysRevD.95.062002}.

\bibitem[Barsotti et~al.(2018)Barsotti, Fritschel, Evans, and
  Slawomir]{aLIGO_design_curve:2018}
L.~Barsotti, P.~Fritschel, M.~Evans, and G.~Slawomir.
\newblock Updated advanced ligo sensitivity design curve, February 2018.
\newblock URL \url{https://dcc.ligo.org/LIGO-T1800044/public}.

\bibitem[Mozzon et~al.(2020)Mozzon, Nuttall, Lundgren, Dent, Kumar, and
  Nitz]{PSD_var:2020}
S.~Mozzon, L.~K. Nuttall, A.~Lundgren, T.~Dent, S.~Kumar, and A.~H. Nitz.
\newblock {Dynamic Normalization for Compact Binary Coalescence Searches in
  Non-Stationary Noise}.
\newblock \emph{Class. Quant. Grav.}, 37\penalty0 (21):\penalty0 215014, 2020.
\newblock \doi{10.1088/1361-6382/abac6c}.

\bibitem[Glanzer et~al.(2023{\natexlab{a}})Glanzer, Soni, Spoon, Effler, and
  Gonz\'alez]{Glanzer:2023}
Jane Glanzer, Siddharth Soni, Jaidyn Spoon, Anamaria Effler, and Gabriela
  Gonz\'alez.
\newblock {Noise in the LIGO livingston gravitational wave observatory due to
  trains}.
\newblock \emph{Class. Quant. Grav.}, 40\penalty0 (19):\penalty0 195015,
  2023{\natexlab{a}}.
\newblock \doi{10.1088/1361-6382/acf01f}.

\bibitem[Dickmann et~al.(2018)Dickmann, Kroker, Levin, Nawrodt, and
  Vyatchanin]{thermal_noise:2018}
Johannes Dickmann, Stefanie Kroker, Yuri Levin, Ronny Nawrodt, and Sergey
  Vyatchanin.
\newblock {Thermal noise of beam splitters in laser gravitational wave
  detectors}.
\newblock \emph{Phys. Rev. D}, 98\penalty0 (8):\penalty0 082002, 2018.
\newblock \doi{10.1103/PhysRevD.98.082002}.

\bibitem[Corbitt and Mavalvala(2003)]{quantum_noise:2003}
Thomas Corbitt and Nergis Mavalvala.
\newblock {Quantum noise in gravitational wave interferometers: Overview and
  recent developments}.
\newblock \emph{Proc. SPIE Int. Soc. Opt. Eng.}, 5111:\penalty0 23, 2003.
\newblock \doi{10.1117/12.507487}.

\bibitem[Abbott et~al.(2020{\natexlab{b}})]{Noise_Guide:2020}
Benjamin~P Abbott et~al.
\newblock {A guide to LIGO\textendash{}Virgo detector noise and extraction of
  transient gravitational-wave signals}.
\newblock \emph{Class. Quant. Grav.}, 37\penalty0 (5):\penalty0 055002,
  2020{\natexlab{b}}.
\newblock \doi{10.1088/1361-6382/ab685e}.

\bibitem[Abbott et~al.(2016{\natexlab{c}})]{GW150914_noise:2016}
B.~P. Abbott et~al.
\newblock {Characterization of transient noise in Advanced LIGO relevant to
  gravitational wave signal GW150914}.
\newblock \emph{Class. Quant. Grav.}, 33\penalty0 (13):\penalty0 134001,
  2016{\natexlab{c}}.
\newblock \doi{10.1088/0264-9381/33/13/134001}.

\bibitem[Nuttall(2018)]{Nuttall:2018}
L.~K. Nuttall.
\newblock {Characterizing transient noise in the LIGO detectors}.
\newblock \emph{Phil. Trans. Roy. Soc. Lond. A}, 376\penalty0 (2120):\penalty0
  20170286, 2018.
\newblock \doi{10.1098/rsta.2017.0286}.

\bibitem[Matichard et~al.(2015)]{seismic_isolation:2015}
F.~Matichard et~al.
\newblock {Seismic isolation of Advanced LIGO: Review of strategy,
  instrumentation and performance}.
\newblock \emph{Class. Quant. Grav.}, 32\penalty0 (18):\penalty0 185003, 2015.
\newblock \doi{10.1088/0264-9381/32/18/185003}.

\bibitem[Sutton(2003)]{range_calculation:2003}
P.~Sutton.
\newblock S3 performance of the ligo interferometers as measured by
  sensemonitor, November 2003.
\newblock URL \url{https://dcc.ligo.org/LIGO-T030276/public}.

\bibitem[Ota(2023)]{ota:2023}
Iara Naomi~Nobre Ota.
\newblock \emph{Optimizing the LIGO Summary Pages Pipeline}.
\newblock PhD thesis, International Centre for Theoretical Physics, 12 2023.
\newblock URL \url{https://hdl.handle.net/20.500.11767/139990}.
\newblock Accessed: 2024-09-02.

\bibitem[Macleod et~al.(2024)Macleod, Goetz, Ota, Isi, Massinger, Davis,
  Pitkin, paulaltin, and Nitz]{gwsumm:2024}
Duncan Macleod, Evan Goetz, Iara Ota, Maximiliano Isi, Thomas Massinger, Derek
  Davis, Matt Pitkin, paulaltin, and Alex Nitz.
\newblock gwpy/gwsumm: 2.2.7, July 2024.
\newblock URL \url{https://doi.org/10.5281/zenodo.12796492}.

\bibitem[Mukherjee et~al.(2010)Mukherjee, Obaid, and
  Matkarimov]{GWMimicking:2010}
S.~Mukherjee, R.~Obaid, and B.~Matkarimov.
\newblock {Classification of glitch waveforms in gravitational wave detector
  characterization}.
\newblock \emph{J. Phys. Conf. Ser.}, 243:\penalty0 012006, 2010.
\newblock \doi{10.1088/1742-6596/243/1/012006}.

\bibitem[Glanzer et~al.(2023{\natexlab{b}})]{gravityspy:2023}
J.~Glanzer et~al.
\newblock {Data quality up to the third observing run of advanced LIGO: Gravity
  Spy glitch classifications}.
\newblock \emph{Class. Quant. Grav.}, 40\penalty0 (6):\penalty0 065004,
  2023{\natexlab{b}}.
\newblock \doi{10.1088/1361-6382/acb633}.

\bibitem[Zevin et~al.(2024)]{gravityspy:2024}
Michael Zevin et~al.
\newblock {Gravity Spy: lessons learned and a path forward}.
\newblock \emph{Eur. Phys. J. Plus}, 139\penalty0 (1):\penalty0 100, 2024.
\newblock \doi{10.1140/epjp/s13360-023-04795-4}.

\bibitem[Chatterji et~al.(2004)Chatterji, Blackburn, Martin, and
  Katsavounidis]{qscan:2004}
S.~Chatterji, L.~Blackburn, G.~Martin, and E.~Katsavounidis.
\newblock {Multiresolution techniques for the detection of gravitational-wave
  bursts}.
\newblock \emph{Class. Quant. Grav.}, 21:\penalty0 S1809--S1818, 2004.
\newblock \doi{10.1088/0264-9381/21/20/024}.

\bibitem[Cabero et~al.(2019)]{blips:2019}
Miriam Cabero et~al.
\newblock {Blip glitches in Advanced LIGO data}.
\newblock \emph{Class. Quant. Grav.}, 36\penalty0 (15):\penalty0 15, 2019.
\newblock \doi{10.1088/1361-6382/ab2e14}.

\bibitem[Tolley et~al.(2023)Tolley, Cabourn~Davies, Harry, and
  Lundgren]{ArchEnemy:2023}
Arthur~E. Tolley, Gareth~S. Cabourn~Davies, Ian~W. Harry, and Andrew~P.
  Lundgren.
\newblock {ArchEnemy: removing scattered-light glitches from gravitational wave
  data}.
\newblock \emph{Class. Quant. Grav.}, 40\penalty0 (16):\penalty0 165005, 2023.
\newblock \doi{10.1088/1361-6382/ace22f}.

\bibitem[Merritt et~al.(2021)Merritt, Farr, Hur, Edelman, and
  Doctor]{glitschen:2021}
Jonathan Merritt, Ben Farr, Rachel Hur, Bruce Edelman, and Zoheyr Doctor.
\newblock {Transient glitch mitigation in Advanced LIGO data}, 2021.

\bibitem[Soni et~al.(2020)]{reducing_scattering:2020}
S.~Soni et~al.
\newblock {Reducing scattered light in LIGO's third observing run}.
\newblock \emph{Class. Quant. Grav.}, 38\penalty0 (2):\penalty0 025016, 2020.
\newblock \doi{10.1088/1361-6382/abc906}.

\bibitem[Wu et~al.(2024)Wu, Zevin, Berry, Crowston, \O{}sterlund, Doctor,
  Banagiri, Jackson, Kalogera, and Katsaggelos]{GlitchPlot:2024}
Yunan Wu, Michael Zevin, Christopher P.~L. Berry, Kevin Crowston, Carsten
  \O{}sterlund, Zoheyr Doctor, Sharan Banagiri, Corey~B. Jackson, Vicky
  Kalogera, and Aggelos~K. Katsaggelos.
\newblock {Advancing Glitch Classification in Gravity Spy: Multi-view Fusion
  with Attention-based Machine Learning for Advanced LIGO's Fourth Observing
  Run}, 1 2024.

\bibitem[Zevin et~al.(2017)]{gravityspy:2017}
Michael Zevin et~al.
\newblock {Gravity Spy: Integrating Advanced LIGO Detector Characterization,
  Machine Learning, and Citizen Science}.
\newblock \emph{Class. Quant. Grav.}, 34\penalty0 (6):\penalty0 064003, 2017.
\newblock \doi{10.1088/1361-6382/aa5cea}.

\bibitem[Soni et~al.(2021)]{gravityspy:2021}
S.~Soni et~al.
\newblock {Discovering features in gravitational-wave data through detector
  characterization, citizen science and machine learning}.
\newblock \emph{Class. Quant. Grav.}, 38\penalty0 (19):\penalty0 195016, 2021.
\newblock \doi{10.1088/1361-6382/ac1ccb}.

\bibitem[Cornish and Littenberg(2015)]{BayesWave:2015}
Neil~J. Cornish and Tyson~B. Littenberg.
\newblock {BayesWave: Bayesian Inference for Gravitational Wave Bursts and
  Instrument Glitches}.
\newblock \emph{Class. Quant. Grav.}, 32\penalty0 (13):\penalty0 135012, 2015.
\newblock \doi{10.1088/0264-9381/32/13/135012}.

\bibitem[Bianchi et~al.(2022)Bianchi, Longo, Valdes, Gonz\'alez, and
  Plastino]{gwadaptive:2022}
Stefano Bianchi, Alessandro Longo, Guillermo Valdes, Gabriela Gonz\'alez, and
  Wolfango Plastino.
\newblock {An automated pipeline for scattered light noise characterization}.
\newblock \emph{Class. Quant. Grav.}, 39\penalty0 (19):\penalty0 195005, 2022.
\newblock \doi{10.1088/1361-6382/ac88b0}.

\bibitem[Davis et~al.(2022)Davis, Littenberg, Romero-Shaw, Millhouse, McIver,
  Di~Renzo, and Ashton]{O3_subtraction:2022}
D.~Davis, T.~B. Littenberg, I.~M. Romero-Shaw, M.~Millhouse, J.~McIver,
  F.~Di~Renzo, and G.~Ashton.
\newblock {Subtracting glitches from gravitational-wave detector data during
  the third LIGO-Virgo observing run}.
\newblock \emph{Class. Quant. Grav.}, 39\penalty0 (24):\penalty0 245013, 2022.
\newblock \doi{10.1088/1361-6382/aca238}.

\bibitem[Powell et~al.(2017)Powell, Torres-Forn\'e, Lynch, Trifir\`o, Cuoco,
  Cavagli\`a, Heng, and Font]{Powell:2016}
Jade Powell, Alejandro Torres-Forn\'e, Ryan Lynch, Daniele Trifir\`o, Elena
  Cuoco, Marco Cavagli\`a, Ik~Siong Heng, and Jos\'e~A. Font.
\newblock {Classification methods for noise transients in advanced
  gravitational-wave detectors II: performance tests on Advanced LIGO data}.
\newblock \emph{Class. Quant. Grav.}, 34\penalty0 (3):\penalty0 034002, 2017.
\newblock \doi{10.1088/1361-6382/34/3/034002}.

\bibitem[Bondarescu et~al.(2023)Bondarescu, Lundgren, and
  Macas]{antiglitch:2023}
Ruxandra Bondarescu, Andrew Lundgren, and Ronaldas Macas.
\newblock {Quasiphysical model for removing short glitches from LIGO and Virgo
  data}.
\newblock \emph{Phys. Rev. D}, 108\penalty0 (12):\penalty0 122004, 2023.
\newblock \doi{10.1103/PhysRevD.108.122004}.

\bibitem[Essick et~al.(2020)Essick, Godwin, Hanna, Blackburn, and
  Katsavounidis]{iDQ:2020}
Reed Essick, Patrick Godwin, Chad Hanna, Lindy Blackburn, and Erik
  Katsavounidis.
\newblock {iDQ: Statistical Inference of Non-Gaussian Noise with Auxiliary
  Degrees of Freedom in Gravitational-Wave Detectors}.
\newblock \emph{Machine Learning: Science and Technology}, 2\penalty0
  (1):\penalty0 015004, 5 2020.
\newblock \doi{10.1088/2632-2153/abab5f}.

\bibitem[Abbott et~al.(2018)]{DQ_vetoes:2017}
B~P Abbott et~al.
\newblock {Effects of data quality vetoes on a search for compact binary
  coalescences in Advanced LIGO\textquoteright{}s first observing run}.
\newblock \emph{Class. Quant. Grav.}, 35\penalty0 (6):\penalty0 065010, 2018.
\newblock \doi{10.1088/1361-6382/aaaafa}.

\bibitem[Davis et~al.(2021)]{O2O3_DetChar:2021}
Derek Davis et~al.
\newblock {LIGO detector characterization in the second and third observing
  runs}.
\newblock \emph{Class. Quant. Grav.}, 38\penalty0 (13):\penalty0 135014, 2021.
\newblock \doi{10.1088/1361-6382/abfd85}.

\bibitem[Acernese et~al.(2023)]{VirgoDetChar:2023}
F.~Acernese et~al.
\newblock {Virgo detector characterization and data quality: results from the
  O3 run}.
\newblock \emph{Class. Quant. Grav.}, 40\penalty0 (18):\penalty0 185006, 2023.
\newblock \doi{10.1088/1361-6382/acd92d}.

\bibitem[Zooniverse(2024)]{zooniverse}
Zooniverse.
\newblock Zooniverse: People-powered research, 2024.
\newblock URL \url{https://www.zooniverse.org/}.
\newblock Accessed: 2024-09-22.

\bibitem[Buikema et~al.(2020)]{O3_sensitivity:2020}
Aaron Buikema et~al.
\newblock {Sensitivity and performance of the Advanced LIGO detectors in the
  third observing run}.
\newblock \emph{Phys. Rev. D}, 102\penalty0 (6):\penalty0 062003, 2020.
\newblock \doi{10.1103/PhysRevD.102.062003}.

\bibitem[{Cabourn Davies} and Harry(2022)]{PyCBC_singles:2022}
Gareth~S. {Cabourn Davies} and Ian~W. Harry.
\newblock {Establishing significance of gravitational-wave signals from a
  single observatory in the PyCBC offline search}.
\newblock \emph{Class. Quant. Grav.}, 39\penalty0 (21):\penalty0 215012, 2022.
\newblock \doi{10.1088/1361-6382/ac8862}.

\bibitem[Goetz(2021)]{gwdetchar_tools:2021}
Evan Goetz.
\newblock gwdetchar: Gravitational-wave detection characterization tools,
  November 2021.
\newblock URL \url{https://doi.org/10.5281/zenodo.5722057}.

\bibitem[Accadia et~al.(2010)]{TAccadia:2010}
T.~Accadia et~al.
\newblock {Noise from scattered light in Virgo's second science run data}.
\newblock \emph{Class. Quant. Grav.}, 27:\penalty0 194011, 2010.
\newblock \doi{10.1088/0264-9381/27/19/194011}.

\bibitem[Valdes et~al.(2017)Valdes, O'Reilly, and Diaz]{HilbertHuang:2017}
Guillermo Valdes, Brian O'Reilly, and Mario Diaz.
\newblock {A Hilbert\textendash{}Huang transform method for scattering
  identification in LIGO}.
\newblock \emph{Class. Quant. Grav.}, 34\penalty0 (23):\penalty0 235009, 2017.
\newblock \doi{10.1088/1361-6382/aa8e6b}.

\bibitem[Longo et~al.(2020)Longo, Bianchi, Plastino, Arnaud, Chiummo, Fiori,
  Swinkels, and Was]{tvf-EMD:2020}
Alessandro Longo, Stefano Bianchi, Wolfango Plastino, Nicolas Arnaud, Antonino
  Chiummo, Irene Fiori, Bas Swinkels, and Michal Was.
\newblock {Scattered light noise characterisation at the Virgo interferometer
  with tvf-EMD adaptive algorithm}.
\newblock \emph{Class. Quant. Grav.}, 37\penalty0 (14):\penalty0 145011, 2020.
\newblock \doi{10.1088/1361-6382/ab9719}.

\bibitem[Longo et~al.(2022)Longo, Bianchi, Valdes, Arnaud, and
  Plastino]{Scattering_Monitoring:2022}
Alessandro Longo, Stefano Bianchi, Guillermo Valdes, Nicolas Arnaud, and
  Wolfango Plastino.
\newblock {Daily monitoring of scattered light noise due to microseismic
  variability at the Virgo interferometer}.
\newblock \emph{Class. Quant. Grav.}, 39\penalty0 (3):\penalty0 035001, 2022.
\newblock \doi{10.1088/1361-6382/ac4117}.

\bibitem[Was et~al.(2021)Was, Gouaty, and Bonnand]{Was_Subtract:2021}
Michal Was, Romain Gouaty, and Romain Bonnand.
\newblock {End benches scattered light modelling and subtraction in advanced
  Virgo}.
\newblock \emph{Class. Quant. Grav.}, 38\penalty0 (7):\penalty0 075020, 2021.
\newblock \doi{10.1088/1361-6382/abe759}.

\bibitem[Abbott et~al.(2021{\natexlab{d}})]{GWOSC:2021}
Rich Abbott et~al.
\newblock {Open data from the first and second observing runs of Advanced LIGO
  and Advanced Virgo}.
\newblock \emph{SoftwareX}, 13:\penalty0 100658, 2021{\natexlab{d}}.
\newblock \doi{10.1016/j.softx.2021.100658}.

\bibitem[Ashton et~al.(2019)]{BILBY:2019}
Gregory Ashton et~al.
\newblock {BILBY: A user-friendly Bayesian inference library for
  gravitational-wave astronomy}.
\newblock \emph{Astrophys. J. Suppl.}, 241\penalty0 (2):\penalty0 27, 2019.
\newblock \doi{10.3847/1538-4365/ab06fc}.

\bibitem[Udall and Davis(2023)]{Udall:2023}
Rhiannon Udall and Derek Davis.
\newblock {Bayesian modeling of scattered light in the LIGO interferometers}.
\newblock \emph{Appl. Phys. Lett.}, 122\penalty0 (9):\penalty0 094103, 2023.
\newblock \doi{10.1063/5.0136896}.

\bibitem[Nitz et~al.(2018)Nitz, Dal~Canton, Davis, and Reyes]{PyCBC_Live:2018}
Alexander~H. Nitz, Tito Dal~Canton, Derek Davis, and Steven Reyes.
\newblock {Rapid detection of gravitational waves from compact binary mergers
  with PyCBC Live}.
\newblock \emph{Phys. Rev. D}, 98\penalty0 (2):\penalty0 024050, 2018.
\newblock \doi{10.1103/PhysRevD.98.024050}.

\bibitem[Chatziioannou et~al.(2024)Chatziioannou, Dent, Fishbach, Ohme,
  P\"urrer, Raymond, and Veitch]{Chatziioannou:2024}
K.~Chatziioannou, T.~Dent, M.~Fishbach, F.~Ohme, M.~P\"urrer, V.~Raymond, and
  J.~Veitch.
\newblock {Compact binary coalescences: gravitational-wave astronomy with
  ground-based detectors}, 9 2024.

\bibitem[Kumar and Dent(2024)]{PyCBC_focussed_bbh:2024}
Praveen Kumar and Thomas Dent.
\newblock {Optimized Search for a Binary Black Hole Merger Population in
  LIGO-Virgo O3 Data}, 3 2024.

\bibitem[Collaboration(2024{\natexlab{a}})]{ldas_caltech:2024}
LIGO~Scientific Collaboration.
\newblock Ligo data analysis system (ldas) at caltech, 2024{\natexlab{a}}.
\newblock URL
  \url{https://computing.docs.ligo.org/guide/computing-centres/cit/}.
\newblock Accessed: 2024-09-23.

\bibitem[Ghez et~al.(2000)Ghez, Morris, Becklin, Kremenek, and
  Tanner]{Ghez:2000}
A.~Ghez, M.~Morris, E.~E. Becklin, T.~Kremenek, and A.~Tanner.
\newblock {The Accelerations of stars orbiting the Milky Way's central black
  hole}.
\newblock \emph{Nature}, 407:\penalty0 349, 2000.
\newblock \doi{10.1038/35030032}.

\bibitem[Metzger(2017)]{Kilonovae:2017}
Brian~D. Metzger.
\newblock {Kilonovae}.
\newblock \emph{Living Rev. Rel.}, 20\penalty0 (1):\penalty0 3, 2017.
\newblock \doi{10.1007/s41114-017-0006-z}.

\bibitem[Cowperthwaite et~al.(2017)]{kilonova_lightcurve:2017}
P.~S. Cowperthwaite et~al.
\newblock {The Electromagnetic Counterpart of the Binary Neutron Star Merger
  LIGO/Virgo GW170817. II. UV, Optical, and Near-infrared Light Curves and
  Comparison to Kilonova Models}.
\newblock \emph{Astrophys. J. Lett.}, 848\penalty0 (2):\penalty0 L17, 2017.
\newblock \doi{10.3847/2041-8213/aa8fc7}.

\bibitem[Neronov(2019)]{multi_mess_astro:2019}
Andrii Neronov.
\newblock {Introduction to multi-messenger astronomy}.
\newblock \emph{J. Phys. Conf. Ser.}, 1263\penalty0 (1):\penalty0 012001, 2019.
\newblock \doi{10.1088/1742-6596/1263/1/012001}.

\bibitem[Schutz(1986)]{Schutz:1986}
Bernard~F. Schutz.
\newblock {Determining the Hubble Constant from Gravitational Wave
  Observations}.
\newblock \emph{Nature}, 323:\penalty0 310--311, 1986.
\newblock \doi{10.1038/323310a0}.

\bibitem[Hubble(1929)]{hubble:1929}
Edwin Hubble.
\newblock A relation between distance and radial velocity among extra-galactic
  nebulae.
\newblock \emph{Proceedings of the National Academy of Sciences of the United
  States of America}, 15\penalty0 (3):\penalty0 168--173, 1929.
\newblock \doi{10.1073/pnas.15.3.168}.

\bibitem[Spergel et~al.(2003)Spergel, Verde, Peiris, Komatsu, Nolta, Bennett,
  Halpern, Hinshaw, Jarosik, Kogut, Limon, Meyer, Page, Tucker, Weiland,
  Wollack, and Wright]{WMAP_H0:2003}
D.~N. Spergel, L.~Verde, H.~V. Peiris, E.~Komatsu, M.~R. Nolta, C.~L. Bennett,
  M.~Halpern, G.~Hinshaw, N.~Jarosik, A.~Kogut, M.~Limon, S.~S. Meyer, L.~Page,
  G.~S. Tucker, J.~L. Weiland, E.~Wollack, and E.~L. Wright.
\newblock First-year wilkinson microwave anisotropy probe (wmap) observations:
  Determination of cosmological parameters.
\newblock \emph{The Astrophysical Journal Supplement Series}, 148:\penalty0
  175--194, 2003.
\newblock \doi{10.1086/377226}.

\bibitem[Kessler et~al.(2009)Kessler, Becker, Cinabro, Vanderplas, Frieman,
  Marriner, Davis, Dilday, Holtzman, Jha, Lampeitl, Sako, Smith, Zheng, Nichol,
  Bassett, Bender, Depoy, Doi, Elson, Filippenko, Foley, Garnavich, Hopp,
  Ihara, Ketzeback, Kollatschny, Konishi, Marshall, McMillan, Miknaitis,
  Morokuma, Mörtsell, Pan, Prieto, Richmond, Riess, Romani, Schneider,
  Sollerman, Takanashi, Tokita, van~der Heyden, Wheeler, Yasuda, and
  York]{BAO_H0:2009}
Richard Kessler, Andrew~C. Becker, David Cinabro, Jake Vanderplas, Joshua~A.
  Frieman, John Marriner, Tamara~M. Davis, Benjamin Dilday, Jon Holtzman,
  Saurabh~W. Jha, Hubert Lampeitl, Masao Sako, Mathew Smith, Chen Zheng,
  Robert~C. Nichol, Bruce Bassett, Ralf Bender, Darren~L. Depoy, Mamoru Doi,
  Ed~Elson, Alexei~V. Filippenko, Ryan~J. Foley, Peter~M. Garnavich, Ulrich
  Hopp, Yutaka Ihara, William Ketzeback, W.~Kollatschny, Kohki Konishi,
  Jennifer~L. Marshall, Russet~J. McMillan, Gajus Miknaitis, Tomoki Morokuma,
  Edvard Mörtsell, Kaike Pan, Jose~Luis Prieto, Michael~W. Richmond, Adam~G.
  Riess, Roger Romani, Donald~P. Schneider, Jesper Sollerman, Naohiro
  Takanashi, Kouichi Tokita, Kurt van~der Heyden, J.~C. Wheeler, Naoki Yasuda,
  and Donald York.
\newblock First-year sloan digital sky survey-ii supernova results: Hubble
  diagram and cosmological parameters.
\newblock \emph{The Astrophysical Journal Supplement Series}, 185\penalty0
  (1):\penalty0 32–84, October 2009.
\newblock ISSN 1538-4365.
\newblock \doi{10.1088/0067-0049/185/1/32}.
\newblock URL \url{http://dx.doi.org/10.1088/0067-0049/185/1/32}.

\bibitem[Freedman et~al.(2001{\natexlab{a}})Freedman, Madore, Gibson,
  Ferrarese, Kelson, Graham, Mould, Worthey, Jr., Ford, Hoessel, Illingworth,
  Macri, and Saha]{Cepheids_H0:2001}
Wendy~L. Freedman, Barry~F. Madore, Brad~K. Gibson, Laura Ferrarese, Daniel~D.
  Kelson, John~A. Graham, Jeremy~R. Mould, John R.~P. Worthey, Ronald
  C.~Kennicutt Jr., Hans~C. Ford, John~A. Hoessel, John G.~S. Illingworth,
  Lucas~M. Macri, and Brad~E. Saha.
\newblock The hubble space telescope key project on the extragalactic distance
  scale: The final results.
\newblock \emph{The Astrophysical Journal}, 553:\penalty0 47--72,
  2001{\natexlab{a}}.
\newblock \doi{10.1086/320638}.

\bibitem[Riess et~al.(1998)]{TypeIa_H0:1998}
Adam~G. Riess et~al.
\newblock {Observational evidence from supernovae for an accelerating universe
  and a cosmological constant}.
\newblock \emph{Astron. J.}, 116:\penalty0 1009--1038, 1998.
\newblock \doi{10.1086/300499}.

\bibitem[Di~Valentino et~al.(2021)]{H0_tension:2020}
Eleonora Di~Valentino et~al.
\newblock {Snowmass2021 - Letter of interest cosmology intertwined II: The
  hubble constant tension}.
\newblock \emph{Astropart. Phys.}, 131:\penalty0 102605, 2021.
\newblock \doi{10.1016/j.astropartphys.2021.102605}.

\bibitem[Chassande-Mottin et~al.(2019)Chassande-Mottin, Leyde, Mastrogiovanni,
  and Steer]{inclin_degen_2:2019}
E.~Chassande-Mottin, K.~Leyde, S.~Mastrogiovanni, and D.~A. Steer.
\newblock {Gravitational wave observations, distance measurement uncertainties,
  and cosmology}.
\newblock \emph{Phys. Rev. D}, 100\penalty0 (8):\penalty0 083514, 2019.
\newblock \doi{10.1103/PhysRevD.100.083514}.

\bibitem[Holz et~al.(2018)Holz, Hughes, and Schutz]{inclin_degen:2018}
Daniel~E. Holz, Scott~A. Hughes, and Bernard~F. Schutz.
\newblock {Measuring cosmic distances with standard sirens}.
\newblock \emph{Physics Today}, 71\penalty0 (12):\penalty0 34--40, 12 2018.
\newblock ISSN 0031-9228.
\newblock \doi{10.1063/PT.3.4090}.
\newblock URL \url{https://doi.org/10.1063/PT.3.4090}.

\bibitem[Jaranowski and Krolak(2009)]{Jaranowski:2009}
Piotr Jaranowski and Andrzej Krolak.
\newblock \emph{Analysis of Gravitational-Wave Data}.
\newblock Cambridge Monographs on Particle Physics, Nuclear Physics and
  Cosmology. Cambridge University Press, 2009.

\bibitem[{NASA}(2023)]{gcn_circulars:2024}
{NASA}.
\newblock Gamma-ray coordinates network (gcn) circulars, 2023.
\newblock URL \url{https://gcn.nasa.gov/circulars}.
\newblock Accessed: 2024-09-06.

\bibitem[Abbott et~al.(2017{\natexlab{b}})]{gw170817_joint:2017}
B.~P. Abbott et~al.
\newblock {Gravitational Waves and Gamma-rays from a Binary Neutron Star
  Merger: GW170817 and GRB 170817A}.
\newblock \emph{Astrophys. J. Lett.}, 848\penalty0 (2):\penalty0 L13,
  2017{\natexlab{b}}.
\newblock \doi{10.3847/2041-8213/aa920c}.

\bibitem[Thompson and Wilson-Hodge(2022)]{Fermi:2022}
David~J. Thompson and Colleen~A. Wilson-Hodge.
\newblock {Fermi Gamma-ray Space Telescope}, 10 2022.

\bibitem[Kocevski et~al.(2017)Kocevski, Omodei, and
  Vianello]{Fermi_GW170817:2017}
Daniel Kocevski, Nicola Omodei, and Giacomo Vianello.
\newblock {Fermi-LAT observations of the LIGO/Virgo event GW170817}, 10 2017.

\bibitem[Winkler et~al.(2003)Winkler, Courvoisier, Di~Cocco, Gehrels, Giménez,
  Grebenev, Hermsen, Mas-Hesse, Lebrun, Lund, Palumbo, Paul, Roques, Schnopper,
  Schönfelder, Sunyaev, Teegarden, Ubertini, Vedrenne, and
  Dean]{INTEGRAL:2003}
C.~Winkler, T.J.-L. Courvoisier, G.~Di~Cocco, N.~Gehrels, A.~Giménez,
  S.~Grebenev, W.~Hermsen, J.M. Mas-Hesse, F.~Lebrun, N.~Lund, G.G.C. Palumbo,
  J.~Paul, J.-P. Roques, H.~Schnopper, V.~Schönfelder, R.~Sunyaev,
  B.~Teegarden, P.~Ubertini, G.~Vedrenne, and A.J. Dean.
\newblock {The INTEGRAL mission}.
\newblock \emph{Astronomy and Astrophysics}, 411:\penalty0 L1--L6, 2003.
\newblock \doi{10.1051/0004-6361:20031288}.

\bibitem[Savchenko et~al.(2017)]{INTEGRAL_GW170817:2017}
V.~Savchenko et~al.
\newblock {INTEGRAL Detection of the First Prompt Gamma-Ray Signal Coincident
  with the Gravitational-wave Event GW170817}.
\newblock \emph{Astrophys. J. Lett.}, 848\penalty0 (2):\penalty0 L15, 2017.
\newblock \doi{10.3847/2041-8213/aa8f94}.

\bibitem[Freedman et~al.(2001{\natexlab{b}})]{HST:2000}
W.~L. Freedman et~al.
\newblock {Final results from the Hubble Space Telescope key project to measure
  the Hubble constant}.
\newblock \emph{Astrophys. J.}, 553:\penalty0 47--72, 2001{\natexlab{b}}.
\newblock \doi{10.1086/320638}.

\bibitem[Kilpatrick et~al.(2022)]{HST_GW170817:2021}
Charles~D. Kilpatrick et~al.
\newblock {Hubble Space Telescope Observations of GW170817: Complete Light
  Curves and the Properties of the Galaxy Merger of NGC 4993}.
\newblock \emph{Astrophys. J.}, 926\penalty0 (1):\penalty0 49, 2022.
\newblock \doi{10.3847/1538-4357/ac3e59}.

\bibitem[Sutherland et~al.(2015)Sutherland, Emerson, Dalton, Atad-Ettedgui,
  Beard, Bennett, Bezawada, Born, Caldwell, Clark, Craig, Henry, Jeffers,
  Little, McPherson, Murray, Stewart, Stobie, Terrett, Ward, Whalley, and
  Woodhouse]{VISTA:2015}
Will Sutherland, Jim Emerson, Gavin Dalton, Eli Atad-Ettedgui, Steven Beard,
  Richard Bennett, Naidu Bezawada, Andrew Born, Martin Caldwell, Paul Clark,
  Simon Craig, David Henry, Paul Jeffers, Bryan Little, Alistair McPherson,
  John Murray, Malcolm Stewart, Brian Stobie, David Terrett, Kim Ward, Martin
  Whalley, and Guy Woodhouse.
\newblock {The Visible and Infrared Survey Telescope for Astronomy (VISTA):
  Design, technical overview, and performance}.
\newblock \emph{Astron. Astrophys.}, 575:\penalty0 A25, 2015.
\newblock \doi{10.1051/0004-6361/201424973}.

\bibitem[Tanvir et~al.(2017)]{VISTA_GW170817:2017}
N.~R. Tanvir et~al.
\newblock {The Emergence of a Lanthanide-Rich Kilonova Following the Merger of
  Two Neutron Stars}.
\newblock \emph{Astrophys. J. Lett.}, 848\penalty0 (2):\penalty0 L27, 2017.
\newblock \doi{10.3847/2041-8213/aa90b6}.

\bibitem[Gehrels et~al.(2004)]{Swift:2004}
N.~Gehrels et~al.
\newblock {The Swift Gamma-Ray Burst Mission}.
\newblock \emph{Astrophys. J.}, 611:\penalty0 1005--1020, 2004.
\newblock \doi{10.1086/422091}.
\newblock [Erratum: Astrophys.J. 621, 558 (2005)].

\bibitem[Evans et~al.(2017)]{Swift_GW170817:2017}
P.~A. Evans et~al.
\newblock {Swift and NuSTAR observations of GW170817: detection of a blue
  kilonova}.
\newblock \emph{Science}, 358:\penalty0 1565, 2017.
\newblock \doi{10.1126/science.aap9580}.

\bibitem[Haggard et~al.(2017)Haggard, Nynka, Ruan, Kalogera, Bradley~Cenko,
  Evans, and Kennea]{Chandra_GW170817:2017}
Daryl Haggard, Melania Nynka, John~J. Ruan, Vicky Kalogera, S.~Bradley~Cenko,
  Phil Evans, and Jamie~A. Kennea.
\newblock {A Deep Chandra X-ray Study of Neutron Star Coalescence GW170817}.
\newblock \emph{Astrophys. J. Lett.}, 848\penalty0 (2):\penalty0 L25, 2017.
\newblock \doi{10.3847/2041-8213/aa8ede}.

\bibitem[Lacy et~al.(2020)]{VLA:2019}
M.~Lacy et~al.
\newblock {The Karl G. Jansky Very Large Array Sky Survey (VLASS). Science case
  and survey design}.
\newblock \emph{Publ. Astron. Soc. Pac.}, 132\penalty0 (1009):\penalty0 035001,
  2020.
\newblock \doi{10.1088/1538-3873/ab63eb}.

\bibitem[Hallinan et~al.(2017)]{VLA_GW170817:2017}
G.~Hallinan et~al.
\newblock {A Radio Counterpart to a Neutron Star Merger}.
\newblock \emph{Science}, 358:\penalty0 1579, 2017.
\newblock \doi{10.1126/science.aap9855}.

\bibitem[Nicolaci~da Costa et~al.(1998)]{NGC4993:1998}
Luiz Nicolaci~da Costa et~al.
\newblock {The Southern Sky Redshift Survey}.
\newblock \emph{Astron. J.}, 116:\penalty0 1--7, 1998.
\newblock \doi{10.1086/300410}.

\bibitem[Abbott et~al.(2017{\natexlab{c}})]{GW170817_H0:2017}
B.~P. Abbott et~al.
\newblock {A gravitational-wave standard siren measurement of the Hubble
  constant}.
\newblock \emph{Nature}, 551\penalty0 (7678):\penalty0 85--88,
  2017{\natexlab{c}}.
\newblock \doi{10.1038/nature24471}.

\bibitem[Ade et~al.(2016)]{Planck_H0:2015}
P.~A.~R. Ade et~al.
\newblock {Planck 2015 results. XIII. Cosmological parameters}.
\newblock \emph{Astron. Astrophys.}, 594:\penalty0 A13, 2016.
\newblock \doi{10.1051/0004-6361/201525830}.

\bibitem[Riess et~al.(2016)]{Riess_H0:2016}
Adam~G. Riess et~al.
\newblock {A 2.4\% Determination of the Local Value of the Hubble Constant}.
\newblock \emph{Astrophys. J.}, 826\penalty0 (1):\penalty0 56, 2016.
\newblock \doi{10.3847/0004-637X/826/1/56}.

\bibitem[Palmese et~al.(2023)Palmese, Bom, Mucesh, and Hartley]{Palmese:2021}
Antonella Palmese, Clecio~R. Bom, Sunil Mucesh, and William~G. Hartley.
\newblock {A Standard Siren Measurement of the Hubble Constant Using
  Gravitational-wave Events from the First Three LIGO/Virgo Observing Runs and
  the DESI Legacy Survey}.
\newblock \emph{Astrophys. J.}, 943\penalty0 (1):\penalty0 56, 2023.
\newblock \doi{10.3847/1538-4357/aca6e3}.

\bibitem[Harry and Noller(2022)]{Harry_speed_of_gravity:2022}
Ian Harry and Johannes Noller.
\newblock {Probing the speed of gravity with LVK, LISA, and joint
  observations}.
\newblock \emph{Gen. Rel. Grav.}, 54\penalty0 (10):\penalty0 133, 2022.
\newblock \doi{10.1007/s10714-022-03016-0}.

\bibitem[Baker et~al.(2022)]{Baker_speed_of_gravity:2022}
Tessa Baker et~al.
\newblock {Measuring the propagation speed of gravitational waves with LISA}.
\newblock \emph{JCAP}, 08\penalty0 (08):\penalty0 031, 2022.
\newblock \doi{10.1088/1475-7516/2022/08/031}.

\bibitem[Soares-Santos et~al.(2019)]{DES:2019}
M.~Soares-Santos et~al.
\newblock {First Measurement of the Hubble Constant from a Dark Standard Siren
  using the Dark Energy Survey Galaxies and the LIGO/Virgo
  Binary\textendash{}Black-hole Merger GW170814}.
\newblock \emph{Astrophys. J. Lett.}, 876\penalty0 (1):\penalty0 L7, 2019.
\newblock \doi{10.3847/2041-8213/ab14f1}.

\bibitem[Dalang and Baker(2024)]{Dalang_dark_sirens:2023}
Charles Dalang and Tessa Baker.
\newblock {The clustering of dark sirens' invisible host galaxies}.
\newblock \emph{JCAP}, 02:\penalty0 024, 2024.
\newblock \doi{10.1088/1475-7516/2024/02/024}.

\bibitem[Gompertz et~al.(2020)]{GOTO:2020}
B.~P. Gompertz et~al.
\newblock {Searching for Electromagnetic Counterparts to Gravitational-wave
  Merger Events with the Prototype Gravitational-wave Optical Transient
  Observer (GOTO-4)}.
\newblock \emph{Mon. Not. Roy. Astron. Soc.}, 497\penalty0 (1):\penalty0
  726--738, 2020.
\newblock \doi{10.1093/mnras/staa1845}.

\bibitem[{LIGO Scientific Collaboration and Virgo
  Collaboration}(2024{\natexlab{b}})]{ligo_gracedb:2024}
{LIGO Scientific Collaboration and Virgo Collaboration}.
\newblock {Gravitational Wave Candidate Event Database (GraceDB)}.
\newblock \url{https://gracedb.ligo.org/}, 2024{\natexlab{b}}.
\newblock URL \url{https://gracedb.ligo.org/}.
\newblock Accessed: 2024-09-08.

\bibitem[Abbott et~al.(2016{\natexlab{d}})]{ligo_prospects:2016}
B.~P. Abbott et~al.
\newblock {Prospects for observing and localizing gravitational-wave transients
  with Advanced LIGO, Advanced Virgo and KAGRA}.
\newblock \emph{Living Rev. Rel.}, 19:\penalty0 1, 2016{\natexlab{d}}.
\newblock \doi{10.1007/s41114-020-00026-9}.

\bibitem[Kasliwal et~al.(2017)]{gw170817_skymap:2017}
M.~M. Kasliwal et~al.
\newblock {Illuminating Gravitational Waves: A Concordant Picture of Photons
  from a Neutron Star Merger}.
\newblock \emph{Science}, 358:\penalty0 1559, 2017.
\newblock \doi{10.1126/science.aap9455}.

\bibitem[Singer and Price(2016)]{BAYESTAR:2016}
Leo~P. Singer and Larry~R. Price.
\newblock {Rapid Bayesian position reconstruction for gravitational-wave
  transients}.
\newblock \emph{Phys. Rev. D}, 93\penalty0 (2):\penalty0 024013, 2016.
\newblock \doi{10.1103/PhysRevD.93.024013}.

\bibitem[Nitz et~al.(2020{\natexlab{b}})Nitz, Sch\"afer, and
  Dal~Canton]{PyCBC_earlywarning:2020}
Alexander~H. Nitz, Marlin Sch\"afer, and Tito Dal~Canton.
\newblock {Gravitational-wave Merger Forecasting: Scenarios for the Early
  Detection and Localization of Compact-binary Mergers with Ground-based
  Observatories}.
\newblock \emph{Astrophys. J. Lett.}, 902:\penalty0 L29, 2020{\natexlab{b}}.
\newblock \doi{10.3847/2041-8213/abbc10}.

\bibitem[Pratten et~al.(2020)Pratten, Husa, Garcia-Quiros, Colleoni,
  Ramos-Buades, Estelles, and Jaume]{IMRPhenomXAS:2020}
Geraint Pratten, Sascha Husa, Cecilio Garcia-Quiros, Marta Colleoni, Antoni
  Ramos-Buades, Hector Estelles, and Rafel Jaume.
\newblock {Setting the cornerstone for a family of models for gravitational
  waves from compact binaries: The dominant harmonic for nonprecessing
  quasicircular black holes}.
\newblock \emph{Phys. Rev. D}, 102\penalty0 (6):\penalty0 064001, 2020.
\newblock \doi{10.1103/PhysRevD.102.064001}.

\bibitem[Fairhurst(2011)]{Fairhurst:2010}
Stephen Fairhurst.
\newblock {Source localization with an advanced gravitational wave detector
  network}.
\newblock \emph{Class. Quant. Grav.}, 28:\penalty0 105021, 2011.
\newblock \doi{10.1088/0264-9381/28/10/105021}.

\bibitem[Collaboration(2024{\natexlab{b}})]{gracedb_superevent_selection}
LIGO~Scientific Collaboration.
\newblock Selection of the preferred event - {LIGO}/{Virgo} public alerts user
  guide, 2024{\natexlab{b}}.
\newblock URL
  \url{https://emfollow.docs.ligo.org/userguide/analysis/superevents.html#selection-of-the-preferred-event}.
\newblock Accessed: 2024-09-24.

\bibitem[sup(2024)]{superevent_S240924a}
Superevent {S240924a}, 2024.
\newblock URL \url{https://gracedb.ligo.org/superevents/S240924a/}.
\newblock Accessed: 2024-09-24.

\bibitem[Virtanen et~al.(2020)]{SciPy:2020}
Pauli Virtanen et~al.
\newblock {SciPy 1.0--Fundamental Algorithms for Scientific Computing in
  Python}.
\newblock \emph{Nature Meth.}, 17:\penalty0 261, 2020.
\newblock \doi{10.1038/s41592-019-0686-2}.

\bibitem[Storn and Price(1997)]{DE:1997}
Rainer Storn and Kenneth Price.
\newblock Differential evolution - a simple and efficient heuristic for global
  optimization over continuous spaces.
\newblock \emph{Journal of Global Optimization}, 11\penalty0 (4):\penalty0
  341--359, 1997.

\bibitem[{Harry, Ian and Dal Canton, Tito}(2019)]{pycbc_pull_request_2659}
{Harry, Ian and Dal Canton, Tito}.
\newblock Pull request \#2659: [optimize snr time series in pycbc live].
\newblock \url{https://github.com/gwastro/pycbc/pull/2659}, 2019.
\newblock Accessed: 2024-09-25.

\bibitem[Cokelaer(2007)]{Cokelaer:2007}
Thomas Cokelaer.
\newblock {Gravitational waves from inspiralling compact binaries: Hexagonal
  template placement and its efficiency in detecting physical signals}.
\newblock \emph{Phys. Rev. D}, 76:\penalty0 102004, 2007.
\newblock \doi{10.1103/PhysRevD.76.102004}.

\bibitem[Endres et~al.(2018)Endres, Sandrock, and Focke]{shgo:2018}
Stefan Endres, Carl Sandrock, and Walter Focke.
\newblock A simplicial homology algorithm for lipschitz optimisation.
\newblock \emph{Journal of Global Optimization}, 72, 10 2018.
\newblock \doi{10.1007/s10898-018-0645-y}.

\bibitem[Kennedy and Eberhart(1995)]{pso:1995}
James Kennedy and Russell Eberhart.
\newblock Particle swarm optimization, 1995.

\bibitem[Srivastava et~al.(2018)Srivastava, Rajesh~Nayak, and
  Bose]{pso_search_1:2018}
Varun Srivastava, K.~Rajesh~Nayak, and Sukanta Bose.
\newblock {Toward low-latency coincident precessing and coherent aligned-spin
  gravitational-wave searches of compact binary coalescences with particle
  swarm optimization}, 11 2018.

\bibitem[Pal and Nayak(2024)]{pso_search_2:2023}
Souradeep Pal and K.~Rajesh Nayak.
\newblock {Swarm-intelligent search for gravitational waves from eccentric
  binary mergers}.
\newblock \emph{Phys. Rev. D}, 110\penalty0 (4):\penalty0 042003, 2024.
\newblock \doi{10.1103/PhysRevD.110.042003}.

\bibitem[{Akiba} et~al.(2019){Akiba}, {Sano}, {Yanase}, {Ohta}, and
  {Koyama}]{optuna:2019}
Takuya {Akiba}, Shotaro {Sano}, Toshihiko {Yanase}, Takeru {Ohta}, and Masanori
  {Koyama}.
\newblock {Optuna: A Next-generation Hyperparameter Optimization Framework}.
\newblock \emph{arXiv e-prints}, art. arXiv:1907.10902, July 2019.
\newblock \doi{10.48550/arXiv.1907.10902}.

\end{thebibliography}
}





\end{document}